# GUIDE TO MATHEMATICAL CONCEPTS OF QUANTUM THEORY

**Teiko Heinosaari** [1]
*Department of Physics, University of Turku, FI-20014, Finland*

**Mario Ziman** [2]
*Research Center for Quantum Information, Slovak Academy of Sciences,*
*Dúbravská cesta 9, 84511 Bratislava, Slovakia*

Quantum Theory is one of the pillars of modern science developed over the last hundred years. In this review paper we introduce, step by step, the quantum theory understood as a mathematical model describing quantum experiments. We start with splitting the experiment into two parts: a preparation process and a measurement process leading to a registration of a particular outcome. These two ingredients of the experiment are represented by states and effects, respectively. Further, the whole picture of quantum measurement will be developed and concepts of observables, instruments and measurement models representing the three different descriptions on experiments will be introduced. In the second stage, we enrich the model of the experiment by introducing the concept of quantum channel describing the system changes between preparations and measurements. At the very end we review the elementary properties of quantum entanglement. The text contains many examples and exercise covering also many topics from quantum information theory and quantum measurement theory. The goal is to give a mathematically clear and self-containing explanation of the main concepts of the modern language of quantum theory.

---

[1]E-mail address: heinosaari@gmail.com
[2]E-mail address: ziman@savba.sk

# Contents

# 1   Preface

Dear reader, we are happy that you decided to use this Guide to discover the hidden world of quantum theory. Before you start, let us warn you. For many people the quantum world is strange and puzzling, full of paradoxes and giving only a little sense. Still more than 70 years after its birth, quantum mechanics gives raise to vigorous debates.

Our senses, trained in classical world, are insensitive to beauty of quantum. Due to this incapability, the quantum world was hidden for ages to human minds. Nowadays we use artificial devices to reveal the hidden quantum nature. And one of the main goal of this Guide is to describe ways how we understand and model the quantum observations.

Even if our intuition is very limited, we are able to create an abstract picture of what is going on. This is the power of mathematics - we have access to the details of quantum theory although we cannot see or touch quantum objects. The knowledge of quantum theory also gives us possibility to train our physical intuition and hence get acquainted with quantum phenomena.

We mostly avoid discussion of philosophical consequences and reasons of quantum theory. Not because those issues would not be interesting, but because we think that proper understanding can be achieved only by knowing the mathematical structure first. Instead, we shall spend more spacetime on mathematical proofs. The abstract mathematical and physical concepts are illustrated in numerous examples. The examples are essential part of this Guide and it is not recommendable to skip them.

This Guide is based on the lectures that the authors gave in 2007/08 in Research Center for Quantum Information, Bratislava (`http://www.quniverse.sk`). The audience consisted mostly of PhD students studying various subfields of quantum theory. The aim of the lectures was to introduce a common language for the core part of quantum theory.

Based on this history, this Guide is not meant to be the first textbook in quantum theory. We expect that the reader has already some training in quantum mechanics, such as one or two university courses. Perhaps she now wants to get more systematic picture of different components occurring in quantum mechanics, such as states, observables and channels. And these are exactly the targets aimed in this Guide.

## 2 Hilbert space refresher

Quantum theory, in it is conventional formulation, builds on the theory of Hilbert spaces and operators. In this chapter we go through the basic material which is central for the rest of the chapters. Our treatment is intended as a refresher and reminder - it is assumed that the reader is already familiar with some of the concepts and elementary results, at least in the case of finite dimensional inner product spaces. Proofs are given if they are seen instructive or illustrative. Good references for this chapter are, for instance, [27], [69], [73]. They also contain those proofs that we skip here.

### 2.1 Hilbert spaces

#### 2.1.1 Definition and examples

Let $\mathcal{H}$ be a complex vector space. We recall that a complex valued function $\langle \cdot \,|\, \cdot \rangle$ on $\mathcal{H} \times \mathcal{H}$ is an *inner product* if it satisfies the following three conditions for all $\varphi, \psi, \phi \in \mathcal{H}$ and $c \in \mathbb{C}$:

- $\langle \varphi \,|\, c\psi + \phi \rangle = c \langle \varphi \,|\, \psi \rangle + \langle \varphi \,|\, \phi \rangle$   (linear in the second argument),

- $\overline{\langle \varphi \,|\, \psi \rangle} = \langle \psi \,|\, \varphi \rangle$   (conjugate symmetric),

- $\langle \psi \,|\, \psi \rangle > 0$   if   $\psi \neq 0$   (positive definite).

Notice that the linearity and conjugate symmetry imply that $\langle 0 \,|\, \psi \rangle = \langle \psi \,|\, 0 \rangle = 0$.

A complex vector space with an inner product defined on it is an *inner product space*. An alternative name for an inner product is a *scalar product*, and then one naturally speaks about scalar product spaces. A word of warning: unlike in all quantum mechanics textbooks, in most functional analysis textbooks inner products are linear in the first argument.

**Example 1.** Let $\mathbb{C}^d$ denote the set of all $d$-tuples of complex numbers. For two vectors $\psi = (\psi_1, \ldots, \psi_n)$ and $\varphi = (\varphi_1, \ldots, \varphi_n)$, the inner product $\langle \psi \,|\, \varphi \rangle$ is defined to be

$$\langle \psi \,|\, \varphi \rangle = \sum_{j=1}^{d} \bar{\psi}_j \varphi_j \,. \tag{2.1}$$

Here $\bar{\psi}_j$ denotes the complex conjugate of $\psi_j$. There are also other inner products on $\mathbb{C}^d$, but when referring to $\mathbb{C}^d$ we always assume that the inner product is the one defined in (2.1).

An isomorphism is, generally speaking, a structure preserving bijection. Hence, in the context of inner product spaces we have the following definition.

**Definition 1.** Two inner product spaces $\mathcal{H}_1$ and $\mathcal{H}_2$ are *isomorphic* if there is a bijective linear mapping $U : \mathcal{H}_1 \to \mathcal{H}_2$ such that

$$\langle U\psi \,|\, U\varphi \rangle = \langle \psi \,|\, \varphi \rangle \tag{2.2}$$

for all $\psi, \varphi \in \mathcal{H}$. The mapping $U$ is an *isomorphism*.

Notice that a linear mapping $U$ satisfying (2.2) is automatically injective: if $U\psi = 0$, then $\langle\,\psi\,|\,\psi\,\rangle = \langle\,U\psi\,|\,U\psi\,\rangle = 0$ and hence $\psi = 0$. Therefore, to check whether a given linear mapping $U$ is an isomorphism it is enough to verify that $U$ is surjective and satisfies the condition (2.2).

Two vectors $\psi, \varphi \in \mathcal{H}$ are called *orthogonal* if $\langle\,\psi\,|\,\varphi\,\rangle = 0$. In this case we denote $\psi \perp \varphi$. A set $X \subset \mathcal{H}$ is called *an orthogonal set* of vectors if any two distinct vectors belonging to $X$ are orthogonal.

**Definition 2.** Let $\mathcal{H}$ be an inner product space. If for any positive integer $d$ there exists an orthogonal set of $d$ vectors, then $\mathcal{H}$ is *infinite dimensional*. Otherwise $\mathcal{H}$ is *finite dimensional*.

We recall the following characterization of finite dimensional inner product spaces.

**Proposition 1.** If $\mathcal{H}$ is a finite dimensional inner product space, then there is a positive integer $d$ such that:

- there are $d$ nonzero orthogonal vectors;

- for $d' > d$, any set of $d'$ nonzero vectors contains non-orthogonal vectors.

The number $d$ is called the *dimension* of $\mathcal{H}$. A finite dimensional inner product space of dimension $d$ is isomorphic to $\mathbb{C}^d$.

Not all inner product spaces are finite dimensional. The following example, which is also used later, illustrates this fact.

**Example 2.** We denote by $\mathbb{N}$ the set of natural numbers, including 0. Let $\ell^2(\mathbb{N})$ be the set of functions $f : \mathbb{N} \to \mathbb{C}$ such that the sum $\sum_{j=0}^{\infty} |f(j)|^2$ is finite. The formula

$$\langle\,f\,|\,g\,\rangle = \sum_{j=0}^{\infty} \overline{f(j)} g(j)$$

defines an inner product on $\ell^2(\mathbb{N})$. For $k \in \mathbb{N}$, let $\delta_k$ be the Kronecker function defined by

$$\delta_k(j) = \left\{ \begin{array}{ll} 1 & \text{if } j = k, \\ 0 & \text{if } j \neq k. \end{array} \right.$$

The inner product space $\ell^2(\mathbb{N})$ is infinite dimensional since the Kronecker functions $\delta_0, \delta_1, \ldots$ are orthogonal.

Every inner product space $\mathcal{H}$ is a normed space with the norm defined as

$$\|\psi\| \equiv \langle\,\psi\,|\,\psi\,\rangle^{\frac{1}{2}}. \tag{2.3}$$

Thus, it makes sense to speak about topological properties of $\mathcal{H}$. We recall that a normed space is

- *complete* if every Cauchy sequence is convergent;

- *separable* if it has a countable dense subset.

Every finite dimensional inner product space is complete and separable. For infinite dimensional inner product spaces this is not true, and we have the following definition.

**Definition 3.** An inner product space which is complete with respect to the norm (2.3) is a *Hilbert space*.

An orthogonal set $X \subset \mathcal{H}$ is an *orthonormal set* if each vector $\psi \in X$ has unit norm. An *orthonormal basis* for a Hilbert space $\mathcal{H}$ is a maximal orthonormal set; this means that there is no other orthonormal set containing it as a proper subset. A useful criterion for the maximality of an orthonormal set $X \subset \mathcal{H}$ is the following: if $\psi$ is orthogonal to all vectors in $X$, then $\psi = 0$.

It can be proved using Zorn's lemma that every Hilbert space has an orthonormal basis and, moreover, that all orthonormal bases of a given Hilbert space have the same cardinality.

**Example 3.** It can be shown that the inner product space $\ell^2(\mathbb{N})$ is complete and separable. The set $\{\delta_0, \delta_1, \ldots\}$ is an orthonormal basis for $\ell^2(\mathbb{N})$.

The following proposition should be compared to Proposition 1.

**Proposition 2.** A Hilbert space $\mathcal{H}$ is separable if and only if it has a countable orthonormal basis. A separable infinite dimensional Hilbert space is isomorphic to $\ell^2(\mathbb{N})$.

**Example 4.** Let $L^2(\mathbb{R})$ be the set of complex valued measurable functions on $\mathbb{R}$ which satisfy $\int_{\mathbb{R}} |f(x)|^2 \, dx < \infty$. This is a separable infinite dimensional Hilbert space under the inner product

$$\langle f \,|\, g \rangle = \int_{\mathbb{R}} \overline{f(x)} g(x) \, dx \,.$$

A unit vector $\psi$ in $L^2(\mathbb{R})$ is thus a function $\psi : \mathbb{R} \to \mathbb{C}$ satisfying $\int_{\mathbb{R}} |\psi(x)|^2 \, dx = 1$.

From now on, all Hilbert spaces that we deal with are assumed to be separable (hence, either finite dimensional or countably infinite dimensional).

### 2.1.2 Basic properties

An elementary but extremely important result for inner product spaces is *Cauchy-Schwarz inequality*: if $\varphi, \psi \in \mathcal{H}$, then

$$|\langle \varphi \,|\, \psi \rangle|^2 \leq \langle \varphi \,|\, \varphi \rangle \, \langle \psi \,|\, \psi \rangle \,. \tag{2.4}$$

Moreover, equality in (2.4) occurs if and only if $\varphi$ and $\psi$ are linearly dependent. Cauchy-Schwarz inequality will be constantly used in our calculations. Two other useful formulas are stated in the following exercises.

**Exercise 1.** (*Parallelogram law*) Let $\psi$ and $\varphi$ be vectors in an inner product space $\mathcal{H}$. Prove that the following equality, known as the *parallelogram law*, holds:

$$\|\psi + \varphi\|^2 + \|\psi - \varphi\|^2 = 2 \, \|\psi\|^2 + 2 \, \|\varphi\|^2 \,.$$

It is interesting to note that the converse is also true: a normed linear space is an inner product space if the norm satisfies the parallelogram law. (For a proof of this latter fact, see e.g. Theorem 6.1.5. in [36].)

**Exercise 2.** (*Pythagorean formula*) Let $\psi$ and $\varphi$ be orthogonal vectors in an inner product space $\mathcal{H}$. Prove that the following equality, known as the *Pythagorean formula*, holds:

$$\|\psi + \varphi\|^2 = \|\psi\|^2 + \|\varphi\|^2 \, .$$

In the case of a finite dimensional Hilbert space, it is often useful to understand an orthonormal basis as a list of vectors rather than just a set, i.e., we have ordered the elements of the orthonormal basis. There is then a unique correspondence with the vectors and the $d$-tuples of complex numbers. Similarly, in the case of a separable infinite dimensional Hilbert space we take an orthonormal basis to mean a sequence of orthogonal vectors (rather than just a set) whenever this is convenient.

Let $\mathcal{H}$ be either finite or separable infinite dimensional Hilbert space and let $\{\varphi_k\}_{k=0}^d$ be an orthonormal basis for $\mathcal{H}$. The *basis expansion* of a vector $\psi$ is

$$\psi = \sum_{k=0}^d \langle\, \varphi_k \mid \psi \,\rangle \, \varphi_k \, . \tag{2.5}$$

If $d < \infty$, then the basis expansion is just of finite sum. In the case of an infinite dimensional Hilbert space, the basis expansion is a convergent infinite series.

The coefficients $\langle\, \varphi_k \mid \psi \,\rangle$ in (2.5) are called the *Fourier coefficients* of $\psi$ with respect to the orthonormal basis $\{\varphi_k\}_{k=0}^d$. The norm of $\psi$ is given by the *Parseval's formula*:

$$\|\psi\|^2 = \sum_{k=0}^d |\langle\, \varphi_k \mid \psi \,\rangle|^2 \, . \tag{2.6}$$

### 2.2   Operators on Hilbert spaces

#### 2.2.1   $C^*$-algebra of bounded operators

**Definition 4.** We call a linear mapping $T : \mathcal{H} \to \mathcal{H}$ an *operator*. An operator $T$ is *bounded* if there exists a number $t \geq 0$ such that

$$\|T\psi\| \leq t \, \|\psi\| \quad \text{for all } \psi \in \mathcal{H} \, .$$

We denote by $\mathcal{L}(\mathcal{H})$ the set of bounded operators on $\mathcal{H}$.

It is a basic result in functional analysis that an operator is bounded if and only if it is continuous. For a bounded operator $T$, we use the following notations:

$$
\begin{aligned}
\ker T &= \{\psi \in \mathcal{H} : T\psi = 0\} & &(kernel)\,; \\
\operatorname{ran} T &= \{\psi \in \mathcal{H} : \psi = T\varphi \text{ for some } \varphi \in \mathcal{H}\} & &(range)\,; \\
\operatorname{supp} T &= \{\psi \in \mathcal{H} : \psi \perp \ker T\} \equiv (\ker T)^\perp & &(support)\,.
\end{aligned}
\tag{2.7}
$$

Note that in a finite dimensional Hilbert space $\mathcal{H}$ every operator is bounded. Even more, every linear mapping $T : X \to \mathcal{H}$ defined on a linear subspace $X \subset \mathcal{H}$ has an extension to a bounded operator $\widetilde{T} : \mathcal{H} \to \mathcal{H}$. In an infinite dimensional Hilbert space this is no longer true. In Example 5 we demonstrate the concept of an unbounded operator.

**Example 5.** (*Unbounded operator*) For each $f \in \ell^2(\mathbb{N})$, define a function $Nf : \mathbb{N} \to \mathbb{C}$ by formula

$$(Nf)(n) = nf(n) \,.$$

It may happen that $Nf$ is not a vector in $\ell^2(\mathbb{N})$. For instance, let

$$f(n) = \left\{ \begin{array}{ll} 0 & \text{if } n = 0, \\ \frac{1}{n} & \text{if } n > 0. \end{array} \right.$$

Then $f \in \ell^2(\mathbb{N})$ but $Nf \notin \ell^2(\mathbb{N})$. The set

$$\mathcal{D}(N) = \{ f \in \ell^2(\mathbb{N}) : Nf \in \ell^2(\mathbb{N}) \}$$

is a linear subspace of $\ell^2(\mathbb{N})$ and $N$ is a linear mapping from $\mathcal{D}(N)$ into $\ell^2(\mathbb{N})$. For each $k \in \mathbb{N}$, we have $N\delta_k = k\delta_k$ and hence $\|N\delta_k\| = k \|\delta_k\|$. This shows that there is no bounded operator $\widetilde{N}$ on $\ell^2(\mathbb{N})$ which would be an extension of $N$.

The set $\mathcal{L}(\mathcal{H})$ is a vector space. Namely, two operators can be added in the usual way and the scalar multiplication with a complex number $c$ is also defined in a natural way,

$$(S + T)\psi = S\psi + T\psi \,,$$
$$(cT)\psi = c(T\psi) \,.$$

The vector space $\mathcal{L}(\mathcal{H})$ is a normed space when we define a norm by formula

$$\|T\| := \sup_{\|\psi\|=1} \|T\psi\| \,. \tag{2.8}$$

This norm on $\mathcal{L}(\mathcal{H})$ is called the *operator norm*. One can show that $\mathcal{L}(\mathcal{H})$ is complete in the operator norm topology. Let us note that complete normed vector spaces are called *Banach spaces*, hence, $\mathcal{L}(\mathcal{H})$ is a Banach space.

It follows directly from the definition of the operator norm, Eq. (2.8), that if $T \in \mathcal{L}(\mathcal{H})$, then for every $\psi \in \mathcal{H}$ we have

$$\|T\psi\| \leq \|T\| \, \|\psi\| \,. \tag{2.9}$$

Together with Cauchy-Schwarz inequality this implies that for every $\varphi, \psi \in \mathcal{H}$,

$$|\langle \, \varphi \, | \, T\psi \, \rangle| \leq \|\varphi\| \, \|\psi\| \, \|T\| \,. \tag{2.10}$$

Both (2.9) and (2.10) are very useful inequalities.

We can multiply two operators by forming their composition. If $S, T \in \mathcal{L}(\mathcal{H})$, then by using inequality (2.9) twice we get

$$\sup_{\|\psi\|=1} \|ST\psi\| \leq \|S\| \sup_{\|\psi\|=1} \|T\psi\| = \|S\| \|T\| \sup_{\|\psi\|=1} \|\psi\| = \|S\| \|T\| \ .$$

Thus, we have

$$\|ST\| \leq \|S\| \|T\| \ . \tag{2.11}$$

This shows, in particular, that the product operator $ST$ is bounded and hence $ST \in \mathcal{L}(\mathcal{H})$. In other words, $\mathcal{L}(\mathcal{H})$ is an algebra. Moreover, inequality (2.11) implies that multiplication in $\mathcal{L}(\mathcal{H})$ is separately continuous in each variable.

For each bounded operator $T$, we can define the *adjoint operator* $T^*$ by the formula

$$\langle \varphi \,|\, T^*\psi \rangle = \langle T\varphi \,|\, \psi \rangle \ , \tag{2.12}$$

required to hold for all $\varphi, \psi \in \mathcal{H}$. The mapping $T \mapsto T^*$ is conjugate linear, i.e.,

$$(cT + S)^* = c^*T^* + S^* \ .$$

Moreover, if $S, T \in \mathcal{L}(\mathcal{H})$, then

$$(ST)^* = T^*S^* \ , \tag{2.13}$$
$$(T^*)^* = T \ . \tag{2.14}$$

**Exercise 3.** Prove the properties (2.13) and (2.14) directly from the defining condition (2.12).

**Proposition 3.** A bounded operator $T$ and its adjoint $T^*$ satisfy

$$\|T\| = \|T^*\| = \|T^*T\|^{\frac{1}{2}} \ . \tag{2.15}$$

*Proof.* For $\psi \in \mathcal{H}$, $\|\psi\| = 1$, we get

$$\|T\psi\|^2 = \left| \|T\psi\|^2 \right| = |\langle T\psi \,|\, T\psi \rangle| = |\langle \psi \,|\, T^*T\psi \rangle| \overset{(2.10)}{\leq} \|\psi\|^2 \|T^*T\| =$$
$$= \|T^*T\| \overset{(2.11)}{\leq} \|T^*\| \|T\| \ ,$$

which implies that

$$\|T\|^2 \leq \|T^*T\| \leq \|T^*\| \|T\| \ . \tag{2.16}$$

This shows, first of all, that $\|T\| \leq \|T^*\|$. Substituting $T^*$ instead of $T$ into this inequality and using the identity $(T^*)^* = T$ we also get $\|T^*\| \leq \|T\|$. Therefore, $\|T\| = \|T^*\|$. Using this fact in (2.16) proves the claim. $\qquad \square$

We can summarize the previous discussion by saying that $\mathcal{L}(\mathcal{H})$ is a $C^*$-*algebra*. This means that

- $\mathcal{L}(\mathcal{H})$ is an algebra;

- $\mathcal{L}(\mathcal{H})$ is a complete normed space (i.e. Banach space);

- the adjoint mapping $T \mapsto T^*$ on $\mathcal{L}(\mathcal{H})$ is conjugate linear and satisfies (2.13) and (2.14);

- the operator norm on $\mathcal{L}(\mathcal{H})$ satisfies (2.11) and (2.15).

These are the basic properties of $\mathcal{L}(\mathcal{H})$ and also the most often used facts for proving other relevant things.

**Example 6.** (*Shift operators*) Let us again continue the discussion on the Hilbert space $\ell^2(\mathbb{N})$. It is often convenient to write an element $\zeta \in \ell^2(\mathbb{N})$ as

$$\zeta = (\zeta_0, \zeta_1, \zeta_2, \dots),$$

where $\zeta_i \equiv \zeta(i)$. Let $A : \ell^2(\mathbb{N}) \to \ell^2(\mathbb{N})$ be the *shift operator* defined by

$$A(\zeta_0, \zeta_1, \dots) = (0, \zeta_0, \zeta_1, \dots).$$

We have

$$\|A\zeta\| = \sum_{j=0}^{\infty} |\zeta_j|^2 = \|\zeta\|,$$

and therefore $A$ is bounded and $\|A\| = 1$. To calculate the adjoint operator $A^*$, let $\zeta = (\zeta_0, \zeta_1, \dots), \eta = (\eta_0, \eta_1, \dots) \in \ell^2(\mathbb{N})$. The defining condition (2.12) for $A^*$ gives

$$\langle\, \eta \,|\, A^*\zeta \,\rangle = \langle\, A\eta \,|\, \zeta \,\rangle = \sum_{j=0}^{\infty} \bar{\eta}_j \zeta_{j+1}.$$

As the vector $\eta$ is arbitrary, we conclude that

$$A^*(\zeta_0, \zeta_1, \zeta_2, \dots) = (\zeta_1, \zeta_2, \dots).$$

The operator $A^*$ is called the *backward shift operator*.

We will later need the following simple facts.

**Exercise 4.** Let $T \in \mathcal{L}(\mathcal{H})$ and $\phi, \psi \in \mathcal{H}$. Verify the following identity, know as *Polarization Identity*:

$$\langle\, \phi \,|\, T\psi \,\rangle = \frac{1}{4} \sum_{k=0}^{3} i^k \, \langle\, \psi + i^k\phi \,|\, T(\psi + i^k\phi) \,\rangle. \tag{2.17}$$

(Hint: expand the right hand side of (2.17).)

**Proposition 4.** Let $S, T \in \mathcal{L}(\mathcal{H})$. If $\langle\, \psi \,|\, S\psi \,\rangle = \langle\, \psi \,|\, T\psi \,\rangle$ for every $\psi \in \mathcal{H}$, then $S = T$.

*Proof.* Assume that $\langle\,\psi\,|\,S\psi\,\rangle = \langle\,\psi\,|\,T\psi\,\rangle$ for every $\psi \in \mathcal{H}$. By Polarization Identity (2.17), we have $\langle\,\phi\,|\,S\psi\,\rangle = \langle\,\phi\,|\,T\psi\,\rangle$ for every $\psi, \phi \in \mathcal{H}$. Fix a vector $\psi \in \mathcal{H}$ and choose an orthonormal basis $\{\phi_k\}$ for $\mathcal{H}$. We then have $\langle\,\phi_k\,|\,S\psi\,\rangle = \langle\,\phi_k\,|\,T\psi\,\rangle$ for every $k = 0, 1, \ldots$. This means that the basis expansions of the vectors $S\psi$ and $T\psi$ are the same, hence $S\psi = T\psi$. As this is true for every $\psi \in \mathcal{H}$, we conclude that $S = T$. $\qquad\square$

**Definition 5.** Let $T$ be a bounded operator.

- A number $\lambda \in \mathbb{C}$ is an *eigenvalue* of $T$ if there exists a vector $\psi \in \mathcal{H} \setminus \{0\}$ such that $T\psi = \lambda\psi$. The vector $\psi$ is *eigenvector* of $T$ associated with the eigenvalue $\lambda$.

- A number $\lambda \in \mathbb{C}$ is in the *spectrum* of $T$ if the inverse mapping of the operator $\lambda I - T$ does not exist.

It is clear that all of the eigenvalues of $T$ are in the spectrum. However, spectrum may also contain other numbers than just the eigenvalues. In a finite dimensional Hilbert space the eigenvalues of $T$ are solutions of the equation $\det(T - \lambda I) = 0$. In particular, every operator has eigenvalues. This is no longer true in infinite dimensional Hilbert space, as we demonstrate in the following example.

**Example 7.** The shift operator $A$ defined in Example 6 does not have any eigenvalues. Indeed, assume that $\psi = \sum_k c_k \delta_k$ would be an eigenvector of $A$ with an eigenvalue $\lambda$. Since $A\delta_k = \delta_{k+1}$, we get the following set of equations:

$$
\begin{aligned}
\lambda c_0 &= 0, \\
\lambda c_1 &= c_0, \\
\lambda c_2 &= c_1, \\
&\;\;\vdots
\end{aligned}
$$

This can happen only if $c_0 = c_1 = \ldots = 0$, hence $\psi = 0$.

The eigenvalues and the spectrum play an important role in the analysis of operators. However, we are avoiding the machinery of spectral analysis, and for our purposes and the knowledge of elementary properties of eigenvalues and eigenvectors is sufficient.

**Exercise 5.** Prove the following: if $\psi_1$ and $\psi_2$ are eigenvectors of a bounded operator $T$ and they correspond to the same eigenvalue $\lambda$, then also every linear combination of $\psi_1$ and $\psi_2$ is an eigenvector associated with the same eigenvalue $\lambda$.

### 2.2.2    Partially ordered vector space of selfadjoint operators

**Definition 6.** An operator $T \in \mathcal{L}(\mathcal{H})$ is *selfadjoint* if $T = T^*$. We denote by $\mathcal{L}_s(\mathcal{H})$ the set of selfadjoint operators on $\mathcal{H}$.

Since the adjoint mapping $T \mapsto T^*$ is linear with respect to real linear combinations, it follows that the set $\mathcal{L}_s(\mathcal{H})$ is a real vector space. We denote by $O$ and $I$ the null operator and the identity operator, respectively; they are defined as $O\psi = 0$ and $I\psi = \psi$ for every $\psi \in \mathcal{H}$. Both $O$ and $I$ are clearly selfadjoint operators.

**Proposition 5.** An operator $T \in \mathcal{L}(\mathcal{H})$ is selfadjoint if and only if $\langle\, \psi \,|\, T\psi \,\rangle$ is a real number for every $\psi \in \mathcal{H}$.

*Proof.* By Proposition 4, an operator $T$ is selfadjoint iff $\langle\, \psi \,|\, T\psi \,\rangle = \langle\, \psi \,|\, T^*\psi \,\rangle$ for every $\psi \in \mathcal{H}$. But $\langle\, \psi \,|\, T^*\psi \,\rangle = \langle\, T\psi \,|\, \psi \,\rangle$, so the previous condition is the same as to require that $\langle\, \psi \,|\, T\psi \,\rangle = \overline{\langle\, \psi \,|\, T\psi \,\rangle}$ for every $\psi \in \mathcal{H}$. $\qquad\square$

It is not immediate from the definition that selfadjoint operators (except real multiples of the identity operator) exist. A moments reflection shows, however, that there are plenty of them. Linear (complex) combinations of selfadjoint operators actually span $\mathcal{L}(\mathcal{H})$, as illustrated in the following example.

**Example 8.** (*Real and imaginary parts of an operator*) Any operator $T \in \mathcal{L}(\mathcal{H})$ can be written as a sum of two selfadjoint operators. Indeed, denote $T_R = \frac{1}{2}(T + T^*)$ and $T_I = \frac{1}{2i}(T - T^*)$. These operators are called the real and the imaginary parts of $T$, respectively. It is easy to verify that $T_R$ and $T_I$ are selfadjoint and $T = T_R + iT_I$.

**Definition 7.** An operator $T \in \mathcal{L}(\mathcal{H})$ is *positive* if $\langle\, \psi \,|\, T\psi \,\rangle \geq 0$ for every $\psi \in \mathcal{H}$.

A comparison of Definition 7 to Proposition 5 shows that positive operators are selfadjoint.

**Exercise 6.** Prove the following: if a selfadjoint operator $T$ has an eigenvalue $\lambda$, then $\lambda$ is a real number. Moreover, if $T$ is positive, then $\lambda$ is a positive number.

The concept of positivity defines a partial ordering in $\mathcal{L}_s(\mathcal{H})$ in a natural way - this is spelled out in the following definition.

**Definition 8.** Let $S, T \in \mathcal{L}_s(\mathcal{H})$. We denote $S \geq T$ if the operator $S - T$ is positive.

Notice that an operator $T$ is positive exactly when $T \geq O$. It is straightforward to verify that the relation $\geq$ in Definition 8 is a partial ordering. This relation has also some further properties, which connect the order structure and the vector space structure of $\mathcal{L}_s(\mathcal{H})$. Namely, let $T_1, T_2, T_3 \in \mathcal{L}_s(\mathcal{H})$ and $\alpha \in \mathbb{R}, \alpha \geq 0$. It follows directly from the definition of positivity that

- if $T_1 \geq T_2$, then $T_1 + T_3 \geq T_2 + T_3$ .

- if $T_1 \geq T_2$, then $\alpha T_1 \geq \alpha T_2$ .

These properties mean that the relation $\geq$ makes $\mathcal{L}_s(\mathcal{H})$ a *partially ordered vector space*.

The following two results illustrate the partial ordering relation of selfadjoint operators. They are also useful later.

**Exercise 7.** Let $T \in \mathcal{L}_s(\mathcal{H})$ and $T \neq O$. Show that $-I \leq \|T\|^{-1} T \leq I$.

**Proposition 6.** Let $T \in \mathcal{L}_s(\mathcal{H})$. If $O \leq T \leq I$, then $O \leq T^2 \leq T$.

*Proof.* Let $\psi \in \mathcal{H}$. Then

$$\langle\, \psi \,|\, T^2\psi \,\rangle = \left\langle\, T^{\frac{1}{2}}\psi \,\middle|\, TT^{\frac{1}{2}}\psi \,\right\rangle \leq \left\langle\, T^{\frac{1}{2}}\psi \,\middle|\, T^{\frac{1}{2}}\psi \,\right\rangle = \langle\, \psi \,|\, T\psi \,\rangle \ .$$

Hence, $T^2 \leq T$. On the other hand,

$$\langle\, \psi \,|\, T^2 \psi \,\rangle = \langle\, T\psi \,|\, T\psi \,\rangle = \|T\psi\|^2 \geq 0 \,,$$

and thus $T \geq O$. $\hfill\square$

We need later the following fact, also known as the *square root lemma*. It is a standard result and can be found in any functional analysis book.

**Theorem 1.** (*Square root lemma*) Let $T \in \mathcal{L}(\mathcal{H})$ be a positive operator. There is a unique positive operator, denoted by $T^{\frac{1}{2}}$, satisfying $(T^{\frac{1}{2}})^2 = T$. The operator $T^{\frac{1}{2}}$ (also denoted by $\sqrt{T}$) is called the *square root* of $T$. It has the following properties:

(a) If $S \in \mathcal{L}(\mathcal{H})$ and $ST = TS$, then $ST^{\frac{1}{2}} = T^{\frac{1}{2}}S$.

(b) If $T$ is invertible, then also $T^{\frac{1}{2}}$ is invertible and $(T^{\frac{1}{2}})^{-1} = (T^{-1})^{\frac{1}{2}}$.

The square root lemma is important in many situations. In the following we derive some useful consequences.

**Proposition 7.** Let $T \in \mathcal{L}(\mathcal{H})$ be a positive operator and $\psi \in \mathcal{H}$. If $\langle\, \psi \,|\, T\psi \,\rangle = 0$, then $T\psi = 0$.

*Proof.* Suppose that $\langle\, \psi \,|\, T\psi \,\rangle = 0$. Then

$$0 = \langle\, \psi \,|\, T\psi \,\rangle = \langle\, T^{\frac{1}{2}}\psi \,|\, T^{\frac{1}{2}}\psi \,\rangle = \left\| T^{\frac{1}{2}}\psi \right\| \,.$$

Hence, $T^{\frac{1}{2}}\psi = 0$. Applying the operator $T^{\frac{1}{2}}$ on both sides of this equality gives $T\psi = 0$. $\hfill\square$

For each $T \in \mathcal{L}(\mathcal{H})$, the operator $T^*T$ is positive. Indeed, we have

$$\langle\, \psi \,|\, T^*T\psi \,\rangle = \langle\, T\psi \,|\, T\psi \,\rangle = \|T\psi\|^2 \geq 0 \,.$$

With this observation we ready for the following definition.

**Definition 9.** Let $T \in \mathcal{L}(\mathcal{H})$. We denote $|T| := (T^*T)^{\frac{1}{2}}$.

**Theorem 2.** (*Polar decomposition*) Let $T \in \mathcal{L}(\mathcal{H})$. There exists an operator $V \in \mathcal{L}(\mathcal{H})$ such that

- $T = V\,|T|$ ;

- $\|V\psi\| = \|\psi\|$ for every $\psi \in (\ker V)^{\perp} = \{\phi \in \mathcal{H} : \langle\, \phi \,|\, \varphi \,\rangle = 0 \,\forall \varphi \in \ker V\}$.

The second condition in Theorem 2 means that $V$ is a *partial isometry*.

### 2.2.3   Group of unitary operators

In Section 2.1.1 we defined the concepts of isomorphic inner product spaces and isomorphism. Any Hilbert space $\mathcal{H}$ is, trivially, isomorphic with itself; one isomorphism is the identity mapping $I : \psi \mapsto \psi$. There are also other isomorphisms on a given Hilbert spaces and they play an important role in various different situations.

**Proposition 8.** A linear mapping $U : \mathcal{H} \to \mathcal{H}$ is an isomorphism if and only if $U$ is bounded and

$$UU^* = U^*U = I . \tag{2.18}$$

*Proof.* Assume that $U$ is an isomorphism, hence satisfying Eq. (2.2). It follows that $\|U\psi\| = \|\psi\|$, and thus, $U$ is a bounded operator with $\|U\| = 1$. Comparing (2.2) with the defining condition (2.12) for $U^*$, we see that

$$\langle\, U\varphi \,|\, U\psi \,\rangle = \langle\, \varphi \,|\, U^*U\psi \,\rangle = \langle\, \varphi \,|\, \psi \,\rangle \tag{2.19}$$

for every $\varphi, \psi \in \mathcal{H}$. Therefore, $U^*U = I$. The inverse mapping $U^{-1}$ is also an isomorphism and a similar reasoning as before leads to $UU^* = I$. Hence, $U$ satisfies (2.18).

On the other hand, a bounded operator $U$ satisfying (2.18) is an isomorphism, as is seen from (2.19). □

It is common to rename the isomorphisms on $\mathcal{H}$ in the following way.

**Definition 10.** An operator $U \in \mathcal{L}(\mathcal{H})$ is *unitary* if it satisfies (2.18). We denote by $\mathcal{U}(\mathcal{H})$ the set of unitary operators on $\mathcal{H}$.

**Example 9.** (*Eigenvalues of a unitary operator*) In the proof of Proposition 8 we have already seen that $\|U\| = 1$ for a unitary operator $U$. Hence, an eigenvalue $\lambda$ of $U$ satisfies $|\lambda| \leq 1$. It is actually easy to see that all eigenvalues of $U$ must satisfy $|\lambda| = 1$. Namely, suppose that $\psi$ is an eigenvector of $U$, i.e., $U\psi = \lambda\psi$ for some $\lambda \in \mathbb{C}$. Since $|\lambda|^2 \langle\, \psi \,|\, \psi \,\rangle = \langle\, U\psi \,|\, U\psi \,\rangle = \langle\, \psi \,|\, \psi \,\rangle$, we conclude that $|\lambda| = 1$. Thus, all the eigenvalues of $U$ are of the form $\lambda = e^{ia}$, $a \in \mathbb{R}$. Note, however, that a unitary operator in an infinite dimensional Hilbert space need not have eigenvalues at all.

In linear algebra unitary operators are usually introduced as mappings which transform an orthonormal basis to another orthonormal basis. This property is valid also in the infinite dimensional case and we have the following useful result.

**Proposition 9.** Let $\mathcal{H}$ be a Hilbert space and $\{\varphi_j\}$ an orthonormal basis for $\mathcal{H}$. If $U$ is a unitary operator on $\mathcal{H}$, then also $\{U\varphi_j\}$ is an orthonormal basis for $\mathcal{H}$.

*Proof.* It is clear that $\{U\varphi_j\}$ is an orthonormal set. To prove that it is maximal orthonormal set, suppose that $\psi \in \mathcal{H}$ is such that $\langle\, \psi \,|\, U\varphi_j \,\rangle = 0$ for every $j$. This implies that $\langle\, U^*\psi \,|\, \varphi_j \,\rangle = 0$ for every $j$, which means that $U^*\psi = 0$ as $\{\varphi_j\}$ is an orthonormal basis. Since $U^*$ is bijective, we conclude that $\psi = 0$. This shows that $\{U\varphi_j\}$ is maximal. □

The unitary operators form a group with respect to the operator multiplication. The identity operator $I$ is the unit element in the group $\mathcal{U}(\mathcal{H})$. To verify the other group axioms, let $U$ and $V$ be two unitary operators. We get

$$(UV)(UV)^* = UVV^*U^* = I \,,$$

and

$$(UV)^*(UV) = V^*U^*UV = I \,,$$

where we have used Eq. (2.13). Hence, the product $UV$ is unitary. Since $U^{-1} = U^*$ by Eq. (2.18), an application of Eq. (2.14) shows that $U^{-1}$ is unitary also.

**Example 10.** (*Exponential of a selfadjoint operator*) Let $T$ be a bounded operator. For each $k = 0, 1, 2, \ldots$, we denote

$$F_k(T) := \sum_{n=0}^{k} \frac{T^n}{n!} \,, \qquad f_k(T) := \sum_{n=0}^{k} \frac{\|T^n\|}{n!} \,.$$

We have

$$f_k(T) \leq \sum_{n=0}^{k} \frac{\|T\|^n}{n!} \leq \sum_{n=0}^{\infty} \frac{\|T\|^n}{n!} = e^{\|T\|} \,,$$

which shows that the increasing sequence $f_0(T), f_1(T), \ldots$ of real numbers has an upper bound. Therefore, it converges. This means that the series $\sum_{n=0}^{\infty} \frac{T^n}{n!}$ is absolutely convergent. Since $\mathcal{L}(\mathcal{H})$ is a Banach space, every absolutely convergent series converges. We denote by $e^T$ the limit, that is,

$$e^T := \sum_{n=0}^{\infty} \frac{T^n}{n!} = \lim_{k \to \infty} F_k(T) \,. \tag{2.20}$$

The product and adjoint are continuous mappings on $\mathcal{L}(\mathcal{H})$. Hence, it is straightforward to verify the following formulas for $a, b \in \mathbb{C}$ and $T \in \mathcal{L}(\mathcal{H})$,

$$e^{aT} e^{bT} = e^{(a+b)T} \,,$$
$$(e^{aT})^* = e^{\bar{a}T^*} \,.$$

Assume then that $T$ is selfadjoint. We have $(e^{iT})^* = e^{-iT}$, and

$$e^{iT} e^{-iT} = e^{-iT} e^{iT} = e^{O} = I \,.$$

This shows that $e^{iT}$ is a unitary operator.

As we will see later, unitary operators have an important role in quantum formalism. Also a related concept of an antiunitary operator is sometimes needed.

**Definition 11.** A mapping $A : \mathcal{H} \to \mathcal{H}$ is *antiunitary operator* if $A(\psi + \lambda\varphi) = A\psi + \bar{\lambda}A\varphi$ and $\langle\, A\psi \mid A\varphi \,\rangle = \langle\, \varphi \mid \psi \,\rangle$ for all $\psi, \varphi \in \mathcal{H}$ and $\lambda \in \mathbb{C}$.

**Exercise 8.** Show that a product of two antiunitary operators is a unitary operator. Show also that a product of a unitary operator and an antiunitary operator is an antiunitary operator.

There is a prototypical class of antiunitary operators. Fix an orthonormal basis $\{\varphi_j\}$ for $\mathcal{H}$. Every vector $\psi \in \mathcal{H}$ has a unique representation

$$\psi = \sum_j c_j \varphi_j \,.$$

We define

$$J\psi = \sum_j \bar{c}_j \varphi_j \,.$$

It is straightforward to verify that $J$ is an antiunitary operator and it clearly satisfies $J^2 = I$. We call it the *complex conjugate operator* related to the orthonormal basis $\{\varphi_j\}$.

**Proposition 10.** Let $J$ be the complex conjugate operator related to same orthonormal basis $\{\varphi_j\}$. Every antiunitary operator $A$ can be written in the form $A = UJ$, where $U$ is a unitary operator.

*Proof.* Let $A$ be an antiunitary operator. Denote $U = AJ$. By Exercise 8, $U$ is a unitary operator. Moreover, we have $UJ = AJ^2 = A$. $\qquad\square$

### 2.2.4   Orthocomplemented poset of projections

**Definition 12.** A selfadjoint operator $P \in \mathcal{L}_s(\mathcal{H})$ is a *projection* if $P = P^2$. We denote by $\mathcal{P}(\mathcal{H})$ the set of projections.

Projections are positive operators. Namely, if $P$ is a projection and $\psi \in \mathcal{H}$, we have

$$\langle\, \psi \,|\, P\psi \,\rangle = \langle\, \psi \,|\, P^2\psi \,\rangle = \langle\, P\psi \,|\, P\psi \,\rangle = \|P\psi\|^2 \geq 0 \,.$$

We discuss some other elementary properties of projections below. Let us first, however, consider a prototypical example of a projection.

**Example 11.** (*One-dimensional projection*) Let $\eta \in \mathcal{H}$ be a unit vector. We define an operator $P_\eta$ on $\mathcal{H}$ as

$$P_\eta\psi = \langle\, \eta \,|\, \psi \,\rangle\,\eta \,. \tag{2.21}$$

It is straightforward to check that $P_\eta^* = P_\eta$ and $P_\eta^2 = P_\eta$. Therefore, $P_\eta$ is a projection. It is clear that the range of $P$ is the one-dimensional subspace $\mathbb{C}\eta = \{c\eta \mid c \in \mathbb{C}\}$. For this reason $P_\eta$ is called one-dimensional projection.

**Proposition 11.** Let $P$ be a projection and $O \neq P \neq I$. Then

(a) $\|P\| = 1$.

(b) The eigenvalues of $P$ are 0 and 1.

(c) For every $\psi \in \mathcal{H}$, there are vectors $\psi_0, \psi_1 \in \mathcal{H}$ such that

$$P\psi_0 = 0 , \qquad P\psi_1 = \psi_1 , \qquad \text{and} \qquad \psi = \psi_0 + \psi_1 .$$

*Proof.* (a) An application of Eq. (2.15) gives

$$\|P\| = \|P^2\| = \|P^*P\| = \|P\|^2 .$$

Since $\|P\| \neq 0$ by the assumption, we conclude that $\|P\| = 1$.

(b) First of all, suppose that $\psi \in \mathcal{H}$ is an eigenvector of $P$ with eigenvalue $\lambda$. Then $P\psi = \lambda\psi$ and

$$P\psi = P^2\psi = \lambda P\psi = \lambda^2\psi .$$

Hence, $\lambda^2 = \lambda$, which means that either $\lambda = 0$ or $\lambda = 1$. To prove that $P$ has both of these eigenvalues, choose vectors $\psi, \phi \in \mathcal{H}$ such that $P\psi \neq 0$ and $P\phi \neq \phi$ (this can be done since $O \neq P \neq I$). Then $P\psi$ is an eigenvector of $P$ with eigenvalue 1 and $(I - P)\phi$ is an eigenvector of $P$ with eigenvalue 0.

(c) Denote $\psi_0 = (I - P)\psi$ and $\psi_1 = P\psi$. These vectors have the required properties. □

**Proposition 12.** Let $P$ be a projection and $\psi \in \mathcal{H}$. The following conditions are equivalent:

(i) $\psi \in \operatorname{ran} P$ ;

(ii) $P\psi = \psi$ ;

(iii) $\|P\psi\| = \|\psi\|$ .

*Proof.* It is trivial that (ii)⇒(iii). In the following we prove that (i)⇒(ii) and (iii)⇒(i).

Assume that (i) holds. This means that there is a vector $\varphi \in \mathcal{H}$ such that $P\varphi = \psi$. But then

$$P\psi = PP\varphi = P\varphi = \psi ,$$

and hence, (ii) holds. We conclude that (i)⇒(ii).

Finally, assume that (iii) holds. We show that (i) follows. Let us first notice that $\psi$ can written as a sum

$$\psi = P\psi + (I - P)\psi . \tag{2.22}$$

The vectors $P\psi$ and $(I - P)\psi$ are orthogonal since

$$\langle\, P\psi \,|\, (I - P)\psi \,\rangle = \langle\, \psi \,|\, P(I - P)\psi \,\rangle = 0 .$$

Therefore, we can apply Pythagoras theorem to (2.22), and we get

$$\|\psi\|^2 = \|P\psi\|^2 + \|(I - P)\psi\|^2 .$$

Our assumption (iii) now implies that $\|(I - P)\psi\| = 0$, where it follows that $(I - P)\psi = 0$. Thus, $P\psi = \psi$. □

**Proposition 13.** Let $P$ and $Q$ be projections. The following conditions are equivalent:

  (i) $P \geq Q$ ;

 (ii) $PQ = Q$ ;

(iii) $QP = Q$ ;

 (iv) $QP = PQ = Q$ ;

  (v) $P - Q$ is a projection.

*Proof.* We prove this proposition by showing that (i)$\Rightarrow$(ii)$\Rightarrow$(iii)$\Rightarrow$(iv)$\Rightarrow$(v)$\Rightarrow$(i).

Suppose that (i) holds and fix $\psi \in \mathcal{H}$. Let us first assume that $P \neq O$. Then by Prop. 11 we have $\|P\| = 1$, and an application of (2.9) gives

$$\|PQ\psi\| \leq \|P\| \, \|Q\psi\| = \|Q\psi\| \; .$$

On the other hand, using (i) we get

$$\|PQ\psi\|^2 = \langle \, PQ\psi \mid PQ\psi \, \rangle = \langle \, Q\psi \mid PQ\psi \, \rangle \overset{(i)}{\geq} \langle \, Q\psi \mid QQ\psi \, \rangle = \langle \, Q\psi \mid Q\psi \, \rangle = \|Q\psi\|^2 \; ,$$

and hence, $\|PQ\psi\| \geq \|Q\psi\|$. Therefore, $\|PQ\psi\| = \|Q\psi\|$. This together with Prop. 12 implies that $PQ\psi = Q\psi$. As this is true for any $\psi \in \mathcal{H}$, we conclude that $PQ = Q$ and (ii) holds. If $P = O$, then (i) implies that $O \leq Q \leq O$, which means that $Q = O$. Hence, also in this case (i) implies (ii).

Suppose that (ii) holds. We get

$$Q = Q^* = (PQ)^* = Q^*P^* = QP \, ,$$

and therefore, (iii) holds. In the same way (iii) implies (ii) and they are thus equivalent. This means that (iii) implies (iv).

Assume that (iv) holds. The operator $P - Q$ is selfadjoint as both $P$ and $Q$ are. Using (iv) we get

$$(P - Q)^2 = P - PQ - QP + Q = P - Q \, .$$

Hence, $P - Q$ is a projection, and we conclude that (iv) implies (v).

Finally, assume that (v) holds. As a projection is a positive operator, we have $\langle \, \psi \mid (P - Q)\psi \, \rangle \geq 0$ for every $\psi \in \mathcal{H}$. This means that $\langle \, \psi \mid P\psi \, \rangle \geq \langle \, \psi \mid Q\psi \, \rangle$ for every $\psi \in \mathcal{H}$, and hence, (i) holds. Thus, (v) implies (i). $\qquad\square$

If $P$ is a projection, we denote $P^{\perp} := I - P$ and we call this operator the *complement* of $P$. The operator $P^{\perp}$ is clearly selfadjoint and it is a projection since

$$P^{\perp}P^{\perp} = (I - P)(I - P) = I - P - P + P^2 = I - P = P^{\perp} \; .$$

**Exercise 9.** Verify the following properties of $P^{\perp}$:

- $(P^\perp)^\perp = P$.

- If $Q \leq P$, then $P^\perp \leq Q^\perp$.

**Proposition 14.** Let $P$ and $Q$ be projections. If $Q \leq P$ and $Q \leq P^\perp$, then $Q = O$.

*Proof.* If $Q \leq P$ and $Q \leq P^\perp$, then by Prop. 13 we have $PQ = Q$ and $P^\perp Q = Q$. Adding these two equations together gives $Q = O$. $\qquad\square$

The properties described in Exercise 9 and Proposition 14 mean that the mapping $P \mapsto P^\perp$ on $\mathcal{P}(\mathcal{H})$ is an *orthocomplementation*.

**Proposition 15.** Let $P$ and $Q$ be projections. Then the following conditions are equivalent:

(i) $P^\perp \geq Q$ ;

(ii) $Q^\perp \geq P$ ;

(iii) $PQ = O$ ;

(iv) $QP = O$ ;

(v) $PQ = QP = O$ ;

(vi) $P + Q$ is a projection.

*Proof.* We prove the proposition by showing that (i)$\Rightarrow$(ii)$\Rightarrow$(iii)$\Rightarrow$(iv)$\Rightarrow$(v)$\Rightarrow$(vi)$\Rightarrow$(i). The fact that (i) implies (ii) is a direct consequence of Exercise 9.

Assume then that (ii) holds and fix $\psi \in \mathcal{H}$. Denote $\psi_1 = Q\psi$. We then have

$$\|P\psi_1\|^2 = \langle\, P\psi_1 \mid P\psi_1 \,\rangle = \langle\, \psi_1 \mid P\psi_1 \,\rangle \leq \langle\, \psi_1 \mid Q^\perp \psi_1 \,\rangle = \langle\, \psi_1 \mid Q^\perp Q\psi \,\rangle = 0\,.$$

This implies that $P\psi_1 = 0$, and hence, $PQ\psi = 0$. As $\psi$ was arbitrary vector, we conclude that $PQ = O$. Therefore, (ii) implies (iii).

Suppose that (iii) holds. We get

$$O = O^* = (PQ)^* = Q^*P^* = QP\,,$$

and therefore, (iv) holds. In the same way (iv) implies (iii) and they are thus equivalent. This means that (iv) implies (v).

Assume that (v) holds. Then

$$(P + Q)^2 = P^2 + PQ + QP + Q^2 = P + Q\,.$$

Thus, $P + Q$ is a projection. Therefore, (v) implies (vi).

Finally, assume that (vi) holds. This implies that $P + Q \leq I$. Hence, (i) holds. $\qquad\square$

Two projections $P$ and $Q$ satisfying one (and hence all) of the conditions in Proposition 15 are called *orthogonal*.

**Example 12.** (*Orthogonal one-dimensional projections*) Let us continue from Example 11. Let $\eta, \phi \in \mathcal{H}$ be two unit vectors and $P_\eta$, $P_\phi$ the corresponding one-dimensional projections. For a vector $\psi \in \mathcal{H}$, we get

$$P_\phi P_\eta \psi = \langle \eta \,|\, \psi \rangle \langle \phi \,|\, \eta \rangle \, \phi \,. \tag{2.23}$$

This shows that $P_\phi P_\eta = O$ if $\langle \phi \,|\, \eta \rangle = 0$. On the other hand, choosing $\psi = \eta$ in (2.23) we see that $P_\phi P_\eta = O$ only if $\langle \phi \,|\, \eta \rangle = 0$. We conclude that two one-dimensional projections are orthogonal if and only if the unit vectors defining them are orthogonal.

As seen from Proposition 15, we can sum two orthogonal projections to get a third projection. In this way, one dimensional projections can be used as building blocks to get other projections. Actually, every projection is either a finite or countably infinite sum of one-dimensional projections. We explain this construction but do not prove the details.

Let $P$ be a projection. As shown in Proposition 12, the range of $P$ consists of its eigenvectors with eigenvalue 1. Hence, ran $P$ is a linear subspace of $\mathcal{H}$. Let us first assume that ran $P$ is finite dimensional with dimension $r$. We choose an orthonormal basis $\{\eta_k\}_{k=1}^r$ for ran $P$. For every $k = 1, \ldots, r$, we then have a one-dimensional projection $P_k \equiv P_{\eta_k}$. For $k \neq l$, the projections $P_k$ and $P_l$ are orthogonal as explained in Example 12. By Proposition 15, the sum $\sum_{k=1}^r P_k$ is a projection, and it can be shown that $\sum_{k=1}^r P_k = P$.

If ran $P$ is infinite dimensional, the same procedure still works. In this case, we first note that the set ran $P$ is closed. Indeed, it follows from Proposition 12 that $\psi \in$ ran $P$ exactly when $P^\perp \psi = 0$. Hence, ran $P$ is the preimage of the closed set $\{0\}$ in the continuous mapping $P^\perp$, which implies that ran $P$ is closed. We conclude that ran $P$ is a closed linear subspace of $\mathcal{H}$, and therefore it has an orthonormal basis $\{\eta_k\}_{k=1}^\infty$. The infinite sum $\sum_{k=1}^\infty P_k$ converges in the weak operator topology (see Sec. 2.3.1) and we have again $\sum_{k=1}^\infty P_k = P$.

### 2.2.5   Ideal of trace class operators

In the finite dimensional Hilbert space $\mathbb{C}^d$, the *trace* of an operator $T$ can be calculated by writing $T$ as a matrix in some orthonormal basis and then summing the diagonal entries of the matrix. This number, denoted by $\mathrm{tr}\,[T]$, does not depend on the chosen orthonormal basis and we thus have

$$\mathrm{tr}\,[T] = \sum_{j=1}^d \langle \varphi_j \,|\, T\varphi_j \rangle \tag{2.24}$$

for any orthonormal basis $\{\varphi_j\}_{j=1}^d$ of $\mathbb{C}^d$. As one knows from linear algebra, $\mathrm{tr}\,[T]$ equals the sum of the eigenvalues of $T$, counting multiplicity.

In an infinite dimensional Hilbert space the trace is still a useful concept but things are not so straightforward as in $\mathbb{C}^d$. Let $\mathcal{H}$ be a separable infinite dimensional Hilbert space and $\{\varphi_j\}_{j=1}^\infty$ an orthonormal basis for $\mathcal{H}$. For any positive operator $T \in \mathcal{L}(\mathcal{H})$, we denote

$$\mathrm{tr}\,[T] = \sum_{j=1}^\infty \langle \varphi_j \,|\, T\varphi_j \rangle \,. \tag{2.25}$$

It may happen that the sum in the right hand side does not converge, in which case we denote $\mathrm{tr}\,[T] = \infty$. The number $\mathrm{tr}\,[T]$ does not depend on the chosen orthonormal basis $\{\varphi_j\}_{j=1}^{\infty}$. Indeed, let $\{\psi_j\}_{j=1}^{\infty}$ another orthonormal basis for $\mathcal{H}$. Then

$$
\begin{aligned}
\sum_{j=1}^{\infty} \langle\, \psi_j \mid T\psi_j \,\rangle &= \sum_{j=1}^{\infty} \left\| T^{\frac{1}{2}}\psi_j \right\|^2 = \sum_{j=1}^{\infty} \left( \sum_{k=1}^{\infty} \left| \left\langle\, \varphi_k \mid T^{\frac{1}{2}}\psi_j \,\right\rangle \right|^2 \right) \\
&= \sum_{k=1}^{\infty} \left( \sum_{j=0}^{\infty} \left| \left\langle\, \psi_j \mid T^{\frac{1}{2}}\varphi_k \,\right\rangle \right|^2 \right) = \sum_{k=1}^{\infty} \left\| T^{\frac{1}{2}}\varphi_k \right\|^2 \\
&= \sum_{k=1}^{\infty} \langle\, \varphi_k \mid T\varphi_k \,\rangle\,.
\end{aligned}
$$

Here we have used Parseval's formula (2.6) twice. Interchanging the order of the sums is allowed since all the terms are non-negative.

**Example 13.** (*Trace of a one-dimensional projection*) Let $\eta \in \mathcal{H}$ be a unit vector and $P_\eta$ the corresponding one-dimensional projection as defined in Example 11. Let us choose an orthonormal basis $\{\varphi_j\}_{j=1}^{\infty}$ for $\mathcal{H}$ such that $\varphi_1 = \eta$. Then

$$
\mathrm{tr}\,[P_\eta] = \sum_{j=1}^{\infty} \langle\, \varphi_j \mid P_\eta\varphi_j \,\rangle = \langle\, \varphi_1 \mid P_\eta\varphi_1 \,\rangle = \langle\, \varphi_1 \mid \varphi_1 \,\rangle = 1\,.
$$

Hence, we conclude that $\mathrm{tr}\,[P_\eta] = 1$.

**Exercise 10.** Let $S, T \in \mathcal{L}(\mathcal{H})$ be positive operators. Prove the following properties of the trace:

(a) $\mathrm{tr}\,[S + T] = \mathrm{tr}\,[S] + \mathrm{tr}\,[T]\,.$

(b) $\mathrm{tr}\,[\alpha T] = \alpha\,\mathrm{tr}\,[T]$ for all $\alpha \geq 0\,.$

(c) $\mathrm{tr}\,[UTU^*] = \mathrm{tr}\,[T]$ for all unitary operators $U\,.$

(Hint: For (a) and (b), use directly the definition (2.25). In the case of (c), Prop. 9 in Section 2.2.3 is useful.)

Our discussion so far concerns only positive operators. To proceed to other kind of operators, we need the following definition.

**Definition 13.** A bounded operator $T$ is a *trace class operator* if $\mathrm{tr}\,[|T|] < \infty$. We denote by $\mathcal{T}(\mathcal{H})$ the set of trace class operators.

If $\mathcal{H}$ is infinite dimensional, the set $\mathcal{T}(\mathcal{H})$ is a proper subset of $\mathcal{L}(\mathcal{H})$. For instance, the identity operator $I$ is positive and thus $\mathrm{tr}\,[|I|] = \mathrm{tr}\,[I] = \infty$. Therefore, $I \notin \mathcal{T}(\mathcal{H})$.

**Proposition 16.** If $T \in \mathcal{T}(\mathcal{H})$ and $\{\varphi_j\}_{j=1}^{\infty}$ is an orthonormal basis for $\mathcal{H}$, then $\sum_{j=1}^{\infty} |\langle\, \varphi_j \mid T\varphi_j \,\rangle|$ $< \infty$. The number

$$
\mathrm{tr}\,[T] := \sum_{j=1}^{\infty} \langle\, \varphi_j \mid T\varphi_j \,\rangle \tag{2.26}
$$

is called the trace of $T$ and it is independent of the chosen orthonormal basis.

All the effort we have done may seem superfluous since, in the end, we define the trace of any trace class operator with the same formula as we used earlier for positive trace class operators. The point is that if an operator $T$ is not a trace class operator, then the trace can be finite in some orthonormal basis but infinite in another one. Hence, we need to check that an operator belongs to the trace class before applying the formula (2.26).

In the following proposition we list some basic properties of trace class operators.

**Proposition 17.**    (a)  The set of trace class operators $\mathcal{T}(\mathcal{H})$ is a vector space and the mapping

$$T \mapsto \operatorname{tr}\left[|T|\right] =: \|T\|_{\mathrm{tr}}$$

is a norm on $\mathcal{T}(\mathcal{H})$.

(b)  Let $T \in \mathcal{T}(\mathcal{H})$ and $S \in \mathcal{L}(\mathcal{H})$. Then $TS, ST \in \mathcal{T}(\mathcal{H})$ and

$$\operatorname{tr}\left[TS\right] = \operatorname{tr}\left[ST\right] .$$

(c)  Let $T \in \mathcal{T}(\mathcal{H})$ and $S \in \mathcal{L}(\mathcal{H})$. Then

$$\|T\| \leq \|T\|_{\mathrm{tr}}$$

and

$$\left|\operatorname{tr}\left[TS\right]\right| \leq \|T\|_{\mathrm{tr}} \|S\| .$$

(d)  The mapping

$$(T, S) \mapsto \operatorname{tr}\left[T^*S\right] =: \langle\, T \mid S \,\rangle_{\mathrm{H\text{-}S}}$$

is an inner product on $\mathcal{T}(\mathcal{H})$.

Item (a) in Proposition 17 says that $\mathcal{T}(\mathcal{H})$ is a normed space. In particular, the triangle inequality

$$\|S + T\|_{\mathrm{tr}} \leq \|S\|_{\mathrm{tr}} + \|T\|_{\mathrm{tr}} \tag{2.27}$$

holds for all $S, T \in \mathcal{T}(\mathcal{H})$. The norm $\|\cdot\|_{\mathrm{tr}}$ is called the *trace norm*.

The important point in item (b) is that $S$ is just a bounded operator and need not be a trace class operator. Hence, even though $\operatorname{tr}\left[S\right]$ is not defined for all bounded operators $S$, it is defined for a product $ST$ whenever $T$ is a trace class operator. This property means that $\mathcal{T}(\mathcal{H})$ is an *ideal* in $\mathcal{L}(\mathcal{H})$.

A useful special case of (c) is obtained when we choose $S = I$. Hence, we get inequality

$$\left|\operatorname{tr}\left[T\right]\right| \leq \operatorname{tr}\left[|T|\right] ,$$

true for all $T \in \mathcal{T}(\mathcal{H})$.

The inner product defined in item (d) is called *Hilbert-Schmidt inner product*. Cauchy-Schwarz inequality can be written in the form

$$|\mathrm{tr}\,[ST]|^2 \le \mathrm{tr}\,[S^*S]\,\mathrm{tr}\,[T^*T]\;. \tag{2.28}$$

Hilbert-Schmidt inner product makes sense also for a larger class of operators than trace class operators, and these operators are called Hilbert-Schmidt operators. The inner product space consisting of Hilbert-Schmidt operators is a Hilbert space, and actually Proposition 17 is usually proved by first discovering the basic properties of Hilbert-Schimdt operators. However, for our purposes it is enough to use trace class operators.

**Example 14.** (*Operator norms in a finite dimensional Hilbert space*) Suppose $\mathcal{H}$ is finite dimensional. In this case, each operator $T$ on $\mathcal{H}$ is bounded and trace class. In particular,

$$\|T\| \quad = \quad \sup_{\|\psi\|=1}\|T\psi\| = \max_j |\lambda_j|\;, \tag{2.29}$$

$$\|T\|_{\mathrm{tr}} \quad = \quad \mathrm{tr}\,[|T|] = \sum_j |\lambda_j|\;, \tag{2.30}$$

where $\lambda_j$ are the eigenvalues of $T$. It is also useful to define one more norm, related to the Hilbert-Schmidt inner product. Hence, we define *Hilbert-Schmidt norm* by formula

$$\|T\|_{\mathrm{H\text{-}S}} := \langle\,T\,|\,T\,\rangle_{\mathrm{H\text{-}S}}^{\frac{1}{2}} = \sqrt{\sum_j |\lambda_j|^2}\;. \tag{2.31}$$

### 2.3   Also these are needed

#### 2.3.1   Weak operator topology

A separable infinite dimensional Hilbert space is, in some sense, the closest infinite dimensional analog of $\mathbb{C}^d$. However, the infinite dimension makes some things bit more delicate. The discussion of this section is redundant in a finite dimensional Hilbert space. Hence, we assume here that $\mathcal{H}$ is a separable infinite dimensional Hilbert space.

The topology in $\mathcal{L}(\mathcal{H})$ determined by the operator norm is too strong for many purposes. Several different topologies appear naturally for the set $\mathcal{L}(\mathcal{H})$. In the perspective of quantum mechanics, the weak operator topology is usually the most relevant. The weak operator topology does not come from a norm and the open sets in this topology are a bit lengthy to define. However, we do not need here an explicit description of the topology but it is enough for our purposes to specify when a sequence converges in the weak operator topology.

**Definition 14.** A sequence $\{T_i\} \subset \mathcal{L}(\mathcal{H})$ converges to a bounded operator $T$ *in the weak operator topology*, or *weakly*, if

$$\lim_i |\langle\,\varphi\,|\,T\psi\,\rangle - \langle\,\varphi\,|\,T_i\psi\,\rangle| = 0 \quad \text{for every } \varphi, \psi \in \mathcal{H}\;. \tag{2.32}$$

One should compare this definition with the fact that a sequence $\{T_i\} \subset \mathcal{L}(\mathcal{H})$ converges to a bounded operator $T$ in the operator norm topology if

$$\lim_i \|T - T_i\| = \lim_i \sup_{\|\psi\|=1} \|(T - T_i)\psi\| = 0\;. \tag{2.33}$$

**Proposition 18.** If a sequence $\{T_i\} \subset \mathcal{L}(\mathcal{H})$ converges to a bounded operator $T$ in the operator norm topology, then it also converges to $T$ in the weak operator topology.

*Proof.* For every $\varphi, \psi \in \mathcal{H}$, we get

$$|\langle\, \varphi \mid T\psi \,\rangle - \langle\, \varphi \mid T_i\psi \,\rangle| = |\langle\, \varphi \mid (T - T_i)\psi \,\rangle| \leq \|\varphi\| \, \|\psi\| \, \|T - T_i\| \ .$$

Here we have applied Ineq. (2.10). This proves the claim.                                          □

The converse implication in Proposition 18 is not generally valid. This is demonstrated in the following example.

**Example 15.** For each $i = 1, 2, \ldots$, we denote $T_i = A^i$, where $A$ is the shift operator defined in Example 6. Hence, $T_i$ acts in the following way:

$$T_i(\zeta_0, \zeta_1, \ldots) = (0, \ldots, 0, \zeta_0, \zeta_1, \ldots) \ .$$

Then $\lim_i T_i = O$ in the weak operator topology. Indeed, if $\eta \in \ell^2(\mathbb{N})$, then

$$\lim_{i \to \infty} |\langle\, \eta \mid T_i\zeta \,\rangle| = \lim_{i \to \infty} \left| \sum_{j=0}^{\infty} \bar{\eta}_{j+i}\zeta_j \right| \leq \|\zeta\|^2 \lim_{i \to \infty} \sum_{j=i}^{\infty} |\eta_j|^2 = 0 \ .$$

On the other hand, we have $\|T_i\zeta\| = \|\zeta\|$, and this implies that the sequence $\{T_i\}$ cannot converge to $O$ in the operator norm topology. It follows from Prop. 18 that $\{T_i\}$ does not converge in the operator norm topology as the only option would be that $\{T_i\}$ converges to $O$.

We conclude from Example 15 that the operator norm topology and the weak operator topology are different. For this reason, one has to specify the topology when convergence of operators is discussed. If not otherwise explicitly stated, we will understand all the formulas in the weak sense, i.e., if a sequence or a sum is said to converge, it is meant to converge with respect to the weak operator topology.

### 2.3.2   Dirac notation

In Section 2.2.4 we defined the one-dimensional projection $P_\eta$ for each unit vector $\eta \in \mathcal{H}$. In the so-called *Dirac notation* this projection is written as

$$P_\eta \equiv |\eta\rangle\langle\eta| \ .$$

Generally, if $\eta, \phi \in \mathcal{H}$, we define the operator $|\eta\rangle\langle\phi|$ as

$$|\eta\rangle\langle\phi| \, \psi = \langle\, \phi \mid \psi \,\rangle \, \eta \ . \tag{2.34}$$

A moment's thought shows that $|\eta\rangle\langle\phi|$ behaves as a "rearranged" inner product. For instance, the following rules apply for every $c \in \mathbb{C}$ and $\eta, \phi \in \mathcal{H}$:

$$|c\eta\rangle\langle\phi| = c\, |\eta\rangle\langle\phi| \, , \qquad |\eta\rangle\langle c\phi| = \bar{c}\, |\eta\rangle\langle\phi| \ .$$

Sometimes even a single vector $\eta$ is written in the form $|\eta\rangle$. With this convention, the definition (2.34) becomes simply a reordering rule of the terms. Moreover, $\langle\phi|$ is taken to denote the linear functional

$$\psi \mapsto \langle\,\phi\,|\,\psi\,\rangle\;.$$

Again, this is consistent with the other Dirac notations.

In his famous textbook [32], Dirac gave nice and intuitive names for his notations. First of all, an inner product $\langle\,\eta\,|\,\phi\,\rangle$ is a *bracket*. Since we can decompose it as a "product" of $\langle\eta|$ and $|\phi\rangle$, Dirac called $\langle\eta|$ a *bra vector* and $|\phi\rangle$ a *ket vector*.

**Exercise 11.** Let $P_\eta$ and $P_\phi$ be two one-dimensional projections. Show that

$$P_\eta P_\phi P_\eta = \mathrm{tr}\,[P_\eta P_\phi]\,P_\eta\;.$$

(Hint: write everything in Dirac notation.)

**Example 16.** (*Resolution of the identity operator*) It is sometimes convenient to write the identity mapping $I$ on $\mathcal{H}$ as a sum

$$I = \sum_{k=0}^{d} |\varphi_k\rangle\langle\,\varphi_k\,|\,, \tag{2.35}$$

where $\{\varphi_k\}_{k=0}^{d}$ is an orthonormal basis for $\mathcal{H}$. If $\mathcal{H}$ is infinite dimensional, then this way of writing should be understood in the weak sense. Hence, it is just a shorthand notation for the following formula, true for all $\psi, \xi \in \mathcal{H}$,

$$\langle\,\psi\,|\,\xi\,\rangle = \sum_{k=0}^{d} \langle\,\psi\,|\,\varphi_k\,\rangle\,\langle\,\varphi_k\,|\,\xi\,\rangle\;. \tag{2.36}$$

### 2.3.3   Linear functionals and dual spaces

In Section 2.2 we introduced five classes of operators. They have the obvious inclusions:

$$\mathcal{P}(\mathcal{H}) \subset \mathcal{L}_s(\mathcal{H}) \subset \mathcal{L}(\mathcal{H})\,, \qquad \mathcal{T}(\mathcal{H}) \subset \mathcal{L}(\mathcal{H})\,, \qquad \mathcal{U}(\mathcal{H}) \subset \mathcal{L}(\mathcal{H})\;.$$

There are also more subtle connections and in this subsection we describe an important relationship between $\mathcal{T}(\mathcal{H})$ and $\mathcal{L}(\mathcal{H})$.

A linear mapping $f$ from a complex vector space $V$ into complex numbers is called a *linear functional*. If the vector space $V$ is normed space, then we denote by $V^*$ the set of all continuous linear functionals. It is called the *dual space* of $V$. The dual space $V^*$ is a vector space itself when the linear structure is defined pointwise, i.e., $(f + cg)(v) = f(v) + cg(v)$ for all $v \in V$ and $c \in \mathbb{C}$. We can also define a norm on $V^*$ by setting

$$\|f\| = \sup_{\|v\|=1} |f(v)|\;.$$

In this way, $V^*$ becomes a normed space.

As we have seen in Section 2.2.5, the set $\mathcal{T}(\mathcal{H})$ of trace class operators is a vector space and the trace norm makes it a normed space. For each $S \in \mathcal{L}(\mathcal{H})$, we define a linear functional $f_S$ on $\mathcal{T}(\mathcal{H})$ by formula

$$f_S(T) = \operatorname{tr}[ST] \ . \tag{2.37}$$

It follows from Proposition 17 that $f_S$ is continuous, and thus, $f_S \in \mathcal{T}(\mathcal{H})^*$. Notice also that

$$f_{S_1} + f_{S_2} = f_{S_1 + S_2} \ .$$

Each bounded operator $S$ determines a vector $f_S$ in the dual space $\mathcal{T}(\mathcal{H})^*$ of $\mathcal{T}(\mathcal{H})$. Also the converse is true, namely, each continuous linear functional on $\mathcal{T}(\mathcal{H})$ is of the form $f_S$ for some $S \in \mathcal{L}(\mathcal{H})$. For a proof of the following result, see e.g. [28].

**Theorem 3.** The mapping $S \mapsto f_S$ is a linear isometric bijection from $\mathcal{L}(\mathcal{H})$ to $\mathcal{T}(\mathcal{H})^*$.

In other words, Theorem 3 states that the dual space $\mathcal{T}(\mathcal{H})^*$ of $\mathcal{T}(\mathcal{H})$ can be identified with $\mathcal{L}(\mathcal{H})$, and the identification is given by formula (2.37).

A linear functional $f : \mathcal{T}(\mathcal{H}) \to \mathbb{C}$ is called *positive* if $f(T) \geq 0$ whenever $T \geq O$.

**Proposition 19.** Let $S \in \mathcal{L}(\mathcal{H})$ and $f_S$ as in (2.37). Then $S$ is positive if and only if $f_S$ is positive.

**Exercise 12.** Prove half of Proposition 19: if $f_S$ is positive, then $S$ in positive. (Hint: calculate $f_S(P_\psi)$ for a one-dimensional projection $P_\psi$ and recall that projections are positive.)

### 2.3.4  Tensor product

Tensor product is a way to create a new Hilbert space out of two (or more). There are several different constructions leading to the same thing, and in everyday calculations it is not necessary to remember all the details but mainly some simple computational rules. We therefore first describe tensor product spaces in an informal way without going yet to a precise construction.

Let $\mathcal{H}$ and $\mathcal{K}$ be two finite dimensional inner product spaces. We can form a new inner product space $\mathcal{H} \otimes \mathcal{K}$, called the *tensor product* of $\mathcal{H}$ and $\mathcal{K}$ in the following way. Elements of $\mathcal{H} \otimes \mathcal{K}$ are expressions of the form

$$\sum_{i=1}^{n} \psi_i \otimes \zeta_i \ , \tag{2.38}$$

where $\psi_i \in \mathcal{H}, \zeta_i \in \mathcal{K}$, and $n \in \mathbb{N}$. Symbols $\psi \otimes \zeta$ are assumed to be linear with respect to both arguments, so that

$$c(\psi \otimes \zeta) = (c\psi) \otimes \zeta = \psi \otimes (c\zeta) \ ,$$
$$(\psi_1 + \psi_2) \otimes \zeta = \psi_1 \otimes \zeta + \psi_2 \otimes \zeta \ ,$$
$$\psi \otimes (\zeta_1 + \zeta_2) = \psi \otimes \zeta_1 + \psi \otimes \zeta_2 \ ,$$

for every $\psi_1, \psi_2, \psi \in \mathcal{H}$, $\zeta_1, \zeta_2, \zeta \in \mathcal{K}$, and $c \in \mathbb{C}$. Addition of two elements is defined in a natural way:

$$\sum_{i=1}^{n} \psi_i \otimes \zeta_i + \sum_{i=n+1}^{m} \psi_i \otimes \zeta_i = \sum_{i=1}^{m} \psi_i \otimes \zeta_i \,.$$

Finally, an inner product on $\mathcal{H} \otimes \mathcal{K}$ is defined by setting

$$\langle \psi_1 \otimes \zeta_1 \,|\, \psi_2 \otimes \zeta_2 \rangle = \langle \psi_1 \,|\, \psi_2 \rangle \langle \zeta_1 \,|\, \zeta_2 \rangle \,, \tag{2.39}$$

and then extending by linearity to all elements. In this way, $\mathcal{H} \otimes \mathcal{K}$ becomes an inner product space. If $\{\varphi_i\}$ is an orthonormal basis for $\mathcal{H}$ and $\{\phi_j\}$ for $\mathcal{K}$, then $\{\varphi_i \otimes \phi_j\}$ is an orthonormal basis for $\mathcal{H} \otimes \mathcal{K}$. In particular, the dimension of $\mathcal{H} \otimes \mathcal{K}$ is the product of the dimensions of $\mathcal{H}$ and $\mathcal{K}$.

Similarly as for vectors, two operators $S \in \mathcal{L}(\mathcal{H})$ and $T \in \mathcal{L}(\mathcal{K})$ determine an operator $S \otimes T$ acting in the tensor product space $\mathcal{H} \otimes \mathcal{K}$. If $\psi \in \mathcal{H}$ and $\zeta \in \mathcal{K}$, then

$$S \otimes T \,\psi \otimes \zeta = S\psi \otimes T\zeta \,.$$

This action is then extended to all vectors in $\mathcal{H} \otimes \mathcal{K}$ by linearity.

**Exercise 13.** Let $\mathcal{H}$ be a finite dimensional Hilbert space with an orthonormal basis $\{\varphi_j\}_{j=1}^{d}$. Let us fix the orthonormal basis $\{\varphi_j \otimes \varphi_k\}_{j,k=1}^{d}$ for the tensor product space $\mathcal{H} \otimes \mathcal{H}$. The ordering of the basis vectors is taken to be $\varphi_1 \otimes \varphi_1, \varphi_1 \otimes \varphi_2, \ldots, \varphi_2 \otimes \varphi_1, \ldots$. Prove the following: if $S$ and $T$ are two operators in $\mathcal{H}$ and their matrices in the basis $\{\varphi_j\}_{j=1}^{d}$ are $[s]$ and $[t]$, respectively, then the $d^2 \times d^2$-matrix corresponding to $S \otimes T$ is

$$\begin{bmatrix} s_{11}[t] & s_{12}[t] & \ldots & s_{1d}[t] \\ s_{21}[t] & s_{22}[t] & \ldots & s_{2d}[t] \\ \vdots & \vdots & \ddots & \vdots \\ s_{d1}[t] & s_{d2}[t] & \ldots & s_{dd}[t] \end{bmatrix} \,.$$

If $\mathcal{H}$ and $\mathcal{K}$ are two Hilbert spaces (not necessarily finite dimensional), then the above construction still leads to an inner product space. However, if $\mathcal{H}$ and $\mathcal{K}$ are not both finite dimensional, we need to take the completion of this inner product space to get a Hilbert space. This Hilbert space is then the tensor product space and denoted by $\mathcal{H} \otimes \mathcal{K}$. In this case not all vectors in $\mathcal{H} \otimes \mathcal{K}$ are of finite sums of the form (2.38). However, every vector in $\mathcal{H} \otimes \mathcal{K}$ can be approximated arbitrarily well with vectors of form (2.38).

Let us then take a brief look to an explicit construction of a tensor product space. This is needed when one needs to prove something concerning the structure of the tensor product space. Let $\mathcal{H}$ and $\mathcal{K}$ be two Hilbert spaces. For each $\psi \in \mathcal{H}$ and $\zeta \in \mathcal{K}$, we define a mapping $\psi \otimes \zeta$ from $\mathcal{H} \times \mathcal{K}$ to $\mathbb{C}$ by formula

$$\psi \otimes \zeta(\varphi, \xi) = \langle \varphi \,|\, \psi \rangle \langle \xi \,|\, \zeta \rangle \,.$$

The set $\mathcal{V}$ of all finite linear combinations of such mappings is a linear space, and it becomes an inner product space when we define

$$\langle \psi_1 \otimes \zeta_1 \,|\, \psi_2 \otimes \zeta_2 \rangle = \langle \psi_1 \,|\, \psi_2 \rangle \langle \zeta_1 \,|\, \zeta_2 \rangle \tag{2.40}$$

and extending this by linearity to all elements in $\mathcal{V}$. The tensor product of $\mathcal{H} \otimes \mathcal{K}$ is the completion of $\mathcal{V}$ under the inner product (2.40). For more details, see e.g. [73]. There are also other ways to construct $\mathcal{H} \otimes \mathcal{K}$. One possibility is to use bounded antilinear mappings; see e.g. [35].

**Example 17.** In many cases tensor product spaces can be given a concrete equivalent form. We have earlier in Section 2.1.1 encountered two Hilbert spaces, $\mathbb{C}^d$ and $\ell^2(\mathbb{N})$. Their tensor product space $\ell^2(\mathbb{N}) \otimes \mathbb{C}^d$ is isomorphic with Hilbert space $\ell^2(\mathbb{N}; \mathbb{C}^d)$. This latter space is quite like $\ell^2(\mathbb{N})$ but its elements are functions $f : \mathbb{N} \to \mathbb{C}^d$ satisfying $\sum_i \|f(i)\|^2 < \infty$.

### 3    States and effects

One of the main purpose of a physical theory is to describe events that are observed in experiments. In this chapter we introduce the key physical concepts used in a description of physical experiments and present their mathematical formalization within quantum theory.

#### 3.1    Duality of states and effects

As a general reference for this section, we recommend the book of Kraus [55].

##### 3.1.1    Basic framework

The most basic situation in physics is the following: we have an object system under investigation, and we try to obtain information about it by making an experiment. As a result, measurement outcomes are registered. Statistical theory, such as quantum mechanics, does not predict the individual measurement outcomes but merely the probabilities of the measurement outcomes. Hence, we take the output of an experiment to be a probability distribution on a set $\Omega$ of the measurement outcomes.

It is practical to divide an experiment into a preparation procedure and a measurement. In a given experiment this division may be quite arbitrary, but this is not a problem. We simply assume that there is a collection of possible preparations and a collection of possible measurements, and any pair of a preparation and a measurement can be combined to an experiment. Hence, a preparation specifies a probability distribution for every possible measurement of the system. Two preparation procedures can be superficially quite different and yet lead to the same probability distribution in any chosen measurement. From this point of view, a *state* of the system is an equivalence class of preparation procedures which give the same probability distributions in all measurements. Similarly, an *observable* is an equivalence class of measurements which give the same probability distributions in all preparations. For a pair of a state $\varrho$ and an observable A, we denote by $p_\varrho^A$ the corresponding probability distribution of measurement outcomes.

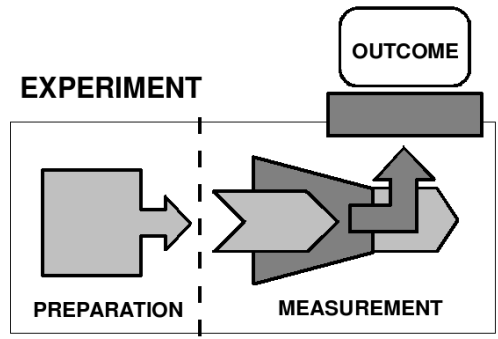

Figure 3.1. Preparation and measurement.

**Example 18.** (*Mixing of the preparation procedures.*) Suppose that we are performing an experiment in which the measurement is fixed, but we are randomly alternating between two preparation procedures with probabilities $\lambda$ and $1 - \lambda$. What is then the predicted probability for the mixture of the preparation procedures? If the two preparation procedures correspond to states $\varrho_1$ and $\varrho_2$, we denote by $\varrho = \lambda \varrho_1 + (1 - \lambda) \varrho_2$ the state corresponding to this mixture.

For example, we can imagine an experiment measuring a circular polarization of the light coming out of a laser source randomly either passing an optically active media, or not. What is the probability that a photon will pass the circular polarization filter? We know the answer for both preparation procedures. Without any doubts the statistics of subensembles of outcomes splitted according to the particular preparation procedures must be in accordance with probabilities $p^{\mathsf{A}}_{\varrho_1}$ and $p^{\mathsf{A}}_{\varrho_2}$, respectively. In particular, let us consider an experimental data of the following form

$$2+, 1+, 2-, 1+, 2-, 1+, 1+, 2+, 2-, 2+, 2+, 1+, 1+, 2-, 1+, 1+, \dots, \tag{3.1}$$

where the number denotes the choice of the preparation procedure and $\pm$ the outcomes of the measurement. Collecting all the outcomes associated with the first preparation procedure we find $p^{\mathsf{A}}_{\varrho_1}(+) = 1$ and similarly, for the second preparation procedure we have $p^{\mathsf{A}}_{\varrho_2}(+) = 1/2$. For the whole sequence we get $p^{\mathsf{A}}_{\varrho}(+) = 3/4$. Since both preparation procedures were chosen with equal probabilities (i.e. $\lambda = 1/2$), we see that the identity $p^{\mathsf{A}}_{\varrho}(+) = \lambda p^{\mathsf{A}}_{\varrho_1}(+) + (1 - \lambda) p^{\mathsf{A}}_{\varrho_2}(+)$ holds. We expect such identity to be valid for a general probabilistic mixture of preparation procedures, thus we require that

$$p^{\mathsf{A}}_{\varrho} = \lambda p^{\mathsf{A}}_{\varrho_1} + (1 - \lambda) p^{\mathsf{A}}_{\varrho_2} \tag{3.2}$$

holds for all observables A.

As illustrated in the previous example, the possibility of randomly mixing two preparation procedures embeds the set of states with a *convex structure*. For states $\varrho_1, \varrho_2$ and a number $\lambda \in [0, 1]$, there exists a unique state $\varrho$, denoted by $\lambda \varrho_1 + (1 - \lambda) \varrho_2$, which is the mixture of $\varrho_1$ and $\varrho_2$ with parts $\lambda$ and $1 - \lambda$. We come to a *basic assumption* demanding that the statistical correspondence between the states and the observables is consistent with the convex structure of the states, that is, if $\varrho_1$ and $\varrho_2$ are states and $0 \leq \lambda \leq 1$, then

$$p^{\mathsf{A}}_{\lambda \varrho_1 + (1 - \lambda) \varrho_2} = \lambda p^{\mathsf{A}}_{\varrho_1} + (1 - \lambda) p^{\mathsf{A}}_{\varrho_2} \tag{3.3}$$

for any observable A.

Let A be an observable and $X$ a subset of possible measurement outcomes of A. We then have a mapping

$$\varrho \mapsto p^{\mathsf{A}}_{\varrho}(X) \tag{3.4}$$

from the set of states to the interval $[0, 1]$. By the basic assumption, this mapping is affine. It describes the statistics of particular measurement outcome, and one can associate these mappings with *statistical events*, also called *effects*.

**Definition 15.** An *effect* is an affine mapping from the set of states to the interval $[0, 1]$ which has the form (3.4) for some observable A and subset $X$ of possible measurement outcomes of A.

Alternatively, an effect can be thought as an equivalence class of pairs $(\mathsf{A}, X)$, where $\mathsf{A}$ is an observable and $X$ is a subset of possible measurement outcomes of $\mathsf{A}$. Two such pairs $(\mathsf{A}, X)$ and $(\mathsf{B}, Y)$ are equivalent if $p_\varrho^{\mathsf{A}}(X) = p_\varrho^{\mathsf{B}}(Y)$ for all states $\varrho$.

**Example 19.** (Identity and zero effects) The *identity effect* $I$ is defined as a mapping assigning probability 1 for every state $\varrho$, i.e., $I(\varrho) = 1$. Physically, it corresponds to a measurement with an outcome that will always click. In a similar way, the *zero effect* $O$ is defined as a mapping assigning probability 0 for every state $\varrho$.

### 3.1.2  Quantum state space

There are several equivalent ways how to mathematically represent the state space of a quantum system. In the most common Hilbert space formulation the states are described by positive trace class operators of trace one, also called density matrices. We denote

$$\mathcal{S}(\mathcal{H}) := \{ \varrho \in \mathcal{T}(\mathcal{H}) \mid \varrho \geq O, \operatorname{tr}[\varrho] = 1 \},$$

and from now on we identify the set of states with $\mathcal{S}(\mathcal{H})$. The set $\mathcal{S}(\mathcal{H})$ is a convex subset of the real vector space $\mathcal{T}_s(\mathcal{H})$ and this gives $\mathcal{S}(\mathcal{H})$ the convex structure described in Section 3.1.1.

An example of a state is a one-dimensional projection. Indeed, we have seen in Section 2.2 that one-dimensional projections are positive and have trace 1. Also recall from Section 2.2.4 that any one-dimensional projection is of the form $|\varphi\rangle\langle\varphi|$ for some unit vector $\varphi \in \mathcal{H}$.

By forming convex combinations of one-dimensional projections we get other states. Actually, the basic fact in $\mathcal{S}(\mathcal{H})$ is that all states are (either finite or infinite) convex combinations of one-dimensional projections. The following theorem is a consequence of the spectral theorem for compact operators; see e.g. [73] for a proof.

**Theorem 4.** A state $\varrho \in \mathcal{S}(\mathcal{H})$ has a *canonical convex decomposition* of the form

$$\varrho = \sum_j \lambda_j P_j \,, \tag{3.5}$$

where $\{\lambda_j\}$ is a sequence of non-negative numbers summing to one and $\{P_j\}$ is an orthonormal sequence of one-dimensional projections. If there are infinitely many nonzero terms, then the sum converges with respect to the trace norm.

The set $\mathcal{S}(\mathcal{H})$ is convex and this corresponds to the possibility of making mixtures as discussed in Subsection 3.1.1. The set $\mathcal{S}(\mathcal{H})$ is actually *σ-convex*, which means that if $\{\varrho_j\}$ is a sequence in $\mathcal{S}(\mathcal{H})$ and $\{\lambda_j\}$ is a sequence of positive numbers summing to one, then the sequence $\sum_j \lambda_j \varrho_j$ converges in $\mathcal{T}(\mathcal{H})$ with respect to the trace norm and the limit belongs to $\mathcal{S}(\mathcal{H})$.

**Definition 16.** An extreme element of the convex set $\mathcal{S}(\mathcal{H})$ is called a *pure state*. Any other element of $\mathcal{S}(\mathcal{H})$ is called a *mixed state*.

As a preparation for the following characterization of pure states, we make a short observation. If $\varrho$ is a state, then the operator $\varrho^2$ is a positive trace class operator. By definition, states satisfy the normalization $\mathrm{tr}\,[\varrho] = 1$. This leads to a bound for the trace of $\varrho^2$. Namely, using the inequalities presented in Proposition 17 in Section 2.2.5 we get

$$\mathrm{tr}\,[\varrho^2] \leq \|\varrho\|\,\mathrm{tr}\,[\varrho] = \|\varrho\| \leq \mathrm{tr}\,[\varrho] = 1\,.$$

Hence, for each state $\varrho$ we can calculate a number $0 < \mathrm{tr}\,[\varrho^2] \leq 1$.

**Proposition 20.** Let $\varrho \in \mathcal{S}(\mathcal{H})$. The following conditions are equivalent:

  (i) $\varrho$ is a pure state;

  (ii) $\varrho$ is a one-dimensional projection;

  (iii) $\mathrm{tr}\,[\varrho^2] = 1$.

*Proof.* We prove the proposition by showing that (i)$\Rightarrow$(ii)$\Rightarrow$(iii)$\Rightarrow$(i),

Suppose that $\varrho$ is a pure state. Then in the canonical decomposition (3.5) there can be only one nonzero number $\lambda_j$, i.e. $\varrho = \lambda P$, where $P$ is a one-dimensional projector. The normalization $\mathrm{tr}\,[\varrho] = \lambda \mathrm{tr}\,[P] = 1$ implies that $\lambda = 1$ and $\varrho = |\varphi\rangle\langle\varphi|$. Thus, (i) implies (ii).

Let us then assume that (ii) holds. By definition, a projection satisfies $\varrho^2 = \varrho$. Therefore, $\mathrm{tr}\,[\varrho^2] = \mathrm{tr}\,[\varrho] = 1$. Thus, (iii) follows.

Finally, assume that (iii) holds. Let us write $\varrho$ in the form $\varrho = \lambda \varrho_1 + (1 - \lambda)\varrho_2$ for some $0 < \lambda < 1$. We need to prove that this is necessarily a trivial convex decomposition, i.e., $\varrho_1 = \varrho_2$. By the assumption we get

$$\begin{aligned}
1 &= \mathrm{tr}\,[\varrho^2] = \lambda^2\,\mathrm{tr}\,[\varrho_1^2] + (1-\lambda)^2\,\mathrm{tr}\,[\varrho_2^2] + 2\lambda(1-\lambda)\,\mathrm{tr}\,[\varrho_1\varrho_2] \\
&\leq \lambda^2 + (1-\lambda)^2 + 2\lambda(1-\lambda)\,|\mathrm{tr}\,[\varrho_1\varrho_2]| \leq 1\,.
\end{aligned}$$

In the last line we have used Cauchy-Schwarz inequality. Since there must be equality in the last inequality, we conclude that $\varrho_1 = c\varrho_2$ for some complex number $c$. As $\mathrm{tr}\,[\varrho_1] = \mathrm{tr}\,[\varrho_2] = 1$, we have $c = 1$ and therefore $\varrho_1 = \varrho_2$. In conclusion, $\varrho$ admits only trivial convex decompositions, which means that it is pure. $\square$

A mixed state $\varrho$ has a canonical decomposition (3.5) as a mixture of pure states. Generally, however, $\varrho$ can be written as a mixture of pure states in many other ways. One can therefore ask for a classification of all decompositions of a given mixed state $\varrho$ to pure states. In the case of finite convex mixtures, this question was answered in [51]. A complete solution, also taking into account $\sigma$-convex mixtures, was then given in [23]. We briefly describe a method presented in [23] to generate convex decompositions. As we describe it here, this procedure does not give all possible convex decompositions but it is sufficient to demonstrate the non-uniqueness of convex decompositions.

Let $\rho \in \mathcal{S}(\mathcal{H})$ be a mixed state. Choose an orthonormal basis $\{\varphi_j\}_{j=1}^d$ for $\mathcal{H}$ (here either $d < \infty$ or $d = \infty$). For each $j$, define $\lambda_j := \left\|\varrho^{\frac{1}{2}}\varphi_j\right\|^2$. These numbers then satisfy the identity

$$\sum_{j=1}^d \lambda_j = \sum_{j=1}^d \left\|\varrho^{\frac{1}{2}}\varphi_j\right\|^2 = \sum_{j=1}^d \langle\,\varphi_j\,|\,\varrho\varphi_j\,\rangle = \mathrm{tr}\,[\varrho] = 1\,.$$

For each $j$ such that $\lambda_j \neq 0$, we denote

$$\phi_j := \lambda_j^{-\frac{1}{2}} \varrho^{\frac{1}{2}} \varphi_j \,. \tag{3.6}$$

The vectors $\phi_j$ are unit vectors and

$$\varrho = \sum_j \lambda_j |\phi_j\rangle\langle\phi_j| \,, \tag{3.7}$$

where the sum contains those terms with $\lambda_j \neq 0$. Indeed, for any $\psi \in \mathcal{H}$, we get

$$
\begin{aligned}
\left\langle \psi \,\middle|\, \left(\sum_j \lambda_j |\phi_j\rangle\langle\phi_j|\right)\psi \right\rangle
&= \sum_{j=1}^{d} \left|\left\langle \psi \,\middle|\, \varrho^{\frac{1}{2}}\varphi_j \right\rangle\right|^2 = \sum_{j=1}^{d} \left|\left\langle \varrho^{\frac{1}{2}}\psi \,\middle|\, \varphi_j \right\rangle\right|^2 \\
&= \left\|\varrho^{\frac{1}{2}}\psi\right\|^2 = \left\langle \psi \,\middle|\, \varrho\psi \right\rangle \,,
\end{aligned}
$$

which shows that (3.7) holds. The decomposition (3.7) is not yet exactly what we want as it may happen that $|\phi_i\rangle\langle\phi_i| = |\phi_j\rangle\langle\phi_j|$ for two different indices $i$ and $j$. However, we can simply sum up these kind of terms and as a result we get a mixture consisting of different pure states.

The decomposition defined in (3.7) shows, in particular, that a pure state $|\phi\rangle\langle\phi|$ can be in some decomposition of a state $\varrho$ if $\phi$ is in the range of the operator $\varrho^{\frac{1}{2}}$. It can be shown that these two conditions are actually equivalent [41].

**Exercise 14.** Conclude from the previous discussion that a mixed state has (uncountably) infinitely many different convex decompositions to pure states.

The following proposition gives a characterization of the ambiguity of finite pure state decompositions of a mixed state [53].

**Proposition 21.** If a mixed state $\varrho$ has a convex decomposition $\varrho = \sum_{j=1}^{n} p_j |\phi_j\rangle\langle\phi_j|$, then all its finite convex decompositions have the form $\varrho = \sum_{k=1}^{m} q_k |\varphi_k\rangle\langle\varphi_k|$, where vectors $\varphi_1, \ldots, \varphi_m \in \mathcal{H}$ and probabilities $q_1, \ldots, q_m$ satisfy the system of equations

$$\sqrt{p_j}\,\phi_j = \sum_k u_{jk}\sqrt{q_k}\,\varphi_k \tag{3.8}$$

with complex numbers $u_{jk}$ defining an $n \times m$ matrix of partial isometry.

*Proof.* Using the identity in Eq.(3.8) the direct calculations gives

$$\sum_j p_j |\phi_j\rangle\langle\phi_j| = \sum_{jk} \sqrt{q_k q_{k'}}\, u_{jk} u_{jk'}^* |\varphi_k\rangle\langle\varphi_{k'}| = \sum_k q_k |\varphi_k\rangle\langle\varphi_k| \,, \tag{3.9}$$

because for the partial isometry $\sum_j u_{jk} u_{jk'}^* = \delta_{kk'}$.

Conversely, suppose $\varrho = \sum_j \lambda_j |\psi_j\rangle\langle\psi_j|$ is the canonical decomposition, i.e. $\lambda_j$ are the eigenvalues of $\varrho$ and $\psi_j$ are the corresponding eigenvectors forming an orthonormal basis of the Hilbert space $\mathcal{H}$ of dimension $d$, thus $\sqrt{p_j}\,\phi_j = \sum_l c_{jl}\sqrt{\lambda_l}\,\psi_l$ and $\sqrt{q_k}\,\varphi_k = \sum_l d_{kl}\sqrt{\lambda_l}\,\psi_l$, where $c_{jl}, d_{jl}$ are suitable complex numbers forming $n \times d$, $m \times d$ matrices, respectively. The identity

$$\varrho = \sum_l \lambda_l |\psi_l\rangle\langle\psi_l| = \sum_j p_j |\phi_j\rangle\langle\phi_j| = \sum_{ll'} \sum_j \sqrt{\lambda_l \lambda_{l'}} c_{jl} c_{jl'}^* |\psi_l\rangle\langle\psi_{l'}| \tag{3.10}$$

requires that $\sum_j c_{jl} c_{jl'}^* = \delta_{ll'}$, i.e. the entries $c_{jl}$ defines the matrix $C$ being a partial isometry. Similarly, the numbers $d_{kl}$ defines a partial isometry $D$ relating vectors $\psi_l$ with vectors $\varphi_k$. Partial isometries being the rectangular matrices can be extended to square unitary matrices, thus they are invertible and $C^*, D^*$ describes the inverse transformations from vectors $\phi_j, \varphi_k$, respectively, to vectors $\psi_l$. Consequently, the $n \times m$ matrix $U = CD^*$ with entries $u_{jk} = \sum_l c_{jl} d_{lk}^*$ gives $\sum_k u_{jk} \sqrt{q_k} \varphi_k = \sqrt{p_j} \phi_j$ as it is stated in the theorem. $\qquad\square$

The set $\mathcal{S}(\mathcal{H})$ is a convex subset of the real normed space $\mathcal{L}_s(\mathcal{H})$. It is instructive to understand the geometry of $\mathcal{S}(\mathcal{H})$ from this topological point of view. We say that a state $\varrho$ belongs to the *boundary* of $\mathcal{S}(\mathcal{H})$ if for each $\epsilon > 0$, there exists an operator $\xi_\epsilon \in \mathcal{L}_s(\mathcal{H})$ such that $\|\varrho - \xi_\epsilon\| < \epsilon$ but $\xi_\epsilon \notin \mathcal{S}(\mathcal{H})$.

**Proposition 22.** If a state $\varrho$ has eigenvalue 0, then it belongs to the boundary of $\mathcal{S}(\mathcal{H})$ .

*Proof.* Assume that $\varrho$ has eigenvalue 0 and let $\varphi \in \mathcal{H}$ be a corresponding eigenvector. Fix $\epsilon > 0$. The selfadjoint operator $\xi_\epsilon := \varrho - \frac{1}{2}\epsilon |\varphi\rangle\langle\varphi|$ is not positive as $\langle\varphi | \xi_\epsilon \varphi\rangle = -\frac{1}{2}\epsilon < 0$. In particular, $\xi_\epsilon \notin \mathcal{S}(\mathcal{H})$. On the other hand,

$$\|\varrho_\epsilon - \varrho\| = \frac{1}{2}\epsilon \, \||\varphi\rangle\langle\varphi|\| = \frac{1}{2}\epsilon < \epsilon \, .$$

Consequently, $\varrho$ belongs to the boundary of $\mathcal{S}(\mathcal{H})$.

$\qquad\square$

Proposition 22 shows, in particular, that all pure states are boundary points of $\mathcal{S}(\mathcal{H})$. However, if $\dim \mathcal{H} \geq 3$, then we can easily define also mixed states which are in the boundary of $\mathcal{S}(\mathcal{H})$. For instance, let $\varphi_1, \varphi_2, \varphi_3$ be three orthogonal unit vectors. Fix a number $0 < \lambda < 1$ and define

$$\varrho = \lambda P_{\varphi_1} + (1 - \lambda) P_{\varphi_2} \, .$$

It is then clear that $\varrho$ is a mixed state. On the other hand,

$$\varrho\varphi_3 = \lambda \langle\varphi_1 | \varphi_3\rangle \varphi_1 + (1 - \lambda) \langle\varphi_2 | \varphi_3\rangle \varphi_2 = 0 \, ,$$

and hence $\varrho$ has eigenvalue 0. Proposition 22 then implies that $\varrho$ is in the boundary of $\mathcal{S}(\mathcal{H})$.

The case $\dim \mathcal{H} = 2$ is an exception and then the boundary points of $\mathcal{S}(\mathcal{H})$ are exactly the pure states; this will become clear in Subsection 3.1.3.

### 3.1.3 Quantum state space for a finite dimensional system

In this subsection we illustrate the concepts of Subsection 3.1.2 in the case of finite dimensional system and we also discuss some further topics which are specific for a finite dimensional situation.

A convenient way to illustrate the state space for a finite dimensional Hilbert space is to adopt the so-called *Bloch representation*. The vector space $\mathcal{L}(\mathcal{H})$ is a Hilbert space endowed with the Hilbert-Schmidt scalar product $\langle A \mid B \rangle_{\text{H-S}} = \text{tr}\,[A^*B]$. For a $d$ dimensional Hilbert space $\mathcal{H}$ the associated complex Hilbert space $\mathcal{L}(\mathcal{H})$ is $d^2$ dimensional. The subspace of selfadjoint operators $\mathcal{L}_s(\mathcal{H})$ is a $d^2$ dimensional real subspace of $\mathcal{L}(\mathcal{H})$. It is convenient to choose the identity operator to be one of the element of an orthogonal basis of selfadjoint operators, $E_0 = I$. In such case the orthogonality implies that the remaining basis operators $E_1, \ldots, E_{d^2-1}$ have trace equal to 0. Moreover, we choose the normalization to be $\langle E_j \mid E_k \rangle_{\text{H-S}} = d\delta_{jk}$. Consequently, a general quantum state can be expressed as

$$\varrho = \frac{1}{d}(I + \vec{r} \cdot \vec{E}), \tag{3.11}$$

where $\vec{r}$ is a $d^2 - 1$ dimensional real vector called *Bloch (state) vector*. Hence, quantum states can be viewed as real vectors and the convex body of the state space is embedded into $\mathbb{R}^{d^2-1}$.

Let us note that general linear combinations of states are not reflected by linear combinations in the Bloch state vectors. Only affine linear combinations of Bloch vectors can be properly interpreted as affine linear combinations of corresponding quantum states. The reason for this fact is that affine linear combinations preserve the trace and Bloch vectors are in one-to-one correspondence with selfadjoint operators of a fixed trace. Consequently, although Bloch vectors are elements of $d^2 - 1$-dimensional real vector space, there are $d^2$ linearly independent states associated with $d^2$ affinely independent Bloch vectors in $\mathbb{R}^{d^2-1}$.

**Example 20.** (*Bloch sphere*) Consider a two-dimensional Hilbert space $\mathcal{H}$ with an orthonormal basis $\{\varphi, \varphi_\perp\}$. The standard operator basis consists of the identity operator $I$ and the Pauli operators

$$
\begin{aligned}
\sigma_x &= |\varphi\rangle\langle\varphi_\perp| + |\varphi_\perp\rangle\langle\varphi|, \\
\sigma_y &= -i|\varphi\rangle\langle\varphi_\perp| + i|\varphi_\perp\rangle\langle\varphi|, \\
\sigma_z &= |\varphi\rangle\langle\varphi| - |\varphi_\perp\rangle\langle\varphi_\perp|.
\end{aligned}
$$

They satisfy the orthogonality relations $\text{tr}\,[\sigma_j\sigma_k] = 2\delta_{jk}$. In this basis a general state can be written as

$$\varrho = \frac{1}{2}(I + \vec{r} \cdot \vec{\sigma}), \tag{3.12}$$

where $\vec{r} \in \mathbb{R}^2$. The eigenvalues of the operator $\varrho$ in (3.12) are $\lambda_\pm = \frac{1}{2}(1 \pm \|\vec{r}\|)$. It follows that the positivity of $\varrho$ is equivalent to the condition $\|\vec{r}\| \leq 1$. Hence, the states of the two-dimensional Hilbert space form a unit sphere in the three-dimensional real vector space of Bloch vectors, the so-called *Bloch sphere*. The pure states correspond to the vectors $\vec{r}$ of unit length. Clearly, the vectors of unit length form the boundary of the unit sphere, hence for two-dimensional Hilbert space the boundary of state space consists of pure states only.

**Exercise 15.** This exercise is related to Example 20. Define pure states $\varrho_x = \frac{1}{2}(I + \sigma_x)$, $\varrho_y = \frac{1}{2}(I + \sigma_y)$, $\varrho_z = \frac{1}{2}(I + \sigma_z)$. Show that they do not form an operator basis. Specify which states can be used to complete the basis. [Hint: Try to express the identity operator $I$ as a linear (not necessarily convex) combinations of $\varrho_x$, $\varrho_y$, $\varrho_z$.]

The item (iii) in Proposition 20 motivates the quantification of so-called *mixedness* of quantum states. The starting point is that we have a numerical function on the set of states that takes certain value if and only if the state is pure. In what follows we introduce two functions commonly used to quantify the mixedness of quantum states.

**Definition 17.** The *purity* $\mathcal{P}(\varrho)$ of a state $\varrho$ is defined as

$$\mathcal{P}(\varrho) := \mathrm{tr}\left[\varrho^2\right] = \sum_j \lambda_j^2 \,,$$

where $\lambda_j$ are the eigenvalues of $\varrho$.

**Definition 18.** The *von Neumann entropy* $S(\varrho)$ of a state $\varrho$ is defined as

$$S(\varrho) := -\mathrm{tr}\left[\varrho \log \varrho\right] = -\sum_j \lambda_j \log \lambda_j \,,$$

where $\lambda_j$ are the nonzero eigenvalues of $\varrho$.

**Proposition 23.** Purity has the following properties:

1. $\mathcal{P}$ is convex, i.e., $\mathcal{P}(p\varrho_1 + (1-p)\varrho_2) \leq p\mathcal{P}(\varrho_1) + (1-p)\mathcal{P}(\varrho_2)$.

2. $\mathcal{P}$ is invariant under unitary conjugation, i.e., $\mathcal{P}(U\varrho U^*) = \mathcal{P}(\varrho)$.

3. $\mathcal{P}(\varrho) = 1$ if and only if $\varrho$ is a pure state.

*Proof.* 1. We prove the claim by showing that the difference $\Delta \equiv p\mathcal{P}(\varrho_1) + (1-p)\mathcal{P}(\varrho_2) - \mathcal{P}(p\varrho_1 + (1-p)\varrho_2)$ is positive. A direct calculation gives

$$\begin{aligned}
\Delta &= p\mathrm{tr}\left[\varrho_1^2\right] + (1-p)\mathrm{tr}\left[\varrho_2^2\right] - p^2\mathrm{tr}\left[\varrho_1^2\right] - (1-p)^2\mathrm{tr}\left[\varrho_2^2\right] - 2p(1-p)\mathrm{tr}\left[\varrho_1\varrho_2\right] \\
&= p(1-p)\mathrm{tr}\left[\varrho_1^2 + \varrho_2^2 - \varrho_1\varrho_2 - \varrho_2\varrho_1\right] = p(1-p)\mathrm{tr}\left[(\varrho_1 - \varrho_2)^2\right] \,. \quad (3.13)
\end{aligned}$$

Since $0 \leq p \leq 1$ and $(\varrho_1 - \varrho_2)^2 \geq O$ it follows that $\Delta$ is positive, thus the purity is convex.

2. If $\varrho$ is a state and $U$ is a unitary operator, we get

$$\mathcal{P}(U\varrho U^*) = \mathrm{tr}\left[U\varrho U^* U\varrho U*\right] = \mathrm{tr}\left[\varrho^2\right] = \mathcal{P}(\varrho) \,.$$

3. This is proved in Prop. 20.                                                                                    □

In the following we list some properties of von Neumann entropy that are needed for our purposes. We omit the proofs which can be found in [57], [70], [65]. See also Exercise 16 below.

**Proposition 24.** von Neumann entropy has the following properties:

1. $S$ is concave, i.e., $S(p\varrho_1 + (1-p)\varrho_2) \geq pS(\varrho_1) + (1-p)S(\varrho_2)$.

2. $S$ is invariant under unitary conjugation, i.e., $S(U\varrho U^*) = S(\varrho)$.

3. $S(\varrho) = 0$ if and only if $\varrho$ is a pure state.

**Exercise 16.** Prove the third property in Prop. 24.

Both purity and von Neumann entropy induce partial orders in the set of states. As shown in [82], these partial orders are different. Namely, there are states $\varrho_1$ and $\varrho_2$ such that $\mathcal{P}(\varrho_1) > \mathcal{P}(\varrho_2)$, but not $S(\varrho_1) < S(\varrho_2)$. However, the following example shows that the maximal element is the same for both of them.

**Example 21.** (*Maximally mixed state*.) According to previous discussion the pure states are the only least mixed states with respect to both purity and von Neumann entropy. We can also ask which states are maximally mixed. It turns out that there exists a unique quantum state called *complete*, or *total mixture* which is the unique maximally mixed state with respect to the both quantities.

States $\varrho$ and $\varrho_U = U\varrho U^*$ have the same value of purity and von Neumann entropy for all unitary operators $U$. Making a convex combination of states $\varrho_k = U_k\varrho U_k^*$ we obtain a state $\varrho' = \sum_k p_k\varrho_k$ with purity $\mathcal{P}(\varrho') \leq \sum_k p_k\mathcal{P}(\varrho_k) = \mathcal{P}(\varrho)$ and entropy $S(\varrho') \geq \sum_k p_kS(\varrho_k) = S(\varrho)$. It follows that with respect to both measures the state $\varrho'$ is more (or equally) mixed as $\varrho$. The maximally mixed state cannot be affected by such mixing procedure, i.e., if $\varrho$ is maximally mixed, then also $\varrho'$ is a maximally mixed state.

Let $\varphi_1, \ldots, \varphi_d$ be the (mutually orthogonal) eigenvectors of the state $\varrho$, so that $\varrho = \sum_j \lambda_j |\varphi_j\rangle\langle\varphi_j|$. Define a shift operator

$$U_{\text{shift}} = \sum_{j=1}^{d} |\varphi_j\rangle\langle\varphi_{j\oplus1}| \tag{3.14}$$

with modulo $d$ summation, i.e., $d \oplus 1 = 1$. Applying the shift operator to $\varrho$ we get a state $\varrho_1 = U_{\text{shift}}\varrho U_{\text{shift}}^* = \sum_j \lambda_j|\varphi_j\rangle\langle\varphi_j|$. Recursively, $\varrho_r = U_{\text{shift}}\varrho_{r-1}U_{\text{shift}}^* = \sum_j \lambda_{j\oplus r}|\varphi_j\rangle\langle\varphi_j|$. Equal probabilistic mixture of states $\varrho_1, \ldots, \varrho_d$ results in the state

$$\varrho' = \frac{1}{d}\sum_r p_r\varrho_r = \frac{1}{d}\sum_{j=1}^{d}\left(\sum_{r=1}^{d}\lambda_{j\oplus r}\right)|\varphi_j\rangle\langle\varphi_j| = \frac{1}{d}I, \tag{3.15}$$

because $\sum_{r=1}^{d}\lambda_{j\oplus r} = \lambda_1 + \cdots + \lambda_d = \text{tr}\,[\varrho] = 1$. Moreover, the state $\frac{1}{d}I$ is not affected by unitary transformations, i.e. $U\frac{1}{d}IU^* = \frac{1}{d}I$. As a result we get that the mixedness of $\frac{1}{d}I$ is larger than the mixedness of arbitrarily state and stability of $\frac{1}{d}I$ under unitary transformations implies that it is indeed the unique maximum of mixedness with respect to both measures.

The maximally mixed state $\frac{1}{d}I$ can be also obtained as the average state over all states. Namely, the group of unitary operators posses an invariant integration, called Haar integral (see e.g. [35] for explanation). Fix a state $\varrho$ and define $\overline{\varrho} = \int U\varrho U^* dU$. It follows from the invariance of Haar integration that the operator $\overline{\varrho}$ commutes with all unitary operators, $[\overline{\varrho}, U] = 0$ for all $U$. By Schur lemma $\overline{\varrho} = cI$. The trace condition fixes the constant $c = 1/d$, where $d = \dim\mathcal{H}$. Thus, $\overline{\varrho} = \frac{1}{d}I$.

We have seen that each quantum state can be represented as a Bloch vector, however, not every Bloch vector is associated with some density operator. The shape of the convex body of the state space in this representation determined by the positivity constraint is not known for general dimension. The picture of Bloch sphere for two-dimensional case is in many respects exceptional and its properties are not valid for larger dimensions. For example, although the pure states are belonging to boundary for any dimension, unless $d = 2$ the boundary always contains mixed states. In fact, each density operator of rank smaller than the dimension of the Hilbert space belongs to the boundary of the state space. By Proposition 20 a state $\varrho$ is pure if

$$1 = \text{tr}\left[\varrho^2\right] = \frac{1}{d^2}\text{tr}\left[(I + \vec{r} \cdot \vec{E})(I + \vec{r} \cdot \vec{E})\right] = \frac{1}{d}(1 + \|\vec{r}\|^2)\,, \tag{3.16}$$

hence $\|\vec{r}\| = \sqrt{d-1}$. This means that the state space is embedded in a sphere of $d^2 - 1$ dimensional real vector space. However, not every operator with $\|\vec{r}\| \leq \sqrt{d-1}$ is positive and the particular form of the state space is more complicated.

Let $\psi, \varphi$ be two unit vectors, $\varrho_\psi, \varrho_\varphi$ the corresponding states and $\vec{r}_\psi, \vec{r}_\varphi$ the associated Bloch vectors. Then

$$|\langle \psi \,|\, \varphi \rangle|^2 = \text{tr}\left[\varrho_\psi \varrho_\varphi\right] = \frac{1}{d}(1 + \vec{r}_\psi \cdot \vec{r}_\varphi)\,. \tag{3.17}$$

Consequently, the orthogonality of vectors $\psi, \varphi$ is reflected by the scalar product $\vec{r}_\psi \cdot \vec{r}_\varphi = -1$ of the associated Bloch vectors. It follows that the angle between the Bloch vectors corresponding to orthogonal pure states is

$$\theta = \arccos\left(\frac{1}{1-d}\right)\,. \tag{3.18}$$

Interestingly, if $\vec{r}_\psi$ (with $\|\vec{r}_\psi\| = \sqrt{d-1}$) defines a pure quantum state, then its antipodal vector $\vec{t} = -\vec{r}_\psi$ does not correspond to any quantum states. In fact, since $\left\|\vec{t}\right\| = \sqrt{d-1}$ the candidate state should be pure. However,

$$\frac{1}{d}(1 + \vec{r}_\psi \cdot \vec{t}) = \frac{1}{d}(1 - d + 1) = \frac{2-d}{d} < 0$$

for $d > 2$, which is in contradiction with the fact that $|\langle \psi \,|\, \varphi \rangle|^2 \geq 0$ for all vectors $\varphi \in \mathcal{H}$. Therefore, the antipodal vector $\vec{t}$ does not correspond to any physical state.

**Example 22.** Fix a unit vector $\varphi$. Consider a traceless selfadjoint operator

$$E = \frac{1}{\sqrt{d-1}}(d \,|\varphi\rangle\langle\varphi| - I)\,, \tag{3.19}$$

which is an element of some orthogonal operator basis consisting of traceless selfadjoint operators $E_1, \ldots, E_{d^2-1}$ such that $\text{tr}\left[E_j^2\right] = d$. Then the operator

$$\varrho = \frac{1}{d}(I + \sqrt{d-1}\,E) = |\varphi\rangle\langle\varphi| \tag{3.20}$$

describes a pure state with the Bloch vector $\vec{r} = (\sqrt{d-1}, 0, \ldots, 0)$. According to previous paragraph an operator

$$A = \frac{1}{d}(I - \sqrt{d-1}\,E) = \frac{2}{d}I - |\varphi\rangle\langle\varphi| \tag{3.21}$$

associated with the antipodal Bloch vector $\vec{t} = -\vec{r}$ is not positive. In fact, it is diagonal and the eigenvalue $(2 - d)/d$ is clearly negative for $d > 2$.

### 3.1.4   From states to effects

In Subsection 3.1.2 we have identified the states of a quantum system with the positive trace class operators of trace one on a Hilbert space $\mathcal{H}$, and this set is denoted by $\mathcal{S}(\mathcal{H})$. An effect, as defined in Subsection 3.1.1, is therefore an affine mapping from $\mathcal{S}(\mathcal{H})$ to $[0, 1]$. However, also effects can be identified with certain specific operators on $\mathcal{H}$.

First of all, an effect $E$ has a unique extension to a positive linear functional $\widetilde{E}$ from $\mathcal{T}(\mathcal{H})$ to $\mathbb{C}$. Indeed, we set $\widetilde{E}(O) := 0$ and for each positive trace class operator $T \neq O$ we define

$$\widetilde{E}(T) := \operatorname{tr}[T] \, E(\operatorname{tr}[T]^{-1} \, T) \, .$$

In this way, $\widetilde{E}$ is defined for all positive trace class operators. Then, for every selfadjoint operator $T \in \mathcal{T}_s(\mathcal{H})$ we define

$$\widetilde{E}(T) := \widetilde{E}(T^+) - \widetilde{E}(T^-) \, ,$$

where $T^+, T^-$ are the positive and negative parts of $T$, respectively. Finally, an operator $T \in \mathcal{T}(\mathcal{H})$ can be written as a sum of two selfadjoint trace class operators $T_R$ and $T_I$ (recall Example 8 in Section 2.2.2). Hence, we can further extend $\widetilde{E}$ to all trace class operators.

Since $\widetilde{E}$ is a positive linear functional on the Banach space $\mathcal{T}(\mathcal{H})$, it is bounded. From the duality relation between the trace class operators and bounded operators (see Section 2.3.3) it follows that there exists a unique operator $\hat{E} \in \mathcal{L}(\mathcal{H})$ such that

$$\widetilde{E}(T) = \operatorname{tr}\left[\hat{E}T\right] \quad \forall T \in \mathcal{T}(\mathcal{H}) \, . \tag{3.22}$$

The effect $E$ takes values between 0 and 1. Hence, for a pure state $P_\psi$ we get

$$0 \leq \operatorname{tr}\left[\hat{E}P_\psi\right] = \left\langle \, \psi \mid \hat{E}\psi \, \right\rangle \leq 1 \, .$$

Equivalently, we can write this condition as

$$0 \leq \left\langle \, \psi \mid \hat{E}\psi \, \right\rangle \leq 1 \, ,$$

required to hold for all unit vectors $\psi \in \mathcal{H}$. Hence, we can identify the effects as selfadjoint operators satisfying

$$O \leq \hat{E} \leq I \, . \tag{3.23}$$

Here these inequalities are operator inequalities in the sense explained in Section 2.2.2. We conclude that effects can be (and will be) identified with the bounded operators satisfying the operator inequalities (3.23). We denote by $\mathcal{E}(\mathcal{H})$ the set of all effects.

**Exercise 17.** Show that every projection is an effect. (Hint: One way is to use Proposition 11 and apply that to the inequality in Exercise 7).

### 3.1.5  From effects to states

In Section 3.1.4 we have seen that if the set of states is chosen to be $\mathcal{S}(\mathcal{H})$, then the mathematical form of effects follows from the basic framework. It is also possible to fix the mathematical form for effects first and take this as a starting point. Here we briefly explain this line of thought.

Let us forget Subsection 3.1.2 for a moment and start again from the general framework of Subsection 3.1.1. In particular, suppose we have not fixed the specific mathematical form of states and effects. Starting from the definition of an effect as an affine mapping on the set of states, we can define a partial binary operation $\boxplus$ on the set of effects. Namely, if $E_1, E_2, E_3$ are effects and for all states $\varrho$ we have

$$E_1(\varrho) + E_2(\varrho) = E_3(\varrho) \,,$$

then we denote $E_1 \boxplus E_2 = E_3$. The operation $E_1 \boxplus E_2$ is not defined for all pairs of effects and for this reason $\boxplus$ is called partial operation. (For instance, $I \boxplus I$ is not defined.)

The partial binary operation $\boxplus$ defines also a partial ordering. Namely, for two effects $E_1$ and $E_3$, we denote $E_1 \leq E_3$ if there exists an effect $E_2$ such that $E_1 \boxplus E_2 = E_3$.

Since effects are affine mappings from the set of states to the interval $[0, 1]$, each state $\varrho$ defines a mapping $f_\varrho$ from the set of effects to the interval $[0, 1]$ by formula

$$f_\varrho(E) := E(\varrho) \,.$$

It is clear from this definition that $f_\varrho$ satisfies the normalization

$$f_\varrho(I) = I(\varrho) = 1 \,, \tag{3.24}$$

where $I$ is the identity effect introduced in Example 19. Moreover, assume that $E_1, E_2$ are effects and $E_1 \boxplus E_2$ exists. Then we get

$$f_\varrho(E_1 \boxplus E_2) = (E_1 \boxplus E_2)(\varrho) = E_1(\varrho) + E_2(\varrho) = f_\varrho(E_1) + f_\varrho(E_2) \,. \tag{3.25}$$

In conclusion, each state determined mapping $f_\varrho$ is a normalized and additive (with respect to $\boxplus$) mapping from the set of effects to the interval $[0, 1]$.

The properties (3.24) and (3.25) are immediate consequences of our definitions. To proceed, we also need to require something more. Namely, let $E_1, E_2, \ldots$ be a sequence of effects such that the sum $E_1 \boxplus \cdots \boxplus E_n$ exists for each $n$. By $E_1 \boxplus E_2 \boxplus \cdots$ we denote the least upper bound of the increasing sequence $E_1, E_1 \boxplus E_2, E_1 \boxplus E_2 \boxplus E_3, \ldots$, if it exists.

**Definition 19.** A mapping $f$ from the set of effects to the interval $[0, 1]$ is a *generalized probability measure* if it satisfies the normalization condition $f(I) = 1$ and

$$f(E_1 \boxplus E_2 \boxplus \cdots) = f(E_1) + f(E_2) + \cdots \tag{3.26}$$

whenever $E_1 \boxplus E_2 \boxplus \cdots$ exists.

We now require that the states correspond to the generalized probability measures on effects. This can be seen as an additional assumption in the basic framework. Note, however, that properties (3.24) and (3.25) already follow from the basic framework, and (3.26) is just a slight generalization of (3.25).

Let us then see the consequences if the set $\mathcal{E}(\mathcal{H}) = \{E \in \mathcal{L}(\mathcal{H}) \mid O \leq E \leq I\}$ is taken to represent the set of effects. The binary operation $\boxplus$ is identified with the usual addition of operators. For the proof of the following result, we refer to [12].

**Proposition 25.** Let $f$ be a generalized probability measure on $\mathcal{E}(\mathcal{H})$. There exists a unique operator $\varrho_f \in \mathcal{S}(\mathcal{H})$ such that

$$f(E) = \mathrm{tr}\,[\varrho_f E] \quad \forall E \in \mathcal{E}(\mathcal{H}). \tag{3.27}$$

We conclude that the choices $\mathcal{S}(\mathcal{H})$ and $\mathcal{E}(\mathcal{H})$ for the mathematical description of states and effects, respectively, are compatible in the sense that already fixing one implies the other. We emphasize that the discussion here and earlier subsections is not in any way a derivation of the Hilbert space structure of quantum mechanics. We have simply shown how the Hilbert space structure fits to the basic framework.

### 3.1.6 Gleason's Theorem

Let us continue the discussion of Subsection 3.1.5 from a slightly different point of view. Recall that effects themselves correspond to measurements with two outcomes. This raises the following question: can we prepare a system in such a way that the measurement outcomes in all two outcome measurements are predictable with probability one? This would mean that we have a generalized probability measure which takes only values 0 and 1. A generalized probability measure satisfying this feature is called *dispersion-free*.

**Exercise 18.** Let $\varrho \in \mathcal{S}(\mathcal{H})$. Show that the related generalized probability measure $f_\varrho$ on $\mathcal{E}(\mathcal{H})$ is not dispersion-free.

Since by Proposition 25 all generalized probability measures on $\mathcal{E}(\mathcal{H})$ are of the form $f_\varrho$ for some state $\varrho \in \mathcal{S}(\mathcal{H})$ we can conclude from Exercise 18 that there are no dispersion-free generalized probability measures on $\mathcal{E}(\mathcal{H})$.

This fact is not unexpected and we can also understand it without Proposition 25. Namely, for each $E \in \mathcal{E}(\mathcal{H})$, also $\frac{1}{2}E \in \mathcal{E}(\mathcal{H})$. If $f$ is a generalized probability measure on $\mathcal{E}(\mathcal{H})$ and $f(E) = 1$, then it follows from (3.26) that $f(\frac{1}{2}E) = \frac{1}{2}$. This argument relates to the fact that $\mathcal{E}(\mathcal{H})$ is a convex set. Hence, perhaps there is some rule of calculating predictions with the probabilities 1 and 0 for those effects which are free from classical randomness, i.e., which are not convex mixtures of other effects.

**Example 23.** (*States and effects in classical mechanics*) This is a good place to think about the mathematical representation of states and effects of classical systems. Briefly, we can say that classical states are probability distributions on a suitable phase space $\Omega$ and classical effects are associated with fuzzy subsets of $\Omega$. In the usual physical situations the phase space $\Omega$ is infinite. For example, the phase space for a moving particle is the six dimensional manifold $\mathbb{R}^3 \times \mathbb{R}^3$ consisting of position and momentum vectors, i.e. $\vec{x} = (\vec{q}, \vec{p}) \in \Omega$.

To keep things simple, we focus only on finite phase spaces which are sufficient for our purposes. In the case of a finite phase space $\Omega$ (say containing $d$ elements), the states can be represented as probability vectors $\vec{p} = (p_1, \ldots, p_d)$ with $0 \leq p_j \leq 1$ and $\sum_j p_j = 1$. The classical effects can be represented by vectors $\vec{e} = (e_1, \ldots, e_d)$ satisfying $0 \leq e_j \leq 1$. The probability of measuring an effect $\vec{e}$ if the system is in the state $\vec{p}$ is $\vec{p} \cdot \vec{e} = \sum_j p_j e_j$.

The extremal elements of the classical state space are the probability vectors $\vec{\delta}_j$ with all entries vanishing except $j$th one. Clearly, each state $\vec{p}$ is a convex combination of some $\vec{\delta}_j$'s. On the other hand, a state $\vec{\delta}_j$ does not have non-trivial convex decomposition to other states. Similarly, the extremal effects are vectors $\vec{e}$ such that each entry $e_j$ is either 0 or 1.

It is now easy to see the that there are no dispersion-free states. However, there are states which are dispersion-free when restricted to extremal effects. Namely, if we have an extremal state $\vec{\delta}_j$ and an extremal effect $\vec{e}$, then the probability $\vec{\delta}_j \cdot \vec{e}$ is either 0 or 1.

To deal with the above queries, we recall the following result of Davies [29].

**Proposition 26.** The extremal elements of the convex set $\mathcal{E}(\mathcal{H})$ are the projections.

*Proof.* Let $P$ be a projection and assume that there are effects $E_1, E_2$ such that

$$P = \lambda E_1 + (1 - \lambda)E_2 \tag{3.28}$$

for some $0 < \lambda < 1$. Suppose $\psi \in \mathcal{H}$ satisfies $P\psi = 0$. Since $E_1$ and $E_2$ are positive operators, we get

$$0 = \langle\, \psi \,|\, P\psi \,\rangle = \lambda \langle\, \psi \,|\, E_1\psi \,\rangle + (1 - \lambda) \langle\, \psi \,|\, E_2\psi \,\rangle \geq \lambda \langle\, \psi \,|\, E_1\psi \,\rangle \geq 0 \,.$$

Hence, $\langle\, \psi \,|\, E_1\psi \,\rangle = 0$. By Prop. 7 in Section 2.2.2 this implies that $E_1\psi = 0$. On the other hand, equation (3.28) gives

$$I - P = \lambda(I - E_1) + (1 - \lambda)(I - E_2) \,, \tag{3.29}$$

and a similar reasoning now shows that $P\psi = \psi$ implies that $E_1\psi = \psi$. By Prop. 11 in Section 2.2.4 every vector in $\mathcal{H}$ can be written as a sum of eigenvectors of $P$. Hence, $E_1$ and $P$ acts identically on all vectors $\psi \in \mathcal{H}$ and we conclude that $E_1 = P$. Thus, $P$ does not have a nontrivial convex decomposition and it is extremal.

Suppose that $A \in \mathcal{E}(\mathcal{H})$ is not a projection, so that $A \neq A^2$. We define $E_1 = A^2 \neq A$ and $E_2 = 2A - A^2 \neq A$. The operator $E_1$ is an effect by Prop. 6 in Section 2.2.2. Similarly, since $I - E_2 = (I - A)^2$, also $I - E_2$ is an effect. This implies that $E_2$ is an effect as well. The equal convex combination of these effects gives

$$\frac{1}{2}(E_1 + E_2) = A \,. \tag{3.30}$$

Thus, $A$ is not an extreme element of $\mathcal{E}(\mathcal{H})$. $\qquad\square$

Motivated by Proposition 26, it seems interesting to study the properties of generalized probability measures on the set of all projections $\mathcal{P}(\mathcal{H})$. In this case the celebrated result of Gleason tells us the following.

**Theorem 5.** (*Gleason's Theorem*) Suppose that $\dim\mathcal{H} \geq 3$. Let $f$ be a generalized probability measure on $\mathcal{P}(\mathcal{H})$. There exists a unique operator $\varrho_f \in \mathcal{S}(\mathcal{H})$ such that

$$f(P) = \operatorname{tr}[\varrho_f P] \quad \forall P \in \mathcal{P}(\mathcal{H}) \,. \tag{3.31}$$

The proof of Gleason's Theorem, first given in [39], is long and non-trivial. We refer to [] for a detailed proof. From Gleason's Theorem we conclude that when $\dim \mathcal{H} \geq 3$, there seems to be no way to escape the probabilistic nature of quantum mechanics. Generalized probability measures on $\mathcal{P}(\mathcal{H})$ are obtained from states with the usual trace formula and they are never dispersion-free. For $\dim \mathcal{H} = 2$ Gleason's Theorem is not valid and the situation is therefore different. This opens up the possibility of constructing generalized probability measures which are dispersion-free, and one example is given below. However, so far these dispersion-free generalized probability measures do not seem to found any physical interpretation.

**Example 24.** (*Dispersion-free generalized probability measures in two-dimensional case.*) The set of projectors on a two-dimensional system consists of $O$, $I$, and operators $P_{\vec{n}} \equiv \frac{1}{2}(I + \vec{n} \cdot \vec{\sigma})$ with $\|\vec{n}\| = 1$. A generalized probability measure can be understand as a function on vectors $\vec{n}$ taking values $f(O) = 0$ and $f(I) = 1$. Let us note that $P_{\vec{n}} + P_{\vec{m}}$ is an effect if and only if $\vec{m} = -\vec{n}$. Therefore, the additivity constraint gives only a single nontrivial relation

$$f(P_{\vec{n}}) + f(P_{-\vec{n}}) = f(I) = 1$$

that should hold for all $\vec{n}$.

Fix a unit vector $\vec{z}$. Define $f(P_{\vec{n}}) = 0$ if $\vec{n} \cdot \vec{z} < 0$ and $f(P_{\vec{n}}) = 1$ otherwise. The property $f(P_{\vec{n}}) + f(P_{-\vec{n}}) = 1$ obviously holds and, therefore, this function is a generalized probability measure. Let us assume that $f$ comes from a state $\varrho_f$ by formula (3.31). As explained in Section 3.1.3, the state $\varrho_f$ has the form $\frac{1}{2}(I + \vec{r} \cdot \vec{\sigma})$ for some vector $\vec{r}$ with $\|\vec{r}\| \leq 1$. By a direct calculation we get

$$f(P_{\vec{n}}) = \frac{1}{4} \mathrm{tr}\left[(I + \vec{r} \cdot \vec{\sigma})(I + \vec{n} \cdot \vec{\sigma})\right] = \frac{1}{2}(1 + \vec{r} \cdot \vec{n}).$$

It is now clear that it is not possible to produce the correct values for $f$. Thus, there is no state $\varrho_f$ which would correspond to $f$. In conclusion, $f$ is an example of dispersion-free generalized measure that cannot be represented by a state $\varrho_f$.

**Exercise 19.** Show that the generalized probability measure $f$ in Example 24 cannot be extend to all effects.

### 3.2   Superposition structure of pure states

#### 3.2.1   Superposition of two pure states

Pure states is an exceptional subset of states not only because they are extremal points of the state space. They have their own special structure called *superposition*, which has no analog in classical mechanics. Superposition is often marked as the very quantum feature. To explain the mathematical formalism of superposition, it is easier first to adopt another description for pure states.

We saw in Subsection 3.1.2 that pure states are described by one-dimensional projections. As noted in Subsection 2.2.4, each unit vector $\eta$ defines a one-dimensional $P_\eta$. However, this correspondence is not injective. Namely, two unit vectors $\eta$ and $\eta'$ define the same projection if there is a complex number $z$ of unit length such that $\eta' = z\eta$. Also the converse holds: if two unit vectors $\eta$ and $\eta'$ differ in some other way than a scalar multiple, then they define different projections. We conclude that pure states can be alternatively described as equivalence classes $[\eta]$ of unit vectors $\eta \in \mathcal{H}$, where the equivalence relation is defined in the following way:

$$\eta \sim \eta' \quad \Leftrightarrow \quad \eta' = z\eta \text{ for some } z \in \mathbb{C}, |z| = 1 \,. \tag{3.32}$$

Forming a superposition of two pure states is, essentially, the same thing as forming a linear combination of two vectors. We only need to take into account the equivalence relation defined above. Hence, let $\psi, \varphi \in \mathcal{H}$ be two linearly independent unit vectors. Choose two non-zero complex numbers $a, b$, and denote

$$\omega = \frac{a\psi + b\varphi}{\|a\psi + b\varphi\|} \,. \tag{3.33}$$

The pure state $[\omega]$ is called a *superposition* of pure states $[\psi]$ and $[\varphi]$. In terms of the corresponding one-dimensional projections $P_\psi, P_\varphi$ and $P_\omega$ the superposition condition (3.33) reads

$$P_\omega = \frac{1}{\|a\psi + b\varphi\|^2} \left( |a|^2 P_\psi + |b|^2 P_\varphi + a^*b|\varphi\rangle\langle\psi| + ab^*|\psi\rangle\langle\varphi| \right) \,. \tag{3.34}$$

Here we have used the Dirac notation, explained in Section 2.3.2.

**Proposition 27.** A pure state $P_1$ (described as a one-dimensional projection) can be written as superposition of a pure state $P_2$ and some other pure state (say $P_3$) if and only if

$$\text{tr} \left[ P_1 P_2 \right] \neq 0 \,. \tag{3.35}$$

Moreover, if (3.35) holds, then $P_3$ can be chosen such that $\text{tr} \left[ P_2 P_3 \right] = 0$.

*Proof.* The fact that (3.35) is a necessary condition for $P_1$ being a superposition of $P_2$ and some other pure state $P_3$ is evident from (3.34). Indeed, take $P_1 = P_\omega$ and $P_2 = P_\varphi$ and multiply the both sides of (3.34) by $P_\omega$. Taking trace gives $\text{tr} \left[ P_\omega P_\varphi \right] = |b|^2 / \|a\psi + b\varphi\|^2$. Hence, $\text{tr} \left[ P_\omega P_\varphi \right] = 0$ implies that $b = 0$.

Let us then assume that (3.35) holds. Choose unit vectors $\omega$ and $\varphi$ such that $P_1 = P_\omega$ and $P_2 = P_\varphi$. If $|\langle \varphi \,|\, \omega \rangle| = 1$, then $P_1 = P_2$ and the statement becomes trivial. Hence, we assume that $|\langle \varphi \,|\, \omega \rangle| < 1$. Denote $\eta = (1 - |\langle \varphi \,|\, \omega \rangle|)^{-1}(\psi - \langle \varphi \,|\, \omega \rangle \varphi)$, so that $\eta$ is a unit vector and $\langle \eta \,|\, \varphi \rangle = 0$. Now

$$\psi = (1 - |\langle \varphi \,|\, \omega \rangle|)\eta + \langle \varphi \,|\, \omega \rangle \varphi \,,$$

showing that $P_1$ is a superposition of $P_2$ and $P_3 = P_\eta$. Moreover, $\langle \eta \,|\, \varphi \rangle = 0$ means that $\text{tr} \left[ P_2 P_3 \right] = 0$. $\qquad\square$

As convexity gives a way to form a new state out of two states, superposition is a way to form a new pure state out of two pure states. However, the superposition structure of pure states is of different character than the convex structure of states. For instance, it can be seen from Proposition 27 that any pure state can be written as a superposition of some other two pure states. Hence, there are no "extremal" pure states in the sense of superposition.

### 3.2.2   Interference

Let $\psi, \varphi \in \mathcal{H}$ be orthogonal unit vectors and let $\omega$ be the superposed vector as in (3.33). The orthogonality of $\psi$ an $\varphi$ implies that $\|a\psi + b\varphi\|^2 = |a|^2 + |b|^2$. The probability related to a measurement of an effect $E$ in the pure state $P_\omega$ is then

$$\operatorname{tr}\left[EP_\omega\right] = \frac{1}{|a|^2 + |b|^2}\left(|a|^2 \operatorname{tr}\left[EP_\psi\right] + |b|^2 \operatorname{tr}\left[EP_\varphi\right] + 2\operatorname{Re}\{a^*b\left\langle\psi\,|\,E\varphi\right\rangle\}\right)\,. \quad (3.36)$$

The first two terms in (3.36) look like a convex combination. Indeed, denote $\lambda = |a|^2 / (|a|^2 + |b|^2)$, in which case $1 - \lambda = |b|^2 / (|a|^2 + |b|^2)$. Let us then define a mixture of $P_\psi$ and $P_\varphi$ as $\varrho = \lambda P_\psi + (1 - \lambda) P_\varphi$. The probability of measuring the effect $E$ in $\varrho$ is

$$\operatorname{tr}\left[E\varrho\right] = \frac{1}{|a|^2 + |b|^2}\left(|a|^2 \operatorname{tr}\left[EP_\psi\right] + |b|^2 \operatorname{tr}\left[EP_\varphi\right]\right)\,. \quad (3.37)$$

Comparing (3.36) with (3.37) we find that the difference between the superposition $P_\omega$ and the mixture $\varrho$ of the orthogonal pure states is given by the *interference term*

$$I_\omega(E) = \frac{2}{|a|^2 + |b|^2}\operatorname{Re}\{a^*b\left\langle\psi\,|\,E\varphi\right\rangle\}\,. \quad (3.38)$$

This number is real and $-1 \leq I_\omega(E) \leq 1$. Loosely speaking we can say that the interference reflects the difference between the superposition (as a purely quantum structure) and the mixture (coming from the general statistical structure) of pure states.

The difference can be observed in experiments, but the interference depends on the particular choice of the effect $E$. Each pure state is a superposition of some other pure states. It is only a question of proper choice of the experiment in which the interference can be seen. For example, if the superposed vectors $\psi, \varphi$ are orthogonal, then $I_\omega(P_\psi) = I_\omega(P_\varphi) = 0$ and $|I_\omega(P_\omega)| = 2|ab|^2 / (|a|^2 + |b|^2)$.

**Example 25.** *Double-slit experiment.* A double-slit experiment is an elegant and simple demonstration of quantum interference and superposition. We assume that reader already met with this experiment. If not, we refer to usual text books on quantum mechanics. We shall only briefly describe the experiment, point out main features, but the discussion is not meant to be complete.

The situation is as follows: a source is producing quantum particles impinging perpendicularly onto a screen with two slits defining the $x$ axis. It is assumed that particles have equal probability to approach any point in some finite area containing the two slits. Thus, there is a nonzero probability that from time to time the particle will pass beyond the screen. In a sense, the screen with slits is filtering the incoming particles. After passing the slits the particles evolve freely and are registered on the second screen. The first screen with slits followed by the free evolution between the screens is considered to be the preparation process of some state. The registration on the second screen is understood as the measurement part of the experiment. For our purposes it is sufficient to consider the one-dimensional version of the problem, in which a probability density distribution $p(x)$ is measured. We consider three different experimental settings:

- Both slits are open and the probability $q(x)$ is measured.

- Lower slit is closed and upper one is open. We measure the probability $p_+(x)$.

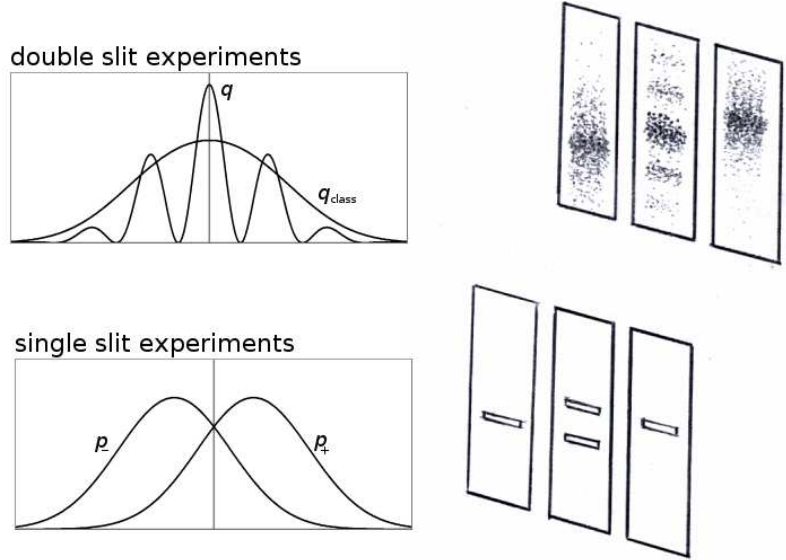

Figure 3.2. Quantum double slit experiment.

- Upper slit is closed and lower one is open. We measure the probability $p_-(x)$.

As it is depicted on Fig.3.2 after the experiments are accomplished we end up with the probability densities $p_\pm(x)$ and $q(x)$. Performing the same experiment with classical particles one will find exactly the same probability densities for $p_\pm(x)$, however, the distribution $q_{\text{class}}(x)$ will be completely different as in quantum case and will satisfy the identity $q_{\text{class}} = \frac{1}{2}[p_+(x) + p_-(x)]$. The interpretation is that the double-slit experiment can be seen as an equal mixture of single-slit experiments, in which with equal probabilities one of the slits is closed. This reasoning is based on the fact that in each run of the double-slit experiment the particle must go through exactly one of the slits. Thus, we can say that the second slit is closed. However, quantumly such "mixture" relation does not hold, $q(x) \neq \frac{1}{2}[p_+(x) + p_-(x)]$, and the "mixture" interpretation fails. In particular, in the quantum case the following relation holds

$$q(x) = \frac{1}{2}[p_+(x) + p_-(x)] + I(x),$$  (3.39)

where $I(x)$ is the so-called *interference term*.

According to classical probability theory a probability of a joint event $a \cup b$ is given as $P(a \cup b) = P(a) + P(b) - P(a \cap b)$, where the last term exhibits the dependencies between the events $a$ and $b$. Let us try to employ this formula to interpret the quantum case. The event $a$ is associated with registration of particle passing the first of the slits in position $x$. Similarly, the event $b$ is identified with the registration of particle passing through the second slit in position $x$. In this probability theory picture the existence of the nonvanishing interference term implies that these two events are independent. That is, the particles although separated in time are not passing the slits independently. But still their behavior can be understood within the structures of probability theory. Unfortunately, the situation is even more complicated. Whereas the probability $P(a \cap b)$ is always positive, the interference term can be both: negative and positive. Thus, the probability theory fails to explain the interference term and the double-slit experiment cannot be related to single-slit experiments within the language of usual probability theory. The quantum particles are passing through pair of slits in a very curious way. The registration of the particles on the second screen does not tell us anything about the slit the particle was passing through. In fact, there is no way how to find such information without disturbing the whole experiment.

The quantum situation is properly described within the quantum theory. As we said the first screen is used to prepare the particular pure state of the particle associated with vectors $\phi, \psi_+, \psi_- \in \mathcal{H} = L^2(\mathbb{R})$, respectively. The probability densities of the position measurement are given as $p_\pm(x) = |\psi_\pm(x)|^2$ and $q(x) = |\phi(x)|^2$. It follows that the state of the particles prepared by the double-slit screen in a state $\phi \in L_2(\mathbb{R})$ is a superposition of the states prepared by single-slit screens, i.e. $\phi(x) = \frac{1}{\sqrt{2}}[\psi_-(x) + \psi_+(x)]$. The calculation

$$|\phi(x)|^2 = \frac{1}{2}|\psi_-(x) + \psi_+(x)|^2 = \frac{1}{2}[p_-(x) + p_+(x)] + \frac{1}{2}[\psi_-(x)^*\psi_+(x) + \psi_+(x)^*\psi_-(x)]$$

gives that $I(x) = \mathrm{Re}[\psi_-(x)^*\psi_+(x)]$.

### 3.3 Automorphism groups

Quantum systems are described by Hilbert spaces, and unitary operators are automorphisms of a Hilbert space. Therefore, one could presume that the group of unitary operators (added perhaps with antiunitary operators) play the role of automorphism group in quantum mechanics. This is almost but not exactly true, and the delicate difference is crucial, for instance, in the description of the spin of a particle. It is therefore instructive to take a closer look on this topic.

This section can be also taken as a compatibility check of the different mathematical structures introduced in the previous sections. Namely, we will see that the basic structures of states, pure states and effects lead all to the same automorphism group.

The discussion of this section follows Chapter 2 of [22], where also many other automorphism groups are treated. We also refer to a series of articles [60], [61], [62] by L. Molnar for a detailed study on this subject.

#### 3.3.1 State automorphisms

As discussed earlier, the characteristic feature for states is the possibility of forming mixtures. This is reflected as the convex structure of $\mathcal{S}(\mathcal{H})$. The automorphisms of $\mathcal{S}(\mathcal{H})$ are therefore taken to be the convex structure preserving bijections.

**Definition 20.** A function $s : \mathcal{S}(\mathcal{H}) \to \mathcal{S}(\mathcal{H})$ is a *state automorphism* if

- $s$ is a bijection;

- $\forall \varrho_1, \varrho_2 \in \mathcal{S}(\mathcal{H}), \lambda \in [0,1] : s(\lambda \varrho_1 + (1-\lambda)\varrho_2) = \lambda s(\varrho_1) + (1-\lambda)s(\varrho_2)$.

We denote by $\mathrm{Aut}(\mathcal{S})$ the set of all state automorphisms.

**Proposition 28.** Let $s$ be a state automorphism. Then also the inverse mapping $s^{-1}$ is state automorphism.

*Proof.* The state automorphism $s$ is, by definition, a bijection on $\mathcal{S}(\mathcal{H})$. Hence, the inverse mapping $s^{-1}$ exists and it is a bijection. For all $\varrho_1, \varrho_2 \in \mathcal{S}(\mathcal{H}), \lambda \in [0,1]$, we get

$$s(\lambda s^{-1}(\varrho_1) + (1-\lambda)s^{-1}(\varrho_2)) = \lambda \varrho_1 + (1-\lambda)\varrho_2 = s(s^{-1}(\lambda \varrho_1 + (1-\lambda)\varrho_2)).$$

Since $s$ is a bijection, it is injective. Therefore,

$$\lambda s^{-1}(\varrho_1) + (1-\lambda)s^{-1}(\varrho_2) = s^{-1}(\lambda \varrho_1 + (1-\lambda)\varrho_2).$$

This shows that $s^{-1}$ is a state automorphism. $\qquad\square$

As a consequence of Proposition 28, we conclude that $\mathrm{Aut}(\mathcal{S})$ is a group.

**Proposition 29.** Let $s$ be a state automorphism. Then a state $\varrho$ is pure if and only if $s(\varrho)$ is pure.

*Proof.* We prove the proposition by showing that a state $\varrho$ is mixed if and only if $s(\varrho)$ is mixed. Let $\varrho$ be a mixed state, so that $\varrho = \lambda \varrho_1 + (1-\lambda)\varrho_2$ for some $\varrho_1 \neq \varrho_2$ and $0 < \lambda < 1$. Then $s(\varrho) = \lambda s(\varrho_1) + (1-\lambda)s(\varrho_2)$ and $s(\varrho_1) \neq s(\varrho_2)$. Thus, $s(\varrho)$ is mixed. Employing the inverse authomorphism $s^{-1}$ we can prove in the same way that if $s(\varrho)$ is mixed the state $s^{-1}s(\varrho) = \varrho$ is mixed. $\qquad\square$

We will need the following technical result.

**Lemma 1.** Let $s \in \mathrm{Aut}(\mathcal{S})$. There exists a unique linear mapping $\hat{s} : \mathcal{T}_s(\mathcal{H}) \to \mathcal{T}_s(\mathcal{H})$ such that $\hat{s}(\varrho) = s(\varrho)$ for every $\varrho \in \mathcal{S}(\mathcal{H})$. Moreover, $\hat{s}$ is positive, trace norm preserving and invertible.

*Proof.* For each nonzero positive operator $T \in \mathcal{T}_s(\mathcal{H})$, we define

$$\tilde{s}(T) := \mathrm{tr}\,[T]\, s(\frac{T}{\mathrm{tr}\,[T]}), \tag{3.40}$$

and we also set $\tilde{s}(O) = O$. For every $\lambda \geq 0$ and $T \in \mathcal{T}_s(\mathcal{H}), T \geq O$, we then have

$$\tilde{s}(\lambda T) = \lambda \mathrm{tr}\,[T]\, s(\frac{\lambda T}{\lambda \mathrm{tr}\,[T]}) = \lambda \tilde{s}(T). \tag{3.41}$$

Let $T_1, T_2$ be nonzero positive operators in $\mathcal{T}_s(\mathcal{H})$. Using the formula

$$T_1 + T_2 = (\mathrm{tr}\,[T_1] + \mathrm{tr}\,[T_2]) \left( \frac{\mathrm{tr}\,[T_1]}{\mathrm{tr}\,[T_1] + \mathrm{tr}\,[T_2]} \frac{T_1}{\mathrm{tr}\,[T_1]} + \frac{\mathrm{tr}\,[T_2]}{\mathrm{tr}\,[T_1] + \mathrm{tr}\,[T_2]} \frac{T_2}{\mathrm{tr}\,[T_2]} \right),$$

Eq. (3.41) and the properties of $s$ we see that $\tilde{s}(T_1 + T_2) = \tilde{s}(T_1) + \tilde{s}(T_2)$. The mapping $\tilde{s}$ also preserves the trace of positive trace class operators since $\mathrm{tr}\,[\tilde{s}(T)] = \mathrm{tr}\,[T]\, \mathrm{tr}\,\left[s(\frac{T}{\mathrm{tr}[T]})\right] = \mathrm{tr}\,[T]$.

For a general $T \in \mathcal{T}_s(\mathcal{H})$, we write $T = T^+ - T^-$, where $T^{\pm} = \frac{1}{2}(|T| \pm T)$ are positive operators. Define

$$\hat{s}(T) := \tilde{s}(T^+) - \tilde{s}(T^-). \tag{3.42}$$

In this way we get a well defined mapping $\hat{s}$ on $\mathcal{T}_s(\mathcal{H})$. Indeed, if $T = T_1 - T_2$ for some positive operators $T_1, T_2 \in \mathcal{T}_s(\mathcal{H})$, then $T^+ + T_2 = T^- + T_1$ and

$$\tilde{s}(T^+) + \tilde{s}(T_2) = \tilde{s}(T^-) + \tilde{s}(T_1)$$

by the additivity of $\tilde{s}$. But this implies that

$$\tilde{s}(T^+) - \tilde{s}(T^-) = \tilde{s}(T_1) - \tilde{s}(T_2),$$

which means that the definition of $\hat{s}$ does not depend on a particular decomposition of $T$.

It follows from the properties of $\tilde{s}$ that $\hat{s}$ is linear and trace preserving. Moreover, $\hat{s}$ is a positive mapping: if $T \geq 0$, then $T^+ = T, T^- = 0$ and therefore $\hat{s}(T) = \tilde{s}(T) \geq O$. A short calculation shows that $\widehat{s^{-1}} = \hat{s}^{-1}$ and hence, $\hat{s}$ is bijective.

Next we show that $\hat{s}$ is trace norm preserving. If $T \in \mathcal{T}_s(\mathcal{H})$, then

$$
\begin{aligned}
\left\| \hat{s}(T) \right\|_{\mathrm{tr}} &= \left\| \hat{s}(T^+ - T^-) \right\|_{\mathrm{tr}} = \left\| \hat{s}(T^+) - \hat{s}(T^-) \right\|_{\mathrm{tr}} \\
&\leq \left\| \hat{s}(T^+) \right\|_{\mathrm{tr}} + \left\| \hat{s}(T^-) \right\|_{\mathrm{tr}} = \mathrm{tr}\left[ \hat{s}(T^+) \right] + \mathrm{tr}\left[ \hat{s}(T^-) \right] \\
&= \mathrm{tr}\left[ T^+ \right] + \mathrm{tr}\left[ T^- \right] = \mathrm{tr}\left[ T^+ + T^- \right] = \mathrm{tr}\left[\mid T \mid\right] = \left\| T \right\|_{\mathrm{tr}}.
\end{aligned}
$$

A similar calculation gives $\left\| \hat{s}^{-1}(T) \right\|_{\mathrm{tr}} \leq \left\| T \right\|_{\mathrm{tr}}$. These inequalities taken together give

$$\left\| T \right\|_{\mathrm{tr}} = \left\| \hat{s}^{-1}(\hat{s}(T)) \right\|_{\mathrm{tr}} \leq \left\| \hat{s}(T) \right\|_{\mathrm{tr}} \leq \left\| T \right\|_{\mathrm{tr}}, \tag{3.43}$$

which shows that $\hat{s}$ is trace norm preserving.

Finally, we prove the uniqueness of the extension of $s$. Let $f : \mathcal{T}_s(\mathcal{H}) \to \mathcal{T}_s(\mathcal{H})$ be a linear mapping on $\mathcal{T}_s(\mathcal{H})$ which is an extension of $s$, i.e., $f(T) = s(T)$ for every $T \in \mathcal{S}(\mathcal{H})$. For $T \in \mathcal{T}_s(\mathcal{H})$, we get

$$
\begin{aligned}
f(T) &= f(T^+ - T^-) = f(T^+) - f(T^-) = \mathrm{tr}\left[ T^+ \right] f\left( \frac{T^+}{\mathrm{tr}\left[ T^+ \right]} \right) - \mathrm{tr}\left[ T^- \right] f\left( \frac{T^-}{\mathrm{tr}\left[ T^- \right]} \right) \\
&= \mathrm{tr}\left[ T^+ \right] s\left( \frac{T^+}{\mathrm{tr}\left[ T^+ \right]} \right) - \mathrm{tr}\left[ T^- \right] s\left( \frac{T^-}{\mathrm{tr}\left[ T^- \right]} \right) \\
&= \hat{s}(T).
\end{aligned}
$$

$\square$

**Proposition 30.** Let $s_1, s_2 \in \mathrm{Aut}(\mathcal{S})$ and assume that $s_1(P) = s_2(P)$ for every $P \in \mathcal{S}^{ext}(\mathcal{H})$. Then $s_1 = s_2$.

*Proof.* Denote $s = s_1 s_2^{-1}$. The assumption means that $s(P) = P$ for every $P \in \mathcal{S}^{ext}(\mathcal{H})$ and we have to prove that $s$ is the identity mapping. For each $\varrho \in \mathcal{S}(\mathcal{H})$, we have a trace norm convergent convex decomposition $\varrho = \sum_i \lambda_i P_i$. Since $s$ is continuous, we get

$$s(\varrho) = s\left( \sum_i \lambda_i P_i \right) = \sum_i \lambda_i s(P_i) = \sum_i \lambda_i P_i = \varrho.$$

This shows that $s$ is the identity mapping.                                                      $\square$

### 3.3.2 Pure state automorphisms

Based on Section 3.2, it seems reasonable to require that automorphisms of pure states preserve the superposition feature. One possibility is thus to define automorphisms in the following way.

**Definition 21.** A function $p : \mathcal{S}^{ext}(\mathcal{H}) \to \mathcal{S}^{ext}(\mathcal{H})$ is a *pure state automorphism* if

- $p$ is a bijection;

- $\mathrm{tr}\,[p(P_1)p(P_2)] = \mathrm{tr}\,[P_1 P_2]$.

We denote by $\mathrm{Aut}(\mathcal{S}^{ext}(\mathcal{H}))$ the set of all pure state automorphisms.

**Lemma 2.** Let $P_1, P_2 \in \mathcal{S}^{ext}(\mathcal{H})$. Then $0 \leq \mathrm{tr}\,[P_1 P_2] \leq 1$ and

$$\|P_1 - P_2\|_{\mathrm{tr}} = 2\sqrt{1 - \mathrm{tr}\,[P_1 P_2]}\,. \tag{3.44}$$

*Proof.* If $P_1 = P_2$, the statement is trivial. Hence, assume that $P_1 \neq P_2$. Choose unit vectors $\psi, \varphi \in \mathcal{H}$ such that $P_1 = P_\psi$ and $P_2 = P_\varphi$. Since $\mathrm{tr}\,[P_1 P_2] = |\langle\, \psi \,|\, \varphi \,\rangle|^2$, it follows from Cauchy-Schwarz inequality that $0 \leq \mathrm{tr}\,[P_1 P_2] \leq 1$.

Let us then calculate the eigenvalues of the operator $P_1 - P_2$. For a vector $\phi \in \mathcal{H}$, we get

$$(P_1 - P_2)\phi = \langle\, \psi \,|\, \phi \,\rangle\, \psi - \langle\, \varphi \,|\, \phi \,\rangle\, \varphi\,.$$

This shows that if $\phi$ is an eigenvector, it must be of the form $\phi = c_1 \psi + c_2 \varphi$ for some $c_1, c_2 \in \mathbb{C}$. Inserting this expression to the eigenvalue equation

$$(P_1 - P_1)\phi = \lambda \phi$$

gives

$$\left\{ \begin{array}{l} \lambda c_1 = c_1 + c_2 \,\langle\, \psi \,|\, \varphi \,\rangle \\ \lambda c_2 = -c_1 - c_1 \,\langle\, \varphi \,|\, \psi \,\rangle\,. \end{array} \right.$$

This leads to two solutions

$$\lambda_\pm = \pm\sqrt{1 - |\langle\, \psi \,|\, \varphi \,\rangle|^2}\,. \tag{3.45}$$

Therefore,

$$\begin{aligned} \|P_1 - P_2\|_{\mathrm{tr}} &= \mathrm{tr}\,[|P_1 - P_2|] = |\lambda_+| + |\lambda_-| \\ &= 2\sqrt{1 - |\langle\, \psi \,|\, \varphi \,\rangle|^2} = 2\sqrt{1 - \mathrm{tr}\,[P_1 P_2]}\,. \end{aligned}$$

$\square$

**Proposition 31.** Let $s$ be a state automorphism. The restriction of $s$ to $\mathcal{S}^{ext}(\mathcal{H})$, denoted by $s\,|_{\mathcal{S}^{ext}(\mathcal{H})}$, is a pure state automorphism.

*Proof.* By Proposition 29, the restriction of $s$ is a mapping from $\mathcal{S}^{ext}(\mathcal{H})$ to $\mathcal{S}^{ext}(\mathcal{H})$. The same is true for the inverse mapping $s^{-1}$, and therefore the restriction of $s$ is a bijection on $\mathcal{S}^{ext}(\mathcal{H})$.

Let $\hat{s}$ be the extension of $s$ to $\mathcal{T}_s(\mathcal{H})$ as in Proposition 1, and let $P_1, P_2 \in \mathcal{S}^{ext}(\mathcal{H})$. Using Lemma 2 we get

$$
\begin{aligned}
2\sqrt{1 - \operatorname{tr}\left[P_1 P_1\right]} &= \|P_1 - P_2\|_{\mathrm{tr}} = \|\hat{s}(P_1 - P_2)\|_{\mathrm{tr}} = \|\hat{s}(P_1) - \hat{s}(P_2)\|_{\mathrm{tr}} \\
&= \|s(P_1) - s(P_2)\|_{\mathrm{tr}} = 2\sqrt{1 - \operatorname{tr}\left[s(P_1)s(P_1)\right]}.
\end{aligned}
$$

Hence, $\operatorname{tr}\left[s(P_1)s(P_2)\right] = \operatorname{tr}\left[P_1 P_2\right]$.                                  $\square$

From Propositions 30 and 31 we conclude that the mapping

$$
s \mapsto s \mid_{\mathcal{S}^{ext}(\mathcal{H})} \tag{3.46}
$$

is an injective mapping from $\operatorname{Aut}(\mathcal{S})$ to $\operatorname{Aut}(\mathcal{S}^{ext}(\mathcal{H}))$.

### 3.3.3  Effect automorphisms

As we have seen in Subsection 3.1.5, the set of effects is endowed with the partial binary operation $\boxplus$. This is the essential structure of the set of effects. Therefore, the effect automorphisms are defined in the following way.

**Definition 22.** A function $e : \mathcal{E}(\mathcal{H}) \to \mathcal{E}(\mathcal{H})$ is an *effect automorphism* if

- $e$ is a bijection;

- for $E_1, E_2 \in \mathcal{E}(\mathcal{H})$, $E_1 \boxplus E_2$ is defined if and only if $e(E_1) \boxplus e(E_2)$ is defined;

- $e(E_1 \boxplus E_2) = e(E_1) \boxplus e(E_2)$.

We denote by $\operatorname{Aut}(\mathcal{E})$ the set of all effect automorphisms.

**Exercise 20.** Let $e \in \operatorname{Aut}(\mathcal{E})$. Show that

(i)  $E_1 \leq E_2$ iff $e(E_1) \leq e(E_2)$;

(ii)  $e(O) = O$ and $e(I) = I$;

(iii)  $e(I - E) = I - e(E)$.

**Proposition 32.** Let $e \in \operatorname{Aut}(\mathcal{E})$. The formula

$$
\operatorname{tr}\left[\varrho e^{-1}(E)\right] = \operatorname{tr}\left[s_e(\varrho)E\right] \quad \varrho \in \mathcal{S}(\mathcal{H}), E \in \mathcal{E}(\mathcal{H}), \tag{3.47}
$$

defines a unique state automorphism $s_e$ and the correspondence $e \mapsto s_e$ is an injective mapping from $\operatorname{Aut}(\mathcal{E})$ to $\operatorname{Aut}(\mathcal{S})$.

*Proof.* Let $\varrho \in \mathcal{S}(\mathcal{H})$. The mapping $E \mapsto \operatorname{tr}\left[\varrho e^{-1}(E)\right]$ is a generalized probability measure, and hence, by Prop. 25 there is a unique state $\varrho'$ such that $\operatorname{tr}\left[\varrho e^{-1}(E)\right] = \operatorname{tr}\left[\varrho' E\right]$ for all $E \in \mathcal{E}(\mathcal{H})$. We define a mapping $s_e$ as $s_e(\varrho) = \varrho'$.                                  $\square$

### 3.3.4  Wigner's Theorem

Let $U$ be a unitary operator on $\mathcal{H}$. We define a mapping $\sigma_U$ on $\mathcal{L}_s(\mathcal{H})$ by

$$\sigma_U(T) = U^*TU\,. \tag{3.48}$$

This mapping preserves the positivity: if $T \geq O$, then $U^*TU \geq O$.

**Exercise 21.** Show that a unitary operator $U$ defines through formula (3.48) (when restricted to proper subsets of $\mathcal{L}_s(\mathcal{H})$)

(a) a state automorphism;

(b) a pure state automorphism;

(c) an effect automorphism.

**Proposition 33.** Let $U$ and $V$ be unitary operators. The following conditions are equivalent:

(i) $\sigma_U = \sigma_V$;

(ii) $\sigma_U(E) = \sigma_V(E)$ for every $E \in \mathcal{E}(\mathcal{H})$;

(iii) $\sigma_U(\varrho) = \sigma_V(\varrho)$ for every $\varrho \in \mathcal{S}(\mathcal{H})$;

(iv) $\sigma_U(P) = \sigma_V(P)$ for every $P \in \mathcal{S}^{ext}(\mathcal{H})$;

(v) $U = tV$ for some $t \in \mathbb{T} = \{z \in \mathbb{C} : |z| = 1\}$.

*Proof.* Trivially, (i)⇒(ii). Moreover, $\mathcal{S}(\mathcal{H})$ is a subset of $\mathcal{E}(\mathcal{H})$ and $\mathcal{S}^{ext}(\mathcal{H})$ is a subset of $\mathcal{S}(\mathcal{H})$, hence (ii)⇒(iii)⇒(iv). It is easy to see that (v)⇒(i). Therefore, to complete the proof we need to show that (iv)⇒(v).

Let us first note that for a one-dimensional projection $P_\psi$, we get $\sigma_U(P_\psi) = P_{U\psi}$. On the other hand, two projections $P_\psi$ and $P_\varphi$ are the same if and only if there is a complex number $z \in \mathbb{T}$ such that $\psi = z\varphi$. Therefore, (iv) implies that there exists a mapping $c$ from the unit vectors of $\mathcal{H}$ to $\mathbb{T}$ such that $U\psi = c(\psi)V\psi$ for all unit vectors $\psi \in \mathcal{H}$. To prove (v), it is enough to show that $c$ is a constant function.

Let $\psi, \varphi \in \mathcal{H}$ and $\psi \neq \varphi$. We then have

$$\langle\,\psi \,|\, \varphi\,\rangle = \langle\, U\psi \,|\, U\varphi\,\rangle = \overline{c(\psi)}c(\varphi)\,\langle\, V\psi \,|\, V\varphi\,\rangle = \overline{c(\psi)}c(\varphi)\,\langle\,\psi \,|\, \varphi\,\rangle\,.$$

Therefore, in the case $\langle\,\psi \,|\, \varphi\,\rangle \neq 0$ we get $c(\psi) = c(\varphi)$.

Suppose then that $\langle\,\psi \,|\, \varphi\,\rangle = 0$. Now $\langle\,\psi \,|\, \phi\,\rangle \neq 0$, where $\phi = \|\psi + \varphi\|^{-1}\,(\psi + \varphi)$. By our earlier observation we conclude that $c(\phi) = c(\psi)$. Thus,

$$\begin{aligned} U\psi + U\varphi &= U(\psi + \varphi) = \|\psi + \varphi\|\,U\phi = \|\psi + \varphi\|\,c(\psi)V\phi = c(\psi)V(\psi + \varphi) \\ &= c(\psi)V\psi + c(\psi)V\varphi = U\psi + c(\psi)V\varphi\,. \end{aligned}$$

Comparing the first and the last expressions we conclude that $U\varphi = c(\psi)V\varphi$, implying that $c(\psi) = c(\varphi)$. This shows that $c$ is a constant function.  $\square$

In the following we are going to need the antiunitary operators introduced in the end of Subsection 2.2.3. An antiunitary operator $A$ defines a mapping $\sigma_A$ on $\mathcal{L}_s(\mathcal{H})$ with the same formula (3.48) as a unitary operator.

**Proposition 34.** Let $A$ be an antiunitary operator. The corresponding mapping $\sigma_A$ on $\mathcal{L}_s(\mathcal{H})$ is linear.

*Proof.* Let $T_1, T_2 \in \mathcal{L}_s(\mathcal{H})$ and $\alpha \in \mathbb{R}$. For all $\varphi, \psi \in \mathcal{H}$, we get

$$
\begin{aligned}
\langle\, \varphi \,|\, \sigma_A(T_1 + \alpha T_2)\psi \,\rangle &= \langle\, \varphi \,|\, A^*(T_1 + \alpha T_2)A\psi \,\rangle = \langle\, A\varphi \,|\, (T_1 + \alpha T_2)A\psi \,\rangle \\
&= \langle\, A\varphi \,|\, T_1 A\psi \,\rangle + \alpha \langle\, A\varphi \,|\, T_2 A\psi \,\rangle \\
&= \langle\, \varphi \,|\, \sigma_A(T_1)\psi \,\rangle + \alpha \langle\, \varphi \,|\, \sigma_A(T_2)\psi \,\rangle
\end{aligned}
$$

It follows that $\sigma_A(T_1 + \alpha T_2) = \sigma_A(T_1) + \alpha\sigma_A(T_2)$, hence $\sigma_A$ is linear.                $\square$

**Exercise 22.** Verify that the results of Exercise 21 and Proposition 33 are true when unitary operators are replaced by antiunitary operators.

**Proposition 35.** Let $U$ be a unitary operator and $A$ an antiunitary operator. Then $\sigma_U \neq \sigma_A$.

*Proof.* We make a counter assumption that $\sigma_U = \sigma_A$. As noted in the proof of Prop. 33, this can be true only if for every unit vector $\psi \in \mathcal{H}$, there is a complex number $c(\psi)$ such that $U\psi = c(\psi)A\psi$. Let $\psi, \varphi \in \mathcal{H}$ be linearly independent unit vectors, and denote $\phi = \|\psi + i\varphi\|^{-1}(\psi + i\varphi)$. Then

$$U(\psi + i\varphi) = U\psi + iU\varphi = c(\psi)A\psi + ic(\varphi)A\varphi\,, \tag{3.49}$$

and, on the other hand,

$$U(\psi + i\varphi) = \|\psi + i\varphi\|\, U\phi = \|\psi + i\varphi\|\, c(\phi)A\psi - i\, \|\psi + i\varphi\|\, c(\phi)A\varphi\,. \tag{3.50}$$

Since $A$ is an antiunitary operator, the vectors $A\psi$ and $A\varphi$ are linearly independent. Comparison of (3.49) and (3.50) therefore gives $c(\psi) = \|\psi + i\varphi\|\, c(\phi)$ and $c(\varphi) = -\|\psi + i\varphi\|\, c(\phi)$. Thus, $c(\psi) = -c(\varphi)$. Repeating the same calculation with the vector $\phi = \|\psi + \varphi\|^{-1}(\psi + \varphi)$ leads to $c(\psi) = c(\varphi)$. This means that $c(\psi) = 0$, which cannot be true as $c(\psi) \in \mathbb{T}$. We conclude that counter assumption is false and $\sigma_U \neq \sigma_A$.                $\square$

For each unitary or antiunitary operator $U$, we denote by $[U]$ the equivalence class consisting of all operators of the form $zU, z \in \mathbb{C} \setminus \{0\}$. As a conclusion from Propositions 33 and 35 the equivalence class $[U]$ consists of those operators defining the same mapping $\sigma_U$. We denote by $\Sigma$ the set of all equivalence classes $[U]$. It is a group when the multiplication is defined in the following way:

$$[U] \cdot [V] = [UV]\,.$$

The fundamental result in the theory of automorphism groups is the following statement, known as *Wigner's Theorem*.

**Theorem 6** (Wigner's Theorem). Let $p \in \mathrm{Aut}(\mathcal{S}^{ext}(\mathcal{H}))$. There is a unitary or antiunitary operator $U$ such that

$$p(P) = \sigma_U(P) \quad \forall P \in \mathcal{S}^{ext}(\mathcal{H}) \,. \tag{3.51}$$

The operator $U$ is unique up to the equivalence class $[U]$.

The proof of this theorem is long and non-trivial; see, for instance, [22]. As a consequence of Wigner's Theorem we conclude that there is an injective mapping from $\mathrm{Aut}(\mathcal{S}^{ext}(\mathcal{H}))$ to $\Sigma$. We thus have the following chain of injective mappings:

$$\Sigma \to \mathrm{Aut}(\mathcal{E}) \to \mathrm{Aut}(\mathcal{S}) \to \mathrm{Aut}(\mathcal{S}^{ext}(\mathcal{H})) \to \Sigma \,. \tag{3.52}$$

The first arrow is easy and it was stated in Exercise 21. The second arrow follows from Proposition 32, while the third arrow follows from Proposition 31. The last arrow is the content of Wigner's Theorem. Let $\sigma \in \Sigma$ and denote by $\sigma' \in \Sigma$ the image of $\sigma$ in the chain of mappings in (3.52). Using formulas (3.46) and (3.47) it is straightforward to verify that $\sigma' = \sigma$. Hence, there is a bijective correspondence between all these sets. Even more, a short inspection shows that all the mappings in (3.52) are group homomorphisms. Therefore, the groups $\mathrm{Aut}(\mathcal{E}), \mathrm{Aut}(\mathcal{S}), \mathrm{Aut}(\mathcal{S}^{ext}(\mathcal{H}))$ and $\Sigma$ are isomorphic.

### 3.4 Composite systems

#### 3.4.1 System vs. subsystems

Suppose that we have two systems $A$ and $B$ and let $\mathcal{H}_A$ and $\mathcal{H}_B$ be the Hilbert spaces used in the description of these systems. To make things simpler, we assume that the systems $A$ and $B$ are of different kind and distinguishable. They could be, for instance, two different modes of an electromagnetic radiation, or electron-proton system. How should we describe the compound system $A + B$?

We assume that $A$ and $B$ are identifiable parts of the compound system $A + B$, and we call them *subsystems* of $A + B$. This assumption means, in particular, that we can separately manipulate the systems $A$ and $B$, i.e., we are able to perform experiments addressing properties of $A$ and $B$ individually. Suppose that we have an effect $E_A$ on $\mathcal{H}_A$ and an effect $E_B$ on $\mathcal{H}_B$. These effects correspond to simple measurements on the systems $A$ and $B$, respectively. Hence, by the assumption we should have an effect $\gamma(E_A, E_B)$ on $\mathcal{H}$ which describes these separate measurements on $A$ and $B$. In other words, we require that there is a mapping $\gamma$ from $\mathcal{E}(\mathcal{H}_A) \times \mathcal{E}(\mathcal{H}_B)$ to $\mathcal{E}(\mathcal{H})$.

In a similar way, we have a mapping $\bar{\gamma}$ from $\mathcal{S}(\mathcal{H}_A) \times \mathcal{S}(\mathcal{H}_B)$ to $\mathcal{S}(\mathcal{H})$. If the measurements and preparations are made separately, the systems are statistically independent and we should have

$$\mathrm{tr}\left[\bar{\gamma}(\varrho_A, \varrho_B)\gamma(E_A, E_B)\right] = \mathrm{tr}\left[\varrho_A E_A\right] \cdot \mathrm{tr}\left[\varrho_B E_B\right] \,. \tag{3.53}$$

The question of the description of the compound system can thus be approached by searching for a suitable Hilbert space $\mathcal{H}$ and mappings $\gamma$ and $\bar{\gamma}$.

Let us make a trial by choosing $\mathcal{H} = \mathcal{H}_A \otimes \mathcal{H}_B$ and setting $\gamma(E_A, E_B) = E_A \otimes E_B$ and $\bar{\gamma}(\varrho_A, \varrho_B) = \varrho_A \otimes \varrho_B$. The properties of the tensor product guarantee that the condition (3.53) is satisfied. This motivates the choice of the tensor product Hilbert space as a description of the compound system. Adding some other natural requirements as (3.53), one can achieve the conclusion that the tensor product structure is actually a unique way to describe compound systems. Here we take this fact as granted and refer to [3] for more details.

A curious thing arises from the above discussion. Namely, an operator $T$ on $\mathcal{H}$ need not be of the tensor product form $T = T_A \otimes T_B$. How should we understand those states and effects of the compound system which are not of the product form? The answer lies on the fact that instead of manipulating subsystems separately, we can make collective measurements and preparations. It may happen that this kind of process cannot be described by separate measurements and preparations. Especially, the states of the subsystems do not, in general, specify the state of the compound system. The states of the two subsystems can be intertwined in a strange way, in which case the compound state is called *entangled*. Entanglement will be our topic in Chapter 7. In this section discuss some basic properties of the tensor product as a description of compound systems.

**Definition 23.** The *partial trace* over the system $A$ is the linear mapping

$$\mathrm{tr}_A : \mathcal{T}(\mathcal{H}_A \otimes \mathcal{H}_B) \to \mathcal{T}(\mathcal{H}_B)$$

satisfying

$$\mathrm{tr}\left[\mathrm{tr}_A[T]E\right] = \mathrm{tr}\left[T(I \otimes E)\right] \tag{3.54}$$

for all $T \in \mathcal{T}(\mathcal{H}_A \otimes \mathcal{H}_B)$ and $E \in \mathcal{L}(\mathcal{H}_B)$. In a similar way we define the partial trace $\mathrm{tr}_B$ over the subsystem $B$.

**Example 26.** Suppose $T \in \mathcal{T}(\mathcal{H}_A \otimes \mathcal{H}_B)$ is of the product form $T = T_A \otimes T_B$. Then the defining condition (3.54) gives

$$\mathrm{tr}\left[\mathrm{tr}_A[T_A \otimes T_B]E\right] = \mathrm{tr}\left[T_A\right]\mathrm{tr}\left[T_B E\right] .$$

Since this holds for every $E \in \mathcal{L}(\mathcal{H}_B)$, we conclude that

$$\mathrm{tr}_A[T_A \otimes T_B] = \mathrm{tr}\left[T_A\right] T_B .$$

Actually, we have not yet shown that the partial trace mapping even exists. To do this, we give a formula to calculate the partial trace. Fix orthonormal bases $\{\psi_i\}$ and $\{\varphi_k\}$ for $\mathcal{H}_A$ and $\mathcal{H}_B$, respectively. If $T \in \mathcal{T}(\mathcal{H}_A \otimes \mathcal{H}_B)$, then

$$\mathrm{tr}_A[T] = \sum_{i,k,l} \langle \psi_i \otimes \varphi_k \,|\, T\psi_i \otimes \varphi_l \rangle \,|\varphi_k\rangle\langle \varphi_l| \tag{3.55}$$

and

$$\mathrm{tr}_B[T] = \sum_{i,j,k} \langle \psi_i \otimes \varphi_k \,|\, T\psi_j \otimes \varphi_k \rangle \,|\psi_i\rangle\langle \psi_j| . \tag{3.56}$$

In some cases one can use directly the defining condition (3.54) to calculate the partial trace, whereas sometimes it is easier to apply formulas (3.55) and (3.56).

**Exercise 23.** Confirm that the operator $\mathrm{tr}_A[T]$ as written in (3.55) satisfies the defining condition (3.54) for every $E \in \mathcal{L}(\mathcal{H}_B)$. (Hint: recall that a possible orthonormal basis for $\mathcal{H}_A \otimes \mathcal{H}_B$ is $\{\psi_i \otimes \varphi_k\}$.)

**Proposition 36.** Let $T \in \mathcal{T}(\mathcal{H}_A \otimes \mathcal{H}_B)$. Then

(a) $\mathrm{tr}\,[T] = \mathrm{tr}\,[\mathrm{tr}_A[T]] = \mathrm{tr}\,[\mathrm{tr}_B[T]]$.

(b) $T \geq O$ implies $\mathrm{tr}_A[T] \geq O$ and $\mathrm{tr}_B[T] \geq O$.

(c) $T \in \mathcal{S}(\mathcal{H}_A \otimes \mathcal{H}_B)$ implies $\mathrm{tr}_A[T] \in \mathcal{S}(\mathcal{H}_A)$ and $\mathrm{tr}_B[T] \in \mathcal{S}(\mathcal{H}_B)$.

*Proof.* (a) Choose $E = I$ in (3.54).

(b) Assume that $T \geq O$. Fix a unit vector $\eta \in \mathcal{H}_B$ and choose $E = P_\eta$ in (3.54). We then get

$$
\begin{aligned}
\langle \eta \,|\, \mathrm{tr}_A[T]\eta \rangle &= \mathrm{tr}\,[\mathrm{tr}_A[T]P_\eta] = \mathrm{tr}\,[T(I \otimes P_\eta)] = \mathrm{tr}\,[(I \otimes P_\eta)T(I \otimes P_\eta)] \\
&= \mathrm{tr}\,\left[(T^{\frac{1}{2}}(I \otimes P_\eta))^*(T^{\frac{1}{2}}(I \otimes P_\eta))\right] \geq 0 \,.
\end{aligned}
$$

This shows that $\mathrm{tr}_A[T] \geq O$.

(c) This follows from (a) and (b). $\qquad\qquad\square$

The effects of the form $E \otimes I$ are interpreted as experiments measuring the properties of the subsystem $A$ only. Since the identity $\mathrm{tr}\,[(E \otimes I)\varrho] = \mathrm{tr}\,[\mathrm{tr}_B[\varrho]E]$ holds for a given state of the composite system $\varrho$ and all effects $E \in \mathcal{E}(\mathcal{H}_A)$ defined on the subsystem $A$, it is natural to identify the state $\varrho_A \equiv \mathrm{tr}_B[\varrho]$ with the state of the subsystem $A$. That is, for arbitrary state $\varrho$ of the composite system, its subsystems are described by well defined states $\varrho_A$ and $\varrho_B$ allowing to predict the probabilities for each possible effect of the subsystems.

**Definition 24.** Let $\varrho \in \mathcal{S}(\mathcal{H}_A \otimes \mathcal{H}_B)$ be a state of the composite bipartite system $A+B$. Then the states $\varrho_A \equiv \mathrm{tr}_B[\varrho]$ and $\varrho_B \equiv \mathrm{tr}_A[\varrho]$ describe the individual subsystems $A$ and $B$, respectively. The states $\varrho_A$ and $\varrho_B$ are called *reduced states*, and $\varrho_{AB} \equiv \varrho$ is a *joint state*.

Assuming that subsystems $A$ and $B$ are described by the states $\varrho_A$ and $\varrho_B$, respectively, we can ask the following question: what are the possible joint states $\varrho_{AB}$ of the composite system? Clearly, one possibility is to have $\varrho_{AB} = \varrho_A \otimes \varrho_B$. Generally, however, there are also other possible choices. This means that the knowledge of the states of the subsystems $A$ and $B$ does not specify the state of the composite sysytem.

Let us denote $\Gamma = \varrho_{AB} - \varrho_A \otimes \varrho_B$. First of all, $\Gamma$ is selfadjoint trace class operator and $\mathrm{tr}\,[\Gamma] = 0$. Moreover, applying the result of Example 26 we see that $\mathrm{tr}_A[\Gamma] = O_B$ and $\mathrm{tr}_B[\Gamma] = O_A$.

**Proposition 37.** If either $\varrho_A$ or $\varrho_B$ is pure, then the joint state $\varrho_{AB}$ is necessarily of the product form $\varrho_{AB} = \varrho_A \otimes \varrho_B$.

*Proof.* Suppose that $\varrho_A = |\varphi\rangle\langle\varphi|$. The positivity of $\varrho_{AB} = |\varphi\rangle\langle\varphi| \otimes \varrho_B + \Gamma$ requires that $\langle\varphi_\perp \otimes \phi\,|\,\Gamma(\varphi_\perp \otimes \phi)\rangle \geq 0$ for all $\phi \in \mathcal{H}_B$ and $\varphi_\perp \in \mathcal{H}_A$ such that $\langle\varphi\,|\,\varphi_\perp\rangle = 0$. Consider orthonormal bases $\varphi_1 \equiv \varphi, \varphi_2, \varphi_3, \cdots \in \mathcal{H}_A$, and $\phi_1, \phi_2, \cdots \in \mathcal{H}_B$. Using the identity $0 = \langle\varphi_\perp\,|\,O\varphi_\perp\rangle = \langle\varphi_\perp\,|\,\mathrm{tr}_B[\Gamma]\varphi_\perp\rangle = \sum_k \langle\varphi_\perp \otimes \phi_k\,|\,\Gamma(\varphi_\perp \otimes \phi_k)\rangle$ and positivity $\langle\varphi_\perp \otimes \phi\,|\,\Gamma(\varphi_\perp \otimes \phi)\rangle \geq 0$ we can conclude that $\langle\varphi_\perp \otimes \phi\,|\,\Gamma(\varphi_\perp \otimes \phi)\rangle = 0$ for all $\varphi_\perp \otimes \phi \in \mathcal{H}_A \otimes \mathcal{H}_B$. In order to show that $\langle\varphi \otimes \phi\,|\,\Gamma(\varphi \otimes \phi)\rangle = 0$ for all $\phi \in \mathcal{H}_B$ we use the identity $0 = \langle\phi\,|\,\mathrm{tr}_A\Gamma\phi\rangle = \sum_j \langle\varphi_j \otimes \phi\,|\,\Gamma(\varphi_j \otimes \phi)\rangle = \langle\varphi \otimes \phi\,|\,\Gamma(\varphi \otimes \phi)\rangle$. In summary, we found that $\Gamma$ vanishes on all product vectors and therefore it is identically zero, i.e. $\Gamma = O$. This proves the proposition. $\qquad\square$

**Proposition 38.** Suppose that $\dim \mathcal{H}_A = \dim \mathcal{H}_B < \infty$. If for all unitary operators $U : \mathcal{H}_A \to \mathcal{H}_B$ the relation $\varrho_A \neq U\varrho_B U^*$ holds, then the joint state $\varrho \in \mathcal{S}(\mathcal{H}_A \otimes \mathcal{H}_B)$ is not pure.

This proposition is a direct consequence of the Schmidt decomposition theorem (see Theorem 14) that will be formulated and proved in Chapter 7. It is essentially saying that if the joint system is in a pure state, then the reduced states $\varrho_A$ and $\varrho_B$ must have identical the nonvanishing part of spectra, hence they are related by a unitary channel, or partial isometry if $\mathcal{H}_A \neq \mathcal{H}_B$.

### 3.4.2   State purification

Mixed states are convex mixtures of pure states. There is also another way how mixed states can be seen as arising from pure states. As we have seen the reduced states of composite systems give different physical interpretation to the nature of mixed states. The reduced states are defined via the partial trace mapping. An inverse procedure to reduction of one of the subsystem is an addition of an extra system, at least into our description. For this purpose we introduce the so called *ancillary system* represented by some additional Hilbert space $\mathcal{H}_{\mathrm{anc}}$. In the following definition we define the inverse mapping to partial trace.

**Definition 25.** Let $\varrho_A$ be a state of a system $A$. A pure state $P_\psi = |\psi\rangle\langle\psi|$ associated with some unit vector $\psi \in \mathcal{H} \otimes \mathcal{H}_{\mathrm{anc}}$ is a *purification* of $\varrho_A$ if

$$\mathrm{tr}_{\mathrm{anc}} P_\psi = \varrho_A \,. \tag{3.57}$$

Let us note that the size of the ancillary system is not limited. In a sense, the purification is an inverse mapping to partial trace, however, it is highly nonunique.

In the following result we describe a method to construct a purification of a given mixed state. This shows, in particular, that every mixed state has purifications.

**Proposition 39.** Consider a convex decomposition of a mixed state $\varrho \in \mathcal{S}(\mathcal{H})$ into pure states $\eta_1, \ldots, \eta_n \in \mathcal{H}$

$$\varrho = \sum_j p_j |\eta_j\rangle\langle\eta_j| \,. \tag{3.58}$$

Define a vector

$$\psi = \sum_j \sqrt{p_j} \eta_j \otimes \phi_j \,, \tag{3.59}$$

where the vectors $\phi_1, \ldots, \phi_n$ form some orthonormal basis of $\mathcal{H}_{\mathrm{anc}}$ of dimension $n$. A state $P_\psi = |\psi\rangle\langle\psi|$ is a purification of $\varrho$.

*Proof.* By direct calculation we get

$$
\begin{aligned}
\mathrm{tr}_{\mathrm{anc}} P_\psi &= \sum_{j,j'} \sqrt{p_j p_{j'}} \sum_k \langle\, \phi_k \mid \phi_j \,\rangle \, \langle\, \phi_{j'} \mid \phi_k \,\rangle \, |\eta_j\rangle\langle\eta_{j'}| \\
&= \sum_j p_j |\eta_j\rangle\langle\eta_j| = \varrho \,.
\end{aligned}
$$

$\square$

A purification associated with the canonical convex decomposition of $\varrho$ is called *canonical purification*. In particular, $\varrho = \sum_j \lambda_j |\varphi_j\rangle\langle\varphi_j|$, where $\lambda_j$ are eigenvalues and $\varphi_j$ are eigenvectors. Then

$$
\psi_{\mathrm{can}} = \sum_j \sqrt{\lambda_j}\, \varphi_j \otimes \phi_j \tag{3.60}
$$

is the canonical purification.

**Proposition 40.** If a vector $\psi \in \mathcal{H} \otimes \mathcal{H}_{\mathrm{anc}}$ defines a purification of $\varrho \in \mathcal{S}(\mathcal{H})$, then $\dim \mathcal{H}_{\mathrm{anc}} \geq \dim \mathrm{supp}(\varrho)$.

*Proof.* Consider a purification $\psi \in \mathcal{H} \otimes \mathcal{H}_{\mathrm{anc}}$ such that $\dim \mathcal{H}_{\mathrm{anc}} < \dim \mathrm{supp}(\varrho)$. According to Proposition 38 the supports $\mathrm{supp}(\varrho)$ and $\mathrm{supp}(\varrho_{\mathrm{anc}})$ have the same dimension. Consequently, since $\dim \mathrm{supp}(\varrho_{\mathrm{anc}}) \leq \dim \mathcal{H}_{\mathrm{anc}}$ it follows that also $\dim \mathrm{supp}(\varrho_{\mathrm{anc}}) \leq \dim \mathcal{H}_{\mathrm{anc}} < \dim \mathrm{supp}(\varrho)$ which is in contradiction with Proposition 38. Therefore, the dimension of the ancillary system is at least the dimension of the support of the state $\varrho$. $\square$

**Example 27.** (*Purifications of total mixture.*) Consider a finite $d$-dimensional Hilbert space $\mathcal{H}$ and let $\varrho = \frac{1}{d} I$ be the total mixture. Suppose $\psi \in \mathcal{H} \otimes \mathcal{H}_{\mathrm{anc}}$ is a purification of $\varrho$, i.e. $\mathrm{tr}_{\mathrm{anc}} |\psi\rangle\langle\psi| = \frac{1}{d} I$. Then also a state $\psi' = (U \otimes V_{\mathrm{anc}})\psi$ defines a purification of the total mixture for all operators $U$, $V_{\mathrm{anc}}$ acting on $\mathcal{H}, \mathcal{H}_{\mathrm{anc}}$, respectively. In order to see this, it is sufficient to verify that the following identity

$$
\mathrm{tr}_{\mathrm{anc}}(U \otimes V_{\mathrm{anc}})\Omega(U^* \otimes V_{\mathrm{anc}}^*) = U(\mathrm{tr}_{\mathrm{anc}}\Omega)U^* \tag{3.61}
$$

holds for all states $\Omega$. Since in our case $\mathrm{tr}_{\mathrm{anc}} |\psi\rangle\langle\psi| = \frac{1}{d} I$, it follows that

$$
U(\mathrm{tr}_{\mathrm{anc}} |\psi\rangle\langle\psi|)U^* = \frac{1}{d} I \,,
$$

which proves that $\psi'$ are purifications of the total mixture.

**Exercise 24.** Prove the identity (3.61). (Hint: $\Omega = \omega_A \otimes \omega_B + \Gamma$.)

## 4  Observables

### 4.1  Observables as positive operator valued measures

Standard references for the topics of this section are [45] and [15].

#### 4.1.1  Definition and basic properties

Intrinsic randomness of quantum measurements is one of the key features of quantum theory. An experiment produces a sequence of outcomes, each occurring with a certain probability depending on the particular setting of the measuring device and preparation of the measured system. Quantum theory predicts only the probabilities of measurement outcomes.

According to Chapter 3 each observed outcome in quantum theory is associated with a particular effect that determines its probability of occurrence via the trace formula. Therefore, an observable describing an experiment with some number of outcomes is represented by a collection of effects. Moreover, since in each run of the experiment some outcome is recorded, it follows that the sum of probabilities for different outcomes equals to one.

Let us first recall some definitions from probability theory. Let $\Omega$ be a set. A $\sigma$-*algebra* on $\Omega$ is a nonempty collection $\mathcal{F}$ of subsets of $\Omega$ that is closed under complements and countable unions. A set $X \in \mathcal{F}$ is called an *event*. A *probability measure* is a mapping $p : \mathcal{F} \to [0, 1]$ that satisfies the following conditions:

(i)  $p(\emptyset) = 0$ ;

(ii)  $p(\Omega) = 1$ ;

(iii)  $p(\cup_i X_i) = \sum_i p(X_i)$ for any sequence $\{X_i\}$ of disjoint sets in $\mathcal{F}$ .

Consequently, the mathematical description of observables is based on the following generalization of probability measures.

**Definition 26.** A *normalized positive operator valued measure* (POVM) on a measurable space $(\Omega, \mathcal{F})$ is a mapping $\mathsf{A} : \mathcal{F} \to \mathcal{E}(\mathcal{H})$ such that

(i)  $\mathsf{A}(\emptyset) = O$ ;

(ii)  $\mathsf{A}(\Omega) = I$ ;

(iii)  $\mathsf{A}(\cup_i X_i) = \sum_i \mathsf{A}(X_i)$ (in the weak sense) for any sequence $\{X_i\}$ of disjoint sets in $\mathcal{F}$ .

There is an alternative way to look at POVMs which makes the physical interpretation transparent. Let $Prob(\Omega, \mathcal{F})$ be the set of all probability measures on a measurable space $(\Omega, \mathcal{F})$. A POVM A, defined on $(\Omega, \mathcal{F})$, determines a mapping $\Phi_{\mathsf{A}}$ from $\mathcal{S}(\mathcal{H})$ to $Prob(\Omega, \mathcal{F})$ by the formula

$$\Phi_{\mathsf{A}}(\varrho) := \operatorname{tr}[\varrho \mathsf{A}(\cdot)] . \tag{4.1}$$

The mapping $\Phi_{\mathsf{A}}$ is affine: for any $\varrho_1, \varrho_2 \in \mathcal{S}(\mathcal{H})$ and $0 \le \lambda \le 1$, we have

$$\Phi_{\mathsf{A}}(\lambda \varrho_1 + (1 - \lambda)\varrho_2) = \lambda \Phi_{\mathsf{A}}(\varrho_1) + (1 - \lambda)\Phi_{\mathsf{A}}(\varrho_2) . \tag{4.2}$$

Conversely, it follows from our earlier discussion on effects in Section 3.1 that a mapping $\Phi : \mathcal{S}(\mathcal{H}) \to Prob(\Omega, \mathcal{F})$ satisfying (4.2) comes from a unique POVM A through formula (4.1).

**Definition 27.** From now on, we will identify observables with POVMs. For an observable A defined on a measurable space $(\Omega, \mathcal{F})$, we say that $\Omega$ is *the set of (measurement) outcomes*, and $(\Omega, \mathcal{F})$ is the *outcome space* of A.

**Example 28.** (*Ideal Stern-Gerlach apparatus*) The Stern-Gerlach apparatus is a device measuring a spin component. In this measurement, a beam of particles passes between the poles of a magnet. The magnetic force depends upon the spin state, and in an idealized picture the beam is splitted into well separate parts. Therefore, the value of the spin component can be inferred from the observed deflection of the beam. In practice this method is useful only for neutral particles since for charged particles the Lorentz force obscures the deflection.

Sending a spin-$\frac{1}{2}$ particle through the Stern-Gerlach apparatus (oriented in the direction $\vec{n}$), it is deflected by an angle $\pm\theta$ with respect to the axis of the apparatus. After that, the particle is detected in one of two detectors labeled by values $\uparrow$ and $\downarrow$. The corresponding effects are denoted by $E_\uparrow$ and $E_\downarrow$. As we assume that each particle in the beam is detected, we have $E_\uparrow + E_\downarrow = I$. The set of measurement outcomes is thus $\Omega = \{\uparrow, \downarrow\}$ and the observable A describing this measurement is defined as $\mathsf{A}(\{\uparrow\}) = E_\uparrow$, $\mathsf{A}(\{\downarrow\}) = E_\downarrow$.

In the ideal situation these effects are the projections $E_\uparrow = \frac{1}{2}(I + \vec{n} \cdot \vec{\sigma})$ and $E_\downarrow = \frac{1}{2}(I - \vec{n} \cdot \vec{\sigma})$. For a general state $\varrho = \frac{1}{2}(I + \vec{r} \cdot \vec{\sigma})$, the probabilities to detect either $\uparrow$ or $\downarrow$ are then $\frac{1}{2}(1 + \vec{r} \cdot \vec{n})$ and $\frac{1}{2}(1 - \vec{r} \cdot \vec{n})$, respectively.

In the following proposition we go through some basic mathematical properties of observables, which are useful later.

**Proposition 41.** Let A be an observable with an outcome space $(\Omega, \mathcal{F})$ and let $X, Y \in \mathcal{F}$.

(a) If $X \subseteq Y$, then $\mathsf{A}(X) \leq \mathsf{A}(Y)$.

(b) If $X \subseteq Y$ and $\mathsf{A}(Y) = O$, then $\mathsf{A}(X) = O$.

(c) $\mathsf{A}(X \cup Y) + \mathsf{A}(X \cap Y) = \mathsf{A}(X) + \mathsf{A}(Y)$.

*Proof.* (a) Since $X \subseteq Y$, we can write $Y$ as a disjoint union $Y = X \cup (Y \smallsetminus X)$. We then have $\mathsf{A}(Y) = \mathsf{A}(X) + \mathsf{A}(Y \smallsetminus X)$, which implies that $\mathsf{A}(X) \leq \mathsf{A}(Y)$.

(b) This is a direct consequence of (a) since $O \leq \mathsf{A}(X)$ by the definition of an observable.

(c) Since $X \subseteq X \cup Y$, we can write $X \cup Y$ as a disjoint union $X \cup Y = X \cup Z$, where $Z \equiv (X \cup Y) \smallsetminus X$. This gives $\mathsf{A}(X \cup Y) = \mathsf{A}(X) + \mathsf{A}(Z)$. Adding $\mathsf{A}(X \cap Y)$ to both sides of this equation and noticing that $Y$ is a disjoint union of $X \cap Y$ and $Z$, we get (c). □

Let A be an observable with an outcome space $(\Omega, \mathcal{F})$. We denote by ran A the range of A, that is,

$$\text{ran } \mathsf{A} := \{\mathsf{A}(X) \mid X \in \mathcal{F}\} \subseteq \mathcal{E}(\mathcal{H}).$$

Since observables are specific kind of set functions, we expect that their ranges have also some specific properties or limitations. A useful concept in studying the structure of observables from this point of view is *coexistence* of effects.

**Definition 28.** A set of effects $\mathcal{C} \subseteq \mathcal{E}(\mathcal{H})$ is *coexistent* if there exists an observable A such that $\mathcal{C} \subseteq \operatorname{ran} A$.

Hence, a measurement of a coexistent set of effects can be realized by measuring a single observable. It is easy to invent examples of both coexistent and non-coexistent sets of effects. For instance, a set $\{A_1, A_2, \ldots, A_n\}$ satisfying $\sum_{i=1}^n A_i \leq I$ is coexistent (see Subsection 4.1.2). On the other hand, a set $\{P, A\}$ of a projection $P$ and an effect $A$ is coexistent only if $P$ and $A$ commute. Generally, however, a method to check whether a set $\mathcal{C} \subseteq \mathcal{E}(\mathcal{H})$ is coexistent or not is not known. Even if $\mathcal{C}$ consists of two effects only, a complete characterization of coexistent sets $\mathcal{C}$ is known just in the simplest case of two dimensional Hilbert space [19, 77, 85]. For more details on coexistence and some related concepts, we refer to survey article [56].

**Example 29.** (*Coexistence of commutative effects*) Let $A, B \in \mathcal{E}(\mathcal{H})$. If $AB = BA$, then the set $\{A, B\}$ is coexistent. To prove this claim, we need to find an observable A such that $\{A, B\} \subseteq \operatorname{ran} A$. A possible choice for A is provided by choosing the outcome set to be $\Omega = \{1, 2, 3, 4\}$ and defining

$$\mathsf{A}(\{1\}) = AB\,, \quad \mathsf{A}(\{2\}) = (I-A)B\,, \quad \mathsf{A}(\{3\}) = A(I-B)\,, \quad \mathsf{A}(\{4\}) = (I-A)(I-B)\,.$$

This defines an observable. For instance, the operator $AB$ is an effect since $AB = A^{\frac{1}{2}}BA^{\frac{1}{2}}$. Moreover, it is clear that $\mathsf{A}(\Omega) = I$. We have $\mathsf{A}(\{1,3\}) = A$ and $\mathsf{A}(\{1,2\}) = B$. Therefore, the set $\{A, B\}$ is coexistent.

### 4.1.2   Discrete observables

Let $\Omega = \{x_1, x_2, \ldots\}$ be a countable set (i.e. either finite or countably infinite). In this case we always choose the $\sigma$-algebra $\mathcal{F}$ to be the collection of all subsets of $\Omega$, i.e., $\mathcal{F}$ is the power set $2^\Omega$. An observable A with the outcome space $(\Omega, 2^\Omega)$ is now completely defined by the effects $\mathsf{A}(x_j) \equiv \mathsf{A}(\{x_j\})$. Notice that formally we have to write $\{x\}$ for the set consisting of the single element $x$, since an observable is defined on events and not on outcomes. However, in the case of a countable outcome space it this convenient to drop the brackets.

For an arbitrary subset $X \subseteq \Omega$, we have

$$\mathsf{A}(X) = \sum_{x_j \in X} \mathsf{A}(x_j)\,. \tag{4.3}$$

Especially, the normalization condition $\mathsf{A}(\Omega) = I$ reads

$$\sum_{x_j \in \Omega} \mathsf{A}(x_j) = I\,. \tag{4.4}$$

It follows that for a countable outcome space, we can think observable as a mapping

$$x_j \mapsto \mathsf{A}(x_j) \tag{4.5}$$

from $\Omega$ to the set of effects $\mathcal{E}(\mathcal{H})$, satisfying the normalization condition (4.4). The effects $\mathsf{A}(X)$ for other kind of sets $X$ than singleton sets are recovered using (4.3).

We adopt the following definition for a discrete observable.

**Definition 29.** An observable A, defined on an outcome space $(\Omega, \mathcal{F})$, is called *discrete* if there is a countable set $\Omega_0 \in \mathcal{F}$ such that $\mathsf{A}(\Omega_0) = I$.

In particular, all observables with countable outcome space are discrete. On the other hand, the outcome space of a discrete observable can be uncountable (e.g. the real line $\mathbb{R}$). However, this is just a trivial difference as the outcome space can be redefined to be the countable set $\Omega_0$ without changing the essence of the observable. The terminology in Definition 29 is set in that way since it is convenient sometimes to have freedom in the choice of the outcome space and we do not want the property of being discrete to depend on that choice.

**Example 30.** (*Observables with finite outcome set*) Sometimes in the literature one meets with the definition of an observable as a finite collection of effects $E_1, \ldots, E_n$ satisfying $\sum_j E_j = I$. Naturally, one is then restricted to observables with finite outcome sets. This definition should be understood in the sense that we first fix an outcome set of $n$ elements (e.g. $\{1, \ldots, n\}$) and then adopt the definition related to the formula (4.5).

### 4.1.3   Real observables

If the outcome set $\Omega$ is $\mathbb{R}$, then it is not convenient to choose the $\sigma$-algebra $\mathcal{F}$ to be the collection of all subsets of $\mathbb{R}$. Instead, it is common to choose $\mathcal{F}$ to be the Borel $\sigma$-algebra $\mathcal{B}(\mathbb{R})$. This $\sigma$-algebra contains all the sets that are needed in calculations. For instance, all open and closed intervals, their complements and countable unions, belong to $\mathcal{B}(\mathbb{R})$. For our purposes this is just a technical remark and a reader who is not familiar with $\mathcal{B}(\mathbb{R})$ before can simply think it as a $\sigma$-algebra on $\mathbb{R}$ that contains all subsets of $\mathbb{R}$ that she can imagine.

**Definition 30.** An observable A is *real*, or *real valued*, if the outcome set of A is either $\mathbb{R}$ or a subset of $\mathbb{R}$. In this case the $\sigma$-algebra is chosen to be the corresponding Borel $\sigma$-algebra.

Let A be an observable with the outcome space $(\mathbb{R}, \mathcal{B}(\mathbb{R}))$. We can then define the *average value*, or *mean value*, of A in a state $\varrho$ as

$$\langle \mathsf{A} \rangle_\varrho := \int_{\mathbb{R}} x \, \mathrm{tr} \left[ \varrho \mathsf{A}(dx) \right] . \tag{4.6}$$

It may happen that the integral in the right hand side of Eq. (4.6) does not converge, in which case we denote $\langle \mathsf{A} \rangle_\varrho = \infty$.

The *variance* $\Delta_\varrho(\mathsf{A})$ of A in a state $\varrho$ is defined as the average distance of the measured value from the mean value of A, i.e.,

$$\Delta_\varrho(\mathsf{A}) = \sqrt{\int_{\mathbb{R}} (x - \langle \mathsf{A} \rangle_\varrho)^2 \, \mathrm{tr} \left[ \varrho \mathsf{A}(dx) \right]} . \tag{4.7}$$

In a similar way the mean value and the variance can be defined also in the cases when $\Omega$ is a subset of $\mathbb{R}$. If, for instance, $\Omega$ is a countable set of real numbers $x_1, x_2, \ldots$, then we have

$$\langle \mathsf{A} \rangle_\varrho = \sum_{x_j \in \Omega} x_j \, \mathrm{tr} \left[ \varrho \mathsf{A}(x_j) \right] , \tag{4.8}$$

$$\Delta_\varrho(\mathsf{A}) = \sqrt{\sum_j (x_j - \langle \mathsf{A} \rangle_\varrho)^2 \, \mathrm{tr} \left[ \varrho \mathsf{A}(x_j) \right]} . \tag{4.9}$$

### 4.1.4 Mixtures of observables

Similarly like in the case of states, we can perform an experiment which is a statistical mixture of two experiments using the same states but different measurement apparatuses. Hence, having two measurement apparatuses we can design a mixture of them. This corresponds to having a convex combination of observables. For simplicity, we discuss here convex combinations only in the case of observables with finite outcome spaces.

Consider two observables A and B with finite outcome spaces $\Omega_A = \{a_1, \ldots, a_n\}$, $\Omega_B = \{b_1, \ldots, b_m\}$, respectively. We define $\Omega_C = \Omega_A \cup \Omega_B$, and then extend A and B to this new outcome space by setting

$$A(X) \;=\; A(X \cap \Omega_A), \quad B(X) = B(X \cap \Omega_B) \tag{4.10}$$

for all $X \subseteq \Omega_C$. In other words, the extension means that we define $A(b_j) = O$ and $B(a_i) = O$. This extension does not change the probability distributions associated to A and B.

Fix a number $0 < \lambda < 1$. An observable C with the outcome set $\Omega_C$ is now defined as

$$C(X) = \lambda A(X) + (1 - \lambda) B(X) \tag{4.11}$$

for all $X \subseteq \Omega_C$. We denote $C = \lambda A + (1 - \lambda) B$, and say that the observable C is a *convex combination* of observables A and B.

**Example 31.** Consider an experiment in which we are randomly switching between observables A, B, both having only two outcomes. If the relative frequency of having A is $\lambda$, then the mixture corresponds to the convex combination $C = \lambda A + (1 - \lambda) B$, which is given by the following four outcome POVM

$$C: \quad a_1 \mapsto \lambda A(a_1), \qquad\qquad a_2 \mapsto \lambda A(a_2),$$
$$\quad\quad b_1 \mapsto (1 - \lambda) B(b_1), \qquad\quad b_2 \mapsto (1 - \lambda) B(b_2).$$

If the outcome sets $\Omega_A$ and $\Omega_B$ are the same, then we can also have a different kind of mixture than above. Namely, we can have a situation where, after a measurement outcome is registered, we cannot anymore track it down to either A or B. This leads to a different kind of convex combination, illustrated in the following example.

**Example 32.** Let A and B be two observables, both having the outcome space $\Omega_A = \Omega_B = \{a_1, a_2\}$. Consider an experiment in which we try to measure A, but due to some noise or fluctuation we actually measure B in some cases. Suppose that we manage to measure A in $\lambda$-part of the experimental runs, but we do not know when the noise affects the measurement. In this case the convex combination $C = \lambda A + (1 - \lambda) B$ is given by the following two outcome POVM

$$C: \quad a_1 \mapsto \lambda A(a_1) + (1 - \lambda) B(a_1), \qquad a_2 \mapsto \lambda A(a_2) + (1 - \lambda) B(a_2).$$

Depending on the situation, there can also be intermediate cases of the two different convex combinations described in Examples 31 and 32. This is demonstrated in the following example.

**Example 33.** Let A and B be two observables with outcome spaces $\Omega_A = \{a_1, a_2\}$ and $\Omega_B = \{b_1, a_2\}$. Suppose that the measurements are mixed and the second outcomes are identified as one. The convex combination $C = \lambda A + (1 - \lambda) B$ is then the following three outcome observable

$$C: \quad a_1 \mapsto \lambda A(a_1), \qquad b_1 \mapsto (1 - \lambda) B(b_1), \qquad a_2 \mapsto \lambda A(a_2) + (1 - \lambda) B(a_2).$$

We use the notation $C = \lambda A + (1 - \lambda) B$ for all these different types of convex combinations. The explicit form of $C$ is specified when the outcome space $\Omega_C$ is given.

### 4.2 Sharp observables

Sharp observables form a specific class of observables which deserves an additional treatment. One should notice that in some context (especially in older literature) the concept of an observable refers only to what we here call sharp observables.

#### 4.2.1 Definition and basic properties

**Definition 31.** A POVM A is a *normalized projection valued measure* (PVM) if $A(X)$ is a projection for every $X \in \mathcal{F}$. The corresponding observable (identified with A) is called *sharp observable*.

**Example 34.** (*Sharp observable associated to an orthonormal basis*) Let $\mathcal{H}$ be a $d$-dimensional Hilbert space (either $d < \infty$ or $d = \infty$) and $\{\varphi_j\}_{j=1}^d$ an orthonormal basis for $\mathcal{H}$. For each $j = 1, \ldots, d$, we denote $A(j) := |\varphi_j\rangle\langle\varphi_j|$. Thus, $A(j)$ is a one-dimensional projection. Since $\{\varphi_j\}_{j=1}^d$ is an orthonormal basis, the normalization condition $\sum_{j=1}^d A(j) = I$ holds (see Example 16 in Section 2.3.2). The mapping $j \mapsto A(j)$ defines a discrete observable with the outcome set $\{1, \ldots, d\}$. We say that A is the sharp observable associated to the orthonormal basis $\{\varphi_j\}_{j=1}^d$. This type of observables are common in applications and we are going to use them later in various instances.

**Proposition 42.** Let A be an observable with an outcome space $(\Omega, \mathcal{F})$. The following conditions are equivalent.

(i) A is sharp.

(ii) $A(X)A(Y) = A(X \cap Y)$ for every $X, Y \in \mathcal{F}$.

(iii) $A(X)A(\neg X) = O$ for every $X \in \mathcal{F}$.

*Proof.* We prove this proposition by showing that (i)⇒(ii)⇒(iii)⇒(i).

Assume that (i) holds and let $X, Y \in \mathcal{F}$. By Prop. 41a, we have $A(X \cap Y) \leq A(X) \leq A(X \cup Y)$. As all these operators are projections, Prop. 13 implies that $A(X)A(X \cap Y) = A(X \cap Y)$ and $A(X)A(X \cup Y) = A(X)$. Hence, multiplying the equation in Prop. 41c by $A(X)$ gives $A(X) + A(X \cap Y) = A(X) + A(X)A(Y)$, where (ii) follows.

Assume then that (ii) holds and let $X \in \mathcal{F}$. Choosing $Y = \neg X$ in (ii), we get

$$A(X \cap \neg X) = A(\emptyset) = O.$$

Finally, assume that (iii) holds and let $X \in \mathcal{F}$. We then get

$$O = \mathsf{A}(X)\mathsf{A}(\neg X) = \mathsf{A}(X)\left(I - \mathsf{A}(X)\right) = \mathsf{A}(X) - \mathsf{A}(X)^2 \,,$$

and hence, $\mathsf{A}(X) = \mathsf{A}(X)^2$. This means that $\mathsf{A}$ is sharp.                    □

**Proposition 43.** The range of a sharp observable $\mathsf{A}$ consists of mutually commuting projections.

*Proof.* Let $\mathsf{A}(X)$ and $\mathsf{A}(Y)$ be projections from the range of $\mathsf{A}$. Condition (ii) in Prop. 42 implies that

$$\mathsf{A}(X)\mathsf{A}(Y) = \mathsf{A}(X \cap Y) = \mathsf{A}(Y \cap X) = \mathsf{A}(Y)\mathsf{A}(X) \,,$$

which proves the claim.                    □

Sharp observables form a distinct class among all observables. However, it is not easy to point out operationally meaningful criterion which is valid only for sharp observables. For instance, one characteristic property of any sharp observable $\mathsf{A}$ is that for every $X \in \mathcal{F}$ with $\mathsf{A}(X) \neq O$, there exists a state $\varrho$ such that $\mathrm{tr}\left[\varrho\mathsf{A}(X)\right] = 1$. However, this condition does not require an observable to be sharp. Namely, it can happen that an observable $\mathsf{B}$ satisfies this criterion without being sharp. The condition $\mathrm{tr}\left[\varrho\mathsf{B}(X)\right] = 1$ can be satisfied if the operator $\mathsf{B}(X)$ has eigenvalue 1.

**Proposition 44.** Let $E$ be an effect and $\varrho$ a pure state. The following conditions are equivalent:

  (i)  $\mathrm{tr}\left[\varrho E\right] = 1$ ;

  (ii)  $E\varrho = \varrho$ ;

  (iii)  $\varrho \leq E$ .

*Proof.* We prove this proposition by showing that (i)⇒(ii)⇒(iii)⇒(i). In the following, we fix a unit vector $\psi \in \mathcal{H}$ such that $\varrho = |\psi\rangle\langle\psi|$.

Suppose that (i) holds. Then

$$1 = \mathrm{tr}\left[\varrho E\right] = \langle\,\psi\,|\,E\psi\,\rangle = |\langle\,\psi\,|\,E\psi\,\rangle| \leq \|\psi\|\,\|E\psi\| \leq \|E\| \leq 1 \,,$$

where the first inequality follows from the Cauchy-Schwarz inequality and the second from ineq. (2.9). This shows that there is actually an equality in the first inequality, which can be true only if the vectors $E\psi$ and $\psi$ are collinear, i.e., $E\psi = c\psi$ for some $c \in \mathbb{C}$. Moreover, $1 = \langle\,\psi\,|\,E\psi\,\rangle = c$ and hence, $E\psi = \psi$. Thus, $E\varrho = \varrho$ and (ii) holds.

Suppose that (ii) holds. Since

$$\varrho E = \varrho^* E^* = (E\varrho)^* = \varrho^* = \varrho = E\varrho \,,$$

we conclude that the operators $E$ and $\varrho$ commute. By Theorem 1, also the square-root, $E^{\frac{1}{2}}$, commutes with $\varrho$. Hence, (ii) implies that $\varrho = E^{\frac{1}{2}}\varrho E^{\frac{1}{2}}$. For every vector $\phi \in \mathcal{H}$, we then get

$$\langle\,\phi\,|\,\varrho\phi\,\rangle = \left\langle\,E^{\frac{1}{2}}\phi\,|\,\varrho E^{\frac{1}{2}}\phi\,\right\rangle \leq \left\|E^{\frac{1}{2}}\phi\right\|\,\left\|\varrho E^{\frac{1}{2}}\phi\right\| \leq \|\varrho\|\,\left\|E^{\frac{1}{2}}\phi\right\|^2 \leq \langle\,\phi\,|\,E\phi\,\rangle \,.$$

The first inequality is an application of the Cauchy-Schwarz inequality, in the second we have used ineq. (2.9), and the third follows from the fact that $\|\varrho\| = 1$. This shows that $\varrho \leq E$.

Finally, suppose that (iii) holds. Then

$$\mathrm{tr}\,[\varrho E] = \langle\,\psi\,|\,E\psi\,\rangle \geq \langle\,\psi\,|\,\varrho\psi\,\rangle = 1\,.$$

Hence, (i) holds.                                                                                       □

**Proposition 45.** Let $\mathcal{H}$ be a finite dimensional Hilbert space of dimension $d$. If A is a sharp observable on $\mathcal{H}$, then A is discrete and the outcome set $\Omega_\mathsf{A}$ of A has at most $d$ elements such that $\mathsf{A}(j) \neq O$.

*Proof.* By Prop. 42 two projections $\mathsf{A}(X)$ and $\mathsf{A}(Y)$ are orthogonal whenever $X \cap Y = \emptyset$. Hence, to prove the claim it is enough to show that there are at most $d$ pairwise orthogonal projections.

Let us make a counter assumption that $P_1, \ldots, P_r$ are pairwise orthogonal projections and $r \geq d+1$. Each projection $P_j$ has an eigenvector $\varphi_j$ with eigenvalue 1. If $i \neq j$, then the vectors $\varphi_i$ and $\varphi_j$ are orthogonal. Namely,

$$\langle\,\varphi_i\,|\,\varphi_j\,\rangle = \langle\,P_i\varphi_i\,|\,P_j\varphi_j\,\rangle = \langle\,\varphi_i\,|\,P_iP_j\varphi_j\,\rangle = 0\,.$$

This means that there are $r$ orthogonal vectors in $\mathcal{H}$. This cannot be true as the dimension of $\mathcal{H}$ is $d$ and hence, we conclude that the counter assumption is false.                        □

The statement in Proposition 45 is not true if the dimension of $\mathcal{H}$ is infinite. The prototypical instance of a non-discrete sharp observable is the canonical position observable on $\mathbb{R}$, which we recall in the following example.

**Example 35.** (*The canonical position observable*) Let $\mathcal{H} = L^2(\mathbb{R})$. The *canonical position observable* Q has the outcome space $(\mathbb{R}, \mathcal{B}(\mathbb{R}))$ and it is defined as

$$\mathsf{Q}(X)\psi(x) = \chi_X(x)\psi(x)\,,$$

where $\chi_X$ is the characteristic function of a set $X$. For a pure state $\varrho = |\psi\rangle\langle\psi|$, we thus have

$$\mathrm{tr}\,[\varrho\mathsf{Q}(X)] = \langle\,\psi\,|\,\mathsf{Q}(X)\psi\,\rangle = \int_X |\psi(x)|^2\,dx\,.$$

This is the probability that a particle in the state $\varrho$ is localized within the set $X$. Notice that for a one element set $X = \{x\}$, this probability is zero for every state $\varrho$. It follows that Q is not a discrete observable.

### 4.2.2 Sharp observables and selfadjoint operators

In elementary courses on quantum mechanics, observables are defined and understood as selfadjoint operators. Let us briefly clarify the difference between this formalism and Definition 27 in Section 4.1.

The spectral theorem for selfadjoint operators says that for each (bounded or unbounded) selfadjoint operator $A$, there is a unique normalized projection valued measure $\mathsf{A}$ on the Borel space $(\mathbb{R}, \mathcal{B}(\mathbb{R}))$ such that

$$A = \int_{\mathbb{R}} x \, \mathsf{A}(dx) \, . \tag{4.12}$$

This symbolic expression means that for every vector $\psi$ in the domain of $A$, we have

$$\langle \, \psi \, | \, A\psi \, \rangle = \int_{\mathbb{R}} x \, \langle \, \psi \, | \, \mathsf{A}(dx)\psi \, \rangle \, . \tag{4.13}$$

In the typical instances of unbounded operators the domain is chosen to be the natural domain $\{\psi \in \mathcal{H} \mid A\psi \in \mathcal{H}\}$.

On the other hand, each normalized projection valued measure $\mathsf{A}$ on $(\mathbb{R}, \mathcal{B}(\mathbb{R}))$ determines a unique selfadjoint operator $A$ through formula (4.12). Thus, in the language introduced in earlier sections, selfadjoint operators give an alternative description for real sharp observables. We refer to [73] for an explanation of the spectral theorem.

Note that Eq.(4.13) implies that the mean value of $\mathsf{A}$ in a state $\varrho$ can be expressed as

$$\langle \mathsf{A} \rangle_\varrho = \operatorname{tr} \left[ \varrho A \right] =: \langle A \rangle_\varrho \tag{4.14}$$

for all states $\varrho$. Taking into account that

$$\langle \, \psi \, | \, A^2\psi \, \rangle = \int_{\mathbb{R}} x^2 \, \langle \, \psi \, | \, \mathsf{A}(dx)\psi \, \rangle \, ,$$

it follows that the variance of $\mathsf{A}$ in $\varrho$ can be written in terms of the operator $A$ in the following way:

$$\Delta_\varrho(\mathsf{A}) = \sqrt{\operatorname{tr} \left[ \varrho A^2 \right] - \langle A \rangle_\varrho^2} = \sqrt{\langle A^2 \rangle_\varrho - \langle A \rangle_\varrho^2} \, . \tag{4.15}$$

We conclude that the selfadjoint operator $A$ gives a convenient way to calculate mean values and variances. In some cases other statistical information is not even needed.

As we have earlier seen, the probabilistic structure of quantum mechanics leads in a natural way to POVMs as a correct formalization of observables. PVMs describe just one specific class of observables, namely sharp observables. It is sometimes simple and convenient to use selfadjoint operators to describe real sharp observables. However, it is untenably restrictive to use only selfadjoint operators and neglect other kind of observables.

**Example 36.** (*Sharp qubit observables*) Let us consider a qubit system, i.e., a two dimensional Hilbert space $\mathcal{H}$. Any sharp observable $\mathsf{A}$ on $\mathcal{H}$ can have only two outcomes (say $\pm 1$). The corresponding operators $\mathsf{A}(1)$ and $\mathsf{A}(-1)$ are necessarily one dimensional projections.

As explained in Example 20, the states of the qubit system can be described by the Bloch sphere vectors. In particular, each unit vector $\vec{a}$ in the Bloch sphere determines a one dimensional projection $P[\vec{a}] := \frac{1}{2}(I + \vec{a} \cdot \vec{\sigma})$. Thus, the vector $\vec{a}$ defines a sharp two outcome observable $\mathsf{A}$ by

$$\mathsf{A}(1) = P[\vec{a}], \quad \mathsf{A}(-1) = I - P[\vec{a}] = P[-\vec{a}] \, .$$

The selfadjoint observable $A$ corresponding to the sharp observable $\mathsf{A}$ through the spectral formula (4.12) is

$$A = \sum_{j=\pm 1} j\mathsf{A}(j) = P[\vec{a}] - (I - P[\vec{a}]) = \vec{a} \cdot \vec{\sigma}\,.$$

**Exercise 25.** Referring to Example 36, calculate the mean value and the variance of $\mathsf{A}$ in a state $\varrho = \frac{1}{2}(I + \vec{r} \cdot \vec{\sigma})$.

In the case of a finite dimensional Hilbert space $\mathcal{H}$, the spectral theorem states, essentially, that every selfadjoint operator is diagonalizable. Let us shortly recall the diagonalization procedure. First of all, the spectrum of a selfadjoint operator $A \in \mathcal{L}_s(\mathcal{H})$ consists of eigenvalues $\lambda$ satisfying the identity $\det[A - \lambda I] = 0$. The eigenvectors corresponding to an eigenvalue $\lambda$ are obtained by solving the equation $A\psi = \lambda\psi$. For a given eigenvalue $\lambda$, the associated eigenvectors form a linear subspace $\mathcal{H}_\lambda \subset \mathcal{H}$. For different eigenvalues, the subspaces $\mathcal{H}_\lambda$ are mutually orthogonal. The dimension $d_\lambda$ of $\mathcal{H}_\lambda$ is called the degeneracy of the eigenvalue $\lambda$. Let $P_\lambda : \mathcal{H} \to \mathcal{H}$ be the projection mapping $\mathcal{H}$ onto $\mathcal{H}_\lambda$. Then the self-adjoint operator $A$ can be expressed in the diagonal (spectral) form as $A = \sum_j \lambda_j P_{\lambda_j}$. The eigenvalues of $A$ define the outcome space $\Omega$ and the normalized projection valued measure associated with the operator $A$ is given as $\mathsf{A}(\{\lambda_j\}) = P_{\lambda_j}$.

The spectral theorem in an infinite dimensional Hilbert space has two complications. First of all, eigenvalues are not generally enough to give a spectral representation of a selfadjoint operator. One needs a more general concept of spectrum, which leads to continuous spectral representations. The second point is that not all PVMs correspond to bounded selfadjoint operators - unbounded operators are unavoidable in this context. We do not go into details of these topic, but just discuss some common examples.

**Example 37.** (*Position operator*) Let us consider again the canonical position observable $\mathsf{Q}$ introduced in Example 35. The *position operator* $Q$ is formally defined as

$$Q = \int_{\mathbb{R}} x\mathsf{Q}(dx)\,.$$

However, $Q$ is an unbounded operator. For a vector $\psi$ belonging to the domain of $Q$, we get

$$\langle\,\psi\,|\,Q\psi\,\rangle = \int x\,\langle\,\psi\,|\,\mathsf{Q}(dx)\psi\,\rangle = \int x\,|\psi(x)|^2\,dx\,.$$

Thus, the action of the position operator $Q$ is given by

$$Q\psi(x) = x\psi(x)\,. \tag{4.16}$$

**Exercise 26.** (*Momentum operator*) The *canonical momentum observable* $\mathsf{P}$ is Fourier connected to the canonical position observable $\mathsf{Q}$. Hence, for any $X \in \mathcal{B}(\mathbb{R})$ we have

$$\mathsf{P}(X) = \mathcal{F}^{-1}\mathsf{Q}(X)\mathcal{F}\,,$$

where $\mathcal{F}$ is the unitary extension of the Fourier transform to $L^2(\mathbb{R})$. In particular, for a pure state $\varrho = |\psi\rangle\langle\psi|$ we have

$$\mathrm{tr}\,[\varrho\mathsf{P}(X)] = \int_X |(\mathcal{F}\psi)(x)|^2\,dx\,.$$

The corresponding (unbounded) selfadjoint operator $P = \mathcal{F}^{-1}Q\mathcal{F}$ is called the *momentum operator*. Show that for a function $\psi$ belonging to the Schwartz space (smooth functions of rapid decrease), the momentum operator $P$ acts by

$$P\psi(x) = -i\hbar \frac{\partial}{\partial x}\psi(x),$$

where $\hbar$ is the *Planck constant*. (Hint: Start from (4.16) and recall how the multiplication behaves in the Fourier transform.)

**Example 38.** (*Energy operator of hydrogen atom.*) The analysis of the hydrogen atom was one of the first successful demonstrations of elementary principles of quantum theory. In particular, it was suggested by Erwin Schrödinger that the energy is associated with the selfadjoint operator

$$H = -\frac{1}{2m}\vec{P}\cdot\vec{P} - \frac{e^2}{4\pi\epsilon_0}|\vec{Q}|^{-1}, \tag{4.17}$$

where $m$ is the mass of electron, $e$ is the elementary electron charge and $\epsilon_0$ is the electrical permittivity of space. This operator acts as follows

$$H\psi(\vec{r}) = -\frac{\hbar^2}{2m}\left(\frac{\partial^2}{\partial x^2} + \frac{\partial^2}{\partial y^2} + \frac{\partial^2}{\partial z^2}\right)\psi(\vec{r}) - \frac{e^2}{4\pi\epsilon_0}\frac{1}{|\vec{r}|}\psi(\vec{r}) \tag{4.18}$$

for all $\psi \in L^2(\mathbb{R}^3)$ belonging to domain of $H$. The eigenvalues of $H$ are degenerated and read

$$E_n = -\frac{me^2}{32\pi^2\epsilon_0^2\hbar^2}\frac{1}{n^2} = -13.6\text{eV}\frac{1}{n^2} \tag{4.19}$$

for $n = 1,\ldots,\infty$. Each eigenvalue is associated with eigenvectors $\psi_{nlm} \in L^2(\mathbb{R}^3)$, where $l, m$ are integer numbers such that $0 \leq l < n$ and $-l \leq m \leq l$. It follows that eigenvalue $E_n$ is associated with $2n+1$ dimensional projection $\Pi_n = \sum_{l,m}|\psi_{nlm}\rangle\langle\psi_{nlm}|$. It follows that the energy operator $H$ defines an energy observable of the hydrogen atom

$$E_n \mapsto \mathsf{A}_{\text{energy}}(E_n) = \Pi_n, \tag{4.20}$$

that is sharp and discrete. Let us note that the energy is usually determined by measuring the emission spectrum. The corresponding observable is described in Example 39.

**Exercise 27.** Verify that

$$\psi(\vec{r}) = \frac{1}{\sqrt{\pi a^3}}e^{-|\vec{r}|/a} \tag{4.21}$$

is an eigenvector of $H$ for a specific value of parameter $a$ known as *Bohr radius*.

**Example 39.** (*Spectrum of Hydrogen atom.*) Consider a measurement of the emission spectrum of a hydrogen atom. In an ideal version of such experiment a heated ensemble of hydrogen atoms starts to emit light of some particular frequencies forming the emission spectrum. According to the basic theory, the observed frequencies are related to energy eigenstates of the hydrogen atom via the relation $\hbar\omega_{nm} = E_1(\frac{1}{n^2} - \frac{1}{m^2})$, where $n < m$ and $E_1 = -13.6eV$ is the ground state energy, i.e. the lowest possible energy of the hydrogen atom. The observable is real and discrete, i.e. the outcome space $\Omega$ consists of a countable number of frequencies $\omega_\mu$.

Thus, for each observed frequency $\omega_\mu$ there must exist a pair of natural numbers $n, m$ such that $E_1(m^2 - n^2)/(m^2 n^2) = \hbar \omega_\mu$. From this point of view, it is an interesting question whether $\omega_\mu$ is associated with a unique pair $n, m$. It turns out that the pair $n, m$ is not unique, i.e. there are pairs $(n, m)$ and $(n', m')$ for which $(m^2 - n^2)/(m^2 n^2) = (m'^2 - n'^2)/(m'^2 n'^2)$. Let us denote by $J_\mu$ the set of all pairs $(n, m)$ defining the same frequency $\omega_\mu$. Although the complete characterization of $J_\mu$ is not known to authors of this paper, it is not that essential for the purposes of the presentation of the spectrum observable.

Observing a frequency $\omega_\mu$ we can conclude that the hydrogen atom is in the state $\varrho$ with a support containing at least one eigenstate of the energies $E_m$ for which there exist $n$ such that $(n, m) \in J_\mu$. Therefore, also the effect $F_\mu$ associated with the outcome $\omega_\mu$ is an operator with the support spanned on these eigenstates which guarantees that the outcome $\omega_\mu$ can be observed, i.e. $\mathrm{tr}\,[\varrho F_\mu] \neq 0$. It follows that the emission spectrum observable is described by effects $F_\mu = \sum_{m \in J_\mu} q_{nm} \Pi_m$, where $\Pi_m = \mathsf{A}(\{E_m\})$ are projectors onto subspaces spanned by eigenvectors associated with energies $E_m$ and $q_{nm}$ are the probabilities with which an atom being in the eigenstates associated with $E_m$ is emitting the light quantum of the frequency $\omega_\mu = \omega_{nm}$, i.e. $\sum_n q_{nm} = 1$. It is easy to verify that $\sum_\mu F_\mu = \sum_{nm} q_{nm} \Pi_m = \sum_m \Pi_m = I$. The probabilities $q_{nm}$ depends on particular experimental setup (temperature, photon detectors). Let us note that some of the couples are forbidden due to conservation laws, i.e. some of $q_{nm} = 0$ although the algebraic relation $n < m$ holds.

Let us now give an explicit example that $J_\mu$ can contain more than one element. Consider $n = 5, m = 7$ and $n' = 7, m' = 35$. A direct calculation gives that $(m^2 - n^2)/(m^2 n^2) = 49 - 25/(49.25) = 24/35^2$ and $(m' - n')/(m'n') = (35^2 - 49)/(35^2.49) = 49(25 - 1)/(49.35^2) = 24/35^2$, which proves that these numbers belong to the same set $J_\mu$ associated with $\omega_\mu = \frac{E_1}{\hbar} \frac{24}{35^2}$.

The main aim of this example is to show that even the simplest experiments constitute quite difficult observables. Let us note that the main purpose of this type of spectrum identification experiments is not to analyze the statistics of the sampled atoms, but rather to identify all possible frequencies. This was also the main goal of the Schrodinger solution of the hydrogen atom who showed that the observed frequencies in emission (or absorption) spectrum are in accordance with the eigenvalues of the energy operator.

### 4.3   Informationally complete observables

One of the most essential purposes of measurements is to gain knowledge about a system. Typically, the system is in an unknown state and one tries to find out what the state is. In the best case one can determine the state completely - this leads to the concept of informational completeness [16], [72].

**Definition 32.** A collection $\{\mathsf{A}, \mathsf{B}, \ldots\}$ of observables is *informationally complete* if for every $\varrho_1, \varrho_2 \in \mathcal{S}(\mathcal{H})$,

$$\left. \begin{array}{l} \Phi_{\mathsf{A}}(\varrho_1) = \Phi_{\mathsf{A}}(\varrho_2) \\ \Phi_{\mathsf{B}}(\varrho_1) = \Phi_{\mathsf{B}}(\varrho_2) \\ \quad\vdots \end{array} \right\} \Rightarrow \varrho_1 = \varrho_2\,.$$

In particular, a single observable A is *informationally complete* if for every $\varrho_1, \varrho_2 \in \mathcal{S}(\mathcal{H})$,

$$\Phi_{\mathsf{A}}(\varrho_1) = \Phi_{\mathsf{A}}(\varrho_2) \Rightarrow \varrho_1 = \varrho_2. \tag{4.22}$$

In other words, informational completeness of a collection $\{\mathsf{A}, \mathsf{B}, \ldots\}$ means that for two different states $\varrho_1 \neq \varrho_2$, at least one of the observables gives different probability distributions for them. Therefore, every state can be uniquely determined from the measurement data.

It is not surprising that we can form informationally complete collections of observables. For instance, it follows from Proposition 4 that the set of all two outcome sharp observables is informationally complete. In applications the problem is usually to find a physically realizable collection of observables, to minimize the number of observables, or to optimize the collection with respect to some other criteria.

**Example 40.** (*Informationally complete collection of qubit observables*) Let us consider a qubit system, i.e., a system described by a two dimensional Hilbert space $\mathcal{H}$ (isomorphic to $\mathbb{C}^2$). As we have seen in Example 20, the states of the qubit system can be described by the Bloch sphere vectors. Every unit vector $\vec{a}$ in the Bloch sphere determines a one dimensional projection, and hence a two outcome sharp observable A as explained in Example 36. For a state $\varrho$ corresponding to a Bloch vector $\vec{r}$, we get

$$\mathrm{tr}\left[\varrho \mathsf{A}(1)\right] = \frac{1}{2}(1 + \vec{r} \cdot \vec{a}). \tag{4.23}$$

Let A, B, C be three two outcome sharp observables, determined by the unit vectors $\vec{a}, \vec{b}, \vec{c}$. If the vectors $\vec{a}, \vec{b}, \vec{c} \in \mathbb{R}^3$ are linearly independent, then the collection $\{\mathsf{A}, \mathsf{B}, \mathsf{C}\}$ is informationally complete. Namely, in this case a Bloch vector $\vec{r}$ is uniquely determined from the inner products $\vec{r} \cdot \vec{a}, \vec{r} \cdot \vec{b}, \vec{r} \cdot \vec{c}$. As we see from (4.23), these are obtained from the measurement statistics of A, B and C.

**Exercise 28.** Following Example 40, show that the set $\{\mathsf{A}, \mathsf{B}, \mathsf{C}\}$ is not informationally complete if the vectors $\vec{a}, \vec{b}, \vec{c}$ are linearly dependent.

**Example 41.** (*Pauli problem*) Let $\mathcal{H} = L^2(\mathbb{R})$ and Q and P the canonical position and momentum observables of a spin-0 particle moving on the real line $\mathbb{R}$ (see Example 35 and Exercise 26), i.e.,

$$\langle \psi \mid \mathsf{Q}(X)\psi \rangle = \int_X |\psi(x)|^2 \, dx, \quad \langle \psi \mid \mathsf{P}(Y)\psi \rangle = \int_Y \left|\widehat{\psi}(y)\right|^2 \, dy.$$

Here $\widehat{\psi} = \mathcal{F}\psi$ is the Fourier transform of $\psi$. The set $\{\mathsf{Q}, \mathsf{P}\}$ is not informationally complete. Indeed, the functions $|\psi(\cdot)|$ and $\left|\widehat{\psi}(\cdot)\right|$ do not determine the vector $\psi$ uniquely up to a phase factor. This can be seen, for instance, by choosing $a, b > 0$ and setting

$$\psi_{\pm}(x) = e^{-(a \pm ib)x^2}.$$

The question of the informational completeness of the set $\{\mathsf{Q}, \mathsf{P}\}$ was first posed by Wolfgang Pauli and it is therefore called the Pauli problem. Nowadays the Pauli problem is also used to refer to several variants of the original problem, and some of these variants are still open. For instance, there seems not to be an exhaustive characterization of observables A such that the collection $\{\mathsf{Q}, \mathsf{P}, \mathsf{A}\}$ is informationally complete; see [16] for further discussion.

**Proposition 46.** Consider a Hilbert space $\mathcal{H}$ of dimension $d < \infty$. If an observable A is informationaly complete, then its outcome space $\Omega$ contains at least $d^2$ points.

*Proof.* Let A be an observable with an outcome space $\Omega = \{x_1, \ldots, x_n\}$, where $n < d^2$. Since the real inner product space $\mathcal{L}_s(\mathcal{H})$ of selfadjoint operators is $d^2$-dimensional, there exists a selfadjoint operator $T \neq O$ such that $\mathrm{tr}\left[TA(x_j)\right] = 0$ for all $x_j \in \Omega$. Moreover, since $\sum_j \mathsf{A}(x_j) = I$, we have $\mathrm{tr}\left[T\right] = 0$. Define an operator $\varrho_0 = d^{-1}(I + \|T\|^{-1}\,T)$. As $\varrho_0$ is positive (recall Exercise 7 in Section 2.2.2) and $\mathrm{tr}\left[\varrho_0\right] = 1$, it is a state. Now $\mathrm{tr}\left[\varrho_0 \mathsf{A}(x_j)\right] = \mathrm{tr}\left[\frac{1}{d}I\mathsf{A}(x_j)\right]$ for all $x_j \in \Omega$, meaning that A cannot distinguish between the total mixture $\frac{1}{d}I$ and the state $\varrho_0$. This shows that A cannot be informationally complete.                                                             $\square$

The existence of a single informationally complete observable is not evident from its definition. In the following example we demonstrate that informationally complete observables exist. An informationally complete observable A is called a *minimal informationally complete observable* if the outcome space $\Omega$ has the smallest possible number of elements, $|\Omega| = d^2$.

**Example 42.** (*Minimal informationally complete observable*) Let $\mathcal{H}$ be a finite dimensional Hilbert space with dimension $d$ and denote $\Omega = \{1, \ldots, d\}$. We follow [24] in their construction of an informationally complete observable A. First of all, we fix an orthonormal basis $\{\varphi_j\}_{j=1}^{d}$ for $\mathcal{H}$. For every $j, k \in \Omega$, we denote by $P_{jk}$ the following one-dimensional projections

$$P_{jj} = |\varphi_j\rangle\langle\varphi_j|\,;$$

$$P_{jk} = |\tfrac{1}{\sqrt{2}}(\varphi_j + \varphi_k)\rangle\langle\tfrac{1}{\sqrt{2}}(\varphi_j + \varphi_k)| \quad \text{if } j > k\,;$$

$$P_{jk} = |\tfrac{1}{\sqrt{2}}(\varphi_j + i\varphi_k)\rangle\langle\tfrac{1}{\sqrt{2}}(\varphi_j + i\varphi_k)| \quad \text{if } j < k\,.$$

We denote

$$T = \sum_{j,k=1}^{d} P_{jk}\,. \tag{4.24}$$

Since $T \geq \sum_{j=1}^{d} P_{jj} = I$, the operator $T$ is positive and it has square root $T^{\frac{1}{2}}$. Moreover, it follows from $T \geq I$ that $\det T \geq \det I$, hence $T$ is invertible. By Prop. **??** this implies that the operator $T^{\frac{1}{2}}$ is invertible also. For every $j, k \in \Omega$, we then define

$$\mathsf{A}(j,k) := T^{-\frac{1}{2}} P_{jk} T^{-\frac{1}{2}}\,. \tag{4.25}$$

Each operator $\mathsf{A}(j,k)$ is positive and the normalization condition $\sum_{j,k} \mathsf{A}(j,k) = I$ is satisfied. Hence, A is an observable with the outcome set $\Omega \times \Omega$.

Suppose that $\varrho_1$ and $\varrho_2$ are two states such that the numbers $\mathrm{tr}\left[\varrho_1 P_{jk}\right]$ and $\mathrm{tr}\left[\varrho_2 P_{jk}\right]$ are the same for all indices $j, k \in \Omega$. This implies that $\langle\varphi_j \mid \varrho_1\varphi_k\rangle = \langle\varphi_j \mid \varrho_2\varphi_k\rangle$ for all $j, k \in \Omega$, and thus, $\varrho_1 = \varrho_2$. This reasoning holds true also if $P_{jk}$'s are replaced with $\mathsf{A}(j,k)$'s. Therefore, the observable A is informationally complete and since $|\Omega \times \Omega| = d^2$, it is minimal informationally complete observable. Let us note that the above way of constructing an informationally complete observable works also when the Hilbert space $\mathcal{H}$ is infinite dimensional.

While it may be hard to check if a given observable is informationally complete, there are some simple *necessary* conditions which an observable has to satisfy in order to be informationally complete. Two such criteria are given in the next proposition.

**Proposition 47.** Let A be an informationally complete observable with an outcome space $(\Omega, \mathcal{F})$, and let $X \in \mathcal{F}$.

(a) $\mathsf{A}(X)$ does not have both eigenvalues 0 and 1.

(b) If $\mathsf{A}(X)$ is a non-trivial effect (i.e. not a scalar multiple of the identity operator $I$), then it does not commute with all the other effects in the range of A.

*Proof.* (a) Assume that $\mathsf{A}(X)$ has eigenvalues 0 and 1. Let $\psi_1, \psi_0 \in \mathcal{H}$ be unit vectors such that $\mathsf{A}(X)\psi_1 = \psi_1$ and $\mathsf{A}(X)\psi_0 = 0$. The vectors $\psi_1$ and $\psi_0$ are orthogonal, as

$$\langle \psi_1 \,|\, \psi_0 \rangle = \langle \mathsf{A}(X)\psi_1 \,|\, \psi_0 \rangle = \langle \psi_1 \,|\, \mathsf{A}(X)\psi_0 \rangle = 0 \,.$$

Moreover, we have $\mathsf{A}(\neg X)\psi_1 = (I - \mathsf{A}(X))\,\psi_1 = 0$ and $\mathsf{A}(\neg X)\psi_0 = (I - \mathsf{A}(X))\,\psi_0 = \psi_0$.

Denote $\psi = \frac{1}{\sqrt{2}}\,(\psi_1 + \psi_0)$. For any $Y \in \mathcal{F}$, we then have

$$\mathsf{A}(X \cap Y)\psi = \frac{1}{\sqrt{2}}\mathsf{A}(X \cap Y)\psi_1 \,, \quad \mathsf{A}(\neg X \cap Y)\psi = \frac{1}{\sqrt{2}}\mathsf{A}(X \cap Y)\psi_0 \,,$$

and therefore

$$\langle \psi \,|\, \mathsf{A}(Y)\psi \rangle = \frac{1}{2}\,\langle \psi_1 \,|\, \mathsf{A}(X \cap Y)\psi_1 \rangle + \frac{1}{2}\,\langle \psi_0 \,|\, \mathsf{A}(\neg X \cap Y)\psi_0 \rangle \,.$$

On the other hand, denote $\varrho = \frac{1}{2}|\psi_1\rangle\langle\psi_1| + \frac{1}{2}|\psi_0\rangle\langle\psi_0|$. Then

$$\begin{aligned}
\mathrm{tr}\,[\varrho\mathsf{A}(Y)] &= \mathrm{tr}\,[\varrho\mathsf{A}(X \cap Y)] + \mathrm{tr}\,[\varrho\mathsf{A}(\neg X \cap Y)] \\
&= \frac{1}{2}\,\langle \psi_1 \,|\, \mathsf{A}(X \cap Y)\psi_1 \rangle + \frac{1}{2}\,\langle \psi_0 \,|\, \mathsf{A}(\neg X \cap Y)\psi_0 \rangle \,.
\end{aligned}$$

This shows that $\Phi_{\mathsf{A}}(|\psi\rangle\langle\psi|) = \Phi_{\mathsf{A}}(\varrho)$, which contradicts the fact that A is informationally complete.

(b) Assume that $\mathsf{A}(X)\mathsf{A}(Y) = \mathsf{A}(Y)\mathsf{A}(X)$ for every $Y \in \mathcal{F}$. Let $U \equiv e^{i\mathsf{A}(X)}$ be the unitary operator defined by $\mathsf{A}(X)$ (see Example 10 in Section 2.2.3). Since $\mathsf{A}(X)$ is not a multiple of the identity operator $I$, there is a unit vector $\psi \in \mathcal{H}$ which is not an eigenvector of $\mathsf{A}(X)$. Thus, $\psi$ is not an eigenvector of $U$ either and hence, the vectors $\psi$ and $U\psi$ are not parallel. Choose $\varrho_1 = |\psi\rangle\langle\psi|$ and $\varrho_2 = |U\psi\rangle\langle U\psi|$, in which case $\varrho_1 \neq \varrho_2$.

Let $Y \in \mathcal{F}$. As $\mathsf{A}(X)$ commutes with $\mathsf{A}(Y)$, also $U$ commutes with $\mathsf{A}(Y)$. We then get

$$\mathrm{tr}\,[\varrho_2\mathsf{A}(Y)] = \mathrm{tr}\,[|\psi\rangle\langle\psi|U^*\mathsf{A}(Y)U] = \mathrm{tr}\,[|\psi\rangle\langle\psi|\mathsf{A}(Y)] = \mathrm{tr}\,[\varrho_1\mathsf{A}(Y)] \,,$$

and therefore $\Phi_{\mathsf{A}}(\varrho_1) = \Phi_{\mathsf{A}}(\varrho_2)$. This is in contradiction with the fact that A is informationally complete. $\square$

Proposition 47a implies, especially, that an informationally complete observable does not have any non-trivial projection in its range. Therefore, sharp observables are not informationally complete. Proposition 47b is about the non-commutativity of an informationally complete observable. There are also other kind of results related to the non-commutative nature of an informationally complete observable; we refer to [13] for further discussion.

## 4.4   Identification of quantum states

It is the primary role of observables to statistically identify states, or some of their properties. A typical goal of an experiment can be, for instance, to decide among different alternatives or hypotheses about the states. The difficulties and details of a particular identification problem depends on our apriori knowledge and properties we are interested in. The combination of apriori knowledge and available resources defines the alternatives we can experimentally verify. Depending whether the set of alternatives is finite or infinite, we make a distinction between discrimination and estimation problems. For a general reference on these type of problems, we refer to the classic [43] by Helstrom and to more recent review book [68].

Let us start by briefly mentioning some simple examples.

- *State estimation problem.* In its most difficult version the goal is to answer the question *what is the state?*, provided that no additional information is available. In this case, the whole state space $\mathcal{S}(\mathcal{H})$ forms the set of possible hypotheses. Experiments providing the answer to this problem correspond to informationally complete observables. As explained in Section 4.3, these are observables A for which the mapping $\Phi_A$ is injective.

- *State discrimination problem.* On the other hand, perhaps the simplest version of the quantum identification problem is to identify a state which is known to be either $\varrho_1$ or $\varrho_2$, so that our aim is to distinguish between two alternatives only. If we can obtain the hole measurement data, it is sufficient to design an observable A for which $\Phi_A(\varrho_1) \neq \Phi_A(\varrho_2)$.

These two types of identification problem (as defined above) are based on measurement statistics and we shall refer to them as to *statistical identification problems*. Perhaps surprisingly, non-trivial and meaningful identification schemes exits even if we assume that the resources are only finite in the sense that the unknown preparators are used in the experiment only a finite number of times.

**Example 43.** Imagine that we toss a coin only once and observe the outcome *head*. What can we say about the coin? Without any additional assumption we can only conclude that the probability of getting head is nonzero. However, if we are promised that the coin is either a fair one, or an unfair one having on both sides the same value `tail`, then our conclusion can be much stronger. In fact, we may with certainty confirm that the coin is fair.

This example illustrates that even in probabilistic theories there are situations in which conclusions can be made from a small number of experimental runs. And quantum theory is not an exception. Under certain circumstances and with suitable apriori knowledge, the quantum states may be uniquely identified even from individual measurement outcomes. These kind of tasks which are not based on measurement statistics but only some finite number of measurement outcomes are called *non-statistical identification problems*.

### 4.4.1   Unambiguous discrimination of pure states

The goal of *unambiguous state discrimination* (or *error-free state discrimination*) is to identify an unknown state $\varrho$ out of some set of possible states $\{\varrho_1, \ldots, \varrho_n\}$ without an error. We assume that only a single copy of a system is available and $\eta_j$ is the apriori probability that the unknown state $\varrho$ is $\varrho_j$. The main question is which sets of states can be discriminated in an unambiguous way. This problem was originally introduced by Ivanovic in [52].

In the general scheme of unambiguous state discrimination, we accept the possibility of getting no conclusion. However, when a conclusion is made, it is required to be correct. Opposed to unambiguous state discrimination, there are also probabilistic state discrimination schemes which are shortly discussed in Subsection 4.4.3 . In this subsection we focus on unambiguous discrimination of two pure states. We refer to [25] and [10] for reviews of different state discrimination strategies.

As a special type of unambiguous discrimination, we say that a set of states can be *perfectly discriminated* if the probability of getting a conclusive outcome is 1. Hence, a set of states $\{\varrho_1, \ldots, \varrho_n\}$ can be perfectly discriminated if and only if there exists an observable A with the outcome space $\Omega = \{1, 2, \ldots, n\}$ such that

$$\mathrm{tr}\left[\varrho_1 \mathsf{A}(1)\right] = \mathrm{tr}\left[\varrho_2 \mathsf{A}(2)\right] = \cdots = \mathrm{tr}\left[\varrho_n \mathsf{A}(n)\right] = 1 \,. \tag{4.26}$$

Indeed, in the perfect discrimination one must be able to draw the correct conclusion in every possible measurement outcome. This is exactly the condition (4.26).

**Exercise 29.** Let $\psi_1, \ldots, \psi_n$ be orthogonal unit vectors and $\varrho_1, \ldots, \varrho_n$ the corresponding pure states. Show that the set $\{\varrho_1, \ldots, \varrho_n\}$ can be perfectly discriminated. (Hint: recall Example 34 in Section 4.2.)

**Proposition 48.** Let $\varrho_1$ an $\varrho_2$ be two pure states corresponding to unit vectors $\psi_1$ and $\psi_2$. These states can be perfectly discriminated if and only if they are orthogonal, that is, $\langle\, \psi_1 \,|\, \psi_2 \,\rangle = 0$.

*Proof.* We have already seen in Exercise 29 that two orthogonal pure states can be perfectly discriminated. Let us then show that orthogonality is also a necessary condition.

If $\varrho_1$ and $\varrho_2$ can be perfectly discriminated, then there exists an observable A with the outcome space $\Omega = \{1, 2\}$ such that

$$\mathrm{tr}\left[\varrho_1 \mathsf{A}(1)\right] = \mathrm{tr}\left[\varrho_2 \mathsf{A}(2)\right] = 1 \,.$$

By Prop. 44 in Sec. 4.2, this implies that

$$\mathsf{A}(1)\varrho_1 = \varrho_1 \,, \qquad \mathsf{A}(2)\varrho_2 = \varrho_2 \,.$$

Since $\mathsf{A}(2) = I - \mathsf{A}(1)$, these conditions can be written in the form

$$\mathsf{A}(1)\varrho_1 = \varrho_1 \,, \qquad \mathsf{A}(1)\varrho_2 = 0 \,.$$

Therefore, we get

$$\mathrm{tr}\left[\varrho_1 \varrho_2\right] = \mathrm{tr}\left[\mathsf{A}(1)\varrho_1 \varrho_2\right] = \mathrm{tr}\left[\varrho_1 \varrho_2 \mathsf{A}(1)\right] = \mathrm{tr}\left[\varrho_1 (\mathsf{A}(1)\varrho_2)^*\right] = 0 \,.$$

Since $\mathrm{tr}\left[\varrho_1 \varrho_2\right] = |\langle\, \psi_1 \,|\, \psi_2 \,\rangle|^2$, the claim follows.

$\square$

Suppose that two pure states $\varrho_1, \varrho_2$ are given and that these states are not orthogonal. As the perfect discrimination is not possible, we turn into a more modest question - is it possible to identify the state $\varrho_1$ unambiguously? In comparison with the previous discussion, we are now not interested whether the unknown state is $\varrho_2$ but only the conclusion $\varrho = \varrho_1$ matters. Therefore, the state $\varrho_1$ can be identified within the set $\{\varrho_1, \varrho_2\}$ if and only if there exists an observable A with two outcomes 1 and ? such that

$$\mathrm{tr}\,[\varrho_1 \mathsf{A}(1)] > 0\,, \qquad \mathrm{tr}\,[\varrho_2 \mathsf{A}(1)] = 0\,. \tag{4.27}$$

The effect $\mathsf{A}(?) = I - \mathsf{A}(1)$ corresponds to an *inconclusive* outcome since no unambiguous conclusion is assigned to it.

The success of the state identification is quantified by means of the probability to observe the conclusive outcome 1, i.e.,

$$p_{\mathrm{success}} = \eta\,\mathrm{tr}\,[\varrho_1 \mathsf{A}(1)]\,, \tag{4.28}$$

where $\eta$ is the apriori probability that the unknown state $\varrho$ is $\varrho_1$. The second condition $\mathrm{tr}\,[\varrho_2 \mathsf{A}(1)] = 0$ in (4.27) is equivalent to $\mathrm{tr}\,[\varrho_2 \mathsf{A}(?)] = 1$. By Proposition 44, this implies that $\varrho_2 \leq \mathsf{A}(?)$ or, in other words, $\mathsf{A}(1) \leq I - \varrho_2$. Consequently, the probability $\mathrm{tr}\,[\varrho_1 \mathsf{A}(1)]$ is maximal if we set $\mathsf{A}(1) = I - \varrho_2$. Hence, the best achievable success probability is

$$p_{\mathrm{success}} = \eta(1 - \mathrm{tr}\,[\varrho_1 \varrho_2])\,. \tag{4.29}$$

Notice that $\mathrm{tr}\,[\varrho_1 \varrho_2] = 1$ only when $\varrho_1 = \varrho_2$. Therefore, $p_{\mathrm{success}} > 0$ whenever $\varrho_1 \neq \varrho_2$. We conclude that for any pair of two different pure states, the identification task is possible to carry out with non-zero success probability even with a single copy of the system available.

**Example 44.** *(Unambiguous identification of a mixed state.)* Let $\varrho_1, \varrho_2$ be two states, not necessarily pure. The condition for identification of $\varrho_1$ out of the set $\{\varrho_1, \varrho_2\}$ is still the set of constrains written in (4.27). However, the identification task cannot be performed for all pairs of states. Consider an effect $E$ such that $\mathrm{tr}\,[\varrho_1 E] > 0$ and denote $S_E^\perp = \{\varrho \in \mathcal{S}(\mathcal{H}) : \mathrm{tr}\,[\varrho E] = 0\}$. It follows that the two outcome observable A, defined as $\mathsf{A}(1) = E$, $\mathsf{A}(?) = I - E$, can be employed to identify the state $\varrho_1$ out of the set $\{\varrho_1, \varrho_2\}$, provided that $\varrho_2 \in S_E^\perp$. On the other hand, if for all effects $E$ the inequality $\mathrm{tr}\,[\varrho_1 E] > 0$ implies $\mathrm{tr}\,[\varrho_2 E] > 0$, then the state $\varrho_1$ cannot be identified from the set $\{\varrho_1, \varrho_2\}$. In other words, the unambiguous identification of $\varrho_1$ requires that its support is not smaller than the support of $\varrho_2$. It follows that while for pure states the success probability, given in (4.29), is symmetric with respect to change of $\varrho_1$ and $\varrho_2$, it is not necessarily symmetric for mixed states. For example, if $\mathcal{H}$ is finite dimensional and $\varrho_2$ is the total mixture $\frac{1}{d}I$, then the identification is impossible for all states $\varrho_1$. Thus, no state can be identified from the total mixture. However, if we choose $\varrho_1 = \frac{1}{d}I$, then we see that the total mixture can be identified with non-zero success probability, provided that the support of $\varrho_2$ is not the whole Hilbert space $\mathcal{H}$.

Let us turn back to unambiguous discrimination of two pure states $\varrho_1 = |\psi_1\rangle\langle\psi_1|$ and $\varrho_2 = |\psi_2\rangle\langle\psi_2|$ (assuming that $\varrho_1 \neq \varrho_2$). If the states are not orthogonal, then according to Proposition 48 they cannot be perfectly discriminated. However, as we have seen previously, each pure state can be unambiguously identified from any other pure state. We denote by A and B the observables identifying the states $\varrho_1$ and $\varrho_2$ in the optimal way, respectively. Hence, $\mathsf{A}(1) = I - \varrho_2$ and $\mathsf{B}(1) = I - \varrho_1$. Let us then fix a number $0 < q < 1$ and define an observable $\mathsf{C}$ with the outcome space $\{1, 2, ?\}$ in the following way:

$$\mathsf{C}(1) = q(I - \varrho_2)\,, \quad \mathsf{C}(2) = (1 - q)(I - \varrho_1)\,, \quad \mathsf{C}(?) = q\varrho_2 + (1 - q)\varrho_1\,. \tag{4.30}$$

The observable $\mathsf{C}$ is a convex combination of $\mathsf{A}$ and $\mathsf{B}$ in the sense explained in Example 33 in Subsection 4.1.4. The inconclusive outcomes of $\mathsf{A}$ and $\mathsf{B}$ are identified as a single inconclusive outcome $?$. As $\mathsf{C}$ is a mixture of $\mathsf{A}$ and $\mathsf{B}$ and these are the optimal observables for identifying the states $\varrho_1$ and $\varrho_2$, we expect that $\mathsf{C}$ is able to discriminate these states.

The observable $\mathsf{C}$ leads to the following list of probabilities:

$$p_{\varrho_1}^{\mathsf{C}}(1) = q(1 - \mathrm{tr}\,[\varrho_1\varrho_2]) \neq 0\,, \qquad p_{\varrho_2}^{\mathsf{C}}(1) = 0\,,$$
$$p_{\varrho_1}^{\mathsf{C}}(2) = 0\,, \qquad p_{\varrho_2}^{\mathsf{C}}(2) = (1 - q)(1 - \mathrm{tr}\,[\varrho_1\varrho_2]) \neq 0\,,$$
$$p_{\varrho_1}^{\mathsf{C}}(?) = 1 - q + q\mathrm{tr}\,[\varrho_1\varrho_2] \neq 0\,, \qquad p_{\varrho_2}^{\mathsf{C}}(?) = q + (1 - q)\mathrm{tr}\,[\varrho_1\varrho_2] \neq 0\,.$$

Therefore, if we get the outcome 1, we make the conclusion that $\varrho = \varrho_1$. In a similar way, the outcome 2 leads to the conclusion that $\varrho = \varrho_2$. If the outcome $?$ is obtained, then no conclusion can be made.

We conclude that although the perfect discrimination is not possible, there exists an observable discriminating between the non-orthogonal pure states $\varrho_1$ and $\varrho_2$ provided that an inconclusive result is allowed. As before, the success probability $p_{\text{success}}$ is quantified as the probability of getting a conclusive result. In the above scheme we get

$$\begin{aligned} p_{\text{success}} &= \eta\,\mathrm{tr}\,[\mathsf{C}(1)\varrho_1] + (1 - \eta)\,\mathrm{tr}\,[\mathsf{C}(2)\varrho_2] \\ &= q\eta\,\mathrm{tr}\,[\varrho_2(I - \varrho_1)] + (1 - q)(1 - \eta)\,\mathrm{tr}\,[\varrho_1(I - \varrho_2)] \\ &= [q\eta + (1 - q)(1 - \eta)](1 - \mathrm{tr}\,[\varrho_1\varrho_2])\,. \end{aligned}$$

For equal apriori distribution of the states $\varrho_1$ and $\varrho_2$ (i.e. $\eta = 1/2$), the unambiguous discrimination by $\mathsf{C}$ is successful with the probability

$$p_{\text{success}} = \frac{1}{2}(1 - \mathrm{tr}\,[\varrho_1\varrho_2]) = \frac{1}{2}(1 - |\langle\,\psi_1\,|\,\psi_2\,\rangle|^2)\,. \tag{4.31}$$

Does the above procedure give the best achievable success probability? To find the optimal solution, we are looking for an observable $\mathsf{D}$ with three outcomes $1, 2, ?$ such that

$$\mathrm{tr}\,[\varrho_1\mathsf{D}(2)] = 0\,, \qquad \mathrm{tr}\,[\varrho_2\mathsf{D}(1)] = 0\,, \qquad \mathsf{D}(?) = I - \mathsf{D}(1) - \mathsf{D}(2) \geq O\,, \tag{4.32}$$

and the success probability

$$p_{\text{success}} = \eta\,\mathrm{tr}\,[\varrho_1\mathsf{D}(1)] + (1 - \eta)\,\mathrm{tr}\,[\varrho_2\mathsf{D}(2)]$$

is required to be as large as possible.

Before we come to the the optimal value of $p_{\text{success}}$, we derive an upper bound on success probability.

**Lemma 3.** Let $T$ be a trace class operator $T \in \mathcal{T}(\mathcal{H})$. Then

$$\sup_{U \in \mathcal{U}(\mathcal{H})} |\mathrm{tr}\,[TU]| = \mathrm{tr}\,[|T|] \ . \tag{4.33}$$

*Proof.* We give a proof only in the case $\dim \mathcal{H} < \infty$. Let $U \in \mathcal{U}(\mathcal{H})$. According to Proposition 17 it follows that

$$|\mathrm{tr}\,[TU]| \le \|T\|_{\mathrm{tr}}\,\|U\| = \|T\|_{\mathrm{tr}} = \mathrm{tr}\,[|T|] \ .$$

Hence, $|\mathrm{tr}\,[TU]| \le \mathrm{tr}\,[|T|]$ for all unitary operators $U$.

Let us express the operator $T$ in its polar decomposition as $T = W\,|T|$, where $W$ is unitary. We then get

$$|\mathrm{tr}\,[W^*T]| = |\mathrm{tr}\,[W^*W\,|T|]| = |\mathrm{tr}\,[|T|]| = \mathrm{tr}\,[|T|] \ .$$

This proves the lemma.                                                                          $\square$

**Proposition 49.** ( [34]) Let $\varrho_1$ and $\varrho_2$ be two states, occurring with apriori probabilities $\eta$ and $1 - \eta$. The probability of success has the following upper bound:

$$p_{\mathrm{success}} \le 1 - 2\sqrt{\eta(1-\eta)}\,\mathrm{tr}\,[|\sqrt{\varrho_1}\sqrt{\varrho_2}|] \ . \tag{4.34}$$

*Proof.* The probability of success can be written in the form $p_{\mathrm{success}} = 1 - p_{\mathrm{error}}$, where $p_{\mathrm{error}} = \mathrm{tr}\,[\mathsf{D}(?)(\eta\varrho_1 + (1-\eta)\varrho_2)]$ is the error probability. for which

$$\begin{aligned}
p_{\mathrm{error}}^2 &= \eta^2(\mathrm{tr}\,[\mathsf{D}(?)\varrho_1])^2 + (1-\eta)^2(\mathrm{tr}\,[\mathsf{D}(?)\varrho_2])^2 + 2\eta(1-\eta)\mathrm{tr}\,[\mathsf{D}(?)\varrho_1]\,\mathrm{tr}\,[\mathsf{D}(?)\varrho_2] \\
&\ge 4\eta(1-\eta)\mathrm{tr}\,[\mathsf{D}(?)\varrho_1]\,\mathrm{tr}\,[\mathsf{D}(?)\varrho_2] \ ,
\end{aligned}$$

where we used the inequality $a^2 + b^2 \ge 2ab$ for $a = \eta\mathrm{tr}\,[\mathsf{D}(?)\varrho_1]$ and $b = (1-\eta)\mathrm{tr}\,[\mathsf{D}(?)\varrho_2]$. Using the Cauchy-Schwartz inequality we obtain

$$\begin{aligned}
\mathrm{tr}\,[\mathsf{D}(?)\varrho_1]\,\mathrm{tr}\,[\mathsf{D}(?)\varrho_2] &= \mathrm{tr}\left[U\sqrt{\varrho_1}\sqrt{\mathsf{D}(?)}\sqrt{\mathsf{D}(?)}\sqrt{\varrho_1}U^\dagger\right]\mathrm{tr}\left[\sqrt{\varrho_2}\sqrt{\mathsf{D}(?)}\sqrt{\mathsf{D}(?)}\sqrt{\varrho_2}\right] \\
&\ge (\mathrm{tr}\,[U\sqrt{\varrho_1}\mathsf{D}(?)\sqrt{\varrho_2}])^2
\end{aligned}$$

Since $\mathsf{D}(?) = I - \mathsf{D}(1) - \mathsf{D}(2)$ and by definition of the unambiguous discrimination $\mathsf{D}(1)\varrho_2 = \varrho_1\mathsf{D}(2) = O$, it follows that $\sqrt{\varrho_1}\mathsf{D}(?)\sqrt{\varrho_2} = \sqrt{\varrho_1}\sqrt{\varrho_2}$. Thus,

$$p_{\mathrm{error}} \ge 2\sqrt{\eta(1-\eta)}\mathrm{tr}\,[U\sqrt{\varrho_1}\sqrt{\varrho_2}] \tag{4.35}$$

for all unitary operators $U$. According the Lemma 3 the maximum of the right-hand side of this equation gives the formula

$$p_{\mathrm{error}} \ge 2\sqrt{\eta(1-\eta)}\mathrm{tr}\,[|\sqrt{\varrho_1}\sqrt{\varrho_2}|] \ . \tag{4.36}$$

$\square$

For two pure states distributed with equal apriori probabilities Proposition 49 gives

$$p_{\text{success}} \leq 1 - |\langle\,\psi_1\,|\,\psi_2\,\rangle|\,.$$

Then from Eq.(4.31) it is straightforward to see that observable $\mathsf{C}$ does not saturate this upper bound. Namely, the equation

$$\frac{1}{2}\left(1 - |\langle\,\psi_1\,|\,\psi_2\,\rangle|^2\right) = 1 - |\langle\,\psi_1\,|\,\psi_2\,\rangle|$$

holds only if $|\langle\,\psi_1\,|\,\psi_2\,\rangle| = 1$, but this would mean that $\varrho_1 = \varrho_2$.

In what follows we shall define a family of observables and show that for particular choice of parameters there is an observable saturating the bound (4.34).

The conditions in Eq.(4.32) implies that $\mathsf{D}(2) \leq I - \varrho_1$ and $\mathsf{D}(1) \leq I - \varrho_2$. Let us denote by $Q$ a projection onto the support of an operator $(I - \varrho_1)(I - \varrho_2)$. Since $\text{tr}\,[Q\varrho_1] = \text{tr}\,[Q\varrho_2] = 0$ it follows that the support of $Q$ is irrelevant for discrimination, because no effect $F \leq Q$ can be observed. It is straightforward to see that $Q$ has rank $d-2$ and therefore only a two-dimensional subspace spanned on vectors $\psi_1, \psi_2$ is of interest for us. Let us denote by $Q^\perp = I - Q$ the projector onto this two-dimensional subspace. That is, we can assume $\mathsf{D}(1) \leq I - Q$ and $\mathsf{D}(2) \leq I - Q = Q^\perp$. Consider the following ansatz

$$\mathsf{D}(1) = c(Q^\perp - \varrho_2)\,, \qquad \mathsf{D}(2) = c(Q^\perp - \varrho_1)\,, \qquad \mathsf{D}(?) = I - \mathsf{D}(1) - \mathsf{D}(2)\,. \quad (4.37)$$

Using this type of observable for $\eta = 1/2$ the success probability reads

$$
\begin{aligned}
p_{\text{success}} &= \frac{1}{2}\left(\text{tr}\,[\varrho_1\mathsf{D}(1)] + \text{tr}\,[\varrho_2\mathsf{D}(2)]\right) = \frac{c}{2}\left(\text{tr}\,\left[Q^\perp(\varrho_1 + \varrho_2)\right] + 2\text{tr}\,[\varrho_1\varrho_2]\right) \\
&= c(1 - \text{tr}\,[\varrho_1\varrho_2])\,,
\end{aligned}
$$

where the factor $c$ takes the largest possible value for which the operator $\mathsf{D}(?) = I - \mathsf{D}(1) - \mathsf{D}(2)$ is still an effect, i.e., $\mathsf{D}(?) \geq O$.

From the definition of $\mathsf{D}(?) = Q + Q^\perp(1 - 2c) + c(\varrho_1 + \varrho_2)$ it follows that interesting part of $\mathsf{D}(?)$ is effectively two-dimensional. In particular, $\mathsf{D}(?)$ coincides with the identity operator on the $Q$ subspace and the positivity constraints can be violated only by operator $Q^\perp(1 - 2c) + c(\varrho_1 + \varrho_2)$, where $Q^\perp$ acts as the identity on the relevant two-dimensional subspace, i.e. $Q^\perp = \varrho_1 + \varrho_1^\perp$. Here $\varrho_1^\perp$ is a one-dimensional projection onto a state $\psi_1^\perp$ orthogonal to $\psi_1$, but belonging to linear span of vectors $\psi_1, \psi_2$, thus $\langle\,\psi_1^\perp\,|\,\psi_2\,\rangle \neq 0$. projection $\varrho_2$ can be expressed as a superposition of pure states $\varrho_1, \varrho_1^\perp$

$$\varrho_2 = |\alpha|^2\varrho_1 + |\beta|^2\varrho_1^\perp + \alpha\beta^*|\psi_1\rangle\langle\psi_1^\perp| + \alpha^*\beta|\psi_1^\perp\rangle\langle\psi_1|\,, \qquad (4.38)$$

where $|\alpha|^2 + |\beta|^2 = 1$ Inserting this into the formula for relevant part of $\mathsf{D}(?)$ we obtain the operator

$$\mathsf{D}(?)^{\text{rel}} = (1 - c|\beta|^2)\varrho_1 + (1 - c(1 + |\alpha|^2))\varrho_1^\perp + c\alpha\beta^*|\psi_1\rangle\langle\psi_1^\perp| + c\alpha^*\beta|\psi_1^\perp\rangle\langle\psi_1|\,. \quad (4.39)$$

It has positive eigenvalues if and only if its determinant is positive, i.e.

$$
\begin{aligned}
0 &\leq 1 + c^2|\beta|^2(1 + |\alpha|^2) - c(1 + |\alpha|^2 + |\beta|^2) - c^2|\alpha|^2|\beta|^2 \\
&\leq 1 + c^2|\beta|^2 - 2c = \left(c - \frac{1 - |\alpha|}{|\beta|^2}\right)\left(c - \frac{1 + |\alpha|}{|\beta|^2}\right)
\end{aligned}
$$

Since $0 \leq c \leq 1$ it follows that the optimal choice is $c = (1 - |\alpha|)/|\beta|^2 = (1 + |\alpha|)^{-1}$. Since $|\alpha|^2 = |\langle \psi_1 | \psi_2 \rangle|^2 = \mathrm{tr}\,[\varrho_1 \varrho_2]$ we obtain the following formula for success probability

$$
\begin{aligned}
p_{\mathrm{success}} &= c(1 - \mathrm{tr}\,[\varrho_1 \varrho_2]) = \frac{1 - \sqrt{\mathrm{tr}\,[\varrho_1 \varrho_2]}}{1 - \mathrm{tr}\,[\varrho_1 \varrho_2]}(1 - \mathrm{tr}\,[\varrho_1 \varrho_2]) \\
&= 1 - \sqrt{\mathrm{tr}\,[\varrho_1 \varrho_2]} = 1 - |\langle \psi_1 | \psi_2 \rangle|\,.
\end{aligned} \tag{4.40}
$$

Comparing with Proposition 49 we see that the bound is saturated and the optimal observable D depends on the scalar product of pure states $\psi_1, \psi_2$. In particular

$$
\mathsf{D}_{\mathrm{opt}}(1) = \frac{1}{1 + |\langle \psi_1 | \psi_2 \rangle|}(Q^\perp - |\psi_2\rangle\langle\psi_2|)\,, \tag{4.41}
$$

$$
\mathsf{D}_{\mathrm{opt}}(2) = \frac{1}{1 + |\langle \psi_1 | \psi_2 \rangle|}(Q^\perp - |\psi_1\rangle\langle\psi_1|)\,, \tag{4.42}
$$

$$
\mathsf{D}_{\mathrm{opt}}(?) = I - \mathsf{D}_{\mathrm{opt}}(1) - \mathsf{D}_{\mathrm{opt}}(2)\,, \tag{4.43}
$$

where $Q^\perp$ is a projector onto the linear subspace spanned by vectors $\psi_1, \psi_2$.

For a comparison, let us briefly take a look at the unambiguous discrimination of mixed states.

**Proposition 50.** Let $\varrho_1$ and $\varrho_2$ be two states which can be unambiguously discriminated. Then neither $\mathrm{supp}\varrho_1 \not\subseteq \mathrm{supp}\varrho_2$ nor $\mathrm{supp}\varrho_2 \not\subseteq \mathrm{supp}\varrho_1$.

*Proof.* Suppose that $\mathrm{supp}\varrho_1 \subseteq \mathrm{supp}\varrho_2$. Then for any effect $E$, the condition $\mathrm{tr}\,[\varrho_1 E] \neq 0$ implies that $\mathrm{tr}\,[\varrho_2 E] \neq 0$. Thus, we can never unambiguously conclude that the unknown state is $\varrho_1$. Similarly, in the case $\mathrm{supp}\varrho_2 \subseteq \mathrm{supp}\varrho_1$ we can never conclude that the unknown state is $\varrho_2$. Therefore, the condition on the supports is a necessary condition that $\varrho_1$ and $\varrho_2$ can be unambiguously discriminated. $\square$

The following result is a simple consequence of Proposition 50.

**Proposition 51.** Let $\mathcal{H}$ be a two dimensional Hilbert space. Two different states $\varrho_1, \varrho_2 \in \mathcal{S}(\mathcal{H})$ can be unambiguously discriminated if and only if they are both pure.

*Proof.* We have already seen earlier that two different pure states can be unambiguously discriminated. To prove the other implication, let us assume that one of the states, say $\varrho_1$, is mixed. As we have seen in Subsection 3.1.3, the state $\varrho_1$ can be written in the form $\varrho = \frac{1}{2}(I + \vec{r} \cdot \vec{\sigma})$, where $\|\vec{r}\| < 1$. Let us define a pure state $\omega = \frac{1}{2}(I + \vec{n} \cdot \vec{\sigma})$ with $\vec{n} = \frac{\vec{r}}{\|\vec{r}\|}$. Then

$$
\varrho_1 = (1 - \|\vec{r}\|)\frac{1}{2}I + \|\vec{r}\|\,\omega\,,
$$

which shows that $\varrho_1$ can be written as a convex combination of the total mixture and the pure state $\omega$. Since $\mathrm{supp}\frac{1}{2}I = \mathcal{H}$, it follows that $\mathrm{supp}\varrho_1 = \mathcal{H}$. Therefore, according to Prop. 50 it is not possible to unambiguously discriminate $\varrho_1$ and $\varrho_2$. $\square$

**Example 45.** *(Key distribution protocol B92)* A nice application of the unambiguous state discrimination scheme was presented by Ch. Bennett in [4]. The key distribution protocol known as B92 works in the following way. Alice randomly prepares a system in one of two pure states $\varrho_0, \varrho_1$ and sends the system to Bob. Bob then performs a measurement of observable D, hence unambiguously discriminating between the given pair of states. Repeating the experiment $n$ times Alice preparations defines a string of $n$ random bits $\vec{x} = (x_1, \ldots, x_n)$ such that $x_j \in \{0, 1\}$. Bob's measurements defines a string $\vec{y} = (y_1, \ldots, y_n)$ with $y_j \in \{0, 1, ?\}$, where ? is associated with the inconclusive outcome ?. Due to the unambiguity of the conclusive results, the strings are perfectly correlated whenever $y_j \neq ?$. Therefore, Bob announces the positions in which he found inconclusive outcomes and both of them simply erase these entries from their strings. After that Alice's and Bob's reduced strings match perfectly and represent a secret key shared between Alice and Bob.

The security of B92 protocol is based on the nonorthogonality of states. For orthogonal states the correlations between the strings are perfect without any public communication, however, in this case the whole communication can be observed by a third party without being detected. Orthogonal pure states can be discriminated perfectly. Reducing the orthogonality we also reduce the abilities to eavesdrop and the anonymity of the eavesdropper. The presence of an adversary and security of the key is verified if part of the shared key is released and compared. This verification stage of the protocol is based on the same principles as the one that will be discussed in Section 6.3.2, where we discussed the more practical key distribution protocol called BB84.

Let us note that this scheme, although very elegant, is not very practical due to its fragility with respect to noise. In fact, for arbitrarily small noise the unambiguity of conclusion is lost since the no-error conditions $\mathrm{tr}\,[\varrho_2 \mathsf{D}(1)] = \mathrm{tr}\,[\varrho_1 \mathsf{D}(2)] = 0$ no longer hold. The noise transformers the pure states $\varrho_1, \varrho_2$ into mixed states $\varrho_1', \varrho_2'$. For a two-dimensional system, Proposition 51 implies that $\varrho_1', \varrho_2'$ cannot be unambiguously discriminated.

### 4.4.2    How close are two states?

There are several approaches how to introduce the concept of distance between quantum states. One possibility is to adopt some operational definition based on statistical distinguishability of states. The information acquired in measurements comes from observed probabilities. Therefore, it is natural to use these probabilities to introduce the notion of distance between two states. Loosely speaking, the idea is that two states are considered to be close to each other if for all observables it is very difficult to distinguish the associated probabilities.

A probability distribution defined on some finite sample space $\Omega$ can be understood as a real vector $\vec{p}$ with positive entries summing up to one. Given two such probability vectors $\vec{p}$ and $\vec{q}$, their difference is commonly evaluated by means of the following functions

$$A(\vec{p}, \vec{q}) \;:=\; \max_{j \in \Omega} |p_j - q_j|, \tag{4.44}$$

$$D(\vec{p}, \vec{q}) \;:=\; \frac{1}{2} \sum_{j \in \Omega} |p_j - q_j| \quad (\textit{Kolmogorov distance}), \tag{4.45}$$

$$F(\vec{p}, \vec{q}) \;:=\; \sum_{j \in \Omega} \sqrt{p_j q_j} \quad (\textit{Bhattacharyya coefficient}). \tag{4.46}$$

In the quantum case states are associated with generalized probability measures defined on the effect algebra $\mathcal{E}(\mathcal{H})$, i.e., assigning a probability $p(E)$ for every effect $E$ and probability distribution $\vec{p}_A$ for every observable A. We can use both, individual effects and observables, to define statistical distances between a pair of quantum states. In what follows we shall exploit the success probability of state discrimination to quantify the distance between quantum states.

### 4.4.3    Minimum-error state discrimination and the trace distance

Consider an experiment where an observable A is measured only once leading to an outcome $x_j$. Our goal is to assign a state (either $\varrho_1$ or $\varrho_2$) to each measured outcome $x_j$. Let us denote by $p_j = \text{tr}\,[\varrho_1 A(x_j)]$ and $q_j = \text{tr}\,[\varrho_2 A(x_j)]$ the probabilities of outcomes $x_j$ for the states $\varrho_1$ and $\varrho_2$.

The effects $A(x_j)$ forming the observable A can be sorted according to related conclusions into two subsets. Identifying the individual outcomes with the same conclusion we end up with a two outcome observable described by effects $C_1, C_2 = I - C_1$. These effects are associated with the conclusions $\varrho_? = \varrho_1, \varrho_? = \varrho_2$. Usually, if $p_j > q_j$ then the result $x_j$ is associated with the conclusion $\varrho_? = \varrho_1$. Similarly, if $p_j < q_j$, the conclusion is $\varrho_? = \varrho_2$. If for an outcome $x_j$ the predicted probabilities coincide (i.e. $p_j = q_j$), then each time we observe this outcome we randomly choose one of the conclusions. For simplicity, we assume that that both states are equally probable. In such case the effect $A(x_j)$ is splitted into two parts $\frac{1}{2}A(x_j) + \frac{1}{2}A(x_j) = A(x_j)$, and the parts are associated with the conclusions $\varrho_? = \varrho_1, \varrho_? = \varrho_2$, respectively.

The conclusions are not always correct in a sense that we can conclude that the state to be identified is $\varrho_1$ even if it is $\varrho_2$. The error probability is given by the formula

$$p_{\text{error}} = \frac{1}{2}\text{tr}\,[C_1 \varrho_2 + C_2 \varrho_1] = \frac{1}{2}\left(1 + \text{tr}\,[C_1(\varrho_2 - \varrho_1)]\right). \tag{4.47}$$

Naturally, we are then interested to find out the optimal observable which minimizes the error probability. We call this task the *minimum-error state discrimination*.

In order to minimize the error we must choose the effect $C_1$ to be a projection onto the eigenvectors of the operator $\varrho_2 - \varrho_1$ associated with negative eigenvalues. Since $\text{tr}\,[\varrho_2 - \varrho_1] = 0$ it follows that for the selfadjoint operator $\varrho_2 - \varrho_1$ its positive eigenvalues gives the same sum as its negative eigenvalues. This implies that the optimal error probability is symmetric with respect to exchange of $\varrho_1, \varrho_2$ and we can write

$$\sup_{C_1} \text{tr}\,[C_1(\varrho_2 - \varrho_1)] = \sum_{j:\lambda_j < 0} \lambda_j = -\frac{1}{2}(\sum_j |\lambda_j|) = -\frac{1}{2}\text{tr}\,[|\varrho_2 - \varrho_1|], \tag{4.48}$$

where $\lambda_j$ are eigenvalues of $\varrho_2 - \varrho_1$. In summary, for the optimal error probability of discrimination of states $\varrho_1, \varrho_2$ we get

$$p_{\text{error}} = \frac{1}{2}(1 - \frac{1}{2}\text{tr}\,[|\varrho_1 - \varrho_2|]). \tag{4.49}$$

The error probability is associated to the distance in the following sense. The smaller the error probability, the larger is the distance between the states. If $p_{\text{error}} = 1/2$, then the states are necessarily the same. It turns out that the function

$$D(\varrho_1, \varrho_2) = \frac{1}{2}\text{tr}\,[|\varrho_1 - \varrho_2|] \tag{4.50}$$

can be used to measure the distance between the states $\varrho_1$ and $\varrho_2$. In fact, (up to a factor) it is induced by the operator's trace-norm. The above formula gives a clear operational meaning for the trace norm evaluating the minimum-error state discrimination. We refer to this distance as to *trace distance*.

### 4.4.4   Fidelity

An alternative state discrimination task is the one in which the conclusions are either error-free, or no conclusion is made. This is known as the unambiguous discrimination problem and it was discussed in Subsection 4.4.1. In this case the success is quantified by a probability of inconclusive result (associated with the effect $E_?$) and reads

$$p_{\text{error}} = \frac{1}{2}\text{tr}\left[E_?(\varrho_1 + \varrho_2)\right],  \tag{4.51}$$

where $\varrho_1, \varrho_2$ are known states and $E_? = I - E_1 - E_2$ such that $\text{tr}\left[E_1\varrho_2\right] = \text{tr}\left[E_2\varrho_1\right] = 0$. Again, we assume that states are apriori distributed with the same probability. A general solution is not known, but for pure states $\varrho_1, \varrho_2$ we found in Subsection 4.4.1 that the optimal value is

$$p_{\text{error}} = |\langle \psi_1 \,|\, \psi_2 \rangle|,  \tag{4.52}$$

where $\psi_1, \psi_2 \in \mathcal{H}$ are the vectors corresponding to states $\varrho_1, \varrho_2$, respectively.

According to Proposition 49 the quantity $\text{tr}\left[|\sqrt{\varrho_1}\sqrt{\varrho_2}|\right] = \text{tr}\left[\sqrt{\sqrt{\varrho_1}\varrho_2\sqrt{\varrho_1}}\right]$ provides an upper bound on optimal success probability for unambiguous discrimination of a pair of states $\varrho_1, \varrho_2$. This gives an operational meaning to another common way how to measure the distance between quantum states.

**Definition 33.** A *fidelity* of quantum states $\varrho_1$ and $\varrho_2$ is defined as

$$F(\varrho_1, \varrho_2) = \text{tr}\left[\sqrt{\sqrt{\varrho_1}\varrho_2\sqrt{\varrho_1}}\right].  \tag{4.53}$$

**Proposition 52.** (*Basic properties of fidelity.*) For all states $\varrho_1, \varrho_2$ the following statements holds:

1. $F(\varrho_1, \varrho_2) = 1$ if and only if $\varrho_1 = \varrho_2$.

2. $0 \leq F(\varrho_1, \varrho_2) \leq 1$

3. Fidelity is invariant under unitary conjugation, i.e. $F(U\varrho_1 U^*, U\varrho_2 U^*) = F(\varrho_1, \varrho_2)$ for all unitary operators $U$.

**Exercise 30.** Prove the properties of the fidelity listed in Proposition 52.

### 4.5   Relations between observables

In previous sections we have discussed some properties and qualities of observables. A supplementary point of view is obtained when we compare observables together. This means that we study relative properties of observables. In this section we concentrate on three preorderings in the set of all observables. There are also several other, related but different, preorderings. Other preorderings are investigated, for instance, in [11] and [42].

### 4.5.1   State distinction and determination

Let A be an observable and $\varrho_1$ and $\varrho_2$ two different states. If

$$\Phi_\mathsf{A}(\varrho_1) \neq \Phi_\mathsf{A}(\varrho_2)\,,$$

then a measurement of A makes a distinction between the states $\varrho_1$ and $\varrho_2$. Namely, as the probability distributions are different, we see a difference in the measurement statistics correspond to $\varrho_1$ and $\varrho_2$. On the other hand, if

$$\Phi_\mathsf{A}(\varrho_1) = \Phi_\mathsf{A}(\varrho_2)\,,$$

then A cannot make any distinction between $\varrho_1$ and $\varrho_2$. This motivates the following definition.

**Definition 34.** Let A and B be observables. If for all states $\varrho_1, \varrho_2 \in \mathcal{S}(\mathcal{H})$,

$$\Phi_\mathsf{A}(\varrho_1) = \Phi_\mathsf{A}(\varrho_2) \Rightarrow \Phi_\mathsf{B}(\varrho_1) = \Phi_\mathsf{B}(\varrho_2)\,, \tag{4.54}$$

then we denote $\mathsf{B} \preccurlyeq_i \mathsf{A}$, and say that the *state distinction power* of A is greater than or equal to B. If $\mathsf{B} \preccurlyeq_i \mathsf{A} \preccurlyeq_i \mathsf{B}$, we say that A and B are *informationally equivalent*, and denote $\mathsf{A} \overset{i}{\sim} \mathsf{B}$.

Comparing the above definition with the definition of an informationally complete observable (Subsection 4.3), one immediately notice that if A is informationally complete, then $\mathsf{B} \preccurlyeq_i \mathsf{A}$ for any observable B. Moreover, if A is informationally complete and $\mathsf{A} \overset{i}{\sim} \mathsf{B}$, then also B is informationally complete.

**Exercise 31.** Confirm that $\preccurlyeq_i$ is a preorder and $\overset{i}{\sim}$ is an equivalence relation.

It can also happen that A gives a unique probability distribution for some state $\varrho_1$. This means that for all $\varrho \in \mathcal{S}(\mathcal{H})$, we have

$$\Phi_\mathsf{A}(\varrho_1) = \Phi_\mathsf{A}(\varrho) \Rightarrow \varrho_1 = \varrho\,.$$

In this case, we say that the state $\varrho_1$ is *determined* by A. We denote by $\mathcal{D}_\mathsf{A}$ the set of states determined by A. This leads to another way how to compare observables.

**Definition 35.** Let A and B be observables. If $\mathcal{D}_\mathsf{B} \subseteq \mathcal{D}_\mathsf{A}$, then we denote $\mathsf{B} \preccurlyeq_d \mathsf{A}$, and say that the *state determination power* of A is greater than or equal to B. If $\mathsf{B} \preccurlyeq_d \mathsf{A} \preccurlyeq_d \mathsf{B}$, we denote $\mathsf{A} \overset{d}{\sim} \mathsf{B}$.

With these concepts, we can rephrase the defining condition for an informationally complete observable A as $\mathcal{D}_\mathsf{A} = \mathcal{S}(\mathcal{H})$. Similarly as in the case of state distinction power, we see that if A is informationally complete, then $\mathsf{B} \preccurlyeq_d \mathsf{A}$ for any observable B and $\mathsf{A} \overset{d}{\sim} \mathsf{B}$ only if B is informationally complete.

**Exercise 32.** Confirm that $\preccurlyeq_d$ is a preorder and $\overset{d}{\sim}$ is an equivalence relation.

**Proposition 53.** For two observables A and B, the condition $\mathsf{B} \preccurlyeq_i \mathsf{A}$ implies that $\mathsf{B} \preccurlyeq_d \mathsf{A}$.

*Proof.* Assume that $\mathsf{B} \preccurlyeq_i \mathsf{A}$ and $\varrho_1 \in \mathcal{D}_\mathsf{B}$. For every $\varrho \in \mathcal{S}(\mathcal{H})$, we then have

$$\Phi_\mathsf{A}(\varrho_1) = \Phi_\mathsf{A}(\varrho) \;\Rightarrow\; \Phi_\mathsf{B}(\varrho_1) = \Phi_\mathsf{B}(\varrho) \;\Rightarrow\; \varrho_1 = \varrho \,.$$

This shows that $\varrho_1 \in \mathcal{D}_\mathsf{A}$, and therefore $\mathcal{D}_\mathsf{B} \subseteq \mathcal{D}_\mathsf{A}$. $\hfill\square$

In general, $\mathsf{B} \preccurlyeq_d \mathsf{A}$ does not imply that $\mathsf{B} \preccurlyeq_i \mathsf{A}$. This becomes clear in the following discussion.

**Example 46.** An observable $\mathsf{A}$ is *trivial* if it does not distinguish any pair of states, i.e.,

$$\Phi_\mathsf{A}(\varrho_1) = \Phi_\mathsf{A}(\varrho_2) \quad \forall \varrho_1, \varrho_2 \in \mathcal{S}(\mathcal{H}) \,. \tag{4.55}$$

Condition (4.55) is equivalent with the fact that each effect $\mathsf{A}(X)$ is a multiple of the identity operator $I$. If $\mathsf{A}$ is a trivial observable, then clearly $\mathsf{A} \preccurlyeq_i \mathsf{B}$ for any other observable $\mathsf{B}$. Moreover, if $\mathsf{B} \preccurlyeq_i \mathsf{A}$, then also $\mathsf{B}$ is a trivial observable.

**Proposition 54.** Let $\mathsf{A}$ be a sharp observable with the outcome set $\Omega = \{x_1, x_2\}$. If neither $\mathsf{A}(x_1)$ nor $\mathsf{A}(x_2)$ is a one-dimensional projection, then $\mathcal{D}_\mathsf{A} = \emptyset$.

*Proof.* First of all, if $\mathsf{A}(x_1) = O$ or $\mathsf{A}(x_1) = I$, then it is clear that $\mathcal{D}_\mathsf{A} = \emptyset$. Hence, we assume that $O \neq \mathsf{A}(x_1) \neq I$.

For a state $\varrho$, the probability distribution $\Phi_\mathsf{A}(\varrho)$ is uniquely characterized by the number $\mathrm{tr}\,[\varrho\mathsf{A}(x_1)]$, since $\mathrm{tr}\,[\varrho\mathsf{A}(x_2)] = 1 - \mathrm{tr}\,[\varrho\mathsf{A}(x_1)]$. Hence, to prove the statement, it is enough to show that for each number $0 \leq p \leq 1$ there are (at least) two states $\varrho$ and $\varrho'$ such that $p = \mathrm{tr}\,[\varrho\mathsf{A}(x_1)] = \mathrm{tr}\,[\varrho'\mathsf{A}(x_1)]$.

Suppose that $\mathsf{A}(x_1)$ and $\mathsf{A}(x_2)$ are not one-dimensional projections. We then can write $\mathsf{A}(x_1)$ as a sum $\mathsf{A}(x_1) = \sum_{k=1}^{r} P_k$ of $r \geq 2$ orthogonal one-dimensional projections (see Subsection 2.2.4). Similarly, we write $\mathsf{A}(x_2) = \sum_{k=1}^{s} Q_k$. For any number $0 \leq p \leq 1$, we denote $\varrho = pP_1 + (1-p)Q_1$ and $\varrho' = pP_2 + (1-p)Q_2$. Then $\mathrm{tr}\,[\varrho\mathsf{A}(x_1)] = \mathrm{tr}\,[\varrho'\mathsf{A}(x_1)] = p$. $\hfill\square$

Proposition 54 shows, in particular, that if $\mathsf{A}$ is a two-outcome sharp observable and neither $\mathsf{A}(1)$ nor $\mathsf{A}(2)$ is a one-dimensional projection, then $\mathsf{A} \overset{d}{\sim} \mathsf{B}$ with any trivial observable $\mathsf{B}$. However, we do not have $\mathsf{A} \overset{i}{\sim} \mathsf{B}$ since $\mathsf{A}$ is not trivial. This demonstrates that the preorderings $\preccurlyeq_i$ and $\preccurlyeq_d$ are different.

**Exercise 33.** Let $\mathsf{A}$ be a sharp observable with the outcome set $\Omega = \{x_1, x_2\}$ and assume that $\mathsf{A}(x_1)$ is a one-dimensional projection. Show that the pure state $\varrho = \mathsf{A}(x_1)$ is determined by $\mathsf{A}$.

### 4.5.2 Coarse-graining

Coarse-graining means, generally speaking, a reduction in the statistical description of a system. The statistical information related to an observable $\mathsf{A}$ is most directly encoded in the mapping $\Phi_\mathsf{A}$. Hence, the coarse-graining relation for observables can be formulated in the following way.

**Definition 36.** Let $\mathsf{A}$ and $\mathsf{B}$ be observables. We say that $\mathsf{B}$ is a *coarse-graining* of $\mathsf{A}$, and denote $\mathsf{B} \preccurlyeq_c \mathsf{A}$, if there exists an affine mapping $V : Prob(\Omega_\mathsf{A}) \to Prob(\Omega_\mathsf{B})$ such that

$$\Phi_\mathsf{B} = V \circ \Phi_\mathsf{A} \,. \tag{4.56}$$

If $\mathsf{B} \preccurlyeq_c \mathsf{A} \preccurlyeq_c \mathsf{B}$, we denote $\mathsf{A} \overset{c}{\sim} \mathsf{B}$.

**Exercise 34.** Show that $\preccurlyeq_c$ is a preorder and $\stackrel{c}{\sim}$ is an equivalence relation.

The idea of the coarse-graining relation is that the additional mapping $V$ reduces the details of the statistical description. Typically, $V$ induces some loss of information, which can be due, for instance, to the non-injectivity of $V$. This reduction in the statistical description is also manifested in the following simple result.

**Proposition 55.** If B $\preccurlyeq_c$ A, then B $\preccurlyeq_i$ A.

*Proof.* Assume that B $\preccurlyeq_c$ A and $\varrho_1, \varrho_2$ are states such that $\Phi_\mathsf{A}(\varrho_1) = \Phi_\mathsf{A}(\varrho_2)$. Then

$$\Phi_\mathsf{B}(\varrho_1) = V \circ \Phi_\mathsf{A}(\varrho_1) = V \circ \Phi_\mathsf{A}(\varrho_2) = \Phi_\mathsf{B}(\varrho_2)\,.$$

This shows B $\preccurlyeq_i$ A.                                                                                      $\square$

In general, B $\preccurlyeq_i$ A does not imply that B $\preccurlyeq_c$ A. This comes clear in Subsection 4.6.2.

**Example 47.** Let A be a trivial observable defined by a probability measure $p$, i.e., $\mathsf{A}(X) = p(X)I$ (see Example 46). For any observable B, we then have A $\preccurlyeq_c$ B. Indeed, let $V$ be a mapping $V(m) = p$ for every probability measure $m$. Then $\Phi_\mathsf{A} = V \circ \Phi_\mathsf{B}$. Moreover, if B $\preccurlyeq_c$ A, then also B is a trivial observable.

Let $\Omega_\mathsf{A} = \{a_1, \ldots, a_k\}$ and $\Omega_\mathsf{B} = \{b_1, \ldots, b_l\}$ be finite sets, and let $V$ be an affine mapping from $Prob(\Omega_\mathsf{A})$ to $Prob(\Omega_\mathsf{B})$. A probability measure on $\Omega_\mathsf{A}$ is a convex combination of the point measures $\delta_{a_1}, \ldots, \delta_{a_k}$, and similarly for $\Omega_\mathsf{B}$. Thus, we can write

$$
\begin{aligned}
V(\delta_{a_1}) &= \nu_{11}\delta_{b_1} + \ldots + \nu_{1l}\delta_{b_l} \\
&\vdots \\
V(\delta_{a_k}) &= \nu_{k1}\delta_{b_1} + \ldots + \nu_{kl}\delta_{b_l}
\end{aligned}
$$

The properties of $V$ imply that $0 \leq \nu_{ij} \leq 1$ and $\sum_j \nu_{ij} = 1$. In other words, $(\nu_{ij})$ is a *stochastic matrix*. Condition (4.56) can now be written in the form

$$\mathsf{B}(b_j) = \sum_{i=1}^{k} \nu_{ij}\mathsf{A}(a_i)\,. \tag{4.57}$$

Note that this kind of construction can be written also if $\Omega_\mathsf{A}$ and $\Omega_\mathsf{B}$ are countable (but not necessarily finite) sets.

**Example 48.** (*Simple qubit observables*) In Example 36 in Section 4.2 we discussed sharp qubit observable A defined by a unit vector $\vec{a} \in \mathbb{R}^3$. The projections related to this observable are given by $\mathsf{A}(\pm 1) = \frac{1}{2}(I \pm \vec{a} \cdot \vec{\sigma})$. A stochastic $2 \times 2$ -matrix $(\nu_{ij})$ is completely determined by its two entries $\nu_{11}$ and $\nu_{21}$ as the normalization requirement fixes the other entries. We conclude that the most general two outcome observable (also called *simple*) B which is a coarse-graining of A is given by

$$\mathsf{B}(1) = \nu_{11}\mathsf{A}(1) + \nu_{21}\mathsf{A}(-1) = \frac{1}{2}\left((\nu_{11} + \nu_{21})I + (\nu_{11} - \nu_{21})\vec{a} \cdot \vec{\sigma}\right)\,.$$

We can write $\mathsf{B}(1)$ in the form

$$\mathsf{B}(1) = \frac{1}{2}\left(\beta I + \vec{b}\cdot\vec{\sigma}\right),\tag{4.58}$$

where $\beta = \nu_{11} + \nu_{21}$ and $\vec{b} = (\nu_{11} - \nu_{21})\vec{a}$. From the fact that $0 \leq \nu_{ij} \leq 1$ it follows that in this new parametrization

$$\left\|\vec{b}\right\| \leq \beta \leq 2 - \left\|\vec{b}\right\|.\tag{4.59}$$

On the other hand, every effect in $\mathbb{C}^2$ has the form (4.58) for some parameters $\beta$ and $\vec{b}$ satisfying (4.59). Indeed, the inequalities in (4.59) are equivalent to $O \leq \mathsf{B}(1) \leq I$. This leads to the conclusion that every simple qubit observable is a coarse-graining of some sharp qubit observable.

### 4.6    Example: photon counting observables

In this section we demonstrate some of the concepts studied earlier in a quantum optical context. We give only a very brief explanation of the quantum optical description of a single mode electromagnetic field. Not more than the basic concepts are needed, but a reader who is not familiar with this field before may wish to consult e.g. [37] or [81].

#### 4.6.1    Single mode electromagnetic field and number observable

In the usual quantum optical description, photon is a single excitation of a single mode of the electromagnetic field. Let $\mathcal{H}$ be an infinite dimensional Hilbert space. We fix an orthonormal basis $\{|0\rangle, |1\rangle, \ldots\}$ for $\mathcal{H}$, and we denote $\zeta_k = |k\rangle\langle k|$. The state $\zeta_0$ represents the vacuum state of the electromagnetic field, while the state $\zeta_k$ is taken to represent the excited state of the field containing $k$ photons. The states $\zeta_0, \zeta_1, \ldots$ are called the *number states*.

The *number observable* $\mathsf{N}$ is the sharp observable associated to the orthonormal basis $\{|0\rangle, |1\rangle, \ldots\}$ and it has $\mathbb{N} = \{0, 1, \ldots\}$ as its outcome space (recall Example 34 in Section 4.2). Hence, the effects of $\mathsf{N}$ are defined as $\mathsf{N}(n) = |n\rangle\langle n|$. In the above quantum optical setting $\mathsf{N}$ describes the ideal photon counting observable. Namely, for a number state $\zeta_k$, the related probability distribution in a $\mathsf{N}$-measurement is

$$p_{\zeta_k}^{\mathsf{N}}(n) = \mathrm{tr}\left[\zeta_k \mathsf{N}(n)\right] = \mathrm{tr}\left[|k\rangle\langle k||n\rangle\langle n|\right] = \delta_{k,n}.$$

The *number operator* $N$ is the (unbounded) selfadjoint operator corresponding to $\mathsf{N}$, that is,

$$N = \sum_{n=0}^{\infty} n\mathsf{N}(n).$$

Thus, $\mathrm{tr}\left[\rho N\right]$ gives the expectation value $\langle\mathsf{N}\rangle_\varrho$ of $\mathsf{N}$ in a state $\rho$.

To illustrate this framework further, we recall that for every $z \in \mathbb{C}$, a *coherent state* $|z\rangle$ is defined as

$$|z\rangle = e^{-|z|^2/2}\sum_{n=0}^{\infty}\frac{z^n}{\sqrt{n!}}|n\rangle.$$

Coherent states describe electromagnetic field which is generated by a laser. The number $|z|$ is proportional to the energy of the field. Hence, we expect that larger $|z|$ implies more photons. Indeed, the expectation value of $\mathsf{N}$ in the state $|z\rangle$ is

$$\langle \mathsf{N} \rangle_{|z\rangle} = \langle\, z \,|\, N \,|\, z \,\rangle = |z|^2 \;.$$

**Exercise 35.** Calculate the probability distribution of $\mathsf{N}$ in a coherent state $|z\rangle$. Confirm that it is a Poisson distribution.

### 4.6.2   Non-ideal photon counting observables

A non-ideal photodetector does not detect all the photons hitting the detector. Let us assume that a photon hitting the detector is detected with probability $\epsilon$, $0 \leq \epsilon \leq 1$. Therefore, if the electromagnetic field is in the number state $\zeta_k$, we expect that the probability $p(n|k)$ that the detector clicks $n$ times is

$$\begin{aligned}
p(n|k) &= \binom{k}{n}\epsilon^n(1-\epsilon)^{k-n} && \text{if } n \leq k\,, \\
p(n|k) &= 0 && \text{if } n > k\,.
\end{aligned} \tag{4.60}$$

With the usual convention that $0^0 = 1$ and $0^n = 0$ for $n \geq 1$ these probabilities makes sense also in the extreme cases $\epsilon = 0$ and $\epsilon = 1$.

We want to describe this kind of non-ideal photodetector as a coarse-graining of the number observable. For the number state $\zeta_k$, the number observable $\mathsf{N}$ gives the point probability measure $\delta_k$. Hence, a coarse-graining mapping $V$ maps $\delta_k$ to the probability distribution $p(\cdot|k)$. As explained in the end of Subsection 4.5.2, $V$ can be written as a matrix $(\nu_{kn})$ with respect to the point measures. A comparison to (4.60) shows that we need to set

$$\nu_{kn} = \left\{ \begin{array}{ll}
0 & \text{if } k < n, \\
\binom{k}{n}\epsilon^n(1-\epsilon)^{k-n} & \text{if } k \geq n.
\end{array} \right.$$

**Exercise 36.** Show that $(\nu_{kn})$ is a stochastic $\mathbb{N} \times \mathbb{N}$ -matrix.

We conclude that a photodetector with efficiency $\epsilon$ may be described by a discrete observable $\mathsf{N}^\epsilon$ defined by

$$\mathsf{N}^\epsilon(n) = \sum_{k=0}^{\infty} \nu_{kn}\mathsf{N}(k) = \sum_{k=n}^{\infty} \binom{k}{n}\epsilon^n(1-\epsilon)^{k-n}|k\rangle\langle k|, \quad n \in \mathbb{N}\,. \tag{4.61}$$

We call $\mathsf{N}^\epsilon$ a *photon counting observable*. The number observable $\mathsf{N}$ is thus the photon counting observable $\mathsf{N}^1$ with the ideal efficiency $\epsilon = 1$. On the other hand, the photon counting observable $\mathsf{N}^0$ with the worst efficiency $\epsilon = 0$ is a trivial observable.

The photon counting observable $\mathsf{N}^\epsilon$ should be seen as an unsharp or imprecise version of $\mathsf{N}$. Its form does not say anything about the mechanism or source of the imprecision, and $\mathsf{N}^\epsilon$ can arise in different ways. One possible setting which produces this kind of observable is an ideal photodetector which has a beam splitter in front of it. If the transparency of the beam splitter is $\epsilon$, then the observable representing the hole setup is exactly $\mathsf{N}^\epsilon$. We refer to Chapter VII.3 in [15] for a derivation of this result.

By our construction, $\mathsf{N}^\epsilon$ is a coarse-graining of $\mathsf{N}$. As one could expect, the converse is not true. Actually, the set of photon counting observables is totally ordered by the coarse-graining relation and the ordering corresponds to the ordering of the efficiencies. The following two results have been proved in [42].

**Proposition 56.** Let $\epsilon_1, \epsilon_2 \in [0,1]$ and let $\mathsf{N}^{\epsilon_1}$ and $\mathsf{N}^{\epsilon_2}$ be the corresponding photon counting observables. The relation $\mathsf{N}^{\epsilon_1} \preccurlyeq_c \mathsf{N}^{\epsilon_2}$ holds if and only if $\epsilon_1 \leq \epsilon_2$.

*Proof.* Let us first assume that $\mathsf{N}^{\epsilon_1} \preccurlyeq_c \mathsf{N}^{\epsilon_2}$. This means that there exists a stochastic matrix $\mu$ such that

$$\mathsf{N}^{\epsilon_1}(n) = \sum_{k=0}^\infty \mu_{kn} \mathsf{N}^{\epsilon_2}(k), \quad n \in \mathbb{N}.$$

For every $m, n \in \mathbb{N}$, we get

$$\langle m | \mathsf{N}^{\epsilon_1}(n) | m \rangle = \sum_{k=0}^\infty \mu_{kn} \langle m | \mathsf{N}^{\epsilon_2}(k) | m \rangle. \tag{4.62}$$

Substituting (4.61) into both sides of (4.62) we obtain the identity

$$\binom{m}{n} \epsilon_1^n (1-\epsilon_1)^{m-n} = \sum_{k=0}^m \mu_{kn} \binom{m}{k} \epsilon_2^k (1-\epsilon_2)^{m-k}. \tag{4.63}$$

Setting $m = n = 0$ we get $1 = \mu_{00}$. Since $\sum_n \mu_{kn} = 1$ and $\mu_{kn} \geq 0$ it follows that $\mu_{0n} = 0$ for all $n \geq 1$. Let us further set $m = n = 1$. We obtain $\epsilon_1 = \mu_{01}(1-\epsilon_2) + \mu_{11}\epsilon_2$. Since the previous setting implies $\mu_{01} = 0$ we get that $\mu_{11} = \epsilon_1 \epsilon_2^{-1}$. Since $\mu$ is a stochastic matrix, we have $\mu_{11} \leq 1$. This can hold only if $\epsilon_1 \leq \epsilon_2$.

Let us then assume that $\epsilon_1 \leq \epsilon_2$. Define

$$\mu_{kn} = \begin{cases} 0 & \text{if } k < n, \\ \binom{k}{n} \epsilon_1^n \epsilon_2^{-k} (\epsilon_2 - \epsilon_1)^{k-n} & \text{if } k \geq n. \end{cases}$$

Then $\mu$ is a stochastic matrix, and for each $n$ we have

$$
\begin{aligned}
\sum_{k=0}^\infty \mu_{kn} \mathsf{N}^{\epsilon_2}(k) &= \sum_{k=n}^\infty \sum_{m=k}^\infty \binom{k}{n}\binom{m}{k} \epsilon_1^n (\epsilon_2-\epsilon_1)^{k-n}(1-\epsilon_2)^{m-k} |m\rangle\langle m| \\
&= \sum_{m=n}^\infty \epsilon_1^n \left( \sum_{k=n}^m \binom{k}{n}\binom{m}{k} (\epsilon_2-\epsilon_1)^{k-n}(1-\epsilon_2)^{m-k} \right) |m\rangle\langle m| \\
&= \sum_{m=n}^\infty \binom{m}{n} \epsilon_1^n (1-\epsilon_1)^{m-n} |m\rangle\langle m| = \mathsf{N}^{\epsilon_1}(n).
\end{aligned}
$$

Thus, $\mathsf{N}^{\epsilon_1} \preccurlyeq_c \mathsf{N}^{\epsilon_2}$.                                                      $\square$

**Proposition 57.** Let $\epsilon \neq 0$. Then $\mathsf{N}^\epsilon \overset{i}{\sim} \mathsf{N}$.

*Proof.* As the claim is trivial in the case $\epsilon = 1$, we assume that $0 < \epsilon < 1$. Moreover, since $\mathsf{N}^\epsilon \preccurlyeq_c \mathsf{N}$, we have $\mathsf{N}^\epsilon \preccurlyeq_i \mathsf{N}$ by Proposition 55 in Subsec. 4.5.2. To prove that $\mathsf{N} \preccurlyeq_i \mathsf{N}^\epsilon$, let $\varrho_1, \varrho_2 \in \mathcal{S}(\mathcal{H})$ and assume that $\Phi_{\mathsf{N}^\epsilon}(\varrho_1) = \Phi_{\mathsf{N}^\epsilon}(\varrho_2)$. By (4.61) this means that for every $n \in \mathbb{N}$,

$$\sum_{k=n}^\infty \binom{k}{n} \epsilon^n (1-\epsilon)^k \langle k \,|\, \varrho_1 \,|\, k \rangle = \sum_{k=n}^\infty \binom{k}{n} \epsilon^n (1-\epsilon)^k \langle k \,|\, \varrho_2 \,|\, k \rangle \,. \qquad (4.64)$$

We can erase $\epsilon^n$ from both sides and move terms to one side. Hence, (4.64) is equivalent to the condition

$$\sum_{k=n}^\infty \binom{k}{n} (1-\epsilon)^k \langle k \,|\, \varrho_1 - \varrho_2 \,|\, k \rangle = 0 \,. \qquad (4.65)$$

Denote $a_k = (1-\epsilon)^k \langle k \,|\, \varrho_1 - \varrho_2 \,|\, k \rangle$ for every $k \in \mathbb{N}$. Since $|a_k| \leq (1-\epsilon)^k$, the formula

$$f(z) := \sum_{k=0}^\infty a_k z^k$$

defines an analytic function in the region $|z| < (1-\epsilon)^{-1}$. The $n$th derivative of $f$ is

$$f^{(n)}(z) = \sum_{k=n}^\infty k(k-1)\cdots(k-n+1) a_k z^{k-n} = \frac{1}{n!} \sum_{k=n}^\infty \binom{k}{n} a_k z^{k-n} \,,$$

and thus, (4.65) implies that $f^{(n)}(1) = 0$ for every $n \in \mathbb{N}$. Therefore, $f \equiv 0$, and hence $a_k = 0$ for every $k \in \mathbb{N}$. This means that $\langle k \,|\, \varrho_1 \,|\, k \rangle = \langle k \,|\, \varrho_2 \,|\, k \rangle$ for every $k \in \mathbb{N}$. We conclude that $\Phi_{\mathsf{N}}(\varrho_1) = \Phi_{\mathsf{N}}(\varrho_2)$, and therefore, $\mathsf{N} \preccurlyeq_i \mathsf{N}^\epsilon$. $\qquad \square$

It is perhaps surprising that ideal and non-ideal photon counting observables are informationally equivalent. Notice that this also implies that they have the same state determination power. However, an important difference between these observables is that the induced probability distributions possess different statistical properties. Roughly speaking the difference is in the statistical distance between the induced probabilities. In a sense, the probabilities induced by the sharp photon counting observable are better distinguishable than the probabilities induced by the non-ideal observables. This is exactly what the procedure of the coarse-graining is doing - a coarse-grained probability distribution is fuzzier and statistically noisier than the original.

**Exercise 37.** Calculate the probability distributions of $\mathsf{N}$ and $\mathsf{N}^\epsilon$ ($0 < \epsilon < 1$) in the number states $\zeta_1$ and $\zeta_2$. In Section 4.4.2 we introduced functions $A$ and $D$ to quantify the distance between probability distributions. Calculate the distances between the above probability distributions related to the states $\zeta_1$ and $\zeta_2$, and conclude that $\mathsf{N}$ leads to better separated probability distributions than $\mathsf{N}^\epsilon$.

## 5    Operations and channels

In Chapters 3 and 4 we assumed a model according to which an experiment is splitted into two parts - preparation and measurement. This led us to the associated concepts of states and observables, respectively. Based on this picture we can think about two types of apparatuses that can be abstractly characterized as follows:

- *preparators* are devices producing states, i.e. no quantum input required, but a quantum output is produced.

- *measurements* are input-output devices that are accepting quantum systems in their input and produce classical output in the form of recorded outcomes described by probability distributions, i.e. measurements do require some quantum input, but no quantum output is expected.

This picture indicates that one type of apparatuses is missing - devices taking quantum input and producing quantum output. Hence,

- *channels* are input-output devices transforming the quantum input into a quantum output.

Such device can be placed in between arbitrary preparation and measurement apparatuses. Channels and another slightly more general concept of operations are the topic of this chapter.

### 5.1    Operations and complete positivity

#### 5.1.1    Definition and basic properties

By a *physical operation* $\mathcal{N}$ we understand the most general action that can be performed on a considered physical system. Let us note that by definition the operations describe processes that are happening between preparation and measurement procedure of the experiment. Therefore, from the point of view of observed probabilities (being defined for a given state and observable), they can be understood either as parts of preparations (inputs and outputs are states), or as parts of measurements (inputs and outputs are effects). Depending on our preferences operations can be formulated either as mappings on states (*Schrodinger picture*), or as mappings on effects (*Heisenberg picture*). In what follows we mostly use the Schrodinger picture.

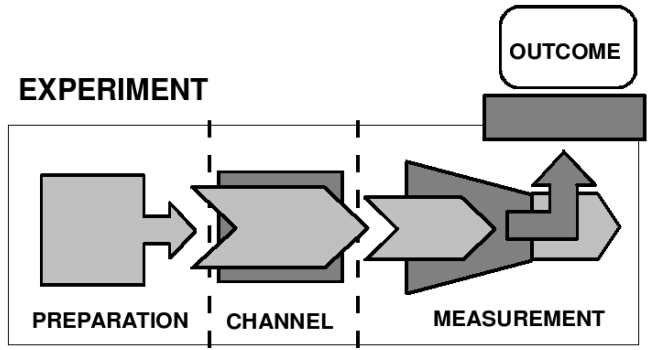

Figure 5.1. A basic framework of experiment including a channel device.

In order to formulate the most general quantum operation it is convenient to introduce a set of unnormalized states. The central notion of quantum theory is the probability rule, i.e., a prescription determining the probability for each pair of a state and an observable. The meaning of unnormalized states is that the predicted probabilities for all (normalized) observables are not necessarily normalized. We require, however, that the probabilities sum up to a number smaller than one, because in this case we can add an empty outcome to achieve the proper normalization formally corresponding to losses of systems in the experiment. Hence, the set of unnormalized states consists of positive trace class operators $\varrho$ with $\mathrm{tr}\,[\varrho] \leq 1$. We denote by $\mathcal{S}^{\mathrm{us}}$ the set of unnormalized states, $\mathcal{S}^{\mathrm{us}} = \{\varrho \in \mathcal{T}(\mathcal{H}) : \varrho \geq O, 0 \leq \mathrm{tr}\,[\varrho] \leq 1\}$.

In the Schrodinger picture a quantum operation $\mathcal{N}$ is defined as a mapping transforming the set of states $\mathcal{S} = \mathcal{S}(\mathcal{H})$ into the set of unnormalized states $\mathcal{S}^{\mathrm{us}}$. The statistical indistinguishability of different convex decompositions of the same state $\varrho$ cannot be affected by any operation as otherwise the state space must be redefined. Hence, operations must be affine, i.e. $\mathcal{N}(\sum_j \lambda_j \varrho_j) = \sum_j \lambda_j \mathcal{N}(\varrho_j)$ for all $\varrho_j \in \mathcal{S}$. This affinity extends by linearity on the linear space of trace class operators $\mathcal{T}(\mathcal{H})$, thus we will think operations as linear maps on $\mathcal{T}(\mathcal{H})$.

**Example 49.** *(No-cloning theorem - weak version.)* A cloning device is machine that produces a clone (a duplicate) of any unknown state. It takes one system as input and give back two of the same type. The other one is a duplicate of the first one in the sense that no experiment would see a difference between them. Hence, a cloning device acts as $\varrho \mapsto \varrho \otimes \varrho$ for all states $\varrho$. However, this kind of operation is not allowed in quantum mechanics, and this fact is known as the *no-cloning theorem*. The no-cloning theorem is a simple consequence of the requirement that operations are linear. Namely, we find the following inconsistency with linearity:

$$\omega = \sum_j \lambda_j \varrho_j \mapsto \omega \otimes \omega = \sum_{i,j} \lambda_i \lambda_j \varrho_i \otimes \varrho_j \neq \sum_j \lambda_j \varrho_j \otimes \varrho_j\,.$$

Later in Example 54 we discuss more sensitive formulation of the no-cloning theorem.

**Definition 37.** A linear mapping $\mathcal{N} : \mathcal{T}(\mathcal{H}) \to \mathcal{T}(\mathcal{H})$ is

  - *positive* if $\mathcal{N}(T) \geq 0$ for all for all $T \geq O$.

  - *trace-preserving* if $\mathrm{tr}\,[\mathcal{N}(T)] = \mathrm{tr}\,[T]$ for all $T \in \mathcal{T}(\mathcal{H})$.

  - *trace-decreasing* if $\mathrm{tr}\,[\mathcal{N}(T)] \leq \mathrm{tr}\,[T]$ for all $T \in \mathcal{T}(\mathcal{H})$.

If an operation $\mathcal{N}$ is understood as a linear mapping on $\mathcal{T}(\mathcal{H})$, then the states are mapped into unnormalized states only if the mapping $\mathcal{N}$ is trace-decreasing and positive. In particular, the requirement $\mathrm{tr}\,[\mathcal{N}(\varrho)] \leq 1$ for all $\varrho \in \mathcal{S}$ implies $\mathrm{tr}\,[\mathcal{N}(T)] \leq \mathrm{tr}\,[T]$ for all $T \in \mathcal{T}(\mathcal{H})$. Similarly $\mathcal{N}(\varrho) \geq O$ for all states implies that $\mathcal{N}(T) \geq O$ for all positive operators $T \geq O$. We thus make the conclusion that:

  • Quantum operation is a trace-decreasing and positive linear mapping on $\mathcal{T}(\mathcal{H})$.

Consider a state of a composite system $A + B$ consisting of two subsystems $A$ and $B$. We can extend the operation $\mathcal{N}_A$ acting only on the subsystem $A$ to a mapping $\mathcal{N}_A \otimes \mathcal{I}_B$ acting on a composite system, where $\mathcal{I}_B$ is the identity mapping defined on subsystem $B$. This should not make any difference as $\mathcal{I}_B$ does nothing.

However, it is a nontrivial assumption that for all possible extensions the transformation $\mathcal{N}_A \otimes \mathcal{I}_B$ transforms states of $\mathcal{S}_{AB} \equiv \mathcal{S}(\mathcal{H}_A \otimes \mathcal{H}_B)$ to unnormalized states on the composite system $A + B$. For a positive trace-decreasing linear mapping $\mathcal{N}$ the joint mapping $\mathcal{N}_A \otimes \mathcal{I}_B$ is trace-decreasing, but its positivity is not guaranteed. If it is positive for all $\varrho_{AB} \in \mathcal{S}_{AB}$ and all extensions $B$, then we say that the map $\mathcal{N}_A$ is *completely positive*. Otherwise the map $\mathcal{N}_A$ is not a physical operation, because the probability interpretation of such mappings leads to negative probabilities.

**Definition 38.** A linear mapping $\mathcal{N}_A : \mathcal{T}(\mathcal{H}_A) \to \mathcal{T}(\mathcal{H}_A)$ is *completely positive* if the mapping $\mathcal{N}_A \otimes \mathcal{I}_B$ is positive on $\mathcal{T}(\mathcal{H}_A \otimes \mathcal{H}_B)$ for all extensions $\mathcal{H}_B$.

As a conclusion, we need to redefine quantum operations in the following way:

- Quantum operation is a completely positive trace-decreasing linear mapping on $\mathcal{T}(\mathcal{H})$.

In the following example we demonstrate that there are positive mappings which are not completely positive.

**Example 50.** *(Partial transposition.)* Let $\{\varphi_i\}$ be an orthonormal basis for a Hilbert space $\mathcal{H}$. Define operators $e_{ij} = |\varphi_i\rangle\langle\varphi_j|$ in $\mathcal{T}(\mathcal{H})$. *Transposition* is a linear mapping $\tau : \mathcal{T}(\mathcal{H}) \to \mathcal{T}(\mathcal{H})$, which acts on the operators $e_{ij}$ as

$$e_{ij} \mapsto \tau(e_{ij}) = e_{ji}. \tag{5.1}$$

Hence, it is a specific permutation of the elements of the operator basis $\{e_{ij}\}$. In matrix representation of operators it is represented by a matrix transposition in the basis $\{\varphi_i\}$.

Let us assume that $d = \dim \mathcal{H} < \infty$. Consider a vector $\psi_+ = \frac{1}{\sqrt{d}} \sum_j \varphi_j \otimes \varphi_j \in \mathcal{H} \otimes \mathcal{H}$. Applying the mapping (*partial transposition*) $\tau_A \otimes \mathcal{I}_B$ onto the state $|\psi_+\rangle\langle\psi_+|$ we obtain an operator $\omega'$ that is not positive. For instance, if $d = 2$, then

$$\omega' = \begin{pmatrix} 1 & 0 & 0 & 0 \\ 0 & 0 & 1 & 0 \\ 0 & 1 & 0 & 0 \\ 0 & 0 & 0 & 1 \end{pmatrix}, \tag{5.2}$$

with eigenvalues $\pm 1$. As a result we see that transposition $\tau$ is an example of positive, but not completely positive map. A practical implication is that such operation is not experimentally implementable.

**Example 51.** *(Quantum NOT gate.)* In classical information theory the NOT operation is one of the elementary single-bit gates changing the original bit value, i.e., $0 \mapsto 1 \mapsto 0$. Two-dimensional quantum systems (known as quantum bits, or qubits) can be used to encode two logical values, hence qubit can store one bit of information. In the quantum language the orthogonality of pure states is replacing the concept of opposite bit values. Quantum NOT gate is defined as machine implementing the transformation

$$\mathcal{E}_{\text{NOT}} : P_\psi \longmapsto P_{\psi_\perp} \,, \tag{5.3}$$

where $\langle \psi_\perp \,|\, \psi \rangle = 0$. Let us note that such definition makes sense only in a two-dimensional Hilbert space, because otherwise the orthogonal vector $\psi_\perp$ is not unique up to complex factor. In particular, if $\psi = a\varphi_0 + b\varphi_1$, then (up to global phase) $\psi_\perp = b^*\varphi_0 - a^*\varphi_1$. It follows that $P_{\psi_\perp} = \sigma_y \tau(P_\psi)\sigma_y$, where $\tau$ is the transposition defined in Example 50. Therefore, $\mathcal{E}_{\text{NOT}}$ is a positive, but not completely positive linear mapping, hence no physical device can implement the quantum NOT gate.

**Exercise 38.** Prove that to verify the positivity, complete positivity, trace-preservity and trace-decreasity it is sufficient to consider the validity only for states. In particular, show that if $\mathcal{N}(\varrho) \geq O$ for all states, then $\mathcal{N}(T) \geq O$ for all positive operators $T$, and if $\text{tr}\,[\mathcal{N}(\varrho)] = 1$ is valid for all states it follows that $\text{tr}\,[\mathcal{N}(T)] = \text{tr}\,[T]$ for all trace class operators $T$.

Quantum operations are interpreted as the most general actions on quantum systems that can be realized experimentally. The usage of unnormalized states in their definition reflects the probabilistic nature of quantum measurements. For some measurements the measured object is available also after the measurement is accomplished and outcome is registered. This kind of scenario will be discussed in the next chapter. In this chapter we pay attention to quantum channels describing *quantum processes* that are placed between preparations and measurements. Let us note that in contrast with state automorphisms (unitary transformations) discussed in Chapter 3 we do not require that quantum channels $\mathcal{E}$ are bijective.

**Definition 39.** A mapping $\mathcal{N} : \mathcal{T}(\mathcal{H}) \to \mathcal{T}(\mathcal{H})$ is an *operation / channel* if it is

1. linear,

2. completely positive,

3. trace-decreasing / trace-preserving.

We denote by $\mathcal{O}$ the set of quantum operations. It is a subset of all completely positive mappings denoted by $\mathcal{M}_{cp}$. A subset $\mathcal{O}_c \subset \mathcal{O}$ containing all trace-preserving quantum operations is the set of all quantum channels.

### 5.1.2 Schrodinger vs. Heisenberg picture

States and effects are dual objects as explained in Chapter 3. Consequently, each quantum operation understood in the Schrodinger picture as a linear mapping $\mathcal{N} : \mathcal{T}(\mathcal{H}) \to \mathcal{T}(\mathcal{H})$ induces a linear transformation $\mathcal{N}_* : \mathcal{L}(\mathcal{H}) \to \mathcal{L}(\mathcal{H})$ defined by the formula

$$\text{tr}\,[\mathcal{N}(T)E] = \text{tr}\,[T\mathcal{N}_*(E)] \,, \tag{5.4}$$

holding for all trace class operators $T \in \mathcal{T}(\mathcal{H})$ and bounded operators $E \in \mathcal{L}(\mathcal{H})$. The mapping $\mathcal{N}_*$ describes the same quantum operation as $\mathcal{N}$ but in the *Heisenberg picture*, i.e., as an operation acting on effects. In this picture the states are not affected, but the effects are. Since $(\mathcal{N} \otimes \mathcal{I})_* = \mathcal{N}_* \otimes \mathcal{I}$ the complete positivity of $\mathcal{N}$ is equivalent to the complete positivity of $\mathcal{N}_*$. The identity

$$\mathrm{tr}\left[\mathcal{N}(T)\right] = \mathrm{tr}\left[\mathcal{N}(T)I\right] = \mathrm{tr}\left[T\mathcal{N}_*(I)\right]$$

implies that an operation $\mathcal{N}$ is trace-decreasing if and only if the dual operation $\mathcal{N}_*$ decreases the identity operator in the sense of partial ordering of positive operators, i.e., $\mathcal{N}_*(I) \leq I$. For channels the trace-preservity of $\mathcal{N}$ is replaced by *unitality* of $\mathcal{N}_*$, namely, $\mathcal{N}_*(I) = I$.

**Definition 40.** A linear mapping $\mathcal{N}_* : \mathcal{L}(\mathcal{H}) \to \mathcal{L}(\mathcal{H})$ is *unital* if $\mathcal{N}_*(I) = I$.

In conclusion, we have seen that a channel $\mathcal{N} : \mathcal{T}(\mathcal{H}) \to \mathcal{T}(\mathcal{H})$ defines a linear mapping $\mathcal{N}_* : \mathcal{L}(\mathcal{H}) \to \mathcal{L}(\mathcal{H})$ which is completely positive and unital.

There is one technical remark that has to be mentioned in the case of infinite dimensional Hilbert spaces. Alternatively, channels can be defined directly in terms of linear mappings on $\mathcal{L}(\mathcal{H})$. In this case, in addition to complete positivity and unitality, one has to add one more requirement, namely, *normality*. We are not going to need this concept, so we refer to [29] for details. Actually, one can usually circumvent this discussion by using operator sum decompositions, which are introduced in Section 5.2.3.

**Example 52.** (*Complete contraction*) Fix a positive trace class operator $F \neq O$. Consider a channel $\mathcal{E}_F$ defined by

$$\mathcal{E}_F(T) = \frac{\mathrm{tr}\left[T\right]}{\mathrm{tr}\left[F\right]} F$$

for all $T \in \mathcal{T}(\mathcal{H})$. Let us specify its action in the Heisenberg picture, i.e., the mapping $\mathcal{E}_{F*} : \mathcal{L}(\mathcal{H}) \to \mathcal{L}(\mathcal{H})$. By definition

$$\mathrm{tr}\left[T\mathcal{E}_{F*}(A)\right] = \mathrm{tr}\left[\mathcal{E}_F(T)A\right] = \frac{\mathrm{tr}\left[T\right]}{\mathrm{tr}\left[F\right]}\mathrm{tr}\left[FA\right] \tag{5.5}$$

holds for all $T \in \mathcal{T}(\mathcal{H})$ and all $A \in \mathcal{L}(\mathcal{H})$. It follows that

$$\mathcal{E}_{F*}(A) = \frac{\mathrm{tr}\left[FA\right]}{\mathrm{tr}\left[F\right]}I \,. \tag{5.6}$$

In summary, the channel $\mathcal{E}_F$ maps in the Schrodinger picture all trace class operators into a one-dimensional linear subspace spanned by the positive trace class operator $F$, whereas in the Heisenberg picture all operators are mapped into one-dimensional subspace spanned on the identity operator $I$. If $\mathcal{E}$ is restricted to states, then $\mathcal{E}_F(\varrho) = \frac{1}{\mathrm{tr}\left[F\right]}F$, i.e., the whole state space is contracted into a single point represented by the state $F/\mathrm{tr}\left[F\right]$.

**Exercise 39.** Show that $\mathcal{I}_* = \mathcal{I}$.

### 5.2  Physical model of quantum channels

Quantum channels were introduced as transformations of quantum states (or effects) satisfying certain mathematical requirements in order to preserve the basic framework of quantum statistics introduced in Chapter 3. No additional assumption (except the mutual compatibility of channels, states and effects) was made. In this section we introduce quantum channels starting from a slightly different point of view.

### 5.2.1 Isolated vs. open systems

For the purposes of system's dynamics physicists are distinguishing among two different types of systems: *isolated* and *open*. We say that a system is isolated if its changes are reversible. Otherwise the system is called open. As it was argued in Chapter 3, the reversibility condition requires that the transformations of the system are either unitary or antiunitary. Although the isolated systems are only mathematical idealizations of the reality, it is a common paradigm in physics that every open system is embedded in a larger isolated system, conventionally called the *environment*. This implies that in principal an evolution can always be reversed, but sufficiently large environment must be taken into account.

Let us denote by $\mathcal{H}_S$ and $\mathcal{H}_E$ the Hilbert spaces of the system and the environment, respectively. Consider a general input state $\omega \in \mathcal{S}(\mathcal{H}_S \otimes \mathcal{H}_E)$ and some unitary, or antiunitary transformation $U : \mathcal{H}_S \otimes \mathcal{H}_E \to \mathcal{H}_S \otimes \mathcal{H}_E$ acting on the total (isolated) system and mapping the state $\omega$ into a state $\omega' = U\omega U^*$. An observer having access to the system $S$ only finds out the evolution

$$\mathrm{tr}_E[\omega] \equiv \varrho \mapsto \varrho' \equiv \mathrm{tr}_E[U\omega U^*] \,. \tag{5.7}$$

We can also think this picture from the other side. Namely, suppose that the system is in a state $\varrho$. We assume that there is a state $\xi$ of the environment such that we can form the compound state $\varrho \otimes \xi$. It is assumed that the state and environment are independent in the sense that we can always prepare the environment into the state $\xi$, independently of $\varrho$. In other words, we assume that the mapping

$$\mathcal{P}_\xi : \mathcal{S}(\mathcal{H}_S) \to \mathcal{S}(\mathcal{H}_S \otimes \mathcal{H}_E) \,, \qquad \mathcal{P}_\xi(\varrho) = \varrho \otimes \xi \tag{5.8}$$

can be realized.

**Proposition 58.** Let $\mathcal{H}_E$ be the Hilbert space describing the environment, $U$ unitary operator on $\mathcal{H} \otimes \mathcal{H}_E$ and $\xi$ a fixed state of the environment. Then the induced mapping

$$\mathcal{E} : \mathcal{S}(\mathcal{H}_S) \to \mathcal{S}(\mathcal{H}_S) \,, \qquad \mathcal{E}(\varrho) = \mathrm{tr}_E[U\varrho \otimes \xi U^*] \tag{5.9}$$

is a channel.

*Proof.* Formula (5.9) can be written as $\mathcal{E} = \mathrm{tr}_E \circ \sigma_U \circ \mathcal{P}_\xi$. The linearity, complete positivity and trace-preservation follows from the properties of mappings $\mathcal{P}_\xi$, $U$ and $\mathrm{tr}_E$. Each of them is linear, completely positive and trace-preserving. $\square$

In Proposition 58 we assumed that $U$ is a unitary operator. Since antiunitary operators are also state isomorphisms, could we prove a similar result using antiunitary operators instead of unitary operators? The answer to this question is negative, as shown in the following result.

**Proposition 59.** Let $A$ be an antiunitary operator acting on $\mathcal{H}$. The mapping $\sigma_A$ is positive, but not completely positive.

*Proof.* As shown in Proposition 10 in Section 2.2.3, the antiunitary operator $A$ can be written as a composition of a unitary operator $U$ and complex conjugate operator $J$ with respect to some basis $\{\varphi_j\}$ of the Hilbert space $\mathcal{H}$. For every unit vector $\psi \in \mathcal{H}$, we get

$$P_{J\psi} = \tau(P_\psi) \,,$$

where $\tau$ is the transposition with respect to the orthonormal basis $\{\varphi_j\}$, as defined in Example 50. Consequently, the complex conjugation of vectors is associated with the transposition of pure states. Since the transposition is positive but not completely positive, it follows that also $\sigma_A$ is positive but not completely positive.                                                                                       $\square$

Our conclusion from Proposition 59 is that if $\sigma_A$ is applied on a subsystem the total system can become unphysical, i.e described by a nonpositive operator. Therefore, we can restrict to unitary mappings when describing the evolution of the closed system $\mathcal{H} \otimes \mathcal{H}_E$.

It may not be so surprising that a combination of the mappings $\mathcal{P}_\xi$, $\sigma_U$ and $\mathrm{tr}_E$ gives a channel. All these three mappings are something that we assume to be physically realizable. However, could there be something more general? Perhaps our construction gives only some special channels and to get other channels we need a more general construction. The final answer to this question in given in Subsection 5.2.2. However, to increase our understanding it is instructive to take a look on these possible generalized constructions.

**Definition 41.** A mapping $\mathcal{P} : \mathcal{S}(\mathcal{H}_S) \to \mathcal{S}(\mathcal{H}_S \otimes \mathcal{H}_E)$ is called a *preparation mapping* if $\mathrm{tr}_E[\mathcal{P}(\varrho)] = \varrho$ for all $\varrho \in \mathcal{H}_S$.

Having the preparation mapping defined, we can express the evolution $\mathcal{E}$ of the system $S$ as a composition of three mappings: preparation $\mathcal{P}$, unitary transformation $\sigma_U$ and partial trace over the environment $\mathrm{tr}_E$,

$$\mathcal{E} = \mathrm{tr}_E \circ \sigma_U \circ \mathcal{P} \,. \tag{5.10}$$

Notice that $\mathcal{P}_\xi$ defined in (5.8) is a preparation mapping with a very special form. With the general concept of preparation mapping, do we get more general class of evolutions than that defined by channels?

**Example 53.** *(Implementation of arbitrary transformation of the state space.)* Suppose the system and the environment are described by the same Hilbert space, i.e., $\mathcal{H}_E = \mathcal{H}_S$. Define a preparation mapping

$$\mathcal{P}_f : \varrho_S \mapsto \varrho_S \otimes f(\varrho_S) \,,$$

where $f : \mathcal{S}(\mathcal{H}_S) \to \mathcal{S}(\mathcal{H}_S)$ is an arbitrary function. Choose $U = V_{\mathrm{SWAP}}$ to be the SWAP transformation defined by the formula

$$V_{\mathrm{SWAP}}(\varrho \otimes \sigma)V_{\mathrm{SWAP}}^* = \sigma \otimes \varrho \,.$$

Applying sequentially all three operations according to Eq.(5.10) we obtain the transformation $\varrho \mapsto f(\varrho)$. In particular,

$$\varrho \overset{\mathcal{P}}{\longmapsto} \varrho \otimes f(\varrho) \overset{\sigma_{V_{\mathrm{SWAP}}}}{\longmapsto} f(\varrho) \otimes \varrho \overset{\mathrm{tr}_E}{\longmapsto} f(\varrho) \,. \tag{5.11}$$

Let us note that no restrictions are applied to function $f$ and in this picture each transformation is possible.

As shown in Example 53, the model under consideration is capable to describe arbitrary state transformation $f : \mathcal{S}(\mathcal{H}) \to \mathcal{S}(\mathcal{H})$. So what goes wrong (if anything)?

The explanation of this conundrum is, once again, that we need to require mappings to be affine. Especially, we require that the preparation mapping $\mathcal{P}$ is affine, i.e., $\mathcal{P}(\sum_j p_j \varrho_j) = \sum_j p_j \mathcal{P}(\varrho_j)$ for all affine combinations of states. Otherwise we shall be able to distinguish different convex decompositions. From the other side a very important assumption allowing us to split experiment into several parts is the *statistical independence* of individual apparatuses in the following sense: an observable describing the measurement apparatus does not depend on the preparation and *vice versa*. In the same way also the description of channels is independent of particular preparations and observables. This assumption requires that the environments affecting the preparations and channels are independent, hence the joint state is uncorrelated, i.e., $\mathcal{P}(\varrho) = \varrho \otimes f(\varrho)$. Applying the affinity constraint it follows that $f$ is necessarily constant, so that there is a fixed state $\xi$ such that $f(\varrho) = \xi$ for all $\varrho$ and $\mathcal{P}(\varrho) = \varrho \otimes \xi$. Hence, with these necessary constrains we are back in preparation mappings of the form (5.8).

### 5.2.2 Stinespring's dilation theorem

It is one of the fundamental facts in the theory of open systems that Proposition 58 has a counterpart. Namely, each channel can be understood as a unitary evolution of an extended system. This result was originally proved in by W.F. Stinespring in [78]. A proof can be also found, for instance, in [28].

**Theorem 7.** *(Stinespring's Theorem.)* Let $\mathcal{E} : \mathcal{T}(\mathcal{H}) \to \mathcal{T}(\mathcal{H})$ be a quantum channel. There exists an environment Hilbert space $\mathcal{H}_E$, unitary operator $U : \mathcal{H} \otimes \mathcal{H}_E \to \mathcal{H} \otimes \mathcal{H}_E$ and an initial state of the environment $\xi \in \mathcal{S}(\mathcal{H}_E)$ such that

$$\mathcal{E}(\varrho) = \mathrm{tr}_E[U(\varrho \otimes \xi)U^*] \tag{5.12}$$

for all $\varrho \in \mathcal{S}(\mathcal{H})$. The triple $\langle \mathcal{H}_E, U, \xi \rangle$ is called a *Stinespring dilation* of the channel $\mathcal{E}$.

Stinespring dilation of a channel $\mathcal{E}$ is not unique. Roughly speaking, this is due to the fact that starting from one Stinespring dilation, we can always choose a bigger environment and in this way produce a new Stinespring dilation. However, Stinespring dilation is unique if an additional requirement of minimality is added.

**Definition 42.** A Stinespring dilation $\langle \mathcal{H}_E, V, \xi \rangle$ is *minimal* if the closure of linear span of the set $\{\psi \in \mathcal{H} \otimes \mathcal{H}_E : \psi = (T \otimes I)V^*(\varphi \otimes \xi) \text{ for some } \varphi \in \mathcal{H}, T \in \mathcal{L}(\mathcal{H})\}$ equals $\mathcal{H} \otimes \mathcal{H}_E$.

In other words, the dilation is minimal if the environment $\mathcal{H}_E$ is smallest possible. In general, for the minimal dilation $\dim \mathcal{H}_E \leq \dim \mathcal{H}$.

**Proposition 60.** Two minimal dilations $V_1, V_2$ of the same quantum channel are related by a unitary transformation $W : \mathcal{H}_{E_2} \to \mathcal{H}_{E_1}$, i.e. $\mathcal{H}_{E_2} = \mathcal{H}_{E_1}$ and $V_1 = (I \otimes W)V_2$.

**Example 54.** *(No cloning theorem for pure states - strong version.)* In Example 49 we pointed out that the cloning transformation $\varrho \mapsto \varrho \otimes \varrho$ is not linear, hence, it cannot be realized experimentally. In this example we show that even stronger statement holds. In particular, a pair of nonorthogonal pure states cannot be cloned. This theorem is of use in many applications in quantum cryptography, because cloning would be an ideal strategy of an adversary.

Consider a pair of pure states $P_\psi = |\psi\rangle\langle\psi|$ and $P_\phi = |\phi\rangle\langle\phi|$. According to quantum model the cloning machine is acting as a unitary channel $\sigma_U$ on a composite system consisting of the system itself (associated with $\mathcal{H}_d$) and some ancilla system described by Hilbert space $\mathcal{H}_{\text{anc}} = \mathcal{H}_d \otimes \mathcal{H}'$. Without loss of generality we may assume that the initial state of the ancillary system is pure, $\xi_{\text{anc}} = |\varphi\rangle\langle\varphi|$. This fact follows from the purification procedures as described in Section 3.4.2. Consequently, the cloning transformation requires

$$P_\psi \otimes |\varphi\rangle\langle\varphi| \quad \mapsto \quad P_\psi \otimes P_\psi \otimes |\psi'\rangle\langle\psi'|,$$
$$P_\phi \otimes |\varphi\rangle\langle\varphi| \quad \mapsto \quad P_\phi \otimes P_\phi \otimes |\phi'\rangle\langle\phi'|,$$

where the purity of states $|\psi'\rangle\langle\psi'|$ and $|\phi'\rangle\langle\phi'|$ is guaranteed by the unitarity of the channel $\sigma_U$. As unitary channels satisfy $\text{tr}\,[\sigma_U[\varrho_1]\sigma_U[\varrho_2]] = \text{tr}\,[\varrho_1\varrho_2]$ for all states $\varrho_1, \varrho_2$, it follows that the identity

$$\text{tr}\,[P_\psi P_\phi] = (\text{tr}\,[P_\psi P_\phi])^2 |\,\langle\,\phi'\,|\,\psi'\,\rangle\,|^2\,, \tag{5.13}$$

must hold for some $\phi', \psi' \in \mathcal{H}'$.

Suppose then that the states $P_\psi$ and $P_\phi$ are different but not orthogonal, so that $0 < \text{tr}\,[P_\psi P_\phi] < 1$. Equation (5.13) now gives

$$1 = \text{tr}\,[P_\psi P_\phi]\,|\,\langle\,\phi'\,|\,\psi'\,\rangle\,|^2\,. \tag{5.14}$$

However, since $\text{tr}\,[P_\psi P_\phi] < 1$ and $|\,\langle\,\phi'\,|\,\psi'\,\rangle\,|^2 \leq 1$, the above condition cannot be satisfied. In summary, we have proved that the cloning of nonorthogonal states cannot be described by means of quantum channels. Hence, the cloning of nonorthogonal states is not a physical transformation.

### 5.2.3   Operator sum decomposition of quantum channels

In this subsection we derive a very convenient alternative form for channels. It is usually helpful especially in checking the complete positivity of a given mapping.

**Proposition 61.** A linear mapping $\mathcal{E} : \mathcal{T}(\mathcal{H}) \to \mathcal{T}(\mathcal{H})$ is an operation if and only if there exists a sequence of bounded operators $A_1, A_2, \dots$ such that $\mathcal{E}(T) = \sum_k A_k T A_k^*$ for every $T \in \mathcal{T}(\mathcal{H})$. In this case, $\mathcal{E}$ is a channel if and only if also the condition $\sum_k A_k^* A_k = I$ holds.

*Proof.* Suppose that $\mathcal{E}(\cdot) = \sum_k A_k \cdot A_k^*$ for some sequence of bounded operators $A_1, A_2, \dots$. Let $\mathcal{H}_{\text{anc}}$ be an ancillary Hilbert space, $T \in \mathcal{T}(\mathcal{H} \otimes \mathcal{H}_{\text{anc}})$ a positive operator, and $\psi \in \mathcal{H} \otimes \mathcal{H}_{\text{anc}}$. Then

$$\langle\,\psi\,|\,(\mathcal{E} \otimes \mathcal{I})(T)\psi\,\rangle = \sum_k \langle\,(A_k^* \otimes I)\psi\,|\,(T)(A_k^* \otimes I)\psi\,\rangle = \sum_k \langle\,\psi_k\,|\,T\psi_k\,\rangle \geq 0\,,$$

showing that $(\mathcal{E} \otimes \mathcal{I})(T) \geq O$, and hence leading to the conclusion that $\mathcal{E}$ is completely positive. In the above calculation we used the positivity of $T$ and we denoted $\psi_k = (A_k^* \otimes I)\psi$. For the trace of $\mathcal{E}(T)$ we get

$$\operatorname{tr}\left[\mathcal{E}(T)\right] = \sum_k \operatorname{tr}\left[A_k T A_k^*\right] = \operatorname{tr}\left[T(\sum_k A_k^* A_k)\right].$$

This proves the second claim.

Next, let us suppose that $\mathcal{E}$ is a channel and $\langle \xi, \mathcal{H} \otimes \mathcal{H}_E, U \rangle$ is its dilation in accordance with the Stinespring theorem. Without loss of generality we can assume $\xi = |\varphi_1\rangle\langle\varphi_1|$, because each mixed state can be purified by adding some ancillary degrees of freedom (see Section 3.4.2). Choose an orthonormal basis $\{\varphi_k\}$ for $\mathcal{H}_E$, and for each $k$ define operator $A_k$ via the identity

$$\langle \psi \,|\, A_k \phi \rangle = \langle \psi \otimes \varphi_k \,|\, U \phi \otimes \varphi_1 \rangle\,,$$

holding for all $\psi, \phi \in \mathcal{H}$. It follows that for all $\psi, \varphi \in \mathcal{H}$ and for all $\varrho \in \mathcal{S}(\mathcal{H})$, we have

$$
\begin{aligned}
\langle \psi \,|\, \mathcal{E}(\varrho)\varphi \rangle &= \langle \psi \,|\, \operatorname{tr}_E[U(\varrho \otimes |\varphi_1\rangle\langle\varphi_1|)U^*]\phi \rangle \\
&= \sum_k \langle \psi \otimes \varphi_k \,|\, U(\varrho \otimes |\varphi_1\rangle\langle\varphi_1|)U^*(\phi \otimes \varphi_k) \rangle \\
&= \sum_k \sum_{ab} \varrho_{ab} \langle U^*(\psi \otimes \varphi_k) \,|\, \psi_a \otimes \varphi_1 \rangle \langle \psi_b \otimes \varphi_1 \,|\, U^*(\phi \otimes \varphi_k) \rangle \\
&= \sum_k \sum_{ab} \varrho_{ab} \langle \psi \,|\, A_k \psi_a \rangle \langle A_k \psi_b \,|\, \phi \rangle \\
&= \sum_k \langle \psi \,|\, A_k \varrho A_k^* \phi \rangle\,,
\end{aligned}
\tag{5.15}
$$

thus $\mathcal{E}(\varrho) = \sum_k A_k \varrho A_k^*$. $\qquad\square$

**Exercise 40.** Prove the following: a mapping $\mathcal{E}_* : \mathcal{L}(\mathcal{H}) \to \mathcal{L}(\mathcal{H})$ describes channel in the Heisenberg picture if and only if there exists a sequence of bounded operators $A_1, A_2, \ldots$ such that $\mathcal{E}_*(T) = \sum_k A_k^* T A_k$ for every $T \in \mathcal{L}(\mathcal{H})$ and $\sum_k A_k^* A_k = I$.

In summary, the expression of quantum channel $\mathcal{E}$ via operators $A_1, \ldots, A_n$ satisfying the normalization $\sum_k A_k^* A_k = I$ is called *operator sum form* (sometimes also *Kraus form* and the operators are called *Kraus operators*). In the Schrodinger picture the channel reads $\mathcal{E}(\varrho) = \sum_k A_k \varrho A_k^*$. In the Heisenberg picture it takes the form $\mathcal{E}_*(T) = \sum_k A_k^* T A_k$. The normalization $\sum_k A_k^* A_k = I$ is picture independent.

**Example 55.** (*Contraction into the total mixture.*) Suppose that $d = \dim \mathcal{H} < \infty$. The *contraction into the total mixture* is a quantum channel $\mathcal{A}_0$ mapping the whole state space into a single point represented by the total mixture, i.e., $\mathcal{A}_0(\varrho) = \frac{1}{d}I$ for all $\varrho \in \mathcal{S}(\mathcal{H})$. Consider an orthonormal basis $\{\varphi_j\}$ of $\mathcal{H}$ and define operators $E_{jk} = |\varphi_j\rangle\langle\varphi_k|$. Since $\sum_{j,k} E_{jk}^* E_{jk} = dI$ it follows that $T \mapsto \frac{1}{d} E_{jk} T E_{jk}^*$ is a channel. In fact,

$$\mathcal{A}_0(T) = \frac{1}{d} \sum_{jk} E_{jk} T E_{jk}^* = \operatorname{tr}\left[T\right] \frac{1}{d} I\,.
\tag{5.16}$$

Let us note that choosing a different basis of the Hilbert space $\mathcal{H}_E$, or different dilation space $\langle \xi'_E, \mathcal{H}_{E'}, U' \rangle$ associated with $\mathcal{E}$ we obtain different operator sum representation of the quantum channel $\mathcal{E}$. The ambiguity in this type of representation is specified in details in the following proposition.

**Proposition 62.** Two sets of bounded operators $\{A_1, \ldots, A_n\}$ and $\{B_1, \ldots, B_m\}$ define the same completely positive map if and only if

$$A_j = \sum_k u_{jk} B_k \,, \tag{5.17}$$

where coefficients $u_{jk}$ form a partial isometry matrix, i.e. $\sum_k u_{jk}^* u_{jl} = \delta_{kl}$.

*Proof.* The sufficiency can be proved by directly. The identities

$$\sum_j A_j T A_j^* = \sum_j u_{jk} u_{jl}^* B_k T B_l^* = \sum_k B_k T B_k^* \tag{5.18}$$

holds for all $T \in \mathcal{T}(\mathcal{H})$. Next let us consider that $\sum_j A_j T A_j^* = \sum_k B_k T B_k^*$ for all $T \in \mathcal{T}(\mathcal{H})$. If we set $T = |\varphi\rangle\langle\varphi|$, then $\sum_j A_j |\varphi\rangle\langle\varphi| A_j^* = \sum_j |\phi_j\rangle\langle\phi_j| = \sum_k |\psi_k\rangle\langle\psi_k| = \sum_k B_k |\varphi\rangle\langle\varphi| B_k^*$, where $\{\phi_j\}$, $\{\psi_k\}$ are collections of unnormalized vectors. The Proposition 21 implies that $\phi_j = \sum_k u_{jk} \psi_k$, where $u_{jk}$ form a matrix of partial isometry. Consequently, $A_j \varphi = \sum_k u_{jk} B_k \varphi$ for all $\varphi \in \mathcal{H}$ implies that $A_j = \sum_k u_{jk} B_k$.  □

### 5.3 Elementary properties of quantum channels

In this section we introduce some basic properties and concepts used for the characterization of quantum channels. Let us recall the notation: $\mathcal{M}_{cp}$ is the set of completely positive linear mappings, $\mathcal{O}$ is the set of quantum operations, and $\mathcal{O}_c$ stands for the set of quantum channels.

#### 5.3.1 Convexity

The sets $\mathcal{O}_c \subset \mathcal{O} \subset \mathcal{M}_{cp}$ are convex, i.e. $\lambda \mathcal{E}_1 + (1-\lambda)\mathcal{E}_2$ is (trace-decreasing/trace-preserving) completely positive map if $0 \leq \lambda \leq 1$ and $\mathcal{E}_1, \mathcal{E}_2$ are (trace-decreasing/trace-preserving) completely positive maps. As in the case of states we can speak about *extremal channels* and about a *completely mixed channel*. We shall not give a complete characterization of extremal channels, but rather give some examples. In particular, there are two elementary examples of extremal elements of $\mathcal{O}_c$:

- *Unitary channels* $\sigma_U(\cdot) = U \cdot U^*$. The extremality comes from the uniqueness of operator-sum representation of unitary channels. In fact, a nontrivial convex combination of two channels necessary results in the operator-sum representation with more than one Kraus operator.

- *Contractions to pure states* mapping the whole state space into a single pure state $|\psi\rangle\langle\psi|$, i.e. $\mathcal{A}_\psi : \mathcal{S}(\mathcal{H}) \to |\psi\rangle\langle\psi|$. In this case the extremality follows from the extremality of the pure states. In particular, let us assume that $\mathcal{A}_\psi = \lambda \mathcal{E}_1 + (1-\lambda)\mathcal{E}_2$. Then necessarily $|\psi\rangle\langle\psi| = \lambda \mathcal{E}_1(\varrho) + (1-\lambda)\mathcal{E}_2(\varrho)$ for all states $\varrho$. However, due to extremality of $|\psi\rangle\langle\psi|$ it follows that $\mathcal{E}_1(\varrho) = \mathcal{E}_2(\varrho) = |\psi\rangle\langle\psi|$ for all $\varrho$, i.e. $\mathcal{E}_1 = \mathcal{E}_2 = \mathcal{A}_\psi$.

In the rest of this subsection we assume that $\dim \mathcal{H} < \infty$. In the search of completely mixed channel we are motivated by some specific properties of the completely mixed state. In particular, we know that the completely mixed state is the unique state commuting with all the unitary operators and it is also the mixture of all states, i.e., it is the average state. Moreover, let us notice that the total mixture is the only state commuting with all operators. In accordance, we shall pay attention to channels commuting with all unitary channels, i.e. $[\mathcal{E}, \sigma_U] = 0$, or equivalently, $\mathcal{E}(UTU^*) = U\mathcal{E}(T)U^*$ for all $U$.

**Example 56.** (*Average unitary channel.*) Consider a channel

$$\mathcal{A}(T) = \int_{U(d)} UTU^* \, dU \,,$$

where $dU$ is the invariant Haar measure. We then have $[\mathcal{A}(T), U] = 0$ for all unitary operators $U$. Hence, by Schur lemma we can conclude that $\mathcal{A}(T)$ is proportional to identity operator, i.e. $\mathcal{A}(T) = c(T)I$, where $c(T)$ is some linear functional. Since $\mathcal{A}$ is trace-preserving it follows that

$$\mathrm{tr}\,[T] = \mathrm{tr}\,[\mathcal{A}(T)] = c(T)\mathrm{tr}\,[I] = c(T)d \,,$$

thus $c(T) = \mathrm{tr}\,[T]\,/d$. In summary, we showed that

$$\mathcal{A}(T) = \frac{\mathrm{tr}\,[T]}{d} I \equiv \mathcal{A}_0(T) \,, \tag{5.19}$$

which is obviously a channel commuting with all unitary channels.

Further, let us fix a channel $\mathcal{E}$ and define its unitary orbit $O_\mathcal{E}$ as a set of all channels $\mathcal{E}_U = \sigma_U \circ \mathcal{E}$. Using the result of Example 56 we find out that starting from any channel $\mathcal{E}$, the average over all channels $\mathcal{E}_U \in O_\mathcal{E}$ gives

$$\int_{U(d)} U\mathcal{E}(T)U^* \, dU = \frac{\mathrm{tr}\,[\mathcal{E}(T)]}{d} I = \frac{\mathrm{tr}\,[T]}{d} I = \mathcal{A}_0(T) \,.$$

Moreover, if $\sigma_U \circ \mathcal{E} = \sigma_V \circ \mathcal{F}$ for some channels $\mathcal{E}, \mathcal{F}$, then $O_\mathcal{E} = O_\mathcal{F}$, because $\sigma_U = \sigma_V \circ \sigma_W$ for some unitary channel $\sigma_W$. In fact, $\sigma_U \circ \mathcal{E} = \sigma_V \circ \sigma_W \circ \mathcal{E} = \sigma_V \circ \mathcal{F}$ implies $\mathcal{F} = \sigma_W \circ \mathcal{E}$. We can use these observations to argue that $\mathcal{A}_0$ is the average channel and therefore it can be identified as the completely mixed channel. Notice that $\mathcal{A}_0 \circ \mathcal{E} = \mathcal{A}_0$ for all channels $\mathcal{E}$. Let us remind that $\mathcal{A}_0$ maps any state into the completely mixed state.

### 5.3.2   Distances and channels

The channels are by their action changing the distances between the initial and the final states. In Sections 4.4.3 and 4.4.4 we introduced distance measures quantifying how close are quantum states (trace distance and fidelity). These measures for states can be used to quantify the degree of disturbance caused by the action of quantum channel, but also to introduce distance between the channels themselves.

**Proposition 63.** Quantum channels are contractive, i.e., they do not increase the distance between two quantum states. In particular, for the trace distance and the fidelity the following inequalities hold:

$$D(\mathcal{E}(\varrho_1), \mathcal{E}(\varrho_2)) \leq D(\varrho_1, \varrho_2), \tag{5.20}$$

$$F(\mathcal{E}(\varrho_1), \mathcal{E}(\varrho_2)) \geq F(\varrho_1, \varrho_2). \tag{5.21}$$

*Proof.* We prove only the inequality for the trace distance. It is sufficient to use only the fact that channels are trace-preserving, i.e. we show that trace-preseving positive linear maps are contractive. A difference of positive operators is a selfadjoint operator. Consequently, from spectral decomposition it follows that $\varrho_1 - \varrho_2 = P - Q$, where $P, Q$ are positive operators with mutually orthogonal supports, i.e. $\mathrm{tr}\,[|P - Q|] = \mathrm{tr}\,[P] + \mathrm{tr}\,[Q]$. The operator $\varrho_1 - \varrho_2$ is traceless, hence $\mathrm{tr}\,[P] = \mathrm{tr}\,[Q] = \mathrm{tr}\,[\mathcal{E}(P)] = \mathrm{tr}\,[\mathcal{E}(Q)]$. According to Eq. (4.48) $\mathrm{tr}\,[C(\varrho_1 - \varrho_2)] \leq \frac{1}{2}\mathrm{tr}\,[|\varrho_1 - \varrho_2|]$ and let us assume that $C$ is the projector for which $\frac{1}{2}\mathrm{tr}\,[|\mathcal{E}(\varrho_1 - \varrho_2)|] = \mathrm{tr}\,[C\mathcal{E}(\varrho_1 - \varrho_2)]$. Using all these facts we find that

$$
\begin{aligned}
D(\varrho_1, \varrho_2) &= \frac{1}{2}\mathrm{tr}\,[|\varrho_1 - \varrho_2|] = \frac{1}{2}\mathrm{tr}\,[|P - Q|] = \mathrm{tr}\,[P] = \mathrm{tr}\,[\mathcal{E}[P]] \\
&\geq \mathrm{tr}\,[C\mathcal{E}[P]] \geq \mathrm{tr}\,[C(\mathcal{E}[P] - \mathcal{E}[Q])] = \frac{1}{2}\mathrm{tr}\,[\mathcal{E}(\varrho_1 - \varrho_2)] \\
&\geq D(\mathcal{E}[\varrho_1], \mathcal{E}[\varrho_2]),
\end{aligned}
\tag{5.22}
$$

what proves the inequality for the trace distance.                                        □

Depending on our preferences there are several options how to understand the concept of distance between quantum channels. Proposition 63 suggests to quantify the difference of quantum channels by means of their action on state space. For instance, the following functions are quantifying the "distance" of channels from slightly different perspectives catching different features of channel's actions.

1. *Minimal distance:* $D_{\inf}(\mathcal{E}_1, \mathcal{E}_2) = \inf_\varrho \|(\mathcal{E}_1 - \mathcal{E}_2)(\varrho)\|_{\mathrm{tr}}$

2. *Maximal distance:* $D_{\sup}(\mathcal{E}_1, \mathcal{E}_2) = \sup_\varrho \|(\mathcal{E}_1 - \mathcal{E}_2)(\varrho)\|_{\mathrm{tr}}$

3. *Average distance:* $D_{\mathrm{ave}}(\mathcal{E}_1, \mathcal{E}_2) = \int_{\mathcal{S}(\mathcal{H})} \|(\mathcal{E}_1 - \mathcal{E}_2)(\varrho)\|_{\mathrm{tr}}\; d\varrho$

**Example 57.** Suppose that $d = \dim \mathcal{H} < \infty$. Let us calculate the distances between a unitary channel $\sigma_U$ and the contraction channel $\mathcal{A}_0$ to the total mixture. We get

$$D_{\inf}(\sigma_U, \mathcal{A}_0) = \frac{1}{2}\inf_\varrho \left\|\sigma_U(\varrho) - \frac{1}{d}I\right\|_{\mathrm{tr}} = 0, \tag{5.23}$$

$$D_{\sup}(\sigma_U, \mathcal{A}_0) = \frac{1}{2}\sup_\varrho \left\|\sigma_U(\varrho) - \frac{1}{d}I\right\|_{\mathrm{tr}} = \frac{1}{2}\left\||\psi\rangle\langle\psi| - \frac{1}{d}I\right\|_{\mathrm{tr}} = (d-1)/d. \tag{5.24}$$

We used the facts that unitary channels preserve total mixture and have some pure state as the fixed point, i.e., $\sigma_U(|\psi\rangle\langle\psi|) = |\psi\rangle\langle\psi|$. Therefore, the infimum is achieved for the total mixture. On the other hand, the pure states maximize the distance from the total mixture and for all pure states this distance equals $(d-1)/d$.

**Example 58.** (*Quantification of noise.*) For the purposes of quantum state transmission a relevant quantity is the degree of noise introduced by the application of the channel describing the transfer. For these purposes one can adopt any measure used to quantify the difference of two quantum channels by setting one of the channels to be the *noiseless channel* identified with the identity channel. In particular, how noisy the channel is can be quantified by the formula

$$\Delta_{\sup}(\mathcal{E}) = D_{\sup}(\mathcal{E}, \mathcal{I}) = \sup_{\varrho \in \mathcal{S}(\mathcal{H})} \frac{1}{2} \text{tr}\left[|\mathcal{E}(\varrho) - \varrho|\right] . \tag{5.25}$$

This formula gives $\Delta_{\sup}(\mathcal{E}) = 0$ if and only if $\mathcal{E} = \mathcal{I}$. In fact, for arbitrary other channel there is a state $\varrho$ such that $\mathcal{E}(\varrho) \neq \varrho$ which guarantees that $\Delta_{\sup}(\mathcal{E}) > 0$ for $\mathcal{E} \neq \mathcal{I}$. The maximal value of trace distance is 1. This is achieved if a difference of mutually orthogonal pure states are evaluated. That is, if a channel $\mathcal{E}$ maps a pure state $|\psi\rangle\langle\psi|$ into an orthogonal state $|\psi_\perp\rangle\langle\psi_\perp|$, then this channel introduces the maximal possible noise with respect to the considered measure. For example, for the pure state contractions

$$\Delta_{\sup}(\mathcal{A}_\psi) = \sup_\varrho \frac{1}{2} \text{tr}\left[||\psi\rangle\langle\psi| - \varrho|\right] = \text{tr}\left[||\psi\rangle\langle\psi| - |\psi_\perp\rangle\langle\psi_\perp||\right] = 1 . \tag{5.26}$$

Similarly, if $\langle\psi\,|\,U\psi\rangle = 0$ for some vector $\psi$, then for these unitary channels $\Delta_{\sup}(\sigma_U) = 1$, i.e. unitary channels can be as noisy as possible. Surprisingly, the contraction into the total mixture $\mathcal{A}_0 : \mathcal{S}(\mathcal{H}) \mapsto \frac{1}{d}I$ does not introduce the maximal possible noise, because

$$\Delta_{\sup}(\mathcal{A}_0) = \sup_\varrho \frac{1}{2} \text{tr}\left[|\frac{1}{d}I - \varrho|\right] = (d-1)/d . \tag{5.27}$$

For the calculation, see Example 57.

### 5.3.3   Conjugate channels

The Stinespring theorem provides a physical model of a quantum channel $\mathcal{E}$ via its Stinespring dilation $\langle\mathcal{H}_E, U, \xi_E\rangle$. Let us observe that each dilation defines also a mapping from $\mathcal{T}(\mathcal{H})$ to $\mathcal{T}(\mathcal{H}_E)$. Namely, we take the partial trace over the system instead of the environment. This leads to the following definition.

**Definition 43.** A channel $\mathcal{E}' : \mathcal{T}(\mathcal{H}) \to \mathcal{T}(\mathcal{H}_E)$ is called *conjugated (or complementary)* to a channel $\mathcal{E} : \mathcal{T}(\mathcal{H}) \to \mathcal{T}(\mathcal{H})$ if there exist a dilation $\langle\mathcal{H}_E, U, \xi\rangle$ of $\mathcal{E}$ such that

$$\mathcal{E}' = \text{tr}_S \circ \sigma_U \circ \mathcal{P}_\xi . \tag{5.28}$$

Here $\text{tr}_S$ stands for the partial trace of the system and $\mathcal{P}_\xi$ denotes the addition of a factorized ancilla.

The fact that $\mathcal{E}'$ is a channel follows directly from the definition, because the mappings $\text{tr}_S, \sigma_U, \mathcal{P}_\xi$ are channels. Since Stinespring dilation is not unique it follows that a channel $\mathcal{E}$ can have many conjugated channels.

Consider an operator form of the channel $\mathcal{E}$, $\mathcal{E}(\varrho) = \sum_j R_j \varrho R_j^*$ and dilation $\langle\mathcal{H}_E, U, \xi_j\rangle$ with $U = \sum_{jk} A_{jk} \otimes |\varphi_k\rangle\langle\varphi_j|$ such that $A_{j1} = R_j$. The unitarity of $U$ is guaranteed if $\sum_j A_{jk}^* A_{jk'} = \delta_{kk'} I$ and $\sum_k A_{jk}^* A_{j'k} = \delta_{jj'} I$. Setting $\xi_E = |\varphi_1\rangle\langle\varphi_1|$ we obtain for the channel and the conjugate channel the following expressions

$$\mathcal{E}(T) = \sum_j R_j T R_j^*; \tag{5.29}$$

$$\mathcal{E}'(T) = \sum_{jk} \operatorname{tr}\left[R_j T R_k^*\right] |\varphi_j\rangle\langle\varphi_k|. \tag{5.30}$$

Therefore, the ambiguity of $\mathcal{E}'$ is in some sense equivalent to ambiguity in operator sum form.

**Example 59.** *(Conjugate channels to a unitary channel.)* Let $\sigma_U$ be a unitary channel. The operator sum form of $\sigma_U$ is unique. By Eq. (5.30) for the conjugate channel we get

$$\sigma_U'(T) = \operatorname{tr}\left[T\right] |\varphi\rangle\langle\varphi|,$$

where the unit vector $\varphi$ can be chosen arbitrarily. Thus, the contraction $\mathcal{A}_\varphi$ into a pure state $|\varphi\rangle\langle\varphi|$ is a conjugate channel for $\sigma_U$. In order to derive Eq.(5.30) it was assumed that the initial state of the environment is pure. Consider $\xi_E$ is a mixed state with a decomposition $\xi_E = \sum_j \lambda_j |\varphi_j\rangle\langle\varphi_j|$. Due to linearity the associated conjugate channel $\mathcal{E}' = \sum_j \lambda_k \mathcal{E}_j'$, where $\mathcal{E}_j'$ are the conjugated channels corresponding to pure states. Let us note that $\mathcal{E}_j'$ and $\mathcal{E}'$ are, in general, conjugated channels to channels $\mathcal{E}_j$ and $\mathcal{E}$, respectively, but these channels are usually different. However, if $\mathcal{E}_j = \mathcal{E}_k$ for all $j, k$, then $\mathcal{E}'$ is also a conjugated channel to $\mathcal{E} \equiv \mathcal{E}_j$. Since in the considered case for all $\varphi \in \mathcal{H}_E$ the channels $\mathcal{A}_\varphi$ are conjugated to $\sigma_U$, it follows that also the convex combinations of $\mathcal{A}_j$ are conjugated to $\sigma_U$. That is, a complete contraction into arbitrary mixed state $\xi$ represents the conjugated channel $\mathcal{A}_\xi$ to each unitary channel $\sigma_U$. Moreover, since unitary channels have the unique operator sum decomposition, there are no others conjugated channels. As a result we get that channels

$$\sigma_U'[T] = \mathcal{A}_\xi[T] = \operatorname{tr}\left[T\right] \xi \tag{5.31}$$

form the set of conjugate channels to arbitrary unitary channel $\sigma_U$.

An interesting and important consequence of this example is that if the system evolves in a unitary way, no trace of the initial state $\varrho$ is left in the final state of the environment. For unitary channels the environment is completely independent of the system and therefore unitarily evolving systems can be considered to be closed.

### 5.4 Parametrizations of quantum channels

In this subsection all Hilbert spaces are assumed to be finite dimensional and we shall discuss several different representations of quantum channels acting on such finite systems.

#### 5.4.1 Matrix representation

Quantum channels are linear mappings on the vector space of operators $\mathcal{L}(\mathcal{H})$ and therefore they can be represented as matrices if operators are understand as vectors (see Section 3.1.3). Consider an orthogonal operator basis $E_0, \ldots, E_{d^2-1}$ of $d$-dimensional quantum system. A general operator $T = \sum_j t_j E_j$ can be expressed as a vector $\vec{t}$ with coefficients $t_j = \frac{1}{\mathrm{tr}\left[E_j^* E_j\right]} \mathrm{tr}\left[E_j^* T\right]$. Similarly, a linear mapping $\mathcal{E}$ on $\mathcal{L}(\mathcal{H})$ form a matrix with elements

$$\mathcal{E}_{jk} = \frac{1}{\mathrm{tr}\left[E_j^* E_j\right] \mathrm{tr}\left[E_k^* E_k\right]} \mathrm{tr}\left[E_j^* \mathcal{E}[E_k]\right] \tag{5.32}$$

and the action of the quantum channel can be expressed as the standard matrix multiplication of matrix $\mathcal{E}$ and vector $\vec{t}$

$$\mathcal{E} : \vec{t} \mapsto \vec{t}' = \mathcal{E}\vec{t}. \tag{5.33}$$

In this type of parametrization the composition of two channels corresponds to the usual matrix multiplication. A disadvantage of this representation is that the constraint of complete positivity is very difficult to verify and other equivalent forms must be exploited.

**Example 60.** (*Matrix operator basis.*) Let $\{\varphi_j\}$ be an orthonormal basis of the Hilbert space $\mathcal{H}$. The operators $e_{jk} = |\varphi_j\rangle\langle\varphi_k|$ expressed as matrices in the basis $\{\varphi_j\}$ contain only one nonzero entry placed at $j$th row and $k$th column. They are linearly independent and form the operator basis of $\mathcal{L}(\mathcal{H})$. Moreover, $\mathrm{tr}\left[e_{jk} e_{ab}\right] = \delta_{ja}\delta_{kb}$, i.e., this matrix operator basis is orthonormal.

**Example 61.** (*Traceless selfadjoint operator basis.*) It is convenient to choose a basis containing the identity operator (up to a factor), i.e., $\tilde{E}_0 = cI$, and for other elements use the notation $\tilde{E}_j = cE_j$. The orthogonality condition fixes all other elements of the basis to be traceless, $\mathrm{tr}\left[E_j\right] = 0$ for $j \neq 0$. Since the interesting operators are selfadjoint, it is natural to choose operators that are selfadjoint. Moreover, we assume that $\mathrm{tr}\left[E_j^2\right] = \mathrm{tr}\left[I\right] = d$, where $d$ is the dimension of the Hilbert space $\mathcal{H}$, so that $\mathrm{tr}\left[\tilde{E}_j \tilde{E}_k\right] = c^2 d\delta_{jk}$ and $\mathrm{tr}\left[E_j E_k\right] = d\delta_{jk}$. Let us note that selfadjoint trace class operators $\mathcal{T}_s(\mathcal{H})$ form a real vector space. In such case the vectors representing states and matrices corresponding to channels are real. We get the so-called *Bloch vector representation* of states and channels that was already introduced in Chapter 3. The so-called *Bloch vector* $\vec{r}$ associated with a state $\varrho$ is defined via identity

$$\vec{r} = \mathrm{tr}\left[\varrho\vec{E}\right] = \frac{1}{c}\mathrm{tr}\left[c\varrho\vec{\tilde{E}}\right] = \frac{d}{c}\vec{x}, \tag{5.34}$$

hence,

$$\varrho = \sum_j \frac{\mathrm{tr}\left[\tilde{E}_j\varrho\right]}{\mathrm{tr}\left[\tilde{E}_j^2\right]}\tilde{E}_j = \frac{c^2}{c^2 d}\sum_j \mathrm{tr}\left[E_j\varrho\right] E_j = \frac{1}{d}(I + \vec{r}\cdot\vec{E}), \tag{5.35}$$

where $\vec{E}$ is a vector of $d^2 - 1$ operators $E_1, \ldots, E_{d^2-1}$. Similarly, a matrix of a channel $\mathcal{E}$ is described by elements

$$\mathcal{E}_{jk} = \frac{\mathrm{tr}\left[\tilde{E}_j \mathcal{E}[\tilde{E}_k]\right]}{\mathrm{tr}\left[\tilde{E}_j^2\right]\mathrm{tr}\left[\tilde{E}_k^2\right]} = \frac{c^2}{c^4 d^2}\mathrm{tr}\left[E_j \mathcal{E}[E_k]\right] = \frac{1}{c^2 d^2}\mathrm{tr}\left[E_j \mathcal{E}[E_k]\right]. \tag{5.36}$$

This matrix is acting on vectors $\vec{x}$ as follows $x_j \mapsto x'_j = \sum_k \mathcal{E}_{jk} x_k$. Since $\mathcal{E}$ is tracepreserving it follows that $\mathrm{tr}\left[I\mathcal{E}[E_k]\right] = \delta_{0k}$, hence $x'_0 = x_0$. That is, for $j \neq 0$

$$r'_j = \mathcal{E}_{j0} + \sum_{k=1}^{d^2-1} \mathcal{E}_{jk} r_k = \frac{1}{c^2 d^2}\left(\mathrm{tr}\left[E_j \mathcal{E}[I]\right] + \sum_{k=1}^{d^2-1} \mathrm{tr}\left[E_j \mathcal{E}[E_k]\right] r_k\right). \tag{5.37}$$

As a result we get the Bloch vector representation of quantum channels. Defining a vector $\vec{\tau}$ with entries $\tau_j = \mathcal{E}_{j0}$ and a $(d^2-1) \times (d^2-1)$ matrix with entries $T_{jk} = \mathcal{E}_{jk}$ for $j, k \neq 0$ the channels take the *affine form*

$$\vec{r} \mapsto \vec{\tau} + T\vec{r}. \tag{5.38}$$

Setting $c = 1/\sqrt{d}$ the operators $\tilde{E}_j = \frac{1}{\sqrt{d}} E_j$ form an orthonormal operator basis and $\vec{\tau} = \frac{1}{d}\mathrm{tr}\left[\vec{E}\mathcal{E}[I]\right]$, $T_{jk} = \frac{1}{d}\mathrm{tr}[E_j\mathcal{E}[E_k]]$.

**Example 62.** *(Unitary channels.)* Consider an orthonormal traceless selfadjoint operator basis of $d$-dimensional quantum system, $\frac{1}{\sqrt{d}}I, \frac{1}{\sqrt{d}}E_1, \ldots, \frac{1}{\sqrt{d}}E_{d^2-1}$. For unitary channels $\mathcal{E}_U$ the vector $\vec{\tau} = \frac{1}{d}\mathrm{tr}\left[\vec{E}UIU^*\right] = \vec{0}$ vanishes and $[T_U]_{jk} = \frac{1}{d}\mathrm{tr}\left[E_j U E_k U^*\right]$. Moreover,

$$\begin{aligned}
\sum_k [T_U]_{jk}[T_U^T]_{kl} &= \sum_k [T_U]_{jk}[T_U]_{lk} = \frac{1}{d^2}\sum_k (U^*E_jU|E_k)(E_k|U^*E_lU) \\
&= \frac{1}{d}(U^*E_jU|U^*E_lU) = \frac{1}{d}\mathrm{tr}\left[U^*E_jUU^*E_lU\right] = \frac{1}{d}\mathrm{tr}\left[E_jE_l\right] \\
&= \delta_{jl}\,,
\end{aligned} \tag{5.39}$$

and similarly also $T_U^T T_U = I$, hence $T_U$ is an orthogonal matrix. In the above definition we used the notation $(A|B) = \mathrm{tr}\left[A^*B\right]$ and the fact that $\frac{1}{d}\sum_{j=0}^{d^2-1} |E_j)(E_j|$ defines an identity map on operators providing that $\{E_j\}$ is an orthonormal operator basis.

By definition $\det T_U$ is a product of all eigenvalues of $T_U$. Let us denote by $e^{ix_j}$ the eigenvalues of unitary operator $U$ and $U\varphi_j = e^{ix_j}\varphi_j$ determines the eigenvectors $\varphi_j \in \mathcal{H}$. Clearly, the eigenvalues of $T_U$ are $\mu_{jk} = e^{i(x_j - x_k)}$ and eigenvectors are the Bloch vectors corresponding to operators $|\varphi_j\rangle\langle\varphi_k|$. Consequently, $\det T_U = \prod_{j,k} \mu_{jk} = 1$.

As a result we found that in the Bloch representation the unitary channels correspond to special orthogonal rotations $T_U$. However, the inverse is not true except the case $d = 2$. This fact can be easily seen if comparing the number of parameters defining a general unitary transformation on $d$-dimensional complex vector space and the number of parameters specifying a general orthogonal rotation on $d^2 - 1$ dimensional real vector space.

### 5.4.2   $\chi$-matrix representation

On contrary to previous representation, the Kraus operator sum form of quantum channels is explicitly completely positive, what is a very important advantage of this representation. However, it is not unique and there is a unitary freedom in particular choice of Kraus operators. As in the previous case, let us fix some orthonormal operator basis $E_0, \ldots, E_{d^2-1}$ and express each Kraus operator $A_n$ in this operator basis, $A_n = \sum_n a_{nj} E_j$. The identity

$$\mathcal{E}(\varrho) = \sum_n A_n \varrho A_n^* = \sum_{rs} \sum_n a_{nr} a_{ns}^* E_r \varrho E_s^* = \sum_{rs} \chi_{rs} E_r \varrho E_s^* \qquad (5.40)$$

defines the $\chi$-*matrix representation* of the quantum channel $\mathcal{E}$. Let us note that composition of quantum channels does not correspond to multiplication of $\chi$ matrices. The $\chi$-matrix representation reduces the ambiguity of the operator sum representation. Two different Kraus decompositions of the same channel $\mathcal{E}$ lead to the same $\chi$-matrix.

Let us now derive the relation between the two considered representations of quantum channels. A direct calculation shows that

$$\mathcal{E}_{jk} = \text{tr} \left[ E_j^* \mathcal{E}[E_k] \right] = \sum \chi_{rs} \text{tr} \left[ E_j^* E_r E_k E_s^* \right] . \qquad (5.41)$$

Defining an object $M_{jk,rs} = \text{tr} \left[ E_j^* E_r E_k E_s^* \right]$ we get the relation between the two representations

$$\mathcal{E}_{jk} = M_{jk,rs} \chi_{rs} \qquad (5.42)$$

of the same quantum channel $\mathcal{E}$.

**Proposition 64.** If $\mathcal{E}$ is a quantum channel, then the corresponding $\chi$-matrix is positive and $\text{tr} [\chi] = d$.

*Proof.* For a quantum channel $\mathcal{E}(\varrho) = \sum_n A_n \varrho A_n^*$ on a $d$-dimensional quantum channel the corresponding $\chi-$ matrix is acting on $d^2$-dimensional complex vector space. For all complex vectors $\vec{x} = (x_1, \ldots, x_{d^2})$

$$\sum_{r,s} x_r^* \chi_{rs} x_s = \sum_n \sum_r (x_r^* a_{nr}) \sum_s (a_{ns}^* x_s) = \sum_n y_n y_n^* = \sum_n |y_n|^2 \geq 0 , \qquad (5.43)$$

hence, the matrix is positive. A direct calculation gives

$$\begin{aligned} \sum_r \chi_{rr} &= \sum_{r,n} (E_r | A_n)(A_n | E_r) = \sum_n (A_n | \left( \sum_r |E_r)(E_r| \right) |A_n) \\ &= \sum_n (A_n | A_n) = \sum_n \text{tr} \left[ A_n^* A_n \right] = \text{tr} \left[ I \right] = d , \end{aligned}$$

where we used the fact that $\sum_r |E_r)(E_r|$ defines the identity map on operators and the normalization $\sum_n A_n^* A_n = I$.  □

### 5.4.3   Choi-Jamiolkowski isomorphism

We denote by $\mathcal{M}_n$ the matrix algebra of $n \times n$ complex matrices.

**Theorem 8.** *(Choi theorem.)* Let $\mathcal{E} : \mathcal{M}_n \to \mathcal{M}_m$ be a positive linear mapping. Then the following statements are equivalent:

*(i)* $\mathcal{E}$ is $n$-positive, i.e., $\mathcal{E} \otimes \mathcal{I}_n$ is a positive map.

*(ii)* The operator matrix

$$
\Phi_{\mathcal{E}} = \begin{pmatrix}
\mathcal{E}(|\varphi_1\rangle\langle\varphi_1|) & \dots & \mathcal{E}(|\varphi_1\rangle\langle\varphi_n|) \\
\vdots & \ddots & \vdots \\
\mathcal{E}(|\varphi_n\rangle\langle\varphi_1|) & \dots & \mathcal{E}(|\varphi_n\rangle\langle\varphi_n|)
\end{pmatrix}
\tag{5.44}
$$

is positive, where $\varphi_j$ is an orthonormal basis of the $n$-dimensional Hilbert space and where $\mathcal{E}(|\varphi_j\rangle\langle\varphi_k|)$ are elements from $\mathcal{M}_m$. The matrix $\Phi_{\mathcal{E}}$ is called *Choi matrix* of $\mathcal{E}$.

*(iii)* $\mathcal{E}$ is completely positive.

*Proof.* The implication $(iii) \Rightarrow (i)$ follows directly from the definition of the complete positivity of $\mathcal{E}$. Consider a positive matrix $M = \sum_{jk} |\varphi_j \otimes \varphi_j\rangle\langle\varphi_k \otimes \varphi_k| \in \mathcal{M}_n \otimes \mathcal{M}_n$. By positivity of $\mathcal{E} \otimes \mathcal{I}_n$ also the matrix $M' = \mathcal{E} \otimes \mathcal{I}_n[M]$ is positive. Since $M' = \mathcal{E}[|\varphi_j\rangle\langle\varphi_k|] \otimes |\varphi_j\rangle\langle\varphi_k| = \Phi_{\mathcal{E}}$ the positivity of $\Phi_{\mathcal{E}}$ follows, i.e. the implication $(i) \Rightarrow (ii)$ holds.

It remains to prove $(ii) \Rightarrow (iii)$. By positivity $\Phi_{\mathcal{E}} = \sum_l |\psi_l\rangle\langle\psi_l|$, where $\psi_l \in \mathbb{C}^n \otimes \mathbb{C}^m$ are unnormalized eigenvectors. The tensor product can be seen as a tensor sum $\mathbb{C}_1^m \oplus \cdots \oplus \mathbb{C}_n^m = \mathbb{C}^n \otimes \mathbb{C}^m$ and let $P_j : \mathbb{C}^n \otimes \mathbb{C}^m \to \mathbb{C}_j^m$ be a projection onto the $j$th copy of $\mathbb{C}^m$. Then $P_j \Phi_{\mathcal{E}} P_k = \mathcal{E}[|\varphi_j\rangle\langle\varphi_k|] = \sum_l P_j |\psi_l\rangle\langle\psi_l| P_k = \sum_l |P_j \psi_l\rangle\langle P_k \psi_l|$. Define $d^2$ operators $V_r : \mathbb{C}^n \to \mathbb{C}^m$ by their action $V_r \varphi_j = P_j \psi_r$. With the help of these operators $\mathcal{E}[|\varphi_j\rangle\langle\varphi_k|] = \sum_r |P_j \psi_r\rangle\langle P_k \psi_r| = \sum_r |V_r \varphi_j\rangle\langle V_r \varphi_k| = \sum_r V_r |\varphi_j\rangle\langle\varphi_k| V_r^*$. By linearity $\mathcal{E}[A] = \sum_r V_r A V_r^*$ is defined on all $A \in \mathcal{M}_n$, i.e. on all operators defined on $n$-dimensional complex Hilbert space. That is, the mapping $\mathcal{E}$ can be written in operator Kraus form and therefore it is completely positive. This proves the theorem. $\qquad\square$

The Choi theorem provides us with a relatively simple test whether a given linear map $\mathcal{E}$ on $d$-dimensional system is completely positive, or not. In particular, it is sufficient to verify that the Choi matrix is positive. We have seen that $\Phi_{\mathcal{E}} = (\mathcal{E} \otimes \mathcal{I})[M]$, where $M : \mathcal{H} \otimes \mathcal{H} \to \mathcal{H} \otimes \mathcal{H}$ is a specific positive operator. Therefore, $P_+ = (\operatorname{tr}[M])^{-1} M = \frac{1}{d} \sum_{j,k=1}^{d} |\varphi_j \otimes \varphi_k\rangle\langle\varphi_j \otimes \varphi_k|$ is a density operator. Consequently, the complete positivity can be tested by application of $\mathcal{E} \otimes \mathcal{I}$ on a specific quantum state $P_+$. In the following theorem we shall see that this particular state allows us to relate the channels on d-dimensional system with states of $d^2$-dimensional system, i.e. of a composite system of a pair of $d$-dimensional systems. As we shall see in the last chapter on entanglement, this state is an example of maximally entangled state.

**Exercise 41.** Verify that $P_+$ is a pure state.

**Theorem 9.** *(Choi-Jamiolkowski isomorphism.)* Consider a $d$-dimensional quantum system, $\varphi_1, \dots, \varphi_d \in \mathcal{H}$ being its orthonormal basis, and a linear map $\mathcal{E} : \mathcal{L}(\mathcal{H}_d) \to \mathcal{L}(\mathcal{H}_d)$. Define an operator $P_+ = \frac{1}{d} \sum_{jk} |\varphi_j\rangle\langle\varphi_k| \otimes |\varphi_j\rangle\langle\varphi_k|$, i.e. a one-dimensional projector onto a subspace spanned on the state $\psi_+ = \frac{1}{\sqrt{d}} \sum_j \varphi_j \otimes \varphi_j$. A mapping

$$
\mathcal{J} : \mathcal{E} \mapsto \Omega_{\mathcal{E}} = (\mathcal{E} \otimes \mathcal{I})[P_+]
\tag{5.45}
$$

defines an isomorphism between linear maps on $d$-dimensional system and linear operators on $d \times d$-dimensional Hilbert space $\mathcal{L}(\mathcal{H}_d \otimes \mathcal{H}_d)$, a so-called *Choi-Jamiolkowski isomorphism*. The inverse mapping reads

$$
\mathcal{J}^{-1} : \Omega \mapsto \mathcal{E}_\Omega : \mathcal{E}_\Omega[X] = \operatorname{tr}_2[(I \otimes X^T)\Omega],
\tag{5.46}
$$

where $\mathrm{tr}_2$ stands for partial trace over the second Hilbert space. In particular (according to above theorem), the Choi-Jamiolkowski isomorphism maps completely positive maps to positive elements $\mathcal{L}_+(\mathcal{H}_d \otimes \mathcal{H}_d)$ and channels are isomorphic to states $\omega \in \mathcal{S}(\mathcal{H}_d \otimes \mathcal{H}_d)$ such that $d\,\mathrm{tr}_2[\omega] = I_d$.

*Proof.* By definition the mapping $\mathcal{J}$ is linear, i.e. $\mathcal{J}(\mathcal{E}_1 + \lambda \mathcal{E}_2) = \mathcal{J}(\mathcal{E}_1) + \lambda \mathcal{J}(\mathcal{E}_2)$. Similarly, also $\mathcal{J}^{-1}$ is linear, hence, the linear structures of the set of linear maps and set of linear operators $\mathcal{L}(\mathcal{H}_d \otimes \mathcal{H}_d)$ are preserved. It is sufficient to prove that $\mathcal{J}^{-1}$ defined above is indeed an inverse mapping to $\mathcal{J}$, i.e.

$$
\begin{aligned}
\mathcal{E}_{\Omega_\mathcal{E}}[X] &= \mathrm{tr}_2[(I \otimes X^T)\Omega_\mathcal{E}] = \mathrm{tr}_2[(I \otimes X^T)(\mathcal{E} \otimes \mathcal{I})[P_+]] \qquad (5.47)\\
&= \frac{1}{d} \sum_{j,k} \mathrm{tr}_2[(I \otimes X^T)(\mathcal{E}[|\varphi_j\rangle\langle\varphi_k|] \otimes |\varphi_j\rangle\langle\varphi_k|)] \\
&= \frac{1}{d} \sum_{j,k} \mathcal{E}[|\varphi_j\rangle\langle\varphi_k|] \mathrm{tr}\left[X^T|\varphi_j\rangle\langle\varphi_k|\right] = \frac{1}{d} \sum_{j,k} \mathcal{E}[|\varphi_j\rangle\langle\varphi_k|] \left\langle \varphi_k \mid X^T \varphi_j \right\rangle \\
&= \mathcal{E}[\sum_{jk} \left\langle \varphi_j \mid X\varphi_k \right\rangle |\varphi_j\rangle\langle\varphi_k|] = \mathcal{E}[X]
\end{aligned}
$$

for all $X \in \mathcal{L}(\mathcal{H}_d)$, i.e. $\mathcal{E}_{\Omega_\mathcal{E}} = \mathcal{E}$ for all linear maps $\mathcal{E}$. In a similar way (left for reader as exercise) we can prove that $\Omega_{\mathcal{E}_\Omega} = \Omega$ for all $\Omega \in \mathcal{L}(\mathcal{H}_d \otimes \mathcal{H}_d)$, i.e. $\mathcal{J} \circ \mathcal{J}^{-1} = \mathcal{J}^{-1} \circ \mathcal{J} = \mathrm{id}$. $\quad\square$

Let us note that for linear maps $\mathcal{E}$ the Choi matrix $\chi_\mathcal{E}$ and Choi-Jamiolkowski state $\omega_\mathcal{E}$ are closely related via $\Phi_\mathcal{E} = d\Omega_\mathcal{E}$. This fact follows directly from the comparison of the definitions of $\Phi_\mathcal{E}$ and $\Omega_\mathcal{E}$. The following proposition is a different version of the Choi-Jamiolkowski isomorphism holding between the Hilbert spaces of Hilbert-Schmidt operators on $d-$dimensional Hilbert space and vectors from $d^2$-dimensional Hilbert space.

**Proposition 65.** A mapping

$$
A \mapsto \psi_A = (A \otimes I)\psi_+, \quad \psi_+ = \frac{1}{\sqrt{d}} \sum_j \varphi_j \otimes \varphi_j \qquad (5.48)
$$

defines an isomorphism between the Hilbert spaces $\mathcal{L}(\mathcal{H}_d)$ endowed with the Hilbert-Schmidt scalar product $(A|B) = \mathrm{tr}\,[A^*B]$ and $\mathcal{H}_d \otimes \mathcal{H}_d$ with the canonical scalar product, i.e.

$$
d\, \langle \psi_A \mid \psi_B \rangle = \mathrm{tr}\,[A^*B] \ . \qquad (5.49)
$$

*Proof.* The proof is similar to the proof of the Choi-Jamiolkowski isomorphism. Let us fix the orthonormal basis $\varphi_1, \ldots, \varphi_d \in \mathcal{H}_d$ that is used in the definition of $\psi_+$. An operator $A \in \mathcal{L}(\mathcal{H})$ takes the form $A = \sum_{j,k} A_{jk}|\varphi_j\rangle\langle\varphi_k|$ and according to definition it results in vector $\psi_A = (A \otimes I)\psi_+ = \frac{1}{\sqrt{d}} \sum_{j,k}(A\varphi_k) \otimes \varphi_k = \frac{1}{\sqrt{d}} \sum_{j,k} A_{jk}\varphi_j \otimes \varphi_k$. Starting from a state $\psi \in \mathcal{H}_d \otimes \mathcal{H}_d$ such that $\psi = \sum_{jk} \psi_{jk}\varphi_j \otimes \varphi_k$ the inverse mapping uniquely defines an operator $A_\psi = \sqrt{d} \sum_{j,k} \psi_{jk}|\varphi_j\rangle\langle\varphi_k|$. It is left as an exercise for the reader that $A_{\psi_A} = A$ and $\psi_{A_\psi} = \psi$. Moreover, the calculation

$$
\begin{aligned}
\langle \psi_A \mid \psi_B \rangle &= \left\langle (A \otimes I)\psi_+ \mid (B \otimes I)\psi_+ \right\rangle = \langle \psi_+ \mid (A^*B \otimes I)\psi_+ \rangle \\
&= \frac{1}{d} \sum_j \left\langle \varphi_j \mid A^*B\varphi_j \right\rangle = \frac{1}{d} \mathrm{tr}\,[A^*B]
\end{aligned}
$$

shows that the scalar product is preserved as it is required for isomorphism between Hilbert spaces.                                                                                                         □

## 5.5  Classes of channels

### 5.5.1  Strictly contractive channels

By Proposition 63 we know that each channel is contractive, i.e., the trace distance between two states cannot increase under the action of a channel. In this section we investigate the properties of strictly contractive channels.

**Definition 44.**  A channel $\mathcal{E}$ is *strictly contractive* if for all states $\varrho_1, \varrho_2 \in \mathcal{S}(\mathcal{H})$, the inequality

$$\|\mathcal{E}(\varrho_1 - \varrho_2)\|_{\mathrm{tr}} \le k \, \|\varrho_1 - \varrho_2\|_{\mathrm{tr}} \tag{5.50}$$

is valid for some $0 \le k < 1$.

**Example 63.**  (*Depolarizing channels*) Let $\mathcal{H}$ be a finite dimensional Hilbert space. An example of strictly contractive channels is provided by the one-parameter family of *depolarizing channels*. For each $0 \le p \le 1$, we define

$$D_p(\varrho) = p \frac{1}{d} I + (1 - p) \varrho \,.$$

For two states $\varrho_1$ and $\varrho_2$, we get

$$\|D_p(\varrho_1 - \varrho_2)\|_{\mathrm{tr}} = (1 - p) \, \|\varrho_1 - \varrho_2\|_{\mathrm{tr}} \,.$$

Thus, the contractivity factor $k$ in (5.50) equals $1 - p$. This shows that the channel $D_p$ is strictly contractive when $p \ne 0$.

**Proposition 66.**  Consider a pair of quantum channels $\mathcal{E}, \mathcal{F}$ and assume that $\mathcal{E}$ is strictly contractive. The compositions $\mathcal{E} \circ \mathcal{F}, \mathcal{F} \circ \mathcal{E}$, and convex combination $\lambda \mathcal{E} + (1 - \lambda)\mathcal{F}$ (with $0 < \lambda \le 1$) are strictly contractive.

*Proof.*  The strict contractivness of $\mathcal{E} \circ \mathcal{F}$ comes from the following inequality

$$\|\mathcal{E} \circ \mathcal{F}(\varrho_1 - \varrho_2)\|_{\mathrm{tr}} \le \|\mathcal{E}(\varrho_1 - \varrho_2)\|_{\mathrm{tr}} \le k \, \|\varrho_1 - \varrho_2\|_{\mathrm{tr}} \,, \tag{5.51}$$

and similarly for $\mathcal{F} \circ \mathcal{E}$. For a convex combination the following calculation

$$
\begin{aligned}
\|(\lambda \mathcal{E} + (1 - \lambda)\mathcal{F})(\varrho_1 - \varrho_2)\|_{\mathrm{tr}} \quad &\le \quad \lambda \, \|\mathcal{E}(\varrho_1 - \varrho_2)\|_{\mathrm{tr}} + (1 - \lambda) \, \|\mathcal{F}(\varrho_1 - \varrho_2)\|_{\mathrm{tr}} \\
&\le \quad (k\lambda + 1 - \lambda) \, \|\varrho_1 - \varrho_2\|_{\mathrm{tr}} \\
&< \quad \|\varrho_1 - \varrho_2\|_{\mathrm{tr}}
\end{aligned}
$$

implies that $\lambda \mathcal{E} + (1 - \lambda)\mathcal{F}$ is a strictly contractive channel for $0 < \lambda \le 1$.                      □

**Proposition 67.**  Let $\mathcal{E}$ and $\mathcal{F}$ be two channels. Their tensor product $\mathcal{E} \otimes \mathcal{F}$ is strictly contractive only if both $\mathcal{E}$ and $\mathcal{F}$ strictly contractive.

*Proof.* Suppose that $\mathcal{F}$ is not strictly contractive. Hence, there are states $\xi_1$ and $\xi_2$, $\xi_1 \neq \xi_2$, such that

$$\|\mathcal{F}(\xi_1 - \xi_2)\|_{\mathrm{tr}} = \|\xi_1 - \xi_2\|_{\mathrm{tr}} \,.$$

Fix a state $\eta$ and denote $\varrho_1 = \eta \otimes \xi_1$ and $\varrho_2 = \eta \otimes \xi_2$. Then we get

$$
\begin{aligned}
\|\mathcal{E} \otimes \mathcal{F}(\varrho_1 - \varrho_2)\|_{\mathrm{tr}} &= \|\mathcal{E}(\eta) \otimes \mathcal{F}(\xi_1 - \xi_2)\|_{\mathrm{tr}} = \|\mathcal{F}(\xi_1 - \xi_2)\|_{\mathrm{tr}} = \|\xi_1 - \xi_2\|_{\mathrm{tr}} \\
&= \|\varrho_1 - \varrho_2\|_{\mathrm{tr}} \,,
\end{aligned}
$$

which shows that the channel $\mathcal{E} \otimes \mathcal{F}$ is not strictly contractive. □

An interesting consequence of the above proposition is that $\mathcal{E} \otimes \mathcal{I}$ is not strictly contractive even if $\mathcal{E}$ is a strictly contractive.

**Proposition 68.** Every channel can be approximated arbitrarily well with strictly contractive channels.

*Proof.* Fix a state $\xi \in \mathcal{S}(\mathcal{H})$. The mapping $\mathcal{F}_\xi : \varrho \mapsto \xi$ is a strictly contractive channel. Let $\mathcal{E}$ be a channel. For each $n = 1, 2, \ldots$, we define

$$\mathcal{E}_n = \frac{1}{2n} \mathcal{F}_\xi + (1 - \frac{1}{2n}) \mathcal{E} \,. \tag{5.52}$$

Clearly, $\mathcal{E}_n$ is strictly contractive. The calculation

$$\|\mathcal{E} - \mathcal{E}_n\| = \frac{1}{2n} \|\mathcal{E} - \mathcal{F}_\xi\| \leq \frac{1}{n}$$

finishes the proof. □

An important property of strictly contractive channels is stated in the following theorem. It is a direct application of Banach fixed points theorem, and the proof can be found e.g. from Wikipedia.

**Theorem 10.** *(Banach fixed point theorem)* Let $\mathcal{E}$ be a strictly contractive channel. There is a unique state $\xi$ such that $\mathcal{E}(\xi) = \xi$. The state $\xi$ called the *fixed point* $\mathcal{E}$. Moreover, for every state $\varrho$, we have

$$\lim_{n \to \infty} \mathcal{E}^n(\varrho) = \xi \,,$$

where the sequence $\{\mathcal{E}^n(\varrho)\}_{n=1}^{\infty}$ converges in the trace norm.

**Exercise 42.** Find the fixed point of the depolarizing channel $D_p$ (see Example 63).

The following example demonstrates that there are nontrivial channels with more than one fixed point.

**Example 64.** *(Fixed points of unitary channels.)* Let $U$ be a unitary operator having eigenvalues. The eigenvectors $\psi_j$ of $U$ determine pure states $P_j = |\psi_j\rangle\langle\psi_j|$. Since the eigenvalues of $U$ have modulus 1 (see Example 9 in Section 2.2.3), we have $\sigma_U(P_j) = P_j$. Therefore, any operator being a linear combination $\sum_j c_j P_j$ of projectors $P_j$ is a fixed point of $\sigma_U$.

### 5.5.2    Random unitary channels

**Definition 45.** A channel $\mathcal{E}$ is a *random unitary channel* if it is a convex mixture of unitary channels, that is,

$$\mathcal{E}(\varrho) = \sum_j p_j U_j \varrho U_j^*, \tag{5.53}$$

where $0 \leq p_j \leq 1$ and $\sum_j p_j = 1$.

The set of random unitary channels is a convex subset of $\mathcal{O}_c$. It is also closed under composition, i.e., $\mathcal{E}_1 \circ \mathcal{E}_2$ is a random unitary channel when ever $\mathcal{E}_1$ and $\mathcal{E}_2$ are.

A random unitary channel $\mathcal{E}$ can be implemented by a unitary transformation $U = \sum_j U_j \otimes |\varphi_j\rangle\langle\varphi_j|$. Indeed, if the environment is initially prepared in a state $\xi$, then the random unitary channel is implemented

$$\mathcal{E}(\varrho) = \sum_j \langle \varphi_j \,|\, \xi \varphi_j \rangle \, U_j \varrho U_j^*, \tag{5.54}$$

and for conjugate channel we get

$$\mathcal{E}'(\varrho) = \sum \xi_{jk} \mathrm{tr}\left[U_j \varrho U_k^*\right] |\varphi_j\rangle\langle\varphi_k|. \tag{5.55}$$

If $\xi$ is diagonal, i.e. only the entries $\xi_{jj}$ are nonzero, then

$$\mathcal{E}'(\varrho) = \sum_j \xi_{jj} |\varphi_j\rangle\langle\varphi_j| = \xi, \tag{5.56}$$

i.e. one of the conjugated channels is the contraction to a specific fixed state $\xi$.

**Proposition 69.** (*Random orthogonal unitary channels.*) Let $\mathcal{H}$ be a finite dimensional Hilbert space and $U_1, \ldots, U_{d^2}$ unitary operators forming an orthogonal basis of $\mathcal{L}(\mathcal{H})$. Then

$$\mathcal{E}(T) = \frac{1}{d^2} \sum_{j=1}^{d^2} U_j T U_j^* = \frac{1}{d} \mathrm{tr}\left[T\right] I = \mathcal{A}_0(T). \tag{5.57}$$

*Proof.* Let $\psi_+ = \frac{1}{\sqrt{d}} \sum_{l=1}^d \varphi_l \otimes \varphi_l$ and $\varphi_1, \ldots, \varphi_d$ is an orthonormal basis of $\mathcal{H}$. According to Choi-Jamiolkowski isomorphism the image of each unitary transformation $U_j$ is a pure state associated with a vector $\psi_j = (I \otimes U)\psi_+$ satisfying $\langle \psi_j \,|\, \psi_k \rangle = \mathrm{tr}\left[U_j^* U_k\right] = d\delta_{jk}$. Since vectors are are mutually orthogonal and form a complete basis of $\mathcal{H} \otimes \mathcal{H}$ it follows that $\sum_j |\psi_j\rangle\langle\psi_j| = I \otimes I$. Consequently, $\Omega_{\mathcal{E}} = (\mathcal{I} \otimes \mathcal{E})[P_+] = \frac{1}{d^2} \sum_j |\psi_j\rangle\langle\psi_j| = \frac{1}{d^2} I \otimes I = (I \otimes \mathcal{A}_0)[P_+]$. The identity of Choi operators $\Omega_{\mathcal{E}} = \Omega_{\mathcal{A}_0}$ implies $\mathcal{E} = \mathcal{A}_0$, which proves the proposition. $\qquad\square$

**Example 65.** (*Private quantum channels.*) A private quantum channel is a cryptographic communication protocol aiming to transmit quantum states in a secure way [1]. The security is based on the shared classical key represented by identical sequences of random bits in possessions of both communicating parties. The protocol works as follows. Alice wants to transfer states $\varrho_1, \ldots, \varrho_n$ to Bob. The classical key is used to set encoding channels $\mathcal{E}_1, \ldots, \mathcal{E}_n$ used by Alice, and decoding channels $\mathcal{D}_1, \ldots, \mathcal{D}_n$ used by Bob. In each run Bob receives a state $\mathcal{E}_j(\varrho_j)$ and applying the proper decoding channel $\mathcal{D}_j$ he reveals the original state $\varrho_j = \mathcal{D}_j \circ \mathcal{E}_j(\varrho)$.

We assume that $\varrho_j$ can be arbitrary state. Therefore, we must have $\mathcal{D}_j = \mathcal{E}_j^{-1}$, and this implies that $\mathcal{E}_j$ must be unitary. Thus, $\mathcal{E}_j = \sigma_{U_j}$ and $\mathcal{D}_j = \sigma_{U_j^*}$ for some unitary operators $U_j$. The question is how many bits is needed in order to protect a single quantum bit, i.e., how many communication runs $n$ are encoded in the sequence of bits of size $N$.

Let $\{U_1, \ldots, U_m\}$ be a set of unitary operators defining the encoding and decoding channels. A sequence of $N$ random bits forming the key defines a random sequence $U_{j_1}, \ldots, U_{j_n}$ determining the sequence of encoding and decoding channels in $n$ runs, i.e. $U_{j_i} \in \{U_1, \ldots, U_m\}$. For anyone except Bob Alice is randomly applying unitary channels, hence she is implementing a random unitary channel constituting the private quantum channel

$$\mathcal{E}_{\mathrm{PQC}}(\varrho) = \sum_j p_j U_j \varrho U_j^* \,, \tag{5.58}$$

where $p_j$ is the probability of operator $U_j$ in the random sequence $U_{j_1}, \ldots, U_{j_n}$ determined by the classical key. The security is achieved if the output state $\mathcal{E}_{\mathrm{PQC}}(\varrho)$ is independent of $\varrho$, i.e. $\mathcal{E}_{\mathrm{PQC}}(\varrho) = \sum_j p_j U_j \varrho U_j^* = \varrho_0$ for some fixed state $\varrho_0$.

Proposition 69 guarantees the existence of random unitary channel with the required properties. In particular, the private quantum channel with $m = d^2$ mutually orthogonal unitary operators $\{U_j\}$ and $p_j = 1/d^2$ satisfies all requirements. As a result we get that if the encoding and decoding channels are associated with mutually orthogonal unitary operators, then the private quantum channel can be implemented and the required key is of the length $N = n \log m = \log d^2 = 2n \log d$. In conclusion, secure transmission of $d$-dimensional quantum state can be done with $2 \log d$ classical bits.

### 5.5.3 Pure decoherence channels

The concept of decoherence was originally introduced as a process standing behind the disappearance of the interference patterns. In a sense, due to the decoherence the initial superposition of pure states is transforming into a mixture of the (orthogonal) pure states. These pure states defines the so-called *decoherence basis*. Nowadays in the literature the decoherence is sometimes understood as arbitrary nonunitary dynamics. Therefore, we shall refer to the original concept as to *pure decoherence channels*.

**Definition 46.** A channel $\mathcal{E}$ describes *pure decoherence* if its power series $\mathcal{E}, \mathcal{E}^2, \mathcal{E}^3, \ldots$, converges to the channel

$$\mathrm{diag}_b : \varrho \mapsto \mathrm{diag}_b(\varrho) = \sum_b \langle \varphi_j | \varrho | \varphi_j \rangle | \varphi_j \rangle \langle \varphi_j | \,, \tag{5.59}$$

where $b = \{\varphi_1, \ldots, \varphi_d\}$ is the decoherence basis.

If the decoherence basis $b$ is fixed and $\mathcal{E}_1, \mathcal{E}_2$ are pure decoherences with respect to this basis, then the convex combination $\lambda \mathcal{E}_1 + (1 - \lambda)\mathcal{E}_2$ and the composition $\mathcal{E}_1 \circ \mathcal{E}_2$ are pure decoherence channels with the same basis $b$. Moreover, $\mathcal{E}_1$ and $\mathcal{E}_2$ commutes, i.e. $\mathcal{E}_1 \circ \mathcal{E}_2 = \mathcal{E}_2 \circ \mathcal{E}_1$. If the decoherence basis of $\mathcal{E}_1$ is different from the decoherence basis of $\mathcal{E}_2$, then their convex combination and composition are no longer pure decoherence channels. The proofs of these statements are left to readers as exercises.

**Proposition 70.** Consider a channel $\mathcal{E}(\varrho) = \sum_j A_j \varrho A_j^*$ expressed in the operator sum form. It is a pure decoherence channel if and only if the Kraus operators $A_j$ commute with the projections $e_1, \ldots, e_d$ ($e_j = |\varphi_j\rangle\langle\varphi_j|$). It follows that for a pure decoherence channel its Kraus operators $A_j$ mutually commute, i.e., $[A_j, A_k] = 0$ for all $j, k$.

*Proof.* Preservation of operators $e_j$ implies that $\mathcal{E}(T) = T$ for all operators $T$ belonging to subalgebra generated by projectors $e_1, \ldots, e_d$. This subalgebra $\mathcal{A}_b$ is a maximal commutative subalgebra of the algebra of bounded operators $\mathcal{L}(\mathcal{H})$. Consequently, the identity

$$(\mathcal{E}(T) - T)(\mathcal{E}(T) - T)^* = 0$$

holds for all $T \in \mathcal{A}_b$. Using the relations $\mathcal{E}(T)\mathcal{E}(T)^* = TT^* = \mathcal{E}(TT^*)$ and $\sum_j A_j^* A_j = I$ the right side can be rewritten into the identity

$$\sum_j \left[ A_j TT^* A_j^* - A_j TA_j^* T^* - TA_j T^* A_j^* - TA_j^* A_j T^* \right] = \sum_j [T, A_j][T, A_j]^* = 0\,.$$

It holds if and only if $[T, A_j] = 0$ for all $T \in \mathcal{A}_b$ and all $j$. Moreover, because the subalgebra $\mathcal{A}_b$ is the maximal commutative subalgebra, it follows that Kraus operators $A_j$ must mutually commute, too.

On the other hand the identities $[A_j, e_n] = 0$ imply $[A_j, T] = 0$ for all $T \in \mathcal{A}_b$ and, consequently, the above arguments can be reversed to prove that $\mathcal{E}^{\mathcal{B}}(T) = \sum_j A_j TA_j^* = T$ for all $T \in \mathcal{A}_b$ including the projectors $e_1, \ldots, e_d$.                                                   □

**Proposition 71.** (*Dilations of pure decoherence channels.*) If $\mathcal{E}$ is pure decoherence channel then $\mathcal{E}[\varrho] = \mathrm{tr}_{\mathrm{env}} U(\varrho \otimes \omega_{\mathrm{env}}) U^*$ and $U = \sum_j |\varphi_j\rangle\langle\varphi_j| \otimes U_j$ is a controlled-U unitary transformation with system playing the role of the control system. The vectors $\varphi_j$ form the decoherence basis.

*Proof.* The preservation of the decoherence basis elements $\varphi_j$, i.e. the identity $U(\varphi_j \otimes \psi) = \varphi_j \otimes \psi'$, implies that $\psi' = U_j \psi$ in order to preserve the scalar product. Using this condition for all $j$ the operator $U$ is defined on the whole Hilbert space and takes the form of so-called control-ed-U transformation

$$U = \sum_j |\varphi_j\rangle\langle\varphi_k| \otimes U_j\,. \tag{5.60}$$

The decohering system is playing the role of the control system and the environment is the target system.                                                   □

### 5.6    Example: qubit channels

In this section we demonstrate that already for the simplest possible quantum system, namely, qubits, the characterization of the channels is quite a complex task. An important steps were made in [75].

Recall from Section 3.1.3 that the state space of a two-dimensional Hilbert space can be nicely represented as the Bloch ball. Hence, qubit channels can be illustrated as mappings on the Bloch ball. Adopting this point of view, the channels are represented by $4 \times 4$-matrices of an affine form. In particular, let us choose the (unnormalized) operator basis $\sigma_0 = I, \sigma_x, \sigma_y, \sigma_z$. Then the matrix elements $\mathcal{E}_{jk} = \frac{1}{2}\text{tr}\left[\sigma_j \mathcal{E}[\sigma_k]\right]$ form a matrix

$$\mathcal{E} = \begin{pmatrix} 1 & 0 & 0 & 0 \\ \tau_x & T_{xx} & T_{xy} & T_{xz} \\ \tau_y & T_{yx} & T_{yy} & T_{yz} \\ \tau_z & T_{zx} & T_{zy} & T_{zz} \end{pmatrix}, \tag{5.61}$$

and channel is acting as $\vec{r} \mapsto \vec{r}' = T\vec{r} + \vec{\tau}$.

The crucial trick used in the characterization of qubit channels is based on the *singular value decomposition* and the one-to-one relation between the unitary qubit channels and orthogonal rotations in three-dimensional space.

**Lemma 4.** A matrix $T$ can be written in the so-called *singular value decomposition* as $T = Q_1 D_{\text{sv}} Q_2$, where $Q_1, Q_2$ are matrices of orthogonal rotations ($Q_j^T = Q_j^{-1}$ for $j = 1, 2$) and $D_{\text{sv}}$ is a diagonal positive matrix $D_{\text{sv}} = \text{diag}\{\mu_1, \ldots, \mu_d\}$, where $\mu_j$ are the so-called *singular values* of matrix $T$.

*Proof.* From polar decomposition it follows that $T = |T|Q$ for positive matrix $|T|$ and some orthogonal rotation $Q$. The positive matrix is selfadjoint and therefore can be diagonalized, i.e. $|T| = Q_1 D_{\text{sv}} Q_1^T$ by some orthogonal matrix $Q_1$. Defining $Q_2 = Q_1^T Q$ we get the singular value decomposition of $T$. ☐

For a general orthonormal matrix $\det Q = \pm 1$. However, in Example 62 we have shown that unitary channels are associated with special orthogonal matrices $R_U$, i.e. $R_U^{-1} = R_U^T$ and $\det R_U = 1$. Fortunately, for the case of qubit the correspondence is one-to-one, i.e. each special orthogonal matrix is associated with some unitary channel $\sigma_U$. We need to modify the above lemma a bit in order to use it for our purposes. Each three-dimensional orthogonal matrix $Q$ is either a rotation ($\det Q = 1$), or can be written as a product of some rotation $R$ and the inversion $-I$, i.e. $Q = \pm R$, where $\det R = 1$. Thus, $T = Q_1 D_{\text{sv}} Q_2 = R_1(\pm D_{\text{sv}})R_2 = (R_1 R_2)R_2^T(\pm D_{\text{sv}})R_2 = RS$, where $S = R_2^T(\pm D_{\text{sv}})R_2$ is semi-definite (either positive, or negative) and $R$ is a rotation.

The rotation $R_k$ associated with a unitary channel $\sigma_k[\varrho] = \sigma_k \varrho \sigma_k$ forms a diagonal matrix with $[R_k]_{jj} = -1$ if $j \neq k$ and $[R_k]_{kk} = 1$. Combining $R_k$ with $D_{\text{sv}} = \text{diag}\{\mu_1, \mu_2, \mu_3\}$ two of the values $\mu_j$ will change their sign. By definition, if $D_{\text{sv}}$ determines a quantum channel, then $R_k D_{\text{sv}}$ must describe a quantum channel as well. Hence, the semidefinitness of $S$ can be replaced by its self-adjointness. Since every self-adjoint real matrix can be diagonalized by some rotation $R_V$ it follows that $T = RS = RR_V^T DR_V$, where $D = \{\lambda_1, \lambda_2, \lambda_3\}$ is a diagonal matrix composed of eigenvalues of $S$. Let us note that $\lambda_j$ are not singular values of $T$. In fact they are not necessarily positive. However, $|\lambda_j| = \mu_j$ are the singular values of $T$.

As a result we get that $T = R_U D R_V$ for suitable rotations $R_U, R_V$, $D = \text{diag}\{\lambda_1, \lambda_2, \lambda_3\}$, where $|\lambda_1|, |\lambda_2|, |\lambda_3|$ are the singular values of $T$. In such case the action of the channel $\mathcal{E}$ can be written as

$$\vec{r} \mapsto \vec{r}' = R_U D R_V \vec{r} + R_U \vec{t}, \tag{5.62}$$

and consequently each qubit channel $\mathcal{E}$ can be written as

$$\mathcal{E} = \sigma_U \circ \mathcal{D} \circ \sigma_V, \text{ i.e. } \mathcal{E}(\varrho) = U\mathcal{D}(V\varrho V^*)U^*, \tag{5.63}$$

where $\sigma_U(\varrho) = U\varrho U^*$, $\sigma_V(\varrho) = V\varrho V^*$ and $\mathcal{D} : \vec{r} \to \vec{r}' = D\vec{r} + \vec{t}$ with $\vec{t} = R_U^T \vec{\tau}$. That is, arbitrary qubit channel is unitarily equivalent to a channel of the form $\mathcal{D}$ parametrized only by 6 real parameters. Moreover, most of the interesting properties of $\mathcal{E}$ are the properties of $\mathcal{D}$.

The above decomposition of $\mathcal{E}$ via $\mathcal{D}_\mathcal{E} = \mathcal{D}$ is not unique, because signs of any pair of parameters $\lambda_j$ can be changed by permutation, i.e. by applying one of the channels $\sigma_j$. Only the product $\lambda_1 \lambda_2 \lambda_3$ is fixed for each channel $\mathcal{E}$.

**Proposition 72.** Let us denote by $\mathcal{D}_\mathcal{E}$ the "diagonal" channel associated with $\mathcal{E}$. Then the following equivalences hold.

- A channel $\mathcal{E}$ is completely positive if and only if $\mathcal{D}_\mathcal{E}$ is completely positive.

- A channel $\mathcal{E}$ is strictly contractive if and only if $\mathcal{D}_\mathcal{E}$ is strictly contractive.

- A channel $\mathcal{E}$ is unital if and only if $\mathcal{D}_\mathcal{E}$ is unital.

### 5.6.1   Complete positivity constraints

The positivity constraint of $\mathcal{D}$ (and $\mathcal{E}$) requires that $|\lambda_j| \leq 1$, because otherwise the length of the Bloch vector will increase and negative region will be reached. In fact, the image of the pure states (Bloch sphere boundary) under $\mathcal{D}$ defines an ellipsoid

$$\left(\frac{r_1' - t_1}{\lambda_1}\right)^2 + \left(\frac{r_2' - t_2}{\lambda_2}\right)^2 + \left(\frac{r_3' - t_3}{\lambda_3}\right)^2 = 1 \,. \tag{5.64}$$

The interpretation is that each quantum channel transforms the Bloch sphere into an ellipsoid (inside the original Bloch sphere), but the converse is not true. Not all ellipsoids included in Bloch sphere can be achieved by some quantum channel.

Let us consider a channel $\mathcal{E}$ with Kraus operators $A_j = \sum_k a_{jk}\sigma_k$. Then

$$T_{jj} = \sum_k \text{tr}\left[\sigma_j A_k \sigma_j A_k^*\right] = \sum_{k,l,l'} a_{kl} a_{kl'}^* \text{tr}\left[\sigma_j \sigma_l \sigma_j \sigma_{l'}\right] = \sum_l j_l \sum_k a_{kl} a_{kl}^* \tag{5.65}$$

where $j_l = \pm$ depending on the $\sigma_j \sigma_l \sigma_j = \pm \sigma_l$. Defining the positive numbers $q_l = \sum_k a_{kl} a_{kl}^*$ we get

$$
\begin{aligned}
T_{00} &= q_0 + q_1 + q_2 + q_3 = 1 \,, \\
T_{11} &= q_0 + q_1 - q_2 - q_3 \,, \\
T_{22} &= q_0 - q_1 + q_2 - q_3 \,, \\
T_{33} &= q_0 - q_1 - q_2 + q_3 \,.
\end{aligned}
$$

It follows that

$$
\begin{aligned}
T_{11} \pm T_{22} &\leq T_{00} \pm T_{33} \,, \\
-T_{11} \pm T_{22} &\leq T_{00} \mp T_{33} \,,
\end{aligned}
$$

or in a more compact form

$$|T_{11} \pm T_{22}| \leq |1 \pm T_{33}| \,. \tag{5.66}$$

The above inequalities fixes the values of $T_{11}, T_{22}, T_{33}$ to form a tetrahedron with vertices $(1,1,1)$, $(1,-1,-1)$, $(-1,1,-1)$, $(-1,-1,1)$. The derived conditions are necessary, but not sufficient for channel being completely positive. For channels of the form $\mathcal{D}$ this condition reads $|\lambda_1 \pm \lambda_2| \leq |1 \pm \lambda_3|$. Without proof we shall state the sufficient and necessary conditions [75] for complete positivity of channels $\mathcal{D}$.

**Proposition 73.** A channel $\mathcal{D}$ is completely positive if and only if the following inequalities hold

$$(\lambda_1 + \lambda_2)^2 \leq (1 + \lambda_3)^2 - t_3^2 \tag{5.67}$$

$$(\lambda_1 - \lambda_2)^2 \leq (1 - \lambda_3)^2 - t_3^2 \tag{5.68}$$

$$\left[1 - |\vec{\lambda}|^2 - |\vec{t}|^2\right]^2 \geq 4[\lambda_1^2(t_1^2 + t_2^2) + \lambda_2^2(t_2^2 + t_3^2) + \lambda_3^2(t_3^2 + t_1^2) - 2\lambda_1\lambda_2\lambda_3] \tag{5.69}$$

The direct calculation shows that the Choi matrix of $\mathcal{D}$ reads

$$\Phi_{\mathcal{D}} = \frac{1}{2} \begin{pmatrix} 1 + t_3 + \lambda_3 & t_1 - it_2 & 0 & \lambda_1 + \lambda_2 \\ t_1 + it_2 & 1 - t_3 - \lambda_3 & \lambda_1 - \lambda_2 & 0 \\ 0 & \lambda_1 - \lambda_2 & 1 + t_3 - \lambda_3 & t_1 - it_2 \\ \lambda_1 + \lambda_2 & 0 & t_1 + it_2 & 1 - t_3 + \lambda_3 \end{pmatrix}. \tag{5.70}$$

The positivity of this matrix is equivalent to conditions in the above proposition.

### 5.6.2   Unital channels

Let us recall that a channel $\mathcal{E}$ is called unital if $\mathcal{E}(I) = I$. In the Bloch sphere picture the unital channels are those for which $\vec{\tau} = \vec{t} = \vec{0}$, hence the ellipsoid is centered in the total mixture. In the discussion we shall restrict only to matrices $D$, i.e. only three real parameters.

**Example 66.** *(Pauli channels.)* Pauli channel is defined as a convex combination of Pauli operators, i.e.,

$$\mathcal{E}_{\text{Pauli}}(\varrho) = \sum_j q_j \sigma_j \varrho \sigma_j, \tag{5.71}$$

where $0 \leq q_j \leq 1$, $\sum_j q_j = 1$. It is clear that these channels are unital. Since each Pauli unitary channel itself is associated with a diagonal matrix $D$, it follows that arbitrary Pauli channel is also diagonal. In particular, $D_{\text{Pauli}} = \text{diag}\{\lambda_1, \lambda_2, \lambda_3\}$ such that

$$\lambda_1 = q_0 + q_1 - q_2 - q_3,$$
$$\lambda_2 = q_0 - q_1 + q_2 - q_3,$$
$$\lambda_3 = q_0 - q_1 - q_2 + q_3.$$

Due to normalization of probability distribution $\{q_j\}$ it follows that for unital channels the conditions $|\lambda_1 \pm \lambda_2| \leq |1 \pm \lambda_3|$ are both necessary and sufficient for complete positivity of $\mathcal{E}$. Thus, the set of all unital channels is unitarily equivalent to the set of Pauli channels.

The three parameters $\vec{\lambda} = (\lambda_1, \lambda_2, \lambda_2) \in \mathbb{R}^3$ restricted by conditions $|\lambda_1 \pm \lambda_2| \leq |1 \pm \lambda_3|$ form a tetrahedron, i.e. a simplex with extremal points being the four unitary channels corresponding to Pauli operators. In particular, $I \leftrightarrow (1,1,1)$, $\sigma_x \leftrightarrow (1,-1,-1)$, $\sigma_y \leftrightarrow (-1,1,-1)$

and $\sigma_z \leftrightarrow (-1, -1, 1)$. Let us note an interesting fact that all the Pauli channels mutually commute although the Pauli operators are not commuting.

In the case $q_1 = q_2 = q_3 = q$ and $q_0 = 1 - 3q$ the Pauli channels are called *depolarizing channels*. Since $0 \leq q_j \leq 1$ it follows that $0 \leq q \leq 1/3$. After the action of these channels the Bloch sphere is homogenuously contracted into a ball of radius $|1 - 4q|$. Except for $q = 0$ these channels are strictly contractive. For $q = 1/4$ the depolarizing channel maps the whole Bloch ball into the maximally mixed state $\vec{r} = \vec{0}$. If $q = 1/3$, then $\lambda_j = -1/3$ for all $j$ and the resulting channel is known as the *best physical approximation* of quantum NOT gate described in Example 51. Let us note that quantum NOT gate would correspond to transformation $\vec{r} \to -\vec{r}$, i.e. to space inversion $-I$.

If $q_1 = q_2 = 0$, then the channels are called *phase damping channels* and describe the pure decoherence processes. In this case $\lambda_3 = 1$ and $\lambda_1 = \lambda_2 = \lambda$, i.e. the Bloch ball is not contracted in direction $z$, i.e. phase-damping channels are not strictly contractive.

## 6 Measurement models and instruments

Until now we have treated a measurement apparatus as a device taking quantum systems as its inputs and producing measurement outcomes as outputs. The distribution of outcomes is determined by the particular input state. In this level of description, measurement apparatuses are described by observables.

It could happen that the systems are still available after the measurement is performed, and we may try to obtain more information by measuring some other observable. Or perhaps we aim to use the measurement apparatus as a preparator, producing some preferred state. In this kind of situations we need a more detailed description of measurement apparatuses than observables. Instruments and measurement models are suited for this purpose.

### 6.1 Measurement models

A standard reference on measurement models is [18], where measurement models are called premeasurements.

#### 6.1.1 Definition

Let us first think about a typical measurement procedure in an informal way. A measurement procedure starts by coupling the system to a measurement apparatus, or a probe. After some time, the system and the probe are decoupled, and a measurement is carried out on the probe. Due to the coupling stage the system and the probe become correlated and the measurement outcome distribution gives us information about the system. The whole procedure can be interpreted directly as a measurement on the system corresponding to some observable A. The following definition formalizes this idea.

**Definition 47.** Let $A : \mathcal{F} \to \mathcal{L}(\mathcal{H})$ be an observable. A *measurement model (memo)* $\mathcal{M}$ is a quadruple $\mathcal{M} = \langle \mathcal{K}, \varrho_0, \mathcal{V}, F \rangle$, where

- $\mathcal{K}$ is a Hilbert space attached to the probe.

- $\varrho_0$ is a state on $\mathcal{K}$. It is the *initial state* of the probe.

- $\mathcal{V}$ is a channel from $\mathcal{T}(\mathcal{H} \otimes \mathcal{K})$ to $\mathcal{T}(\mathcal{H} \otimes \mathcal{K})$. It describes the *measurement interaction* between the system and the probe.

- $F$ is an observable with the outcome space $(\Omega, \mathcal{F})$ and taking values in $\mathcal{K}$. It is the *pointer observable* describing the measurement on the probe.

If the following *probability reproducibility* condition holds, then $\mathcal{M}$ is a measurement model for the observable A:

$$\operatorname{tr} \left[ \varrho A(X) \right] = \operatorname{tr} \left[ \mathcal{V}(\varrho \otimes \varrho_0)(I \otimes F(X)) \right] \quad \forall X \in \mathcal{B}(\Omega), \varrho \in \mathcal{S}(\mathcal{H}) \,. \tag{6.1}$$

The probability reproducibility condition simply means that $\mathcal{M}$ leads to same measurement outcome probabilities as A. We can also interpret this condition in the opposite order: any quadruple $\langle \mathcal{K}, \varrho_0, \mathcal{V}, F \rangle$ defines a unique observable A by equation (6.1). This is actually what happens when we start to describe some measurement scheme. We don't fix A in the beginning but we calculate it from the given memo.

**Example 67.** *(Stern-Gerlach measurement on a spin-$\frac{1}{2}$ particle.)* Let us take another look on the Stern-Gerlach measurement described in Section 4.1.1. A spin-$\frac{1}{2}$ particle is described by the Hilbert space $\mathbb{C}^2 \otimes L^2(\mathbb{R}^3)$ and can be understood as a bipartite composite system consisting of spatial and spin degrees of freedom. During the Stern-Gerlach experiment an external magnetic field couples the spatial degrees of freedom with the spin degrees of freedom. Depending on the spin orientation, the particle passing the Stern-Gerlach apparatus is deflected either to up or down. The channel $\mathcal{V}$ describing the coupling acts as follows:

$$\varphi_\uparrow \otimes \psi \quad \mapsto \quad \varphi_\uparrow \otimes V_+ \psi \,,$$
$$\varphi_\downarrow \otimes \psi \quad \mapsto \quad \varphi_\downarrow \otimes V_- \psi \,.$$

Here $\{\varphi_\uparrow, \varphi_\downarrow\}$ is an orthonormal basis of $\mathbb{C}^2$, $\psi \in L^2(\mathbb{R}^3)$, and $V_\pm$ are unitary operators defined on $L^2(\mathbb{R}^3)$. Assuming an initial state corresponding to a vector $\varphi \otimes \psi_0 = (a\varphi_\uparrow + b\varphi_\downarrow) \otimes \psi_0$, the transformation due to the measurement coupling is

$$\varphi \otimes \psi_0 \mapsto \omega = a\varphi_\uparrow \otimes \psi_+ + b\varphi_\downarrow \otimes \psi_- \,, \tag{6.2}$$

where $\psi_\pm \equiv V_\pm \psi_0$.

A detector is measuring the presence of the particle in a certain region, hence it is the pointer observable measuring the position of the particle. In Stern-Gerlach measurement a screen is used as a detector of particles in the plane orthogonal to incoming beam of particles. We are interested whether the particle is observed in upper or lower half plane of the screen. Thus, in the ideal case the pointer observable consists of two projections $P_\pm$, and the probabilities $p_\pm$ for observing particles in upper (+) or lower (−) half planes are

$$p_\pm = \mathrm{tr}\,[|\omega\rangle\langle\omega|(I \otimes P_\pm)] = |a|^2 \langle\psi_+|P_\pm|\psi_+\rangle + |b|^2 \langle\psi_-|P_\pm|\psi_-\rangle \,, \tag{6.3}$$

where

$$\langle\psi|P_+|\psi\rangle = \int_{\text{upper h.p.}} |\psi(\vec{r})|^2 \; d\vec{r}, \qquad \langle\psi|P_-|\psi\rangle = \int_{\text{lower h.p.}} |\psi(\vec{r})|^2 \; d\vec{r} \tag{6.4}$$

for every $\psi \in L^2(\mathbb{R}^3)$. Defining the effects $E_\pm$ on $\mathbb{C}^2$ as

$$E_\pm = \langle\psi_+|P_\pm|\psi_+\rangle|\varphi_\uparrow\rangle\langle\varphi_\uparrow| + \langle\psi_-|P_\pm|\psi_-\rangle|\varphi_\downarrow\rangle\langle\varphi_\downarrow| \,, \tag{6.5}$$

we obtain the following expression for the probabilities $p_\pm$ only in terms of the spin state

$$p_\pm = \langle\varphi|E_\pm|\varphi\rangle \,. \tag{6.6}$$

If the measurement outcomes are labeled by $\uparrow$ and $\downarrow$, then the resulting observable A is given by

$$\mathsf{A}(\uparrow) = E_+ \,, \qquad \mathsf{A}(\downarrow) = E_- \,.$$

The sharpness of A depends on the initial state $\psi_0$ and on the unitary operators $V_\pm$. In particular, A is sharp if and only if $\langle\psi_\pm|P_\pm|\psi_\pm\rangle = 1$ and $\langle\psi_\mp|P_\pm|\psi_\mp\rangle = 0$.

In a real experiment the effects $E_+$ and $E_-$ are not projections. For a more realistic description of the Stern-Gerlach measurement, we refer to Chapter VII in [15].

### 6.1.2   Normal measurement models

There is a class of measurement models which is enough for almost all situations. This usually makes the investigation simpler.

**Definition 48.** A measurement model $\mathcal{M} = \langle \mathcal{K}, \varrho_0, \mathcal{V}, \mathsf{F} \rangle$ is *normal* if

- $\varrho_0$ is a pure state;

- $\mathcal{V}$ is a unitary channel, i.e., there is a unitary operator $V$ on $\mathcal{H} \otimes \mathcal{K}$ such that $\mathcal{V}(\cdot) = V \cdot V^*$;

- $\mathsf{F}$ is a sharp observable.

At first instance, normal measurement models may seem to be of very special type. Of course, if we start from a description of some real measurement scheme, these conditions of normality may not be fulfilled. However, from a theoretical perspective we can explain the central role of normal memos.

The fact that we can usually restrict to normal memos is justified by the dilation theorems discussed earlier. Namely, as we have seen in Section 3.4.2, every state can be purified. In the similar way, we have seen in 5.2.2 that every channel has a Stinespring dilation to a unitary channel. Finally, an observable has a dilation to a sharp observable. Hence, by choosing suitably large Hilbert space $\mathcal{K}$, we can extend the given memo to a normal memo.

## 6.2   Instruments

If after a measurement of an observable A the system is still available, one may try to perform some other measurement to gain (potentially) more information about the original state. In fact, we can use the measurement procedure as a state preparator. In both of these cases, we need to know not only the outcome probabilities, but also the influence of the first measurement on the system and, consequently, on later measurements. We need not to know all the details of the measurement model which was used, and often such a detailed description is not available or it is too complicated. The concept of an *instrument* neatly captures the relevant description of the measurement process for the above mentioned intentions. Instruments were first systematically studied by E.B. Davies and his book [29] is still a recommendable source.

### 6.2.1   Definition and Ozawa's Theorem

Let $\mathcal{M} = \langle \mathcal{K}, \varrho_0, \mathcal{V}, \mathsf{F} \rangle$ be a measurement model for an observable A. After the measurement of A is performed, we measure another observable B on the system. Altogether we find a joint probability distribution of the values of A and B, and we denote by $p_\varrho(\mathsf{A} \in X \ \& \ \mathsf{B} \in Y)$ the probability that A-measurement gives an outcome from $X$ and B-measurement gives an outcome from $Y$. The subindex refers to the initial state $\varrho$ of the system.

Since the probability reproducibility condition holds between $\mathcal{M}$ and A, we have

$$p_\varrho(\mathsf{A} \in X \ \& \ \mathsf{B} \in Y) = \operatorname{tr}\left[\mathcal{V}(\varrho \otimes \varrho_0)\mathsf{B}(Y) \otimes \mathsf{F}(X)\right] . \tag{6.7}$$

Let us fix the set $X$ for a moment. It turns out to be useful to define an operator $\mathcal{I}_X^{\mathcal{M}}(\varrho)$ such that

$$p_\varrho(\mathsf{A} \in X \ \& \ \mathsf{B} \in Y) = \operatorname{tr}\left[\mathcal{I}_X^{\mathcal{M}}(\varrho)\mathsf{B}(Y)\right] . \tag{6.8}$$

This is indeed possible, and a comparison of (6.7) and (6.8) indicates that

$$\mathcal{I}_X^{\mathcal{M}}(\varrho) = \mathrm{tr}_{\mathcal{K}} \left[ \mathcal{V}(\varrho \otimes \varrho_0) I \otimes \mathsf{F}(X) \right] .\qquad(6.9)$$

If we require that (6.8) holds for all observables B, then formula (6.9) gives the unique choice for the operator $\mathcal{I}_X^{\mathcal{M}}(\varrho)$.

The mapping $\varrho \mapsto \mathcal{I}_X^{\mathcal{M}}(\varrho)$ has a unique linear extension to $\mathcal{T}(\mathcal{H})$, and therefore we will consider $\mathcal{I}_X^{\mathcal{M}}$ as a linear mapping from $\mathcal{T}(\mathcal{H})$ to $\mathcal{T}(\mathcal{H})$. The following properties are straightforward to verify:

(I1) For each $X$, $\mathcal{I}_X^{\mathcal{M}}$ is an operation;

(I2) If $\varrho \in \mathcal{S}(\mathcal{H})$, then $\mathrm{tr}\left[\mathcal{I}_\Omega^{\mathcal{M}}(\varrho)\right] = 1$ and $\mathcal{I}_\emptyset^{\mathcal{M}}(\varrho) = O$;

(I3) If $\varrho \in \mathcal{S}(\mathcal{H})$ and $\{X_j\}$ is a sequence of mutually disjoint sets, then

$$\mathrm{tr}\left[\mathcal{I}_{\cup_j X_j}^{\mathcal{M}}(\varrho)\right] = \sum_j \mathrm{tr}\left[\mathcal{I}_{X_j}^{\mathcal{M}}(\varrho)\right] .$$

The properties of the mapping $\mathcal{I}^{\mathcal{M}}$ can be abstracted and this leads to following definition of a one kind of generalized measure.

**Definition 49.** A mapping $\mathcal{I}$ from an outcome space $(\Omega, \mathcal{F})$ to the set of operations on $\mathcal{T}(\mathcal{H})$ is called an *instrument* if it satisfies the properties (I1)-(I3).

We have seen that each measurement model $\mathcal{M}$ defines an instrument $\mathcal{I}^{\mathcal{M}}$, and we say that $\mathcal{I}^{\mathcal{M}}$ is the *instrument induced by* $\mathcal{M}$. The following fundamental theorem due to Ozawa [66] says that the converse implication also holds.

**Theorem 11.** (*Ozawa's theorem*) If $\mathcal{I}$ is an instrument, then there exists a measurement model $\mathcal{M} = \langle \mathcal{K}, \varrho_0, \mathcal{V}, \mathsf{F} \rangle$ such that $\mathcal{I} = \mathcal{I}^{\mathcal{M}}$. Moreover, $\mathcal{M}$ can be chosen to be normal.

### 6.2.2  A-compatibility

Let $\mathcal{I}$ be an instrument. It defines a unique observable A by the formula

$$\mathrm{tr}\left[\mathcal{I}_X(\varrho)\right] = \mathrm{tr}\left[\varrho \mathsf{A}(X)\right] \quad \forall X \in \mathcal{F}, \varrho \in \mathcal{S}(\mathcal{H}) .\qquad(6.10)$$

We say that $\mathcal{I}$ satisfying (6.10) is A-*compatible*. This connection simply means that $\mathcal{I}$ and A give the same measurement outcome probabilities. It is useful to think this as an equivalence relation in the set of instruments. For a given observable A, there is hence an equivalence class $[\mathcal{I}]$ of A-compatible instruments. Each particular A-compatible instrument $\mathcal{I}$ describes a certain way to measure A, leading to a certain kind of state transformations. Example 68 below demonstrates that every observable A has (infinitely many) A-compatible instruments.

**Example 68.** (*Trivial instrument*) Let A be an observable. Fix a state $\xi$. Then the formula

$$\mathcal{I}_X(\varrho) = \mathrm{tr}\left[\varrho \mathsf{A}(X)\right] \xi \qquad(6.11)$$

defines an A-compatible instrument. In fact, we have

$$\operatorname{tr}\left[\mathcal{I}_X(\varrho)\right] = \operatorname{tr}\left[\varrho \mathsf{A}(X)\right] \operatorname{tr}\left[\xi\right] = \operatorname{tr}\left[\varrho \mathsf{A}(X)\right] ,$$

showing that (6.10) holds. Instruments of this form are called *trivial instruments*.

Let $\mathcal{M}$ be a memo. It defines an instrument $\mathcal{I}^{\mathcal{M}}$ through formula (6.9). On the other hand, $\mathcal{I}^{\mathcal{M}}$ defines an observable by condition (6.10). Combining these two conditions we get

$$\operatorname{tr}\left[\varrho \mathsf{A}(X)\right] = \operatorname{tr}\left[\mathcal{I}_X^{\mathcal{M}}(\varrho)\right] = \operatorname{tr}\left[\operatorname{tr}_{\mathcal{K}}\left[\mathcal{V}(\varrho \otimes \varrho_0) I \otimes \mathsf{F}(X)\right]\right] = \operatorname{tr}\left[\mathcal{V}(\varrho \otimes \varrho_0) I \otimes \mathsf{F}(X)\right] .$$

This is exactly the probability reproducibility condition (6.1).

We conclude that the three different perspectives can be adopted to describe a measurement apparatus in three different levels of description. These perspectives lead to three different mathematical objects: measurement models, instruments and observables. These different layers of description are connected in the following way:

$$\mathcal{M} \longrightarrow \mathcal{I} \longrightarrow \mathsf{A} , \qquad \mathsf{A} \longrightarrow [\mathcal{I}] , \quad \mathcal{I} \longrightarrow [\mathcal{M}] ,$$

meaning that whereas each memo uniquely defines an instrument and each instrument uniquely defines an observable, the opposite relations are not unique. Each observable defines an equivalence class of instruments and each instrument defines an equivalence class of measurement models.

### 6.2.3   Conditional output states

Let us go back to the setting of Subsection 6.2.1, where we have two observables A and B. We measure first the observable A, and after that we measure the observable B. Altogether we find a joint probability distribution of the values of A and B, and in an initial state $\varrho$ this is given by

$$p_\varrho(\mathsf{A} \in X \ \& \ \mathsf{B} \in Y) = \operatorname{tr}\left[\mathcal{I}_X(\varrho)\mathsf{B}(Y)\right] ,$$

where $\mathcal{I}$ is an A-compatible instrument depending on the way how the A-measurement is performed. The conditional probability $p_\varrho(\mathsf{B} \in Y \mid \mathsf{A} \in X)$ can be thus written in the form

$$p_\varrho(\mathsf{B} \in Y \mid \mathsf{A} \in X) = \frac{p_\varrho(\mathsf{A} \in X \ \& \ \mathsf{B} \in Y)}{p_\varrho(\mathsf{A} \in X)} = \frac{\operatorname{tr}\left[\mathcal{I}_X(\varrho)\mathsf{B}(Y)\right]}{\operatorname{tr}\left[\mathcal{I}_X(\varrho)\right]} \equiv \operatorname{tr}\left[\widetilde{\varrho}_X \mathsf{B}(Y)\right] .$$

The state

$$\widetilde{\varrho}_X = \frac{1}{\operatorname{tr}\left[\mathcal{I}_X(\varrho)\right]} \mathcal{I}_X(\varrho)$$

is called a *conditional output state*.

It is worth to notice two things. First of all, the conditional output state $\widetilde{\varrho}_X$ is defined only when $\operatorname{tr}\left[\varrho \mathsf{A}(X)\right] = \operatorname{tr}\left[\mathcal{I}_X(\varrho)\right] \neq 0$. The reason is simply that we cannot define the conditional probability $p_\varrho(\mathsf{B} \in Y \mid \mathsf{A} \in X)$ if $p_\varrho^{\mathsf{A}}(X) = 0$. As a second point, notice that the mapping $\varrho \mapsto \widetilde{\varrho}_X$ is not, in general, linear. However, it is linear (and hence a channel) when $X = \Omega$, since in this case $\widetilde{\varrho}_\Omega = \mathcal{I}_\Omega(\varrho)$.

If the set $\Omega$ of measurement outcomes is finite or countably infinite, the conditional output state corresponding to an outcome $x$ is

$$\widetilde{\varrho}_x = \frac{1}{\operatorname{tr}\left[\mathcal{I}_x(\varrho)\right]}\mathcal{I}_x(\varrho)\,, \tag{6.12}$$

where we have adopted short hand notations $\widetilde{\varrho}_x \equiv \widetilde{\varrho}_{\{x\}}$ and $\mathcal{I}_x \equiv \mathcal{I}_{\{x\}}$.

**Example 69.** (*Conditional output states for a trivial instrument.*) Let us recall Example 68 and consider again the trivial instrument $\mathcal{I}$ defined in (6.11). We then have

$$\widetilde{\varrho}_X = \frac{1}{\operatorname{tr}\left[\mathcal{I}_X(\varrho)\right]}\mathcal{I}_X(\varrho) = \xi \tag{6.13}$$

for all $X$ and $\varrho$ such that $\operatorname{tr}\left[\mathcal{I}_X(\varrho)\right] \neq 0$. Hence, the conditional output state does not depend on the particular measurement outcome nor the input state.

The reason for the name *trivial* instrument can now be explained. Suppose we make an A-measurement and after that a B-measurement. If the A-measurement is described by the trivial instrument, then we get

$$p_\varrho(\mathsf{B} \in Y \mid \mathsf{A} \in X) = \operatorname{tr}\left[\widetilde{\varrho}_X \mathsf{B}(Y)\right] = \operatorname{tr}\left[\xi \mathsf{B}(Y)\right] = p_\xi^{\mathsf{B}}(Y)\,.$$

This is just the same as measuring the trivial observable $Y \mapsto p_\xi^{\mathsf{B}}(Y)I$ in the state $\varrho$. In other words, B-measurement does not produce any additional information on the input state $\varrho$. All measurements following the A-measurement are "trivialized".

### 6.3    Disturbance caused by a measurement

It is one of the basic lessons of quantum physics that a measurement on a quantum system causes an unavoidable disturbance in the sense that after the measurement system is in a different state than before the measurement. In Subsection 6.3.1 we give one simple and precise formulation of this idea in terms of instruments. We note, however, that this is only one version of the "No information without disturbance" statement. In the following form the result has been stated e.g. in [18].

In Subsection 6.3.2 we take a look on the famous BB84 quantum key distribution protocol to demonstrate how one can sometimes benefit from the unavoidable disturbance in quantum measurements.

#### 6.3.1    No information without disturbance

Is there a way to make a measurement of an observable A without causing any disturbance? This would mean that performing a measurement of A does not affect any other measurements performed later on the system. In particular, knowing the measurement outcome for A cannot make any difference for the measurement outcome distribution of an observable B, that is,

$$p_\varrho(\mathsf{B} \in Y \mid \mathsf{A} \in X) = p_\varrho(\mathsf{B} \in Y)\,.$$

for all $X, Y \in \mathcal{F}$ and $\varrho \in \mathcal{S}(\mathcal{H})$. Since B can be chosen to be any observable, this is to require that

$$\widetilde{\varrho}_X = \varrho$$

for all $X \in \mathcal{F}$ and $\varrho \in \mathcal{S}(\mathcal{H})$. Written in terms of the related A-compatible instrument $\mathcal{I}$, this requirement is

$$\mathcal{I}_X(\varrho) = c_X(\varrho)\, \varrho \quad \forall X \in \mathcal{F}, \varrho \in \mathcal{S}(\mathcal{H}) \,, \tag{6.14}$$

where $c_X(\varrho)$ is a non-negative number, possibly depending on $X$ and $\varrho$.

Due to the linearity of $\mathcal{I}_X$ it follows that the number $c_X(\varrho)$ does not depend on $\varrho$ and we can set $c_X \equiv c_X(\varrho)$. Indeed, let $\varrho_1$ and $\varrho_2$ be two different states. Since $\mathcal{I}_X$ is linear, we get

$$\mathcal{I}_X(\varrho_1 + \varrho_2) = \mathcal{I}_X(\varrho_1) + \mathcal{I}_X(\varrho_2) = c_X(\varrho_1)\, \varrho_1 + c_X(\varrho_2)\, \varrho_2 \,, \tag{6.15}$$

and, on the other hand,

$$\mathcal{I}_X(\varrho_1 + \varrho_2) = c_X(\varrho_1 + \varrho_2)(\varrho_1 + \varrho_2) = c_X(\varrho_1 + \varrho_2)\, \varrho_1 + c_X(\varrho_1 + \varrho_2)\, \varrho_2 \,. \tag{6.16}$$

Comparing these two equations we see that $c_X(\varrho_1) = c_X(\varrho_1 + \varrho_2) = c_X(\varrho_2)$ for all $\varrho_1, \varrho_2$, hence $c_X(\varrho) \equiv c_X$.

Taking trace in both sides of (6.14), we get

$$\operatorname{tr}\left[\varrho\mathsf{A}(X)\right] = \operatorname{tr}\left[\mathcal{I}_X(\varrho)\right] = c_X \,.$$

As this is true for all states $\varrho$, we have

$$c_X = \operatorname{tr}\left[|\varphi\rangle\langle\varphi|\mathsf{A}(X)\right] = \langle\, \varphi \mid \mathsf{A}(X)\varphi\,\rangle$$

for all unit vectors $\varphi \in \mathcal{H}$. Therefore, due to Proposition 4 in Section 2.2 this implies that $\mathsf{A}(X) = c_X I$. Thus, A is necessarily a trivial observable which does not provide any information on the state of the system (see Example 46 in Section 4.5). We conclude that in order to acquire at least some non-trivial information, the measurement must produce some disturbance.

### 6.3.2   BB84 quantum key distribution

The fact that measurements necessarily disturb the observed systems can be exploited in cryptographic protocols as a tool to identify the presence of an eavesdroper. Let us discuss the most profound example of quantum cryptography - the so-called BB84 quantum key distribution protocol originally proposed in 1984 by Ch.Bennett and G.Brassard [5].

The goal of this protocol is to establish (in a secure way) a cryptographic key between two parties, conventionally called Alice and Bob. To explain this protocol, we define the following orthonormal bases for a two dimensional Hilbert space $\mathcal{H}$,

$$a_+ = b_+ = \{\varphi, \varphi_\perp\} = \{\psi_{0,+}, \psi_{1,+}\} \,, \tag{6.17}$$
$$a_\times = b_\times = \{\varphi_+, \varphi_-\} = \{\psi_{0,\times}, \psi_{1,\times}\} \,, \tag{6.18}$$

where $\varphi_\pm = \frac{1}{\sqrt{2}}(\varphi_0 \pm \varphi_1)$. The symbols $a, b$ denotes Alice's and Bob's sides, respectively. The BB84 protocol consists of the following steps:

1. In each run Alice and Bob independently and randomly choose orthonormal bases: either $+$, or $\times$, i.e. they both generate a random bit $a_{j_r}$ and $b_{k_r}$, respectively ($j_r, k_r \in \{+, \times\}$ for each run $r$).

2. Alice generates another random bit $x_r \in \{0, 1\}$ and sends the pure state associated with the vector $\psi_{x_r, j_r}$ to Bob.

3. Bob performs a sharp measurement according to bit value $b_{k_r}$ (recall Example 34) and records an outcome $y_r = \{0, 1\}$ with a conditional probability $p = |\langle \psi_{y_r, b_{k_r}} | \psi_{x_r, a_{j_r}} \rangle|^2$ depending on the state received from Alice.

4. After the previous steps have been repeated long enough (e.g. 100 runs), Alice and Bob compare their choices of bases. If they disagree, i.e. $j_r \neq k_r$, that particular run is ignored, meaning that the bits $x_r$ and $y_r$ are discarded. In this procedure the original bit sequences $\vec{x} = x_1, \cdots x_n$ and $\vec{y} = y_1, \cdots y_n$ are transformed into smaller bit sequences $\vec{x}'$ and $\vec{y}'$.

5. The identical subsequences $\vec{x}' \equiv \vec{y}'$ form the rough key. The goal of the key distribution protocol is achieved. Alice and Bob shares the same sequences of random bit values. In order to prove that this is indeed the case we have to evaluate the probabilities for all cases. Omitting the subindex $r$ we obtain the formula

$$|\langle \psi_{y, b_k} | \psi_{x, a_k} \rangle|^2 = \delta_{a_k, b_k} \delta_{xy} + \frac{1}{2}(1 - \delta_{a_k, b_k}), \qquad (6.19)$$

meaning that if the bits $j_r, k_r$ coincide, i.e. Alice and Bob have chosen the same basis, then the bit values $x, y$ are perfectly correlated. However, if $j_r \neq k_r$, then the values $x, y$ are completely independent. Therefore, these later cases are discarded from the strings $\vec{x}, \vec{y}$ and the postselected subsequences $\vec{x}', \vec{y}'$ fit perfectly.

The goal of BB84 is not to communicate any private information, but only to establish a key that can be afterwards used for the (classical) perfectly secure communication protocol *one time pad* [80]. A potential adversary Eve is always considered to have unlimited resources, but her possibilities are restricted by physical laws. Her only task is to learn the key in a way that no one recognizes her existence. It is not difficult to transfer random string from Alice to Bob, but the problem is to do it in a secure way so that no one else has access to the distributed key. Since nothing important is transferred in key distribution, it is not a big issue if the eavesdropper learns part of the key, as long as her presence can be detected. If this happens, Alice and Bob discard their keys and start to establish a new one. But is this possible?

There is no known classical strategy and the security of BB84 was not discussed yet. Let us note that if classical systems are measured any introduced disturbance can be theoretically undone by the eavesdropper, thus her presence cannot be detected. In the case of BB84 a general strategy of the eavesdrop-er is based on the coupling of her system with the transmitted particles. After that she can wait until Alice and Bob publicly announce the bases they used in each run and measure her system in order to estimate the bit values. From Alice & Bob's point of view Eve performs a specific pointer observable, hence realizing a measurement model of some observable on systems transmitted from Alice to Bob.

However, there is no error-free measurement distinguishing among the four possible states associated with vectors $\psi_{x,a_j}$, because they are not mutually orthogonal. As we have discussed previously some disturbance is necessarily introduced. Consequently, the sent and received states are different and there is a nonvanishing probability that the rough keys $\vec{x}'$, $\vec{y}'$ do not match perfectly. In order to find differences Alice and Bob release random bits from their rough keys and compare publicly their bit values. In the ideal and strict version of the protocol any difference implies that the whole key must be discarded.

For the details of how much key must be released in order to detect the eavesdropper with a very high probability we refer to overview paper by Gisin et al [38]. The proof of unconditional security of BB84 can be found in [59], [76].

There are several alternative proposals for quantum key distribution and also for quantum secret communication. A common idea of all these protocols is that the security, in particular the detection of adversaries, is based on the unavoidable disturbance caused by measurements.

**Example 70.** Let us consider the simplest version of attack on BB84 key distribution protocol. Let us assume that Eve is measuring each transmitted particle independently trying to guess what is the state. One option for her is to randomly switch between the same bases as are used by Alice and Bob. Let us denote by $e_{l_r}$ with $l_r \in \{+, \times\}$ the basis chosen by Eve in $r$th run. Her outcome value $z_r \in \{0, 1\}$ is perfectly correlated with Alice's and Bob's bit values only if her choice coincide with the choices of Alice and Bob, simultaneously. However, a chance that $j_r = k_r = l_r$ is once in four runs, i.e. probability is $1/4$. Hence, she can learn half of the bits.

Let us assume that if the eavesdropper gets outcome $z_r$, then she resend to Bob a state associated with the vector $\psi_{z_r, e_{l_r}}$. Under the considered circumstances this is the best what Eve can do in order to minimize the disturbance. A probability that $j_r = k_r \neq l_r$ is also $1/4$, i.e. the second half of the rough key is disturbed. The conditional probability that the original state and the resend one coincide (if $j_r = k_r \neq l_r$) is $1/2$, hence, a quarter of bits in the rough keys is different. Therefore, it is sufficient to publish $1/4$ of the key to find at least one difference with a very large probability.

## 6.4 Repeatable measurements

Given an observable A, we can ask what kind of A-compatible instruments there exist. It should be perhaps emphasized that there are no preferred or canonical A-compatible instrument. Different instruments simply describe different kind of measurements of A, which influence the system in different ways. We may, of course, want to perform the measurement of A in a certain way and wish to influence the system in some way suitable for our purposes. For instance, one possibility is that we use our measurement to prepare the system. We then require that this preparation procedure makes the measurement outcome in the iterated measurement completely predictable. This leads to the notion of a repeatable measurement. An overview on repeatable measurements is presented in [14]. An extensive discussion on repeatability and some other related concepts can be found in [18].

### 6.4.1 Repeatability

To formulate the above idea of a measurement device as preparator, let A be a discrete observable. Suppose that we perform a measurement of A in a state $\varrho$, and that the measurement is described by an A-compatible instrument $\mathcal{I}$. Let $x$ be a measurement outcome which have nonzero probability to occur, i.e., $p_\varrho^A(x) \neq 0$. The conditional output state corresponding to the measurement outcome $x$ is

$$\widetilde{\varrho}_x = \frac{1}{\text{tr}\left[\mathcal{I}_x(\varrho)\right]} \mathcal{I}_x(\varrho) \,, \tag{6.20}$$

where we use the short hand notation $\mathcal{I}_x \equiv \mathcal{I}_{\{x\}}$.

Assume then that after recording some measurement outcome $x$, we repeat the measurement using the same system. Naturally, the measurement outcome now depends on the instrument $\mathcal{I}$. We require that we get the same measurement outcome $x$ as in the first measurement with probability 1. Hence, we set the condition

$$\text{tr}\left[\widetilde{\varrho}_x A(x)\right] = 1 \,. \tag{6.21}$$

This leads us to the following definition.

**Definition 50.** An instrument $\mathcal{I}$ is *repeatable* if $\text{tr}\left[\widetilde{\varrho}_x A(x)\right] = 1$ holds whenever $\text{tr}\left[\varrho A(x)\right] \neq 0$.

Repeatability is thus a property of instruments and it does not depend on the other details of measurement models. Sometimes we say that a measurement model $\mathcal{M}$ is repeatable - this means that the induced instrument $\mathcal{I}^{\mathcal{M}}$ is repeatable.

**Example 71.** (*Trivial instrument is not repeatable.*) Let us continue with Example 68. Let A be a discrete observable, $\xi$ a state, and $\mathcal{I}$ the trivial instrument determined by $\xi$. For each state $\varrho$ and measurement outcome $x$ such that $\text{tr}\left[\varrho A(x)\right] \neq 0$, we then have $\widetilde{\varrho}_x = \xi$. Thus, if the trivial instrument $\mathcal{I}$ is repeatable, then $\text{tr}\left[\xi A(x)\right] = 1$ whenever $A(x) \neq O$. However, since

$$1 = \text{tr}\left[\xi\right] = \sum_x \text{tr}\left[\xi A(x)\right] = \sum_{x : A(x) \neq O} 1 \,,$$

the repeatability condition can hold only if there is just a single nonzero effect in the range of A. In this case, the observable A is of the following form:

$$A(x) = I \quad \text{for some outcome } x, \quad A(y) = O \quad \text{for all the other outcomes } y \,. \tag{6.22}$$

We conclude that trivial instruments are never repeatable, unless the corresponding observable A is of the banal form (6.22).

The repeatability condition can be written in different but equivalent forms.

**Proposition 74.** For an instrument $\mathcal{I}$, the following conditions are equivalent:

(i) $\mathcal{I}$ is repeatable.

(ii) $\text{tr}\left[\mathcal{I}_x(\mathcal{I}_x(\varrho))\right] = \text{tr}\left[\mathcal{I}_x(\varrho)\right]$ for every outcome $x$ and state $\varrho$.

(ii) $\text{tr}\left[\mathcal{I}_y(\mathcal{I}_x(\varrho))\right] = 0$ for all outcomes $x \neq y$ and every state $\varrho$.

*Proof.* Using equations (6.10) and (6.20), we see that the repeatability condition (6.21) for $\mathcal{I}$ is equivalent to

$$\text{tr}\left[\mathcal{I}_x(\mathcal{I}_x(\varrho))\right] = \text{tr}\left[\mathcal{I}_x(\varrho)\right] \;, \tag{6.23}$$

which has to hold whenever $\text{tr}\left[\mathcal{I}_x(\varrho)\right] \neq 0$. If $\text{tr}\left[\mathcal{I}_x(\varrho)\right] = 0$, then both sides in (6.23) give 0. Therefore, (i) and (ii) are equivalent.

The equivalence of (ii ) and (iii) follows immediately by noticing that

$$\sum_y \text{tr}\left[\mathcal{I}_y(\mathcal{I}_x(\varrho))\right] = \text{tr}\left[\mathcal{I}_\Omega(\mathcal{I}_x(\varrho))\right] = \text{tr}\left[\mathcal{I}_x(\varrho)\right] \;.$$

Here we have used the fact that $\mathcal{I}_\Omega$ is trace preserving.                    □

Which observables admit repeatable instruments? It is easy to see that a necessary condition is that each nonzero effect has eigenvalue 1. Namely, assume that a discrete observable A has an A-compatible repeatable instrument. Let $\mathsf{A}(x) \neq O$. Then there is a state $\varrho$ such that $\text{tr}\left[\varrho \mathsf{A}(x)\right] \neq 0$. Repeatability condition then implies that $\text{tr}\left[\widetilde{\varrho}_x \mathsf{A}(x)\right] = 1$. Hence, the effect $\mathsf{A}(x)$ has eigenvalue 1.

On the other hand, suppose that A is a discrete observable and that each nonzero effect $\mathsf{A}(x)$ has eigenvalue 1. For each outcome $x$ satisfying $\mathsf{A}(x) \neq O$, choose a unit vector $\psi_x \in \mathcal{H}$ such that $\mathsf{A}(x)\psi_x = \psi_x$. For an outcome $y$ with $\mathsf{A}(y) = O$, fix an arbitrary unit vector $\psi_y \in \mathcal{H}$. Then define an instrument $\mathcal{I}$ as

$$\mathcal{I}_x(\varrho) = \text{tr}\left[\varrho \mathsf{A}(x)\right] |\psi_x\rangle\langle\psi_x| \,.$$

This instrument is A-compatible since

$$\text{tr}\left[\mathcal{I}_x(\varrho)\right] = \text{tr}\left[\varrho \mathsf{A}(x)\right] \text{tr}\left[|\psi_x\rangle\langle\psi_x|\right] = \text{tr}\left[\varrho \mathsf{A}(x)\right] \,.$$

Using condition (ii) in Proposition 74 it easy to see that $\mathcal{I}$ is repeatable,

$$\text{tr}\left[\mathcal{I}_x(\mathcal{I}_x(\varrho))\right] = \text{tr}\left[\varrho \mathsf{A}(x)\right] \text{tr}\left[\mathcal{I}_x(|\psi_x\rangle\langle\psi_x|)\right] = \text{tr}\left[\varrho \mathsf{A}(x)\right] \text{tr}\left[|\psi_x\rangle\langle\psi_x|\mathsf{A}(x)\right] = \text{tr}\left[\mathcal{I}_x(\varrho)\right] \,.$$

We have thus reached the following result, first proved in [18].

**Proposition 75.** Let A be a discrete observable. The following conditions are equivalent:

  (i)  There exists an A-compatible repeatable instrument.

  (ii)  If $\mathsf{A}(x) \neq O$, then $\mathsf{A}(x)$ has eigenvalue 1.

As a conclusion, we have seen that repeatable measurements are possible, but they set some requirements for the observable in question.

### 6.4.2  Approximate repeatability

One could formulate the repeatability condition also in the general case and not only for discrete observables. In the general case, the repeatability of an instrument $\mathcal{I}$ means that

$$\text{tr}\left[\mathcal{I}_X(\mathcal{I}_X(\varrho))\right] = \text{tr}\left[\mathcal{I}_X(\varrho)\right] \tag{6.24}$$

for every $X \in \mathcal{F}$ and $\varrho \in \mathcal{S}(\mathcal{H})$. However, according to a result of Ozawa [67], there is no A-compatible repeatable instruments unless A is discrete.

The fact that repeatable instruments do not exists for non-discrete observables may seem controversial with the current quantum technology, where a single particle can be repeatedly localized with high precision. One can make a practical argument that in any real experiment there can be only finite number of possible measurement outcomes. However, this issue can also be explained by modifying the repeatability condition slightly, giving perhaps some additional insight to the problem.

An overview of relaxations to the repeatability condition can be found in [17] and [21]. To demonstrate this kind of approach, we follow the seminal article of Davies and Lewis [30] and describe one possible way to relax the repeatability condition. For simplicity, we assume the outcome space of observables and instruments to be the Borel measure space $(\mathbb{R}, \mathcal{B}(\mathbb{R}))$.

For every $\varepsilon > 0$ and $x \in \mathbb{R}$, we denote by $I_{x;\varepsilon}$ the closed interval centered in $x$ and with the length $\epsilon$, i.e., $I_{x;\varepsilon} = [x - \frac{1}{2}\varepsilon, x + \frac{1}{2}\varepsilon]$. For every $X \subseteq \mathbb{R}$, we then denote

$$X_\varepsilon = \bigcup_{x \in X} I_{x;\varepsilon}\,.$$

Thus, if the diameter of a set $X$ (defined as the least upper bound of the distances between pairs of points in $X$) is $d(X)$, the the diameter of $X_\varepsilon$ is $d(X) + \varepsilon$.

**Definition 51.** Let $\varepsilon > 0$. An instrument $\mathcal{I}$ is $\varepsilon$-*repeatable* if

$$\text{tr}\left[\mathcal{I}_{X_\varepsilon}(\mathcal{I}_X(\varrho))\right] = \text{tr}\left[\mathcal{I}_X(\varrho)\right] \tag{6.25}$$

for every $X \in \mathcal{B}(\mathbb{R})$ and every $\varrho \in \mathcal{S}(\mathcal{H})$.

The condition (6.25) has the following meaning. Assume that an outcome from a subset $X$ is recorded. Then a repeated application of the same measurement gives an outcome from $X_\varepsilon$ with probability 1. When $\varepsilon$ is close to 0, then $\varepsilon$-repeatable instrument is almost like a repeatable one. Formally, the definition of $\varepsilon$-repeatability would make sense also for $\varepsilon = 0$; in this case it would reduce to the usual definition of repeatability.

**Exercise 43.** Show that if and an instrument $\mathcal{I}$ is $\varepsilon$-repeatable, then it is $\varepsilon'$-repeatable for every $\varepsilon' \geq \varepsilon$.

As a conclusion of the previous discussion and Exercise 43, for a given observable A, we are interested to find the smallest possible number $\varepsilon$ such that there exists A-compatible instrument $\mathcal{I}$ which is $\varepsilon$-repeatable.

**Proposition 76.** Let A be an observable and assume there is $\varepsilon > 0$ such that for every open interval $X$ with diameter $d(X) = \varepsilon$, the effect $\mathsf{A}(X)$ has eigenvalue 1. Then there exists a $\varepsilon$-repeatable A-compatible instrument.

*Proof.* For each $n \in \mathbb{Z}$, denote $X_n = [n\varepsilon, (n+1)\varepsilon)$. Then $(X_n)$ is a sequence of mutually disjoint sets and $\cup_n X_n = \mathbb{R}$. From the assumption follows that for each $n \in \mathbb{Z}$, we can choose a pure state $\varrho_n$ such that $\mathrm{tr}\left[\varrho_n \mathsf{A}(X_n)\right] = 1$. The formula

$$\mathcal{I}_X(\varrho) := \sum_n \mathrm{tr}\left[\varrho \mathsf{A}(X \cap X_n)\right] \; \varrho_n$$

defines an A-compatible instrument $\mathcal{I}$. Indeed, to see that each $\mathcal{I}_X$ is completely positive, fix an orthonormal basis $\{\varphi_k\}$ for $\mathcal{H}$. Expanding the trace in this basis, $\mathcal{I}_X$ can be written in the Kraus form

$$\mathcal{I}_X(\varrho) = \sum_{k,n} A_{k,n}\varrho A_{k,n}^* \,,$$

where

$$A_{k,n} := |\psi_n\rangle\langle \mathsf{A}(X \cap X_n)^{\frac{1}{2}}\varphi_k| \,.$$

By Proposition 61 this shows that $\mathcal{I}_X$ is completely positive.

To prove that $\mathcal{I}$ is $\varepsilon$-repeatable, let $X \in \mathcal{B}(\mathbb{R})$ and $\varrho \in \mathcal{S}(\mathcal{H})$. We then get

$$
\begin{aligned}
\mathrm{tr}\left[\mathcal{I}_{X_\varepsilon}\left(\mathcal{I}_X(\varrho)\right)\right] &= \sum_n \sum_k \mathrm{tr}\left[\varrho_n \mathsf{A}(X_\varepsilon \cap X_k)\right] \; \mathrm{tr}\left[\varrho \mathsf{A}(X \cap X_n)\right] \\
&= \sum_n \mathrm{tr}\left[\varrho_n \mathsf{A}(X_\varepsilon)\right] \; \mathrm{tr}\left[\varrho \mathsf{A}(X \cap X_n)\right] \,.
\end{aligned}
$$

If $X \cap X_n \neq \emptyset$, then $X_n \subseteq X_\varepsilon$. This implies that either $\mathrm{tr}\left[\varrho \mathsf{A}(X \cap X_n)\right] = 0$ or $\mathrm{tr}\left[\varrho_n \mathsf{A}(X_\varepsilon)\right] = 1$. Therefore

$$\mathrm{tr}\left[\mathcal{I}_{X_\varepsilon}\left(\mathcal{I}_X(\varrho)\right)\right] = \sum_n \mathrm{tr}\left[\varrho \mathsf{A}(X \cap X_n)\right] = \mathrm{tr}\left[\varrho \mathsf{A}(X)\right] \,.$$

$\square$

**Example 72.** (*Approximate repeatability of the canonical position observable*) Let us recall the canonical position observable $\mathsf{Q}$ introduced in Example 35, Section 4.2. It is a sharp observable and thus each $\mathsf{Q}(X) \neq O$ has eigenvalue 1. On the other hand, $\mathsf{Q}(X) \neq O$ whenever $X$ is a nonempty open interval. We conclude from Proposition 76 that there exists a $\varepsilon$-repeatable $\mathsf{Q}$-compatible instrument for each $\varepsilon > 0$. There is thus no theoretical limit for the precision of repeatability.

### 6.5 Lüders measurements

Lüders measurements are perhaps the most often used type of measurement models and instruments in applications. For this reason we spend some time on this topic.

### 6.5.1  Sharp Lüders measurement and ideality

Let us start with a discussion of a specific kind of measurement model for a discrete sharp observable associated to an orthonormal basis. Let $\{\psi_j\}$ be an orthonormal basis for $\mathcal{H}$ and $\mathsf{A}$ the associated sharp observable, i.e. $\mathsf{A}(j) = |\psi_j\rangle\langle\psi_j|$ (recall Example 34 in Section 4.2).

To construct a measurement model for $\mathsf{A}$, fix a Hilbert space $\mathcal{K}$ with the same dimension as $\mathcal{H}$. Let $\{\phi_j\}$ be an orthonormal basis for $\mathcal{K}$, and choose the pointer observable $\mathsf{F}$ to be the sharp observable associated to this basis, i.e., $\mathsf{F}(j) = |\phi_j\rangle\langle\phi_j|$. Let $V : \mathcal{H} \otimes \mathcal{K} \rightarrow \mathcal{H} \otimes \mathcal{K}$ be a unitary operator such that

$$V(\psi_j \otimes \phi_0) = \psi_j \otimes \phi_j \quad \forall j\,.$$

The initial state $\varrho_0$ of the apparatus is chosen to be $\varrho_0 = |\phi_0\rangle\langle\phi_0|$.

If the system is initially in a state $\psi = \sum c_i \psi_i$, then the composition of the system and the apparatus will be in the state

$$|V(\psi \otimes \phi_0)\rangle\langle V(\psi \otimes \phi_0)| = \sum_{i,j} \bar{c}_i c_j \, |\psi_i\rangle\langle\psi_j| \otimes |\phi_i\rangle\langle\phi_j|\,. \tag{6.26}$$

We then get

$$\mathsf{tr}\left[|V(\psi \otimes \phi_0)\rangle\langle V(\psi \otimes \phi_0)| \ I \otimes \mathsf{F}(k)\right]$$

$$= \ \mathsf{tr}\left[\left(\sum_{i,j} \bar{c}_i c_j |\psi_i\rangle\langle\psi_j| \otimes |\phi_i\rangle\langle\phi_j|\right) \ I \otimes |\phi_k\rangle\langle\phi_k|\right]$$

$$= \ \sum_{i,j} \bar{c}_i c_j \, \mathsf{tr}\left[|\psi_i\rangle\langle\psi_j| \otimes |\phi_i\rangle\langle\phi_j| \ I \otimes |\phi_k\rangle\langle\phi_k|\right]$$

$$= \ \sum_i |c_i|^2 \, \mathsf{tr}\left[|\psi_i\rangle\langle\psi_k| \otimes |\phi_i\rangle\langle\phi_k|\right]$$

$$= \ \sum_i |c_i|^2 \, \left(\sum_{r,s} \langle\psi_r \otimes \phi_s \,|\, \psi_i \otimes \phi_i\rangle \, \langle\psi_k \otimes \phi_k \,|\, \psi_r \otimes \phi_s\rangle\right)$$

$$= \ |c_k|^2 \,.$$

On the other hand, we have

$$\mathsf{tr}\left[|\psi\rangle\langle\psi|\mathsf{A}(k)\right] = |c_k|^2 \,,$$

and the probability reproducibility condition is therefore satisfied. We conclude that $\mathcal{M} = \langle\mathcal{K}, \phi_0, V, \mathsf{F}\rangle$ is a measurement model of the observable $\mathsf{A}$.

Let us then calculate the instrument $\mathcal{I}^{\mathcal{M}}$ corresponding to $\mathcal{M}$. We get

$$
\begin{aligned}
\mathcal{I}^{\mathcal{M}}_k(|\psi\rangle\langle\psi|) &= \operatorname{tr}_{\mathcal{K}}\left[|V(\psi\otimes\phi)\rangle\langle V(\psi\otimes\phi)|(I\otimes\mathsf{F}(k))\right] \\
&= \operatorname{tr}_{\mathcal{K}}\left[\left(\sum_{i,j}\bar{c}_i c_j|\psi_i\rangle\langle\psi_j|\otimes|\phi_i\rangle\langle\phi_j|\right)(I\otimes|\phi_k\rangle\langle\phi_k|)\right] \\
&= \sum_{i,j}\bar{c}_i c_j\,\operatorname{tr}_{\mathcal{K}}\left[(|\psi_i\rangle\langle\psi_j|\otimes|\phi_i\rangle\langle\phi_j|)(I\otimes|\phi_k\rangle\langle\phi_k|)\right] \\
&= \sum_i|c_i|^2\,\operatorname{tr}_{\mathcal{K}}\left[|\psi_i\rangle\langle\psi_k|\otimes|\phi_i\rangle\langle\phi_k|\right] \\
&= \sum_i|c_i|^2\left(\sum_{r,s,t}\langle\psi_r\otimes\phi_s\,|\,\psi_i\otimes\phi_i\,\rangle\langle\psi_k\otimes\phi_k\,|\,\psi_t\otimes\phi_s\,\rangle\,|\psi_r\rangle\langle\psi_t|\right) \\
&= |c_k|^2|\psi_k\rangle\langle\psi_k|,
\end{aligned}
$$

which can be written in the form

$$\mathcal{I}^{\mathcal{M}}_k(\varrho) = P_k\varrho P_k, \tag{6.27}$$

where $P_k = |\psi_k\rangle\langle\psi_k|$. This memo $\mathcal{M}$ and instrument $\mathcal{I}^{\mathcal{M}}$ are called Lüders memo and Lüders instrument of A, respectively.

As a generalization, we make the following definition.

**Definition 52.** The *Lüders instrument* $\mathcal{I}^L$ for a discrete observable A is defined as

$$\mathcal{I}^L_x(\varrho) = \mathsf{A}(x)^{\frac{1}{2}}\varrho\,\mathsf{A}(x)^{\frac{1}{2}}. \tag{6.28}$$

Notice that for a projection $P$, we have $P^{\frac{1}{2}} = P$. Therefore, the instrument $\mathcal{I}^{\mathcal{M}}$ in (6.27) is indeed of the Lüders form (6.28).

**Exercise 44.** Let A be a discrete observable (not necessarily sharp) and define an instrument $\mathcal{I}$ as

$$\mathcal{I}_x(\varrho) = \mathsf{A}(x)\varrho\,\mathsf{A}(x).$$

Find an example showing that $\mathcal{I}$ is not, in general, A-compatible.

It is not possible to perform a measurement of a non-trivial observable without at least some disturbance in forthcoming measurements. Hence, it is interesting to ask what is the effect of this unavoidable disturbance. In the following we concentrate on this question in the case of the Lüders instrument of a discrete sharp observable.

**Exercise 45.** Let A be a sharp discrete sharp observable. Show that the corresponding Lüders instrument $\mathcal{I}^L$ is repeatable.

As seen in Subsection 6.3.1, the non-disturbance condition

$$\widetilde{\varrho}_x = \varrho \tag{6.29}$$

cannot hold for all $x$ and $\varrho$, unless A is trivial. We can still ask if (6.29) would hold for some states $\varrho$.

Let A be a sharp discrete observable. An eigenvector $\psi_x$ of $A(x)$ corresponding to the eigenvalue 1 satisfies $A(x)\psi_x = \psi_x$, and this also implies that $A(x)^{\frac{1}{2}}\psi_x = \psi_x$. Therefore, if $\varrho = |\psi_x\rangle\langle\psi_x|$, then $A(x)^{\frac{1}{2}}\varrho A(x)^{\frac{1}{2}} = \varrho$. Hence, in a Lüders measurement the state $\varrho$ does not change at all. This motivates the following definition.

**Definition 53.** Let A be a sharp discrete observable. An A-compatible instrument $\mathcal{I}$ is *ideal* if for every $x \in \Omega$ and $\varrho \in \mathcal{S}(\mathcal{H})$, the following implication holds:

$$p_\varrho^A(x) = 1 \quad \Rightarrow \quad \mathcal{I}_x(\varrho) = \varrho. \tag{6.30}$$

We already saw that the Lüders instrument of a discrete sharp observable A is ideal. This is actually a unique property of the Lüders instrument. Indeed, we have the following result [18].

**Proposition 77.** Let A be a sharp discrete observable. An A-compatible instrument $\mathcal{I}$ is ideal if and only if $\mathcal{I}$ is Lüders instrument.

### 6.5.2 Lüders theorem

Let A and B be two discrete observables. Suppose that we make a Lüders measurement of A, and after that, we perform a measurement of B. We can then ask whether the measurement of B is disturbed by the measurement of A or not. We consider B not to be disturbed if the measurement outcome probabilities of B-measurement do not depend on whether A has been measured first or not. Hence, if $\mathcal{I}^L$ is the A-compatible Lüders instrument, the non-disturbance condition reads

$$\operatorname{tr}\left[\mathcal{I}_\Omega^L(\varrho)B(y)\right] = \operatorname{tr}\left[\varrho B(y)\right],$$

required to hold for all outcomes $y$ and states $\varrho$.

Assume, for a moment, that A and B commute, i.e., $A(x)B(y) = B(y)A(x)$ for all $x, y$. Recall from Theorem 1 in Section 2.2.2 that this implies $A(x)^{\frac{1}{2}}B(y) = B(y)A(x)^{\frac{1}{2}}$ for all $x, y$. Thus, we get

$$\operatorname{tr}\left[\mathcal{I}_\Omega^L(\varrho)B(y)\right] = \sum_x \operatorname{tr}\left[A(x)^{\frac{1}{2}}\varrho A(x)^{\frac{1}{2}}B(y)\right] = \sum_x \operatorname{tr}\left[A(x)\varrho B(y)\right] = \operatorname{tr}\left[\varrho B(y)\right]. \tag{6.31}$$

Thus, B is not disturbed by the A-measurement.

We can also start from the assumption that the measurement statistics of B are not altered by the measurement of A at all. To proceed in this direction, we suppose that A is sharp. Then the assumption means that

$$\sum_x \operatorname{tr}\left[A(x)\varrho A(x)B(y)\right] = \operatorname{tr}\left[\varrho B(y)\right] \tag{6.32}$$

for every state $\varrho$, so that

$$\sum_x A(x)B(y)A(x) = B(y). \tag{6.33}$$

For each $x'$, we get (by multiplying the both sides of (6.33) by $\mathsf{A}(x')$ on the left)

$$\mathsf{A}(x')\mathsf{B}(y) = \sum_x \mathsf{A}(x')\mathsf{A}(x)\mathsf{B}(y)\mathsf{A}(x) = \mathsf{A}(x')\mathsf{B}(y)\mathsf{A}(x') \tag{6.34}$$

and similarly (by multiplying the both sides of (6.33) by $\mathsf{A}(x')$ on the right)

$$\mathsf{B}(y)\mathsf{A}(x') = \sum_x \mathsf{A}(x)\mathsf{B}(y)\mathsf{A}(x)\mathsf{A}(x') = \mathsf{A}(x')\mathsf{B}(y)\mathsf{A}(x') \, . \tag{6.35}$$

Here we have used the fact that $\mathsf{A}(x')\mathsf{A}(x) = \mathsf{A}(\{x\} \cap \{x'\}) = \delta_{x,x'}\mathsf{A}(x')$; see Proposition 42 in Section 4.2. A comparison of (6.34) and (6.35) shows that A and B commute. We thus arrive to the following conclusion, first observed by Lüders [58].

**Theorem 12.** (Lüders Theorem) Let A and B be discrete observables and suppose that A is sharp. A Lüders measurement of A does not disturb B if and only if A and B commute.

Is there a version of Lüders theorem for a general pair of discrete observables? Namely, do two discrete observables commute if their Lüders measurements do not disturb them mutually? The answer to this question is negative and a counterexample was given in [2]. This means that there are observables A and B such that A and B do not commute but, in spite of that, a Lüders measurement of A does not disturb B.

Even though there is no general Lüders Theorem for all pairs of discrete observables, the statement can be extended to certain classes of observables, such as two-outcome observables. All this kind of results are commonly referred as *generalized Lüders Theorem*. For more about this topic, see [20] and [40].

### 6.5.3 Example: mean king problem

Once upon a time ... Alice (a physicists) was sailing in the ocean. A big storm surprised her and she ended up on an island ruled by king Brutus. Brutus loves cats. One day he learned about a cat being alive and dead at once and he started to hate quantum physicists. Nevertheless, he decided to give Alice her last chance. Here is his deadly challenge:

*I shall take a spin-$\frac{1}{2}$ system and make a measurement of either $\sigma_x$, or $\sigma_y$, or $\sigma_z$, thereby getting an outcome 1 or −1. Your task is to correctly guess the outcome I obtain. You are allowed to prepare the system before I measure it. In the morning, I shall make the choice, perform the measurement and give you back the measured system. For the whole day you will have full access to my royal laboratory. However, the sunset shall close and lock the doors without the possibility to take anything out of the lab. I shall tell you my choice of the measurement and you must immediately make your final guess. You have only one chance.*

What are the chances of Alice to survive this game? Is she going to stay in a similar situation as the Schrödinger's cat? Before continuing the story, let us stress that some details must be specified more precisely. It is implicitly assumed that the instruments describing the king's measurement apparatuses are of the Lüder's form. Hence, the states corresponding to the vectors $|\uparrow_b\rangle$ and $|\downarrow_b\rangle$ are the conditional output states related to the outcomes $\pm 1$, respectively, and $b = x, y, z$ identifies the choice of the measurement $\sigma_b$.

**Exercise 46.** If the system is in a state $|\uparrow_x\rangle$, then the outcome of $\sigma_x$ is 1 with probability 1. Hence, if Alice would know before the choice of Brutus, then the problem would be easy. Show that this strategy fails even in a situation of two possible choices, i.e., show that there is no state $\varrho$ such that the outcome for the observables $\sigma_x$ and $\sigma_y$ can be predicted with probability 1.

As seen in Example 46, the measurement outcome obtained by the king will be random whatever way Alice prepares the system. Therefore, Alice chances to survive must depend on the fact that she will know at the end the king's choice. The question is whether this information comes too late or not.

Our goal is not to analyze the problem in its whole generality, but rather to demonstrate that the problem can be solved and Alice can find the king's outcome with certainty. This problem was originally consider in [79]. Here we present a slightly different solution.

Alice prepares two spin-$\frac{1}{2}$ systems into the singlet state $\sigma = |\psi\rangle\langle\psi|$,

$$|\psi\rangle = \frac{1}{\sqrt{2}}\left(|\uparrow\rangle\otimes|\downarrow\rangle - |\downarrow\rangle\otimes|\uparrow\rangle\right),$$

which has the same form in any orthonormal basis. Alice keeps the first system and the second one she gives to the king. Suppose that the king performs a measurement of $\sigma_b$. The conditional output state corresponding to the measurement outcome $\pm 1$ is

$$\frac{1}{\mathrm{tr}\left[\mathcal{I}_{\pm 1}^L(\sigma)\right]}\mathcal{I}_{\pm 1}^L(\sigma) \equiv |\Phi_{b,\pm}\rangle\langle\Phi_{b,\pm}|,$$

where

$$\Phi_{b,+} = |\downarrow_b\rangle\otimes|\uparrow_b\rangle, \qquad\qquad \Phi_{b,-} = |\uparrow_b\rangle\otimes|\downarrow_b\rangle.$$

Alice can count on the additional information she gets after performing her measurement. Let us define an orthonormal basis $\{\theta_1, \theta_2, \theta_3, \theta_4\}$ of $\mathbb{C}^2 \otimes \mathbb{C}^2$ in the following way:

$$\theta_1 = \frac{1}{\sqrt{2}}|\uparrow_z\rangle\otimes|\downarrow_z\rangle + \frac{1}{2}e^{-i\pi/4}|\uparrow_z\rangle\otimes|\uparrow_z\rangle - \frac{1}{2}e^{i\pi/4}|\downarrow_z\rangle\otimes|\downarrow_z\rangle,$$

$$\theta_2 = \frac{1}{\sqrt{2}}|\uparrow_z\rangle\otimes|\downarrow_z\rangle - \frac{1}{2}e^{-i\pi/4}|\uparrow_z\rangle\otimes|\uparrow_z\rangle + \frac{1}{2}e^{i\pi/4}|\downarrow_z\rangle\otimes|\downarrow_z\rangle,$$

$$\theta_3 = \frac{1}{\sqrt{2}}|\downarrow_z\rangle\otimes|\uparrow_z\rangle + \frac{1}{2}e^{i\pi/4}|\uparrow_z\rangle\otimes|\uparrow_z\rangle - \frac{1}{2}e^{-i\pi/4}|\downarrow_z\rangle\otimes|\downarrow_z\rangle,$$

$$\theta_4 = \frac{1}{\sqrt{2}}|\downarrow_z\rangle\otimes|\uparrow_z\rangle - \frac{1}{2}e^{i\pi/4}|\uparrow_z\rangle\otimes|\uparrow_z\rangle + \frac{1}{2}e^{-i\pi/4}|\downarrow_z\rangle\otimes|\downarrow_z\rangle.$$

Alice's observable A is the sharp observable associated to this basis, i.e., $\mathsf{A}(k) = |\theta_k\rangle\langle\theta_k|$ for $k = 1, 2, 3, 4$.

The following table contains the conditional probability distributions (columns) measured by Alice for all possibilities of the king's measurement choice and outcome:

|   | $x, +1$ | $x, -1$ | $y, +1$ | $y, -1$ | $z, +1$ | $z, -1$ |
|---|---------|---------|---------|---------|---------|---------|
| 1 | 1/2     | 0       | 0       | 1/2     | 0       | 1/2     |
| 2 | 0       | 1/2     | 0       | 1/2     | 0       | 1/2     |
| 3 | 0       | 1/2     | 1/2     | 0       | 1/2     | 0       |
| 4 | 1/2     | 0       | 0       | 1/2     | 1/2     | 0       |

The numbers in this table are the overlaps $|\langle\theta_k | \Phi_{b,\pm}\rangle|^2$. The last two columns are easy to calculate as the vectors $\theta_k$ are given in terms of $|\uparrow_z\rangle$ and $|\downarrow_z\rangle$. To calculate the other columns, one can use the expansions

$$|\uparrow_x\rangle = \frac{1}{\sqrt{2}}\left(|\uparrow_z\rangle + |\downarrow_z\rangle\right), \quad |\downarrow_x\rangle = \frac{1}{\sqrt{2}}\left(|\uparrow_z\rangle - |\downarrow_z\rangle\right),$$

$$|\uparrow_y\rangle = \frac{1}{\sqrt{2}}\left(|\uparrow_z\rangle + i|\downarrow_z\rangle\right), \quad |\downarrow_y\rangle = \frac{1}{\sqrt{2}}\left(|\uparrow_z\rangle - i|\downarrow_z\rangle\right).$$

It is straightforward to verify that if $b$ is announced, the outcomes $1, 2, 3, 4$ uniquely match with the king's results $\pm 1$. In conclusion, Alice can solve the problem set by the king. Let us note that the solution strongly depends on the fact that the measurements are described by Lüder's instruments. Without such assumption the perfect success rate cannot be achieved.

### 6.6 Programmable quantum processors

The concept of programmable quantum processors was originally introduced in [64]. The results presented in this section are based on Refs. [44, 87, 89].

#### 6.6.1 Programs and processors

As we have learned in the beginning of this chapter, a measurement model $\mathcal{M}$ is a quadruple $\mathcal{M} = \langle \mathcal{K}, \varrho_0, \mathcal{V}, \mathsf{F}\rangle$. If we change the probe state $\varrho_0$ to another probe state $\varrho_0'$, we get a new measurement model $\mathcal{M}' = \langle \mathcal{K}, \varrho_0', \mathcal{V}, \mathsf{F}\rangle$. In this way, we can potentially realize different observables just by changing the probe state and keeping other components of the memo fixed. Probe states can be seen as *quantum programs* since they encode different measurement models. The rest of the memo components then define a *quantum processor*, which can be programmed by choosing appropriate probe state.

**Definition 54.** A *(programmable) quantum processor* is a triple $\langle \mathcal{K}, V, \mathsf{F}\rangle$ where

- $\mathcal{K}$ is a Hilbert space attached to the apparatus

- $V$ is a unitary channel from $\mathcal{T}(\mathcal{H}\otimes\mathcal{K})$ to $\mathcal{T}(\mathcal{H}\otimes\mathcal{K})$. It describes interaction between the system and the apparatus.

- $\mathsf{F}$ is a pointer observable.

A programmable quantum processor can be used to realize instruments, observables and channels. The probe state $\varrho_0$ being the quantum program encodes all these objects. All possible observables and instruments realized by a given quantum processor are given by the usual set of formulas

$$\mathrm{tr}\left[\varrho\mathsf{A}(X)\right] = \mathrm{tr}\left[\mathcal{I}_X(\varrho)\right] = \mathrm{tr}\left[V(\varrho\otimes\varrho_0)V^*(I\otimes\mathsf{F}(X))\right], \tag{6.36}$$

when $\varrho_0$ runs through all states of the apparatus. The realizable channels correspond to the non-selective state changes. Hence, they are given by

$$\varrho \mapsto \mathrm{tr}_{\mathcal{K}}\left[V(\varrho\otimes\varrho_0)V^*\right]. \tag{6.37}$$

The pointer observable therefore plays no role in the realization of channels.

### 6.6.2 Programming channels

In this section we consider realization of channels using programmable quantum processors. As we said earlier, the pointer observable F does not play any role and therefore for the sake of this section a programmable quantum processor will be represented by a couple $\langle \mathcal{K}, G \rangle$, where $G$ is a unitary transformation defined on $\mathcal{H} \otimes \mathcal{K}$.

**Proposition 78.** Let $\mathcal{E}_1$ and $\mathcal{E}_2$ be two different channels.

(a) There is a quantum processor realizing $\mathcal{E}_1$ and $\mathcal{E}_2$.

(b) In any realization of $\mathcal{E}_1$ and $\mathcal{E}_2$ by a single quantum processor, the program states $\xi_1, \xi_2$ can be chosen to be pure, i.e. $\xi_1 = |\Xi_1\rangle\langle\Xi_1|$ and $\xi_2 = |\Xi_2\rangle\langle\Xi_2|$, and satisfies the identity

$$\langle \Xi_1 \,|\, \Xi_2 \rangle \; I = \sum_j A_j^* B_j \,, \tag{6.38}$$

where $A_j$ are Kraus operators of $\mathcal{E}_1$ and $B_j$ are Kraus operators of $\mathcal{E}_2$.

Let us note a freedom in the specification of Kraus operators for a given channel $\mathcal{E}$ that plays important role in the above proposition. We may assume that the number of Kraus operators is (countably) infinite by adding suitable number of zero operators $O$. Moreover, for the purposes of the above theorem the Kraus operators are ordered by their index values. For example, a channel $\mathcal{E}(\varrho) = q\varrho + (1-q)U\varrho U^*$ has Kraus operators ordered as in many different ways:

$$\sqrt{q}I \,,\; \sqrt{1-q}U \,,\, O \,,\, O \,,\, \dots,$$
$$\sqrt{1-q}U \,,\; \sqrt{q}I \,,\, O \,,\, O \,,\, \dots,$$
$$\sqrt{q}I \,,\, O \,,\; \sqrt{1-q}U \,,\, O \,,\, \dots,$$
$$O \,,\; \sqrt{1-q}U \,,\; \sqrt{q}I \,,\, O \,,\, \dots, \text{ etc}.$$

All of these Kraus decompositions (although containing the same Kraus operators) are considered to be different for the purposes of the Eq. (6.38).

*Proof.* According to Stinespring theorem (see Section 5.2.2) we known that each channel $\mathcal{E}$ can be implemented as a unitary transformation $V_{\mathcal{E}}$ on an extended quantum system $\mathcal{H} \otimes \mathcal{K}_{\mathcal{E}}$ by preparing the system $\mathcal{K}_{\mathcal{E}}$ in a state $\xi_{\mathcal{E}}$. These object form together a dilation $\langle \mathcal{K}_{\mathcal{E}}, V_{\mathcal{E}}, \xi_{\mathcal{E}} \rangle$ of the channel $\mathcal{E}$. Let $\langle \mathcal{K}_1, V_1, \xi_1 \rangle$ and $\langle \mathcal{K}_2, V_2, \xi_2 \rangle$ be dilations for the channels $\mathcal{E}_1$ and $\mathcal{E}_2$, respectively. Define a unitary transformation $G$ acting on $\mathcal{H} \otimes \mathcal{K}_1 \oplus \mathcal{K}_2$ as follows

$$G = V_1 \oplus |\varphi_2\rangle\langle\varphi_2| + V_2 \oplus |\varphi_1\rangle\langle\varphi_1| \,, \tag{6.39}$$

where $\varphi_j \in \mathcal{K}_j$ and $V_1, V_2$ are defined above. Our claim is that a programmable processor $\langle \mathcal{K}_1 \oplus \mathcal{K}_2, G \rangle$ implements both channels $\mathcal{E}_1, \mathcal{E}_2$. In particular, preparing the program register in the state $\xi_1 \in \mathcal{S}(\mathcal{K}_1 \oplus \mathcal{K}_2)$ results in transformation

$$\varrho \mapsto \text{tr}_{\mathcal{K}_1 \oplus \mathcal{K}_2}[G(\varrho \otimes \xi_1)G^*] = \text{tr}_{\mathcal{K}_1}[V_1(\varrho \otimes \xi_1)V_1^*] = \mathcal{E}_1(\varrho) \,, \tag{6.40}$$

and similarly for program state $\xi_2 \in \mathcal{S}(\mathcal{K}_1 \oplus \mathcal{K}_2)$ encodes the channel $\mathcal{E}_2$.

In order to prove the part (b) of the proposition let us evaluate the action of a programmable processor on a pure input state, i.e. when both system and probe are in the pure state. Let us note that a general unitary transformation acting on a bipartite system $\mathcal{H} \otimes \mathcal{K}$ can be written as $G = \sum_{j,k} E_{jk} \otimes |\varphi_j\rangle\langle\varphi_k|$, where vectors $\varphi_j$ form an orthonormal basis of $\mathcal{K}$ and operators $E_{jk}$ must be chosen so that unitarity is guaranteed (find the conditions). We get

$$G|\psi\rangle \otimes |\Xi\rangle = \sum_{j,k} E_{jk}|\psi\rangle \otimes \langle\,\varphi_k\,|\,\Xi\,\rangle\,|\varphi_j\rangle = \sum_j A_j|\psi\rangle \otimes |\varphi_j\rangle\,, \qquad (6.41)$$

where $A_j = \sum_k E_{jk}\,\langle\,\varphi_k\,|\,\Xi\,\rangle$. Tracing out the probe (program) system we end up with the channel

$$|\psi\rangle\langle\psi| \mapsto \sum_j A_j|\psi\rangle\langle\psi|A_j^*\,, \qquad (6.42)$$

written in the operator sum form. Starting with two different pure states $\xi_1, \xi_2$ we get particular operator sum form for two channels $\mathcal{E}_1, \mathcal{E}_2$ with operators $A_j, B_j$, respectively. Let us calculate the following sum

$$\begin{aligned}
\sum_j A_j^* B_j &= \sum_{j,k,k'} E_{jk}^* E_{jk'}\,\langle\,\Xi_1\,|\,\varphi_k\,\rangle\,\langle\,\varphi_{k'}'\,|\,\Xi_2\,\rangle = I \sum_{k,k'} \langle\,\Xi_1\,|\,\varphi_k\,\rangle\,\delta_{kk'}\,\langle\,\varphi_{k'}'\,|\,\Xi_2\,\rangle \\
&= I\,\langle\Xi_1|\left(\sum_k |\varphi_k\rangle\langle\varphi_k|\right)|\Xi_2\rangle = \langle\,\Xi_1\,|\,\Xi_2\,\rangle\,I\,, \qquad (6.43)
\end{aligned}$$

where we used the completeness of basis $\{\varphi_j\}$ and the normalization of operators $E_{jk}$, in particular, $\sum_j E_{jk}^* E_{jk'} = I\delta_{kk'}$. This proves that the identity holds, but it remains to prove that it is sufficient to consider only pure states. The sufficiency follows from the purification. In particular, if we have two mixed program states $\xi_1, \xi_2$, then the programmable processor $G' = G \otimes I_{\mathrm{anc}}$ implements the same channels, but with pure states associated with vectors $\Xi_1, \Xi_2 \in \mathcal{K} \otimes \mathcal{H}_{\mathrm{anc}}$ such that $\mathrm{tr}_{\mathrm{anc}}[|\Xi_j\rangle\langle\Xi_j|] = \xi_j$.                                     $\square$

**Corollary 1.** Arbitrary pair of channels $\mathcal{E}_1, \mathcal{E}_2$ can be realized on the same programmable processor using orthogonal program states. Moreover, in such case the program space $\dim\mathcal{K} \geq n_1 n_2$, where $n_1, n_2$ are the dimensions of the minimal Stinespring dilations of channels $\mathcal{E}_1, \mathcal{E}_2$, respectively.

Let us note that the (b) part of the proposition allows for a special solution for all pairs of channels. In fact, in order to satisfy the identity we can always set $\langle\,\Xi_1\,|\,\Xi_2\,\rangle = 0$ and reorder the Kraus operators so that $A_j \neq O, B_j = O$ for $j = 1, \dots, m$, $A_j = O, B_j \neq O$ for $m < j \leq n$, and $A_j = B_j = O$ for $j > n$. It means that a pair of channels can be implemented on a programmable processor using a pair of suitable orthogonal states to encode the channels. The dimension of the Hilbert space $\mathcal{K}$ is given by the total number of nonvanishing Kraus operators needed to describe both channels, i.e. $n = \dim\mathcal{K}$ in our construction.

Each programmable processor $\langle\mathcal{K}, G\rangle$ defines a linear mapping

$$\mathcal{G} : \mathcal{S}(\mathcal{K}) \to \mathcal{O}_{\mathrm{chan}}(\mathcal{H}), \quad \xi \mapsto \mathcal{E}_\xi : \mathcal{E}_\xi(\varrho) = \mathrm{tr}_{\mathcal{K}}[G(\varrho \otimes \xi)G^*]\,. \qquad (6.44)$$

Let us note that the number of parameters describing the channels on a system $\mathcal{H}$ is fixed, however, the number of parameters of program states can be made arbitrarily large. Therefore, it seems that we can design a universal quantum processor being able to implement all channels. Formally, for universal programmable processor the mapping $\mathcal{G}$ will be surjective. In order to give an answer let us stress that for the purposes of universality it is sufficient to consider only the realization of unitary channels. Unitary channels on $\mathcal{H} \otimes \mathcal{H}$ system are capable to perform any channel on subsystem $\mathcal{H}$. This is a consequence of the existence of the minimal Stinespring dilation. The next theorem is in a sense surprising result due to Nielsen & Chuang [65].

**Theorem 13.** *(Nonexistence of a universal programmable processor.)* There is no quantum processor realizing all unitary channels.

*Proof.* Applying Proposition 78 we see that in order to implement two unitaries, the program states must be orthogonal, because

$$U_1^* U_2 = cI \; \Leftrightarrow \; U_1 = U_2 e^{i\eta}, \text{ or } c = 0 \,. \tag{6.45}$$

The first option means that $U_1, U_2$ are the same up to a phase factor, which is physically irrelevant, so the orthogonality is the only relevant option. And since the set of all unitary channels is uncountable it follows that the Hilbert space $\mathcal{K}$ cannot be separable. Therefore, no universal quantum processor exists.                                                                                            $\square$

**Example 73.** *(Programming the phase damping channels.)* Consider a set of *phase damping* channels $\mathcal{E}_\eta$ acting on a two-dimensional system as

$$\mathcal{E}_\eta(\varrho) = \eta \varrho + (1-\eta) U_{\vec{n}} \varrho U_{\vec{n}}^* \,, \tag{6.46}$$

where $U_{\vec{n}} = \vec{n} \cdot \vec{\sigma}$. These channels are decoherence channels with the decoherence basis defined by eigenvectors of $U_{\vec{n}}$. Let us fix this unitary transformation and let $\varphi_+, \varphi_-$ be the eigenvectors corresponding to eigenvalues $\pm 1$ of $U_{\vec{n}}$. In turns out that $\langle \varphi_\pm | \varrho | \varphi_\pm \rangle = \langle \varphi_\pm | \mathcal{E}_\eta(\varrho) | \varphi_\pm \rangle$ and

$$\gamma = \frac{|\langle \varphi_\pm | \mathcal{E}_\eta(\varrho) | \varphi_\mp \rangle|}{|\langle \varphi_\pm | \varrho | \varphi_\mp \rangle|} = \sqrt{1 + 4\eta(1+\eta)} \,. \tag{6.47}$$

The parameter $\gamma$ is called a *decoherence rate*. The larger it is the faster is the decoherence process if the powers of the channel are applied.

Our goal is to investigate an implementation of channels $\mathcal{E}_\eta$ on the same programmable processor. In particular, what must be the size of the program space $\mathcal{K}$? A direct calculation shows that the Eq.(6.38) for this particular case is fulfilled

$$\sqrt{\eta_1 \eta_2} I + \sqrt{(1-\eta_1)(1-\eta_2)} U^* U = [\sqrt{\eta_1 \eta_2} + \sqrt{(1-\eta_1)(1-\eta_2)}] I \,. \tag{6.48}$$

It means that a pair of channels $\mathcal{E}_{\eta_1}, \mathcal{E}_{\eta_2}$ can be realized with the program states satisfying $\langle \Xi_1 | \Xi_2 \rangle = \sqrt{\eta_1 \eta_2} + \sqrt{(1-\eta_1)(1-\eta_2)}$. The question is, whether this relation can hold simultaneously for all pairs of channels. For $\eta = 0, 1$ the channels are unitary ($U, I$, respectively) and, of course, in this case the states $|\Xi_I\rangle, |\Xi_U\rangle$ are orthogonal. Moreover, it follows that $\langle \Xi_I | \Xi_\eta \rangle = \sqrt{\eta}$ and $\langle \Xi_U | \Xi_\eta \rangle = \sqrt{1-\eta}$, i.e. $\Xi_\eta$ is a superposition of $\Xi_I$ and $\Xi_U$ with amplitudes $\sqrt{\eta}, \sqrt{1-\eta}$, respectively. That is, the program states

$$\Xi_\eta = \sqrt{\eta} \Xi_I + \sqrt{1-\eta} \, \Xi_U \,. \tag{6.49}$$

encoding the channels $\mathcal{E}_\eta$ satisfy the Eq.(6.38). Indeed, a programmable processor

$$G = I \otimes |\Xi_I\rangle\langle\Xi_I| + U \otimes |\Xi_U\rangle\langle\Xi_U| \tag{6.50}$$

realizes the channels $\mathcal{E}_\eta$ with program states $|\Xi_\eta\rangle\langle\Xi_\eta|$ defined above. As a result we get that phase damping channels can be realized on a programmable processor using only a two-dimensional program space.

**Example 74.** *(Programming the decoherence bases.)* Consider the same set of channels as in the previous example, but now let us fix the damping parameter $\eta$. The problem is whether the set of channels

$$\mathcal{E}_{\vec{n}}(\varrho) = \eta I + (1-\eta) U_{\vec{n}} \varrho U_{\vec{n}}^* \tag{6.51}$$

can be realized on some programmable processor. Let us note that for $\eta = 0$ the problem is equivalent to implementation of unitary channels, which is not possible. On the other hand for $\eta = 1$ the problem is trivial, since the set consists of only single element - the identity channel.

The following calculation for a pair of channels $\mathcal{E}_{\vec{n}}, \mathcal{E}_{\vec{m}}$

$$\eta I + \sqrt{1-\eta} U_{\vec{n}}^* O + \sqrt{1-\eta} O U_{\vec{m}} = \eta I \tag{6.52}$$

shows that unlike for unitary channels in this case the orthogonality of program states is not necessary. In particular, $\langle \Xi_{\vec{n}} | \Xi_{\vec{m}} \rangle = \eta$. Hence, the question is how many vectors with pairwisely the same scalar product there exist in $k$-dimensional program space $\mathcal{K}$.

Let us consider a bit different problem: how many equally overlapping vectors is in $\mathcal{K}$? The overlap is the square of the absolute value of the scalar product. Thus, the solution to this problem gives at least as many vectors as the solution of the original problem. The point is that the number of such states is finite (for finite dimensional space), and moreover, the value of overlap $\eta$ is given by the dimension. The larger the dimension the larger is the value of $\eta$. In the limit of an infinite-dimensional space the vectors coincides, hence the program states implement the identity channel. It turns out that for this type of encoding only finite number of channels $\mathcal{E}_{\vec{n}}$ for a special value of $\eta$ can be realized on single programmable processor. Alternatively, we can use orthogonal states to encode the channels $\mathcal{E}_{\vec{n}}$ and in this case the only constraint is that the number of channels $\mathcal{E}_{\vec{n}}$ must be countable. This is the same situation as for the case of unitary channels.

Out of these two example we can make an interesting physical conclusion. Essentially, we get that the decoherence rate (for a fixed basis) can be adjusted by the state of the environment without affecting the strength of the interaction, i.e. with a fixed programmable processor. However, the decoherence basis cannot be completely controlled by the states of the environment and the interaction (defining the programmable processor) must be adjusted. In general, it is an interesting question which parameters of channels can be adjusted by changing the environment (program) and which require different interactions.

**Corollary 2.** Arbitrary countable set of channels can be always realized on the same programmable quantum processor by using orthogonal program states. Similarly, like in the previous corollary, for the dimension of the program space we have $\dim \mathcal{K} \geq n_1 n_2 \ldots$, where $n_j$ is the dimension of the minimal Stinespring dilation of the channel $\mathcal{E}_j$.

**Exercise 47.** A contraction is a quantum channel that maps the whole state space into a single quantum state, i.e. $\mathcal{A}_\xi : \mathcal{S}(\mathcal{H}) \mapsto \xi$. Find a quantum processor implementing all contractions on $d$-dimensional quantum system. (Hint: SWAP operator)

**Exercise 48.** Show that the maximally mixed state as the program state can encode only unital channels. (Hint: No hint.)

### 6.6.3   Programming channels imperfectly

No universal programmable quantum processor implementing all channels does exist. There are two basic approaches how to weaken the universality requirements.

- *approximate universality* - channels are implemented only approximately and the quantity

$$\epsilon_G(\mathcal{E}) = \max_{\xi \in \mathcal{S}(\mathcal{K})} f(\mathcal{G}(\xi), \mathcal{E}) \tag{6.53}$$

  is used to evaluate the quality of the realization of the channel $\mathcal{E}$ on the programmable processor $G$. Let us note that $\mathcal{G}$ is the mapping associated with $G$ (cf. Eq.(6.44)) and $f(\cdot, \cdot) \in [0, 1]$ measures the fidelity of channels, i.e. $f(\mathcal{E}_1, \mathcal{E}_2) = 1$ if and only if $\mathcal{E}_1 = \mathcal{E}_2$. That is, if $\epsilon(\mathcal{E}) = 1$ we say the channel is implemented perfectly. Let us note that each programmable processor is approximately universal, but the differences are in the quantity $\epsilon(\mathcal{E})$. The overall goal is to find a channel optimizing (in some sense) this quality measure with respect to given resources that are usually quantified by the dimension of the program space $\mathcal{K}$. That is, the particular goal is to find an optimal approximate programmable processor for a given size of the program register $\mathcal{K}$.

- *probabilistic (unambiguous) universality* - channels are successfully implemented with some *success probability* $p_{\text{success}}$ associated with particular *success outcome* of the pointer observable F. Moreover, the realization is unambiguous in a sense, that we know whether the channel is implemented, or not. In this case the programmable processors are exploited to realize selective measurements resulting in transformation

$$\varrho \mapsto \varrho' = \mathcal{T}_\xi(\varrho) = \frac{1}{\text{tr}\left[\mathcal{I}_{\text{success}}^\xi(\varrho)\right]} \mathcal{I}_{\text{success}}^\xi(\varrho) \tag{6.54}$$

  where $\mathcal{I}_{\text{success}}^\xi$ is the completely positive map associated with the success outcome and depending on the program state $\xi$. The probability to find the success outcome is

$$p_{\text{success}} = \text{tr}\left[\mathcal{I}_{\text{success}}^\xi(\varrho)\right] . \tag{6.55}$$

  Our goal is to implement channels, hence we are interested only in those cases when $\mathcal{T}_\xi$ is a valid quantum channel. In particular, it is linear if and only if the probability $p_{\text{success}}$ is independent of $\varrho$, i.e. the completely positive map $\mathcal{I}_{\text{success}}^\xi$ is proportional to a trace-preserving map. Only in such case we say that a channel has been implemented and $p_{\text{success}}$ defines the *success probability* of the implementation of some channel $\mathcal{T}_\xi$.

For a given programmable processor we can define a set of *probabilistic programs* $S_{prob} = \{\xi \in \mathcal{S}(\mathcal{K}) : \mathcal{T}_\xi$ is a channel $\}$ as those states that really implements some channel. For a given programmable processor the success probability can be defined for all channels, i.e. $p_{\text{success}}(\mathcal{E}) = 0$ if there is no $\xi \in S_{prob}$ such that $\mathcal{T}_\xi = \mathcal{E}$. Having this in mind we can say that each programmable processor is universal. Similarly like for the approximate scenario also in this case the goal is to find an optimal probabilistic programmable processor given the size of the program space $\mathcal{K}$. The optimality is defined in terms of the success probability $p_{\text{success}}(\mathcal{E})$.

If for a given processor $\epsilon(\mathcal{E}) = 0$, or $p_{\text{success}}(\mathcal{E}) = 0$ we usually say that the processor is not universal. It is not surprising that in both these regimes (approximate and probabilistic) the universal processors (after this redefinition of universality) do exist, i.e. there do exist programmable processors such that $\min_\mathcal{E} \epsilon(\mathcal{E}) > 0$ and $\min_\mathcal{E} p_{\text{success}}(\mathcal{E}) > 0$. Further, we shall give examples of universal processors implementing unitary channels.

**Example 75.** (*Universal approximate programming of unitary channels.*) As in the case of (deterministic) implementation of channels also for approximate programming the choice of the pointer observable is irrelevant. Let us consider a programmable processor being the controlled-U operation

$$G = \sum_j U_j \otimes |\varphi_j\rangle\langle\varphi_j|, \qquad (6.56)$$

where $\{\varphi_1, \dots, \varphi_k\}$ is an orthonormal basis of the program space $\mathcal{K}$. Our task is to approximately implement all unitary channels. Consider a general program state $\xi$. Then the realized channel is a random unitary channel

$$\mathcal{E}_\xi : \quad \mathcal{E}_\xi(\varrho) = \sum_j \langle\varphi_j|\xi|\varphi_j\rangle U_j \varrho U_j^* = \sum_j q_j U_j \varrho U_j^*. \qquad (6.57)$$

Let us note that $\{q_j\}$ is an arbitrary probability distribution on $k$ elements. The fidelity (one minus distance) between a unitary transformation $U$ and the set of channels $\mathcal{E}_\xi$ vanishes (distance is maximal) if $f(U, U_j) = 0$ for all $j$. Strictly speaking, this holds of $f(\cdot, \cdot)$ is convex. From convexity it also follows that

$$\epsilon(U) = \max_\xi f(U, \mathcal{E}_\xi) = \max_j f(U, U_j). \qquad (6.58)$$

A usual choice for quantifying the fidelity of channels is to use the so-called *process fidelity* being defined as fidelity between the corresponding states under Choi-Jamiolkowski isomorphism, i.e.

$$f(\mathcal{E}_1, \mathcal{E}_2) = \text{tr}\left[\sqrt{\sqrt{\omega_1}\omega_2\sqrt{\omega_1}}\right], \qquad (6.59)$$

where $\omega_j = (\mathcal{E}_j \otimes \mathcal{I})[\psi_+]$. For unitaries we have $f(U_1, U_2) = |\text{tr}[U_1^* U_2]|$. Therefore, the processor $G$ will be universal if and only if the operators $\{U_j\}$ span the whole space of operators, i.e. $\text{span}\{U_j\} = \mathcal{L}(\mathcal{H})$. This gives a lower bound on $\mathcal{K}$ since $k = \dim\mathcal{K} \geq \dim\mathcal{L}(\mathcal{H}) = d^2$. For $k < d^2$ no programmable quantum processor with $G$ being a controlled-U transformation exists.

Let us fix the dimension $k = d^2$. In this case we can find the optimal programmable processor within the family of controlled-U processors [89]. In particular, the choice of $U_1, \dots, U_{d^2}$ that are mutually orthogonal in the Hilbert-Schmidt sense will do the job in the sense of minimum error. In this case

$$\min_U \epsilon(U) = \min_U \max_j |\mathrm{tr}\,[U^* U_j]| = \frac{1}{d}\,, \tag{6.60}$$

where the minimum is achieved by the unitary channel $U = \frac{1}{d}\sum_j U_j$ being the equal linear combination of all the channels $U_j$ defining the programmable processor.

This result is surprising, because every unitary channel $U$ is approximated by one of the unitary channels $U_1, \ldots, U_{d^2}$. It means that effectively only $d^2$ program states $|\varphi_j\rangle\langle\varphi_j|$ are used to approximate all unitary channels. Let us note that since the set of deterministically implemented channels $\{\mathcal{E}_\xi\}$ contains the contraction into the total mixture $\mathcal{E}_0 : \varrho \mapsto \frac{1}{d}I$ encoded in a program state $\xi = \frac{1}{d^2}I \in \mathcal{S}(\mathcal{K})$, it follows that none of the channels is realized with $\epsilon(\mathcal{E}) < 1/d$. In fact,

$$f(\mathcal{E}, \mathcal{E}_0) = \mathrm{tr}\left[\sqrt{\sqrt{\omega_\mathcal{E}}\frac{1}{d^2}I\sqrt{\omega_\mathcal{E}}}\right] = \frac{1}{d}\mathrm{tr}\,[\omega_\mathcal{E}] = \frac{1}{d} \tag{6.61}$$

for arbitrary channel $\mathcal{E}$. That is, the total mixture as the program state approximates all channels with a fixed value of $\epsilon(\mathcal{E}) = 1/d$.

**Example 76.** *(Universal probabilistic programming of unitary channels.)* As in the previous example consider a controlled-U programmable processor

$$G = \sum_j U_j \otimes |\varphi_j\rangle\langle\varphi_j|\,, \tag{6.62}$$

and let us perform a projective pointer observable such that $\mathsf{F}_{\text{success}}$ is a one-dimensional projector, i.e. $\mathsf{F}_{\text{success}} = |\phi\rangle\langle\phi| = F_\phi$ for some $\phi \in \mathcal{K}$. For a pure program state $|\Xi\rangle\langle\Xi|$ the total action of the selective measurement implemented by the programmable quantum processor results in the unnormalized state associated with unnormalized vector

$$
\begin{aligned}
\psi \otimes \Xi \mapsto \Omega_{\text{success}} &= \sum_j [(U_j\psi) \otimes (F_\phi \varphi_j \langle\,\varphi_j\,|\,\Xi\rangle)] \\
&= \sum_j \langle\,\phi\,|\,\varphi_j\,\rangle\,\langle\,\varphi_j\,|\,\Xi\rangle\,U_j\psi \otimes \phi
\end{aligned}
$$

giving the probability

$$p_{\text{success}} = \sum_j \langle\varphi_{j'}|P_\phi|\varphi_j\rangle\langle\varphi_j|P_\Xi|\varphi_{j'}\rangle\langle\psi|U_{j'}^* U_j|\psi\rangle\,. \tag{6.63}$$

If $\phi \in \mathcal{K}$ is chosen such that $\langle\varphi_{j'}|P_\phi|\varphi_j\rangle = c$ for all $j, j'$, then the probability can be written in the form

$$p_{\text{success}} = c\langle\psi|A_\Xi^* A_\Xi|\psi\rangle\,, \tag{6.64}$$

where $A_\Xi = \sum_j \langle\,\varphi_j\,|\,\Xi\,\rangle\,U_j$ and the transformation

$$|\psi\rangle\langle\psi| \mapsto \frac{1}{p_{\text{success}}}c\,A_\Xi|\psi\rangle\langle\psi|A_\Xi^* \tag{6.65}$$

is accomplished. Channels with only single Kraus operator are necessarily unitary. Indeed, if $A_\Xi$ is a unitary operator $U_\Xi$, then $p_{\text{success}} = c$ is independent of the input state $\varrho$. Otherwise the implemented transformation is not a channel.

We left open two questions. What is the value of $c$? For which pure program states $|\Xi\rangle\langle\Xi|$ the operators $A_\Xi$ are unitary? It is straightforward to verify that the vector $\phi = \frac{1}{\sqrt{\dim\mathcal{K}}}\sum_j \varphi_j$ satisfies the equalities $\langle\varphi_{j'}|P_\phi|\varphi_j\rangle = c = 1/\dim\mathcal{K}$ for all values of $j, j'$. If $A_\Xi = U_\Xi$ is unitary the probability of success equals $p_{\text{success}}(U_\Xi) = 1/\dim\mathcal{K}$. That is, for each implemented unitary transformation the success probability is the same.

Can we implement all unitary transformations with nonzero success probability? The answer is yes, we can. The operator $A_\Xi = \sum_j \langle\varphi_j\,|\,\Xi\rangle\,U_j = \sum_j a_j U_j$ is a complex linear combination of unitary operators $\{U_j\}$. The coefficients $a_j$ are normalized in the following sense $\sum_j |a_j|^2 = \langle\Xi\,|\,\Xi\rangle = 1$. We know that each operator can be expressed as a complex linear combination of operator basis elements. For a $d$-dimensional system its operator basis consists of $d^2$ operators. It is possible to choose an operator basis consisting only of unitary operators, i.e. linear span of unitary operators equals to $\mathcal{L}(\mathcal{H})$. However, a general unitary operator basis is not necessarily capable to express all unitary operators as normalized linear combinations. Therefore, suppose the unitary operators forming the operator basis are mutually orthogonal, i.e. $\text{tr}\left[U_j^* U_{j'}\right] = d\delta_{jj'}$ for all $j, j' = 1, \dots, d^2$. In such case we can write $U = \sum_j a_j U_j$ with $a_j = \frac{1}{d}\text{tr}\left[U_j^* U\right]$ and the unitarity implies that

$$1 = \frac{1}{d}\text{tr}\left[U^* U\right] = \frac{1}{d}\sum_{j,j'} a_{j'}^* a_j \text{tr}\left[U_{j'}^* U_j\right] = \frac{1}{d}d\sum_j |a_j|^2 = \sum_j |a_j|^2,$$

i.e. the linear combinations are normalized as it is required.

For example, the unitary operators $U_{rs}$ $(r, s = 0, \dots d-1)$

$$U_{rs} = \sum_{l=0}^{d-1} e^{-i2\pi sl/d}|l \ominus_d r\rangle\langle l| \tag{6.66}$$

form an orthogonal basis, i.e. $\text{tr}\left[U_{rs}^* U_{r's'}\right] = d\delta_{rr'}\delta_{ss'}$. In conclusion, using a programmable processor $\langle\mathcal{K} = \mathcal{H} \otimes \mathcal{H}, G, \mathsf{F}\rangle$ with unitary transformation

$$G = \sum_{r,s=0}^{d-1} U_{rs} \otimes |\varphi_{rs}\rangle\langle\varphi_{rs}| \tag{6.67}$$

and two-valued pointer observable

$$\mathsf{F}_{\text{success}} = |\phi\rangle\langle\phi|, \mathsf{F}_0 = I - \mathsf{F}_{\text{success}}, \tag{6.68}$$

where $\phi = \frac{1}{d}\sum_{r,s} \varphi_{rs} \in \mathcal{H} \otimes \mathcal{H}$, each $d$-dimensional unitary channel $U = \sum_{r,s} a_{rs} U_{rs}$ is implemented with the success probability

$$p_{\text{success}}(U) = 1/d^2 \tag{6.69}$$

by using the program state corresponding to vector $\Xi = \sum_{r,s} a_{rs}\varphi_{rs} \in \mathcal{H} \otimes \mathcal{H}$. That is, a universal probabilistic programmable processors do exist.

### 6.6.4   Programming observables

Each program state $\xi \in \mathcal{S}(\mathcal{K})$ encodes a channel, an observable and an instrument, thus each programmable quantum processor induces mappings from the program state space into a set of channels, observables and instruments, respectively, of the considered data system. In the previous sections we have studied the properties of a processor-induced mapping for channels. We have shown that this mapping cannot be onto, i.e., there is no (deterministic) universal programmable processor concerning the implementation of channels. Is there a universal programmable processor for observables?

Let us start with an assumption that a program state $\xi$ implementing an effect $E$ is pure, i.e., $\xi = |\Xi\rangle\langle\Xi|$ for some unit vector $\Xi \in \mathcal{K}$, and that the pointer observable $\mathsf{F}$ is discrete and sharp, i.e. $\mathsf{F}_j = \sum_{\alpha \in l_j} |\phi_\alpha\rangle\langle\phi_\alpha|$ are mutually orthogonal projectors. The vectors $\{\phi_\alpha\}$ form an orthonormal basis of $\mathcal{K}$ and $l_j$ denotes a subset of indexes defining. The action of a general unitary operator $G$ reads

$$G(\psi \otimes \Xi) = \sum_\alpha B_\alpha(\Xi)\psi \otimes \phi_\alpha \,, \qquad (6.70)$$

where and, formally, $B_\alpha(\Xi) = B_\alpha = \langle\phi_\alpha|G|\Xi\rangle$. A probability to find the outcome $\omega_j$ equals

$$p_j = \langle\psi| \sum_{\alpha \in l_j} B_\alpha^* B_\alpha |\psi\rangle = \langle\psi|E_j(\Xi)|\psi\rangle \,, \qquad (6.71)$$

where the effects $E_j(\Xi) = E_j = \sum_{\alpha \in l_j} B_\alpha^* B_\alpha$ define the implemented observable A. Using a mixed program state $\xi = \sum_k \pi_k |\Xi_k\rangle\langle\Xi_k|$ we perform an observable

$$\mathsf{A}_\xi : \omega_j \mapsto E_j = \sum_k \sum_{\alpha \in l_j} \pi_k B_\alpha^*(\Xi_k) B_\alpha(\Xi_k) \,, \qquad (6.72)$$

i.e. a convex combination of observables encoded by pure program states.

**Proposition 79.** If a pair of sharp observables $\mathsf{A}, \mathsf{B}$ with effects being one-dimensional projectors $P_j, Q_j$, respectively, can be implemented on the same programmable processor $\langle\mathcal{K}, G, \mathsf{F}\rangle$ with sharp pointer observable $\mathsf{F}$, then $\dim\mathcal{K} \geq \dim\mathsf{H} = d$ and the program states $\Xi_\mathsf{A}, \Xi_\mathsf{B}$ are orthogonal. It is assumed that at least one of the projectors $P_1, \ldots, P_d$ is not included among the projectors $Q_1, \ldots, Q_d$, i.e. the sets of projectors are different.

*Proof.* Implementation of a sharp observable $\mathsf{A} : \omega_j \mapsto P_j = |\varphi_j\rangle\langle\varphi_j|$ on a programmable quantum processor requires that the pointer observable have at least $d = \dim\mathcal{H}$ outcomes, i.e. it consists of $d$ orthogonal projectors $\mathsf{F}_j = \sum_{\alpha \in l_j} |\phi_\alpha\rangle\langle\phi_\alpha|$, where $l_j$ denotes a subset of indexes labeling the orthonormal basis of the program system $\mathcal{K}$ such that $|l_1| + \cdots + |l_d| = \dim\mathcal{K}$. According to previous paragraph the unitary operator $G$ acts as

$$G(\psi \otimes \Xi_\mathsf{A}) = \sum_\alpha A_\alpha \psi \otimes \phi_\alpha \,. \qquad (6.73)$$

and the identity $\sum_{\alpha \in l_j} A_\alpha^* A_\alpha = P_j$ holds. Moreover, since $P_j$ are one-dimensional projectors it follows that for each $\alpha \in l_j$ the operator $A_\alpha^* A_\alpha$ is proportional to $P_j$. In particular, $A_\alpha^* A_\alpha =$

$U_\alpha^{(j)} P_j$, where $U_\alpha^{(j)}$ are arbitrary unitary operators. Similarly for the observable B

$$G(\psi \otimes \Xi_\mathsf{B}) = \sum_\alpha B_\alpha \psi \otimes \phi_\alpha \,, \qquad (6.74)$$

and $B_\alpha^* B_\alpha = V_\alpha^{(j)} Q_j$, where $V_\alpha^{(j)}$ are unitary and $Q_j = |\psi_j\rangle\langle\psi_j|$. Calculating the scalar product $\langle\, G(\psi \otimes \Xi_\mathsf{A}) \,|\, G(\psi \otimes \Xi_\mathsf{B}) \,\rangle$ we obtain the identity

$$\langle \Xi_\mathsf{A} \,|\, \Xi_\mathsf{B} \rangle = \langle\psi| \sum_\alpha A_\alpha^* B_\alpha |\psi\rangle \qquad (6.75)$$

holding for all $\psi \in \mathcal{H}$ only if

$$\sum_\alpha A_\alpha^* B_\alpha = \langle \Xi_\mathsf{A} \,|\, \Xi_\mathsf{B} \rangle \, I \,. \qquad (6.76)$$

Defining $W_\alpha^{(j)} = U_\alpha^{(j)*} V_\alpha^{(j)}$ we get

$$
\begin{aligned}
\langle \Xi_\mathsf{A} \,|\, \Xi_\mathsf{B} \rangle \, I &= \sum_j P_j \left( \sum_{\alpha \in l_j} W_\alpha^{(j)} \right) Q_j = \sum_j \left( \sum_{\alpha \in l_j} \langle\varphi_j| W_j^{(\alpha)} |\psi_j\rangle \right) |\varphi_j\rangle\langle\psi_j| \\
&= \sum_{jk} w_j a_{jk} |\varphi_j\rangle\langle\varphi_k| \,,
\end{aligned}
$$

where we used the notation $w_j = \sum_{\alpha \in l_j} \langle\varphi_j| W_j^{(\alpha)} |\psi_j\rangle$ and $\psi_j = \sum_k a_{jk} \varphi_k$. This equality holds only if $w_j a_{jk} = c\delta_{jk}$, i.e. either $w_j = 0$, or $a_{jk} = 0$ for $j \neq k$. The first option implies $\langle \Xi_\mathsf{A} \,|\, \Xi_\mathsf{B} \rangle = 0$. The last option means that bases defining the observables A, B coincide, however we assume that the bases are different. In conclusion, if two sharp observables are realizable on the same programmable processor the program states $\Xi_\mathsf{A}, \Xi_\mathsf{B}$ must be orthogonal. $\qquad \square$

According to Ozawa's representation theorem (Theorem 11 in Section 6.2.1) each instrument can be realized by a normal measurement model. Therefore, if a universal programmable processor implementing all observables would exists, it can be chosen so that the pointer observable F is sharp. As a consequence of the above proposition it follows that a universal programmable processor implementing all sharp observables does not exist.

**Example 77.** Consider a controlled-U programmable quantum processor $\langle \mathcal{K}, G, \mathsf{F} \rangle$ as in Example 76, i.e., $\mathcal{K} = \mathcal{H} \otimes \mathcal{H}$, $G = \sum_j U_j \otimes |\varphi_j\rangle\langle\varphi_j|$ ($U_j$ are mutually orthogonal) and $\mathsf{F}_{\text{success}} = |\phi\rangle\langle\phi|$ with $\phi = \frac{1}{\sqrt{d}} \sum_j \varphi_j$. A pure program state $\Xi = \sum_j a_j \varphi_j$ encodes effects

$$E = \frac{1}{d^2} \sum_{jk} a_j^* a_k U_j^* U_k \qquad (6.77)$$

with $\mathrm{tr}\,[E] = \frac{1}{d^2} d \sum_j |a_j|^2 = 1/d$. It follows that only effects with a given trace can be implemented.

## 7    Entanglement

### 7.1    Composite bipartite systems

In quantum theory the paradigm of a composite quantum system is intimately related with the mathematical concept of tensor product of Hilbert spaces. As we have seen in Section 3.4 the composite quantum system composed of subsystems $A$ and $B$ is associated with a tensor product $\mathcal{H}_{AB} = \mathcal{H}_A \otimes \mathcal{H}_B$. As a result of the combination of the tensor product structure and the principle of superposition the quantum theory is embedded with a phenomenon of *entanglement*.

Our goal is not to give an extensive overview of entanglement theory. For an interested reader we recommend the recent reviews [50] and [71] that covers also the quantum information aspects of entanglement. In this chapter we go through the basic mathematical properties of quantum entanglement. To simplify the discussion, we shall assume that Hilbert spaces are finite dimensional.

#### 7.1.1    Vectors

**Definition 55.** Let $\eta \in \mathcal{H}_A \otimes \mathcal{H}_B$ be a unit vector. It is called

1.  *factorized* if $\eta = \phi \otimes \psi$ for some vectors $\phi \in \mathcal{H}_A$, $\psi \in \mathcal{H}_B$.

2.  *entangled* otherwise.

The following example gives a concrete instance of entangled vectors. In particular, it demonstrates that entangled vectors exists.

**Example 78.** Let $\varphi_1, \varphi_2 \in \mathcal{H}$ be two orthogonal unit vectors and let $\alpha, \beta$ be two non-zero complex numbers such that $|\alpha|^2 + |\beta|^2 = 1$. The vector $\eta = \alpha\varphi_1 \otimes \varphi_1 + \beta\varphi_2 \otimes \varphi_2 \in \mathcal{H} \otimes \mathcal{H}$ is entangled. To see this, choose unit vectors $\varphi_3, \varphi_4, \ldots \in \mathcal{H}$ such that the set $\{\varphi_1, \varphi_2, \varphi_3, \ldots\}$ is an orthonormal basis for $\mathcal{H}$. A general factorized vector from $\mathcal{H} \otimes \mathcal{H}$ takes the form $\phi \otimes \psi = \left(\sum_i c_i \varphi_i\right) \otimes \left(\sum_j d_j \varphi_j\right)$. Comparing this expression with $\eta$ we get conditions

$$c_1 d_1 = \alpha, \quad c_2 d_2 = \beta, \quad c_1 d_2 = c_2 d_1 = 0\,.$$

Multiplying the first two equations and the last two, it follows that $\alpha\beta = 0$. But this is not possible, and therefore $\eta$ is entangled.

Consider orthonormal bases $\{\varphi_i\}_{i=1}^{d_A}$ for $\mathcal{H}_A$ and $\{\psi_j\}_{j=1}^{d_B}$ for $\mathcal{H}_B$. A general factorized vector takes the form

$$\eta = \varphi \otimes \psi = \sum_{i,j} c_i d_j \varphi_i \otimes \psi_j$$

Let us notice that if the orthonormal bases do not contain $\varphi$ and $\psi$, the coefficient $c_i d_j$ related to $\varphi_i \otimes \psi_j$ is nonvanishing at least for some values of $i, j$ satisfying $i \neq j$. On the other hand, if $\varphi \in \{\varphi_i\}$ for some $i = i_0$ or $\psi \in \{\psi_j\}$ for some $j = j_0$, then the superposition is trivial, i.e., $\eta = \varphi \otimes \psi$.

Based on this observation we conclude that a vectors of the form $\sum_{j=1}^{d} a_j \varphi_j \otimes \psi_j$, where $d = \min\{d_A, d_B\}$ are necessarily entangled providing that at least two coefficients $a_j$ are nonzero. Notice that the entangled vector in Example 78 is just a special case of this kind. The following theorem is saying that arbitrary entangled vector can be written in this form if suitable orthonormal bases of $\mathcal{H}_A$ and $\mathcal{H}_B$ are chosen. This result, known as Schmidt decomposition, is very useful in many applications.

**Theorem 14.** (*Schmidt decomposition*) For each vector $\psi \in \mathcal{H}_A \otimes \mathcal{H}_B$ there exists an orthogonal basis $\{e_1, \ldots, e_{d_A}\}$ of $\mathcal{H}_A$ and $\{f_1, \ldots, f_{d_B}\}$ of $\mathcal{H}_B$ such that

$$\psi = \sum_{j=1}^{r_s} \sqrt{\lambda_j}\, e_j \otimes f_j \,, \tag{7.1}$$

where $\lambda_1, \ldots, \lambda_{r_s}$ are decreasingly ordered nonzero positive numbers forming the so-called *Schmidt vector* $\vec{\lambda}_\psi$. The *Schmidt rank* $r_s(\psi) \leq \min\{d_A, d_B\}$ is the total number of nonvanishing elements in $\vec{\lambda}_\psi$.

*Proof.* Consider orthonormal bases $\{\varphi_1, \ldots, \varphi_{d_A}\}$ and $\{\phi_1, \ldots, \phi_{d_B}\}$ of Hilbert spaces $\mathcal{H}_A$ and $\mathcal{H}_B$, respectively. A general vector $\psi \in \mathcal{H}_A \otimes \mathcal{H}_B$ takes the form

$$\psi = \sum_{j=1}^{d_A} \sum_{k=1}^{d_B} x_{jk}\, \varphi_j \otimes \phi_k \,, \tag{7.2}$$

where the complex numbers $x_{jk}$ form a $d_A \times d_B$ matrix $X$. In the polar decomposition $X = UF$, where $U$ is a $d_A \times d_B$ isometry matrix and $F = \sqrt{X^*X}$ is $d_B \times d_B$ positive matrix. Due to positivity of $F$ there exists a unitary matrix $V$ such that $V^*FV = D$ and $D$ is a diagonal matrix of square roots of eigenvalues of the matrix $X^*X$, i.e. $D = \mathrm{diag}\{\sqrt{\lambda_1}, \ldots, \sqrt{\lambda_{d_B}}\}$. Without loss of generality we assume that eigenvalues $\lambda_j$ are decreasingly ordered. Since the nonzero eigenvalues of $d_B \times d_B$ matrix $X^*X$ and $d_A \times d_A$ matrix $XX^*$ coincide it follows that the number of nonvanishing eigenvalues $\lambda_j$ is smaller or equal to the smallest of the Hilbert space dimensions. That is, if $d_A < d_B$ then $\lambda_j = 0$ for all $j > \lambda_A$. Without loss of generality let us assume that $d_B \leq d_A$. Putting altogether we get the so-called *singular value decomposition* of the operator $X$

$$X = UV^*DV = W^*DV \,. \tag{7.3}$$

Writing $x_{jk} = \sum_{l,m=1}^{d_B} w_{lj}^* \sqrt{\lambda_l}\, \delta_{lm} v_{mk}$ the vector $\psi$ takes the form

$$\begin{aligned}
\psi &= \sum_{l=1}^{d_B} \sqrt{\lambda_l} \left( \sum_{j=1}^{d_A} w_{lj}^* \varphi_j \right) \otimes \left( \sum_{k=1}^{d_B} v_{lk} \psi_k \right) \\
&= \sum_{l=1}^{d_B} \sqrt{\lambda_l}\, e_l \otimes f_l \,.
\end{aligned}$$

Since $W$ and $V$ are unitary the vectors $\{e_l\}$ and $\{f_l\}$ form orthonormal sets in $\mathcal{H}_A$ and $\mathcal{H}_B$, respectively. $\qquad\square$

Using Schmidt decomposition we can immediately see some interesting consequences related to our earlier discussions. Suppose that the compound system is in a pure state $|\psi\rangle\langle\psi|$. Then the individual subsystems are described by states

$$\varrho_A = \sum_j \lambda_j |e_j\rangle\langle e_j|, \qquad \varrho_B = \sum_j \lambda_j |f_j\rangle\langle f_j|. \tag{7.4}$$

Especially, the spectra of $\varrho_A$ and $\varrho_B$ differ only in the degree of the degeneracy of the eigenvalue 0 (see Proposition 38 in Section 3.4). Therefore, if the composite system is in a pure state, the states of the subsystems have the same purity and entropy. The Schmidt rank $r_s$ equals to ranks of density operators $\varrho_A, \varrho_B$.

### 7.1.2   Positive operators

**Definition 56.** Let $F \in \mathcal{L}_+(\mathcal{H}_A \otimes \mathcal{H}_B)$ be a positive operator. It is called

1. *factorized* if $F = F_A \otimes F_B$. We denote by $\mathcal{L}_+^{\text{fac}} \subset \mathcal{L}_+(\mathcal{H}_A \otimes \mathcal{H}_B)$ the set of all factorized positive operators.

2. *separable* if $F$ belongs to the convex hull of $\mathcal{L}_+^{\text{fac}}$, i.e. $F = \sum_j p_j F_j^A \otimes F_j^B$. We denote by $\mathcal{L}_+^{\text{sep}}$ the set of all separable operators.

3. *entangled* if $F$ is not separable.

If replacing $\mathcal{L}_+$ by $\mathcal{S}(\mathcal{H})$ (states) or by $\mathcal{E}(\mathcal{H})$ (effects) we get a similar classification of quantum states, or effects, respectively. For instance, the states can be classified in the following way.

**Definition 57.** A state $\varrho \in \mathcal{S}(\mathcal{H}_A \otimes \mathcal{H}_B)$ is called

1. *factorized* if $\varrho = \varrho_A \otimes \varrho_B$. Let us denote by $\mathcal{S}^{\text{fac}}$ the set of factorized states.

2. *separable* if $\varrho \in \overline{\text{co}(\mathcal{S}^{\text{fac}})} \equiv \mathcal{S}^{\text{sep}}$.

3. *entangled* if $\varrho \in \mathcal{S} \setminus \mathcal{S}^{\text{sep}}$.

The following lemma follows directly from the definition of separable states as the convex closure of factorized states.

**Proposition 80.** The set of separable states $\mathcal{S}^{\text{sep}}$ is convex and closed in trace norm topology.

The introduced distinction between separable and entangled states is due to R.Werner [83].

### 7.1.3   Operations and channels

Let us recall some notation from Chapter 5. By $\mathcal{O}$ we denote the set of all quantum operations, i.e., trace-decreasing completely positive linear maps. A set of channels is a subset $\mathcal{O}_c \subset \mathcal{O}$ containing all trace-preserving completely positive linear maps.

**Definition 58.** A quantum channel $\mathcal{E} : \mathcal{L}(\mathcal{H}_A \otimes \mathcal{H}_B) \to \mathcal{L}(\mathcal{H}_A \otimes \mathcal{H}_B)$ is called

1. *local/factorized* if $\mathcal{E} = \mathcal{E}_A \otimes \mathcal{E}_B$ and $\mathcal{E}_A, \mathcal{E}_B$ are channels defined on subsystems $A$ and $B$, respectively. We will denote as $\mathcal{O}_c^{\mathrm{fac}} \subset \mathcal{O}_c$ the set of all local channels.

2. *nonlocal* if $\mathcal{E} \in \mathcal{O}_c \setminus \mathcal{O}_c^{\mathrm{fac}}$.

3. *separable* if $\mathcal{E} = \sum_j \mathcal{F}_j^A \otimes \mathcal{F}_j^B$, where $\mathcal{F}_j^A, \mathcal{F}_j^B$ are local operations. $\mathcal{O}_c^{\mathrm{sep}}$ stands for the set of separable channels. Let us note that setting $\mathcal{F}_j^A, \mathcal{F}_j^B$ to be the local operations is not a mistake. Therefore, the set of separable channels $\mathcal{O}_c^{\mathrm{sep}}$ is not a convex hull of factorized channels.

4. *LOCC* (stands for *local operations and classical communication*) if it can be written as a sequence of local quantum operations assisted by the exchange of classical communication. Let us assume that in the first round Bob performs a measurement described by an instrument $j_1 \mapsto \mathcal{F}_{j_1}^B$ defined on a discrete outcome space $\Omega_1$, i.e. $\sum_{j_1 \in \Omega} \mathcal{F}_{j_1}^B = \mathcal{F}_1^B$ is a quantum channel. The observed outcome $j_1$ is lossly communicated to Alice. In the second round Alice chooses a measurement described by an instrument $j_2 \mapsto \mathcal{F}_{j_2|j_1}^A$ ($j_2 \in \Omega_2$) conditioned on the received information $j_1$ from Bob and communicates the outcome $j_2$ back to Bob. Repeating $n$ rounds of communication Alice and Bob jointly apply an LOCC channel

$$\mathcal{E}_{\mathrm{LOCC}} = \sum \cdots (\mathcal{F}_{j_4|j_3,j_2,j_1}^A \otimes \mathcal{F}_{j_3|j_2,j_1}^B)(\mathcal{F}_{j_2|j_1}^A \otimes \mathcal{F}_{j_1}^B) \,,$$

where $\sum_{j_1} \mathcal{F}_{j_1}^B = \mathcal{F}_1^B$, $\sum_{j_2} \mathcal{F}_{j_2|j_1}^A = \mathcal{F}_{2|j_1}^A$, etc., are channels, i.e. tracepreserving completely positive maps. After the last information exchange either Alice, or Bob performs quantum channels $\mathcal{F}_{j_n,\ldots,j_1}$ that finish the implementation of the LOCC channel $\mathcal{E}_{\mathrm{LOCC}}$. The set of all LOCC operations we shall denote by $\mathcal{O}_c^{\mathrm{LOCC}} \subset \mathcal{O}_c$ and clearly $\mathcal{O}_c^{\mathrm{LOCC}} \subset \mathcal{O}_c^{\mathrm{sep}}$.

5. *entangled* if $\mathcal{E} \in \mathcal{O}_c \setminus \mathcal{O}_c^{\mathrm{LOCC}}$.

**Example 79.** (*One-way LOCC channel.*) Suppose that the communication of Alice and Bob is restricted and only Alice can communicate information to Bob, but not vice versa. Since actions of Alice and Bob commute it is sufficient to consider only a single communication from Alice to Bob. Alice makes a measurement described by an instrument $j \mapsto \mathcal{F}_j^A$ ($\sum_j \mathcal{F}_j^A = \mathcal{E}^A$ is a channel). After Bob receives the value of $j$ he applies a channel $\mathcal{E}_j^B$. In this way we get the so-called *one-way LOCC channel* taking the form

$$\mathcal{E}_{1-LOCC}[\varrho] = \sum_j (\mathcal{F}_j^A \otimes \mathcal{E}_j^B)[\varrho] \,, \tag{7.6}$$

where $\mathcal{E}_j^B$ are channels and $\mathcal{F}_j^A$ are operations forming the instrument.

**Example 80.** (*LOCC preparation of factorized bases.*) Consider a channel $\mathcal{E}$ defined on a composite system of two $d$-dimensional systems associated with Hilbert space $\mathcal{H}_{AB} = \mathcal{H}_d \otimes \mathcal{H}_d$

$$\mathcal{E}[\varrho] = \sum_{j,k} A_{jk} \varrho A_{jk}^* \tag{7.7}$$

where $A_{jk} = |\psi_{jk} \otimes \phi_{jk}\rangle\langle\varphi_j \otimes \varphi_k|$ and vectors $\{\varphi_1, \ldots, \varphi_d\}$ form an orthonormal basis of $\mathcal{H}_d$. The complete positivity constraint $\sum_{jk} A_{jk}^* A_{jk} = I$ requires that vectors $\omega_{jk} = \psi_{jk} \otimes \phi_{jk}$ form an orthonormal basis of $\mathcal{H}_{AB}$. The question is whether such channel is LOCC, or not.

Consider the following LOCC procedure. Let Alice and Bob perform local measurements described by Lüders instruments $j \mapsto \mathcal{F}_j$, where $\mathcal{F}_j(\xi) = |\varphi_j\rangle\langle\varphi_j|\xi|\varphi_j\rangle\langle\varphi_j|$. At this step their actions result in transformation

$$\varrho \mapsto (\mathcal{F}_j^A \otimes \mathcal{F}_k^B)[\varrho] = \varrho_{jk,jk}|\varphi_j \otimes \varphi_k\rangle\langle\varphi_j \otimes \varphi_k| \tag{7.8}$$

providing that Alice got a result $j$ and Bob observed an outcome $k$. In order to complete the LOCC channel Alice and Bob exchange the outcomes they found, thus, a classical communication in both directions is needed. In the last step both of them apply local channels $\mathcal{E}_{jk}^A \otimes \mathcal{E}_{jk}^B$ depending on the particular outcomes $j, k$:

$$\mathcal{E}_{jk}^A \quad : \quad |\varphi_j\rangle\langle\varphi_j| \mapsto |\psi_{jk}\rangle\langle\psi_{jk}|,$$
$$\mathcal{E}_{jk}^B \quad : \quad |\varphi_k\rangle\langle\varphi_k| \mapsto |\phi_{jk}\rangle\langle\phi_{jk}|.$$

Therefore, the final state after all these operations takes the form

$$\varrho' = \sum_{j,k} \varrho_{jk,jk}|\psi_{jk} \otimes \phi_{jk}\rangle\langle\psi_{jk} \otimes \phi_{jk}|, \tag{7.9}$$

which coincides with the action of the channel given by Kraus operators $A_{jk}$. Therefore the sequence of LOCC operations implements the separable channel defined in Eq.(7.7).

It remains to show explicitly that a nontrivial example of the factorized basis $\psi_{jk} \otimes \phi_{jk}$ do exist, i.e. $\psi_{jk} \neq \psi_j$ for all $k$ and $\phi_{jk} \neq \phi_k$ for all $j$. For simplicity, let us assume that the system is three-dimensional. It is easy to verify that the following vectors form an orthonormal basis of $\mathcal{H}_3 \otimes \mathcal{H}_3$

$$\varphi_1 \otimes \varphi_{1+2}, \, \varphi_1 \otimes \varphi_{1-2}, \, \varphi_3 \otimes \varphi_{2+3}, \, \varphi_3 \otimes \varphi_{2-3}, \, \varphi_2 \otimes \varphi_2, \\ \varphi_{2+3} \otimes \varphi_1, \, \varphi_{2-3} \otimes \varphi_1, \, \varphi_{1+2} \otimes \varphi_3, \, \varphi_{1-2} \otimes \varphi_3, \tag{7.10}$$

where we used the short-hand notation $\varphi_{j\pm k} = \frac{1}{\sqrt{2}}(\varphi_j \pm \varphi_k)$.

**Exercise 49.** Convince yourself that vectors forming the basis in Eq. (7.10) cannot be discriminated perfectly if only LOCC strategies are allowed. Find a maximal subset of these vectors that can be perfectly discriminated by means of LOCC.

Let us note that in classical theory all channels are LOCC. In fact, the classical communication is nothing else but an exchange of classical systems. Consider we want to implement a classical transformation on a composite system, one part being in the possession of Alice and second one in the possession of Bob. Alice can send her system to Bob (classical communication). Bob can apply the desired operation locally and send the Alice's system back to Alice. In this way the operation on a joint system is accomplished. Thus, all classical operations are implementable in LOCC manner by using only two information exchanges. The inclusions $\mathcal{O}_c^{\text{fac}} \subset \mathcal{O}_c^{\text{LOCC}} \subset \mathcal{O}_c^{\text{sep}} \subset \mathcal{O}_c$ imply that in quantum case there are channels that are not LOCC, because all the inclusions are strict, i.e. they do not reduce to equalities. Perhaps the most surprising and the most nontrivial relation is that $\mathcal{O}_c^{\text{LOCC}} \neq \mathcal{O}_c^{\text{sep}}$, i.e. not every separable channel can be implemented in the LOCC manner and we can formulate the following proposition.

**Proposition 81.** The set $\mathcal{O}_c^{\mathrm{sep}} \setminus \mathcal{O}_c^{\mathrm{LOCC}}$ is nonempty.

*Proof.* The proof is based on counter-example which was originally reported in [7]. Consider a similar separable channel on $\mathcal{H}_d \otimes \mathcal{H}_d$ composite system as in Example 80, i.e.

$$\mathcal{E}^*[\varrho] = \sum_{j,k} B_{jk} \varrho B_{jk}^* \tag{7.11}$$

with $B_{jk} = A_{jk}^* = |\varphi_j \otimes \varphi_k\rangle\langle\psi_{jk} \otimes \phi_{jk}|$. Thus, the channel $\mathcal{E}^*$ is the adjoint channel to channel $\mathcal{E}$ defined in Example 80. However, unlike $\mathcal{E}$, it turns out that this channel is not LOCC although it is separable. The crucial difference between these two separable channels is that the Lüder's measurement in the basis $\psi_{jk} \otimes \phi_{jk}$ cannot be implemented locally. Therefore, the LOCC procedure exploited for the implementation of $\mathcal{E}$ fails for $\mathcal{E}^*$.

If $\mathcal{E}^*$ would be an LOCC channel, then the elements of the product basis $\psi_{jk} \otimes \phi_{jk}$ will be perfectly distinguishable in a single shot. That is, the sharp observable described by projectors $P_{jk} = |\psi_{jk} \otimes \phi_{jk}\rangle\langle\psi_{jk} \otimes \phi_{jk}|$ can be realized by applying the LOCC realization of $\mathcal{E}^*$ and performing local sharp observables on both sides in basis $\varphi_1, \ldots, \varphi_d \in \mathcal{H}_d$. Therefore, the existence of LOCC realization of the channel $\mathcal{E}^*$ is equivalent to the perfect (single shot) distinguishability of states $\psi_{jk} \otimes \phi_{jk}$ by local operations and classical communications. The proof of impossibility of perfect LOCC discrimination is a bit technical and can be found for example in [31]. $\qquad\square$

Comparing Example 80 with Proposition 81 it seems that we come to a paradox. Alice and Bob can easily prepare the factorized states corresponding to vectors $\psi_{jk} \otimes \phi_{jk}$ by LOCC channel starting from states associated with vectors $\varphi_j \otimes \varphi_k$. However, if they forgot which state is actually prepared, they are no longer able to identify it by means of LOCC. This is a kind of pointwise irreversibility of LOCC operations. The phenomenon described in the proof of Proposition 81 is called *nonlocality without entanglement*.

**Example 81.** (*Twirling channel*) Let Alice randomly chooses some unitary channel $\sigma_U[\cdot] = U \cdot U^*$ and send her choice to Bob. Bob applies the same unitary channel on his part of the system. Without any doubts the resulting transformation constitutes an LOCC channel, which is known under the name *twirling* and reads

$$\mathcal{T}[X] = \int_{U(d)} dU \, U \otimes U X U^* \otimes U^* \,, \tag{7.12}$$

where $dU$ is Haar measure on the group of unitary operators $U(d)$. We prove that on selfadjoint operators $X$ the twirling performs the transformation

$$\mathcal{T}[X] = \frac{\mathrm{tr}\,[X P_+]}{d_+} P_+ + \frac{\mathrm{tr}\,[X P_-]}{d_-} P_- \,. \tag{7.13}$$

The invariance property of Haar measure $dU$ implies that the operator $\mathcal{T}[X]$ commutes with all unitary operators of the type $U \otimes U$, i.e. $[\mathcal{T}[X], U \otimes U] = 0$ for all $U \in U(d)$. If $X$ is selfadjoint, then $\mathcal{T}[X]$ is also selfadjoint. Therefore, in its spectral form it reads $\mathcal{T}[X] = \sum_j x_j P_j$, where $x_j$ are real numbers and $P_j$ are projections. The commutation of $\mathcal{T}[X]$ with unitaries $U \otimes U$ implies that each projector $P_j$ must also commute with all operators of the type $U \otimes U$. Therefore, the corresponding subspaces $\mathcal{H}_j = P_j(\mathcal{H}_d \otimes \mathcal{H}_d) = \{\psi \in \mathcal{H}_d \otimes \mathcal{H}_d$ such that $P_j\psi = \psi\}$ are invariant under the action of operators $U \otimes U$.

It turns out there are only two invariant subspaces of $\mathcal{H}_d \otimes \mathcal{H}_d$: *symmetric* and *antisymmetric* subspace. A vector $\psi \in \mathcal{H}_d \otimes \mathcal{H}_d$ is called symmetric (antisymmetric) if $S\psi = \pm\psi$, respectively, where $S$ is the so-called *swap* operator defined as $S = \sum_{jk} |\varphi_j \otimes \varphi_k\rangle\langle\varphi_k \otimes \varphi_j|$, where the vectors $\varphi_1, \ldots, \varphi_d$ form an orthonormal basis of $\mathcal{H}_d$. The definition of $S$ is independent of the choice of the basis of $\mathcal{H}_d$. Let us define the operators $P_\pm$ being the projectors onto the symmetric and antisymmetric subspaces, respectively. Defining the vectors $\psi_{j\pm k} = \frac{1}{\sqrt{2}}(\varphi_j \otimes \varphi_k \pm \varphi_k \otimes \varphi_j)$ for $j \neq k$ and $\psi_{j+j} = \varphi_j \otimes \varphi_j$ we can write

$$P_\pm = \sum_{jk} |\psi_{j\pm k}\rangle\langle\psi_{j\pm k}|, \tag{7.14}$$

and, equivalently, $P_\pm = \frac{1}{2}(I \pm S)$. Let us note that vectors $\psi_{j\pm k}$ $(j, k = 1, \ldots, d)$ are forming an orthonormal basis of $\mathcal{H}_d \otimes \mathcal{H}_d$. Formally, the vectors $\psi_{j-j}$ are associated with the zero vector $0 \in \mathcal{H}_d \otimes \mathcal{H}_d$. It follows that the dimensions of symmetric and antisymmetric subspaces are $d_\pm = d(d \pm 1)/2$.

As a result of the above discussion we obtain that the operator $\mathcal{T}[X]$ written in its spectral form is a linear combination of mutually orthogonal projectors $P_+$ and $P_-$, i.e.

$$\mathcal{T}[X] = a_+(X)P_+ + a_-(X)P_-. \tag{7.15}$$

Our goal is to verify that $a_\pm(X) = \frac{\mathrm{tr}[XP_\pm]}{d_\pm}$, thus the twirling channel has the form (7.13). In order to verify that Eq. (7.12) and Eq. (7.13) define the same mapping, it is sufficient to verify their actions on elements of arbitrary operator basis. We shall use the operators $E_{j\pm k, m\pm n} = |\psi_{j\pm k}\rangle\langle\psi_{m\pm n}|$ forming an orthonormal operator basis, i.e. each operator $X$ can be written as a linear combination of these basis operators.

Define an orthogonal operator basis containing the operators $P_\pm$. For arbitrary operator $Y$ orthogonal to $P_\pm$, i.e. $\mathrm{tr}[YjP_\pm] = 0$. Therefore, according to Eq. (7.15) $\mathrm{tr}[Y^*\mathcal{T}[X]] = 0$. And consequently, It is sufficient to verify that the values of $\mathrm{tr}[P_\pm \mathcal{T}[E_{j\pm k, m\pm n}]]$ coincide for both expressions of the twirling channel given in Eq. (7.12) and in Eq. (7.13). Direct calculation gives

$$
\begin{aligned}
\mathrm{tr}[P_\pm \mathcal{T}[E_{j\pm k, m\pm n}]] &= \mathrm{tr}\left[P_\pm \int_{U(d)} dU\, U \otimes U E_{j\pm k, m\pm n} U^* \otimes U^*\right] \\
&= \mathrm{tr}\left[E_{j\pm k, m\pm n} \int_{U(d)} dU\, U \otimes U P_\pm U^* \otimes U^*\right] \\
&= \mathrm{tr}[E_{j\pm k, m\pm n} P_\pm]
\end{aligned}
$$

and, simultaneously,

$$
\begin{aligned}
\mathrm{tr}[P_\pm \mathcal{T}[E_{j\pm k, m\pm n}]] &= \frac{\mathrm{tr}[E_{j\pm k, m\pm n} P_+]}{d_+}\mathrm{tr}[P_\pm P_+] + \frac{\mathrm{tr}[E_{j\pm k, m\pm n} P_-]}{d_-}\mathrm{tr}[P_\pm P_-] \\
&= \mathrm{tr}[E_{j\pm k, m\pm n} P_\pm].
\end{aligned}
$$

That is, the action of the twirling channel is described by Eq. (7.13).

## 7.2 Entanglement vs LOCC

In this section we use LOCC channels to give operational meaning to the key concept of this chapter - *entanglement*. Roughly speaking, the entanglement is a property of quantum states exhibiting the "quantumness" of composite quantum systems in comparison with the composite classical systems. As we said earlier, for composite classical systems all the states are trivially related by some LOCC channels since any classical channel is LOCC. In quantum case we shall use the LOCC relation between the states to define an ordering of quantum states.

**Definition 59.** The subset of states

$$O_\varrho^{\text{LOCC}} = \{\varrho' \in \mathcal{S}(\mathcal{H}_A \otimes \mathcal{H}_B) : \varrho' = \mathcal{F}_{\text{LOCC}}(\varrho) \text{ for some } \mathcal{F}_{\text{LOCC}} \in \mathcal{O}_c^{\text{LOCC}}\}$$

is called an *LOCC orbit* of a state $\varrho \in \mathcal{S}(\mathcal{H}_A \otimes \mathcal{H}_B)$.

**Definition 60.** We say that $\varrho$ is *LOCC smaller* than $\xi$ (denoted as $\varrho \leq_{\text{LOCC}} \xi$) if $\varrho \in O_\xi^{\text{LOCC}}$, or equivalently, if there exists an LOCC operation $\mathcal{F}_{\text{LOCC}} : \xi \mapsto \varrho$.

The above relation defines an LOCC-induced partial ordering of the state space.

**Definition 61.** We say that states $\varrho$ and $\xi$ are *LOCC equivalent* (and denote $\varrho \sim_{\text{LOCC}} \xi$) if $\varrho \leq_{\text{LOCC}} \xi$ and $\xi \leq_{\text{LOCC}} \varrho$.

**Proposition 82.** Following statements hold:

1. $\varrho \leq_{\text{LOCC}} \xi$ if and only if $O_\varrho^{\text{LOCC}} \subseteq O_\xi^{\text{LOCC}}$. Consequently, states $\varrho$ and $\xi$ are LOCC equivalent if and only if $O_\varrho^{\text{LOCC}} = O_\xi^{\text{LOCC}}$.

2. Consider a factorized pure state $\omega = |\varphi \otimes \varphi\rangle\langle\varphi \otimes \varphi|$. Then $O_\omega^{\text{LOCC}} = \mathcal{S}^{\text{sep}}$.

3. $\mathcal{S}^{\text{sep}} \subseteq O_\varrho^{\text{LOCC}}$ for arbitrary state $\varrho \in \mathcal{S}(\mathcal{H}_A \otimes \mathcal{H}_B)$.

4. All separable states are mutually LOCC equivalent and, moreover, the set of separable states $\mathcal{S}^{\text{sep}}$ is closed under LOCC channels and $\mathcal{S}^{\text{sep}} = \bigcap_\varrho O_\varrho^{\text{LOCC}}$.

5. The separable states form an LOCC equivalence class of LOCC smallest element, i.e. if $\varrho \leq_{\text{LOCC}} \xi$ for all $\xi \in \mathcal{S}(\mathcal{H}_A \otimes \mathcal{H}_B)$, then the state $\varrho$ is separable.

*Proof.* 1. A general element in $\varrho' \in O_\varrho^{\text{LOCC}}$ can be written as $\varrho' = \mathcal{F}_{\text{LOCC}}(\varrho)$. The relation $\varrho \leq_{\text{LOCC}} \xi$ implies that there exists an LOCC channel $\mathcal{F}_{\text{LOCC}}$ such that $\varrho = \mathcal{F}_{\text{LOCC}}(\xi)$ and, consequently, $\varrho' = \mathcal{F}_{\text{LOCC}} \circ \mathcal{F}_{\text{LOCC}}(\xi)$. It follows that $\varrho' \in O_\xi^{\text{LOCC}}$, i.e. $O_\varrho^{\text{LOCC}} \subset O_\xi^{\text{LOCC}}$. It is straightforward to see that LOCC equivalence requires that the inclusions $O_\varrho^{\text{LOCC}} \subset O_\xi^{\text{LOCC}}$ and $O_\xi^{\text{LOCC}} \subset O_\varrho^{\text{LOCC}}$ holds simultaneously.

2. Consider a general separable state $\xi = \sum_j p_j |\psi_j \otimes \varphi_j\rangle\langle\psi_j \otimes \varphi_j| \in \mathcal{S}^{\text{sep}}$. A mapping $\mathcal{A}_{\psi_j} \otimes \mathcal{A}_{\varphi_j}$, where $\mathcal{A}_\psi : \mathcal{S}(\mathcal{H}) \mapsto |\psi\rangle\langle\psi|$ is a contraction into a pure state, is an LOCC channel and $\omega \to \mathcal{F}_{\text{LOCC}}^{(j)}(\omega) = (\mathcal{A}_{\psi_j} \otimes \mathcal{A}_{\varphi_j})(\omega) = |\psi_j \otimes \varphi_j\rangle\langle\psi_j \otimes \varphi_j|$. Since the set of LOCC channels is convex, it follows that $\mathcal{F}_{\text{LOCC}} = \sum_j p_j \mathcal{F}_{\text{LOCC}}^{(j)}$ is an LOCC channel, too. It follows that $\xi = \mathcal{F}_{\text{LOCC}}(\omega)$, hence $\mathcal{S}^{\text{sep}} \subset O_\omega^{\text{LOCC}}$. LOCC channels form a subset

of separable channels, but a general separable channel applied to the state $\omega$ can produce only separable state. That is, no entangled state $\varrho$ belongs to the LOCC orbit of the state $\omega$, i.e. $O_\omega^{\mathrm{LOCC}} = \mathcal{S}^{\mathrm{sep}}$.

3. Starting with a general state $\varrho \in \mathcal{S}(\mathcal{H}_A \otimes \mathcal{H}_B)$ we can apply the local channel $\mathcal{A}_\varphi \otimes \mathcal{A}_\varphi$ resulting in transformation $\varrho \mapsto |\varphi \otimes \varphi\rangle\langle\varphi \otimes \varphi| \equiv \omega$. Because, $\mathcal{S}^{\mathrm{sep}} = O_\omega^{\mathrm{LOCC}}$ it follows that for each separable state $\xi = \mathcal{F}_{\mathrm{LOCC}}(\omega) = \mathcal{F}_{\mathrm{LOCC}} \circ (\mathcal{A}_\varphi \otimes \mathcal{A}_\varphi)(\varrho)$, i.e. $\xi \leq_{\mathrm{LOCC}} \varrho$. Thus, the statement holds, i.e. $\mathcal{S}^{\mathrm{sep}} \subset O_\varrho^{\mathrm{LOCC}}$ for all $\varrho$.

4. It is a simple consequence of the previous facts that separable states are LOCC equivalent, because we have shown that $\omega \mapsto \xi \mapsto \omega$ for all separable states $\xi$ by means of LOCC channels. It means $O_\omega^{\mathrm{LOCC}} = O_\xi^{\mathrm{LOCC}} = \mathcal{S}^{\mathrm{sep}}$ for all separable states $\xi \in \mathcal{S}^{\mathrm{sep}}$, i.e. the set of states $\mathcal{S}^{\mathrm{sep}}$ forms an LOCC equivalence class. It follows that separable states are closed under LOCC channels and also $\mathcal{S}^{\mathrm{sep}} = \bigcap_\varrho O_\varrho^{\mathrm{LOCC}}$.

5. In the previous point we proved that separable states form an LOCC equivalence class. In point 3 it is shown that separable states are comparable with all states and they are smaller than any other quantum state, i.e. they are the smallest elements. This is their unique property and, thus, it can be used to characterize the separable states as it is done in this point.

$\square$

In less formal words Proposition 82 is saying that the separable (not entangled) states are the LOCC-smallest elements of the state space, hence the LOCC ordering gives some interesting structure only to the set of entangled states. In what follows we characterize also the maximal elements (i.e. those states that are either larger than any other state, or incomparable). As we shall see not only that there do exist the maximal elements, but there exists also an LOCC equivalence class of the greatest elements (i.e. states that are greater than any other state) of the state space. As a consequence of the above proposition we can give operational meaning to definitions of entangled and separable states and also introduce the concept of the maximally entangled states. Corollary 3 can be used as an alternative operational definition of the entanglement.

**Corollary 3.** A state is *entangled* if and only if it does not belong to the equivalence class of smallest elements with respect to LOCC-induced ordering. In other words, a state is entangled if and only if it cannot be prepared from a separable state by means of LOCC channel.

**Definition 62.** If a state $\varrho$ belongs to the LOCC equivalence class of the greatest element (if such element exists), then the state $\varrho$ is called the *maximally entangled state.*

**Proposition 83.** The following statements are equivalent:

1. $\Psi$ is maximally entangled.

2. $O_\Psi = \mathcal{S}(\mathcal{H}_A \otimes \mathcal{H}_B)$.

3. $\Psi = (U \otimes I)\Psi_+(U^* \otimes I)$, where $U$ is unitary, $\Psi_+ = |\psi_+\rangle\langle\psi_+|$,

$$\psi_+ = \frac{1}{\sqrt{d}}\sum_{j=0}^{d-1} \varphi_j \otimes \varphi_j, \tag{7.16}$$

where $\varphi_0, \ldots, \varphi_{d-1}$ are mutually orthogonal unit vectors in $\mathcal{H}_A, \mathcal{H}_B$ and $d = \min\{d_A, d_B\}$.

4. $\Psi$ is a pure state and $\mathrm{Tr}_A \Psi = \mathrm{Tr}_B \Psi = \frac{1}{d} \sum_{j=0}^{d-1} |\varphi_j\rangle\langle\varphi_j|$.

*Proof.* The equivalences $1 \Leftrightarrow 2$ and $3 \Leftrightarrow 4$ are direct consequences of the definitions. Therefore, we shall focus on equivalence relation $2 \Leftrightarrow 3$. Before we start with the proof, let us note that the states $(U \otimes I)\Psi_+(U^* \otimes I)$ and $\Psi_+$ are locally unitary equivalent, hence they are LOCC equivalent. Therefore, we can fix the state shared by Alice and Bob to be $\Psi_+$ defined in Eq. (7.16). For simplicity we shall assume that $d_A = d_B = d$. According to Schmidt decomposition a general vector can be written as $\phi = \sum_j a_j \varphi_j \otimes \varphi_j$. Let us assume that Alice and Bob share the maximally entangled state $\Psi_+$. The Choi-Jamiolkowski isomorphism implies that $\phi = (I \otimes R_\phi)\psi_+$ with $R_\phi = \sqrt{d} \sum_j a_j |\varphi_j\rangle\langle\varphi_j|$. Alice adds a $d$-dimensional ancilla in a state $|\varphi_0\rangle_{A'}\langle\varphi_0|$ and applies an isometry transformation $G : \varphi_0 \otimes \varphi_j \to (U_j \otimes R_\phi)\psi_+$ on systems $A'$ and $A$, where $U_j$ are shift operators, i.e. $U_j\varphi_{j'} = \varphi_{j' \oplus j}$. After Alice applied her transformation the composite system is described by a pure state associated with the vector

$$\psi_{A'AB} = \frac{1}{\sqrt{d}} \sum_{j,j'} a_{j'} \varphi_{j \oplus j'} \otimes \varphi_{j'} \otimes \varphi_j = \frac{1}{\sqrt{d}} \sum_k \varphi_k \otimes \left( \sum_{j,j' : j \oplus j' = k} a_{j'} \varphi_{j'} \otimes \varphi_j \right)$$

Alice performs a projective measurement of the ancilla system associated with the Lüder's instrument $k \mapsto \mathcal{I}_k(\varrho) = \langle\varphi_k|\varrho|\varphi_k\rangle|\varphi_k\rangle\langle\varphi_k|$, hence measuring an outcome $k$ the composite system ends up in

$$\psi_{A'AB}^{(k)} = \varphi_k \otimes \sum_{j,j' : j \oplus j' = k} a_{j'} \varphi_{j'} \otimes \varphi_j . \tag{7.17}$$

Alice sends the outcome she found to Bob who applies the unitary channel $\mathcal{U}_k : \varrho \to U_k^* \varrho U_k$ (the inverse of the shift unitary channel) to obtain a state associated with a unit vector

$$\psi_{AB}^{(k)} = \sum_{j,j' : j \oplus j' = k} a_{j'} \varphi_{j'} \otimes \varphi_{j \ominus k} = \sum_{j'} a_{j'} \varphi_{j'} \otimes \varphi_{j'} = \phi_{AB} \tag{7.18}$$

which is the same for all values of $k$. It follows that each pure state $\Phi = |\phi\rangle\langle\phi|$ can be obtained by an LOCC channel from the state $\Psi_+$. Moreover, since a general mixed state $\varrho_{AB}$ is a mixture of pure states, it follows that the corresponding mixture of LOCC channels can be used to transform $\Psi_+$ to any mixed state $\varrho_{AB}$, thus $O_{\Psi_+} = \mathcal{S}(\mathcal{H}_A \otimes \mathcal{H}_B)$, which proves the equivalence. $\qquad\square$

**Example 82.** Let us consider an explicit example showing that arbitrary pure state can be achieved by an LOCC channel from the state $\psi_+$ for the composite system of two qubits. A general pure state $\Phi = |\phi\rangle\langle\phi|$ with $\phi = a\varphi_0 \otimes \varphi_0 + b\varphi_1 \otimes \varphi_1$ written in its Schmidt form. Add a single qubit ancilla in a pure state $|\varphi_0\rangle\langle\varphi_0|$ and apply the local unitary operator on Alice's side

$$\varphi_0 \otimes \varphi_0 \mapsto a\varphi_0 \otimes \varphi_0 + b\varphi_1 \otimes \varphi_1 \tag{7.19}$$

$$\varphi_0 \otimes \varphi_1 \mapsto a\varphi_1 \otimes \varphi_0 + b\varphi_0 \otimes \varphi_1 \tag{7.20}$$

to get

$$\frac{1}{\sqrt{2}} [\varphi_0 \otimes (a\varphi_0 \otimes \varphi_0 + b\varphi_1 \otimes \varphi_1) + \varphi_1 \otimes (a\varphi_0 \otimes \varphi_1 + b\varphi_1 \otimes \varphi_0)] . \tag{7.21}$$

Alice performs a measurement of a two-valued observable associated with basis vectors $\varphi_0, \varphi_1$ and sends the observed outcome $b$ to Bob. If the received value is $b = 0$, then Bob will do nothing. If $b = 1$, Bob will apply a unitary channel such that $\varphi_1 \mapsto \varphi_0$ and $\varphi_0 \mapsto \varphi_1$ to obtain the joint state associated with the vector $\phi$.

The last proposition guarantees the existence of the maximally entangled states. A general question on the existence of LOCC channel between arbitrary pair of states is a very difficult one. The following theorem completely characterize the LOCC ordering between pure states. A proof of this theorem is quite long and requires several additional lemmas and therefore it is omitted. For an interested reader we refer to [63] and [65].

**Theorem 15.** (*Majorization criterion*) Let $\vec{\lambda}_\psi, \vec{\lambda}_\phi$ are the Schmidt vectors of $\psi, \phi \in \mathcal{H}_A \otimes \mathcal{H}_B$, respectively. Then

$$\Phi \leq_{\text{LOCC}} \Psi \Leftrightarrow \vec{\lambda}_\psi \prec \vec{\lambda}_\phi \Leftrightarrow \sum_{j=1}^n \lambda_\psi^{(j)} \leq \sum_{j=1}^n \lambda_\phi^{(j)} \quad \forall n \leq d = \min\{d_A, d_B\} , \tag{7.22}$$

where $\Psi = |\psi\rangle\langle\psi|$ and $\Phi = |\phi\rangle\langle\phi|$ are the projectors representing the pure states.

There are many interesting and surprising results concerning the quantum entanglement. The following discussion cannot be considered as a complete list of all results and known facts on quantum entanglement.

Let us note that a unitary channel $\sigma_U$ is separable if and only if it is local, i.e. $U = U_A \otimes U_B$. This follows directly from the uniqueness of the operator sum decomposition for unitary channels. Consider a composite system $\mathcal{H}_{AB} = \mathcal{H} \otimes \mathcal{H}$ and define a unitary self-adjoint operator $V_{\text{SWAP}}(\varphi \otimes \psi) = \psi \otimes \varphi$ for all $\varphi, \psi \in \mathcal{H}$. As the name is suggesting this unitary operator determines a unitary channel swapping the systems, i.e. mapping $\varrho_{AB}$ to $\varrho_{BA} = V_{\text{SWAP}} \varrho_{AB} V_{\text{SWAP}}$. By unitarity it is not separable, hence it is not LOCC, because $V_{\text{SWAP}} \neq U_A \otimes U_B$. This is not completely unexpected, because the exchange of quantum systems is exactly the procedure that is not permitted in LOCC concept. However, this observation suggests that states $\varrho_{AB}$ and $\varrho_{BA}$ can be LOCC inequivalent, i.e. the shared entanglement may not be symmetric with respect to the exchange of individual parties. The following result, presented in [46], gives a precise meaning to this fact.

**Proposition 84.** (*Exchange asymmetry of entanglement*.) Consider $\mathcal{H}_{AB} = \mathcal{H} \otimes \mathcal{H}$ and define a unitary self-adjoint operator $V_{\text{SWAP}}(\varphi \otimes \psi) = \psi \otimes \varphi$ for all $\varphi, \psi \in \mathcal{H}$. There are states $\varrho_{AB}$ for which the state $\varrho_{BA} = V_{\text{SWAP}} \varrho_{AB} V_{\text{SWAP}}$ is not LOCC equivalent to $\varrho_{AB}$.

In summary, the exchange of systems cannot be implemented by an LOCC channel even if the state is known. That is, pointwise the transformation $\varrho_{AB} \mapsto \varrho_{BA}$ is not LOCC. Such asymmetry of entanglement can have practical implications. For example, as it is shown in [86], this asymmetry property has an interesting consequence for the so-called super-dense coding (see Example 84) in which the communication capacity is different if Alice and Bob exchange the roles of sender and receiver although the shared state remains the same.

Another interesting and surprising feature of quantum entanglement is that LOCC channels do not preserve the LOCC ordering [54], [88], [90].

**Proposition 85.** LOCC-ordering is not preserved under the action of local channels. In particular,

$$\varrho \leq_{\text{LOCC}} \omega \nRightarrow (\mathcal{E}^A \otimes \mathcal{I})(\varrho) \leq_{\text{LOCC}} (\mathcal{E}^A \otimes \mathcal{I})(\omega), \tag{7.23}$$

where $\mathcal{E}^A$ is a channel applied on subsystem $A$ only.

*Proof.* Consider the following counter-example. Let us define two pure entangled states of a composite system $\mathcal{H} \otimes \mathcal{H}$ of a pair of three-dimensional systems associated with vectors:

$$\psi_1 = (\varphi_2 \otimes \varphi_2 + \varphi_3 \otimes \varphi_3)/\sqrt{2}, \tag{7.24}$$

$$\psi_2 = \sqrt{q}\,\varphi_1 \otimes \varphi_1 + \sqrt{1-q}\,\varphi_2 \otimes \varphi_2, \tag{7.25}$$

where $\varphi_1, \varphi_2, \varphi_3$ form an orthonormal basis of $\mathcal{H}$. Consider a local channel $\mathcal{E}^A$ defined via Kraus operators $M_1 = |\varphi_1\rangle\langle\varphi_1| + |\varphi_2\rangle\langle\varphi_2|$ and $M_2 = |\varphi_3\rangle\langle\varphi_3|$. Applying this local channel to the above pure states we get the states

$$\varrho_1^{\text{out}} = (|\varphi_2 \otimes \varphi_2\rangle\langle\varphi_2 \otimes \varphi_2| + |\varphi_3 \otimes \varphi_3\rangle\langle\varphi_3 \otimes \varphi_3|)/2, \tag{7.26}$$

$$\varrho_2^{\text{out}} = |\psi_2\rangle\langle\psi_2|. \tag{7.27}$$

According to Theorem 15 the input state $|\psi_1\rangle\langle\psi_1|$ can be mapped to $|\psi_2\rangle\langle\psi_2|$ by an LOCC channel, but the inverse is not possible unless $q = 1/2$, i.e. $|\psi_2\rangle\langle\psi_2| <_{\text{LOCC}} |\psi_1\rangle\langle\psi_1|$. However, since $\varrho_1^{\text{out}}$ is separable, the transformation $\varrho_1^{\text{out}} \to \varrho_2^{\text{out}}$ cannot be an LOCC channel, but still the transformation $\varrho_2^{\text{out}} \to \varrho_1^{\text{out}}$ can be LOCC, because the set of separable states is contained in the LOCC orbit of each entangled state. Consequently, $\varrho_1^{\text{out}} > \varrho_2^{\text{out}}$. This example shows that the statement of the proposition holds. $\square$

An interesting question is how generic is the above property. That is, which channels on a subsystem do preserve the LOCC ordering? It is straightforward to see that unitary channels cannot affect the LOCC ordering, because they do not change the amount of entanglement.

**Definition 63.** A channel $\mathcal{E} : \mathcal{T}(\mathcal{H}) \to \mathcal{T}(\mathcal{H})$ is *entanglement-breaking* if for all input states $\omega \in \mathcal{S}(\mathcal{H} \otimes \mathcal{H}_{\text{anc}})$, the output states $\mathcal{E} \otimes \mathcal{I}_{\text{anc}}(\omega)$ are separable for all possible $\mathcal{H}_{\text{anc}}$.

By definition also the entanglement-breaking channels [48] trivially preserves the LOCC ordering, because they completely destroy any entanglement present in the state.

**Proposition 86.** A channel $\mathcal{E}$ is entanglement-breaking if and only if $\mathcal{E}$ is of the form $\mathcal{E}(\varrho) = \sum_j \xi_j \text{tr}[\varrho F_j]$, where $\xi_j$ are states and the positive operators $F_j$ defines a discrete POVM, i.e. $\sum_j F_j = I$.

*Proof.* By definition if $\mathcal{E}$ is entanglement-breaking then the state $\omega = (\mathcal{E} \otimes \mathcal{I})[\Psi_+]$ is separable, i.e. $\omega_{\mathcal{E}} = \sum_n p_n |\varphi_n\rangle\langle\varphi_n| \otimes |\phi_n\rangle\langle\phi_n|$. Consider a map $\mathcal{E}'$ of the form $\mathcal{E}'(\varrho) = \sum_n \xi_n \mathrm{tr}\,[\varrho F_n]$ with $\xi_n = |\phi_n\rangle\langle\phi_n|$ and $F_n = dp_n|\varphi_n\rangle\langle\varphi_n|$, where $d$ is the dimension of the Hilbert space the channel $\mathcal{E}$ acts on. Let us note that $\mathrm{tr}_A\omega_{\mathcal{E}} = \frac{1}{d}I = \sum_n p_n |\phi_n\rangle\langle\phi_n|$, thus the positive operators $F_n$ are properly normalized ($\sum_n F_n = I$) and they form a POVM. Due to Choi-Jamiolkowski isomorphism the state $\omega_{\mathcal{E}}$ determines the channel $\mathcal{E}$ uniquely. Therefore, since $\omega_{\mathcal{E}'} = (\mathcal{E}' \otimes \mathcal{I})[\Psi_+] = \sum_n p_n |\varphi_n\rangle\langle\varphi_n| \otimes |\phi_n\rangle\langle\phi_n| = \omega_{\mathcal{E}}$ we can conclude that $\mathcal{E} = \mathcal{E}'$, hence if $\mathcal{E}$ is entanglement-breaking, then it has the desired form.

Conversely, if $\mathcal{E}$ is defined as $\mathcal{E}(\varrho) = \sum_j \xi_j \mathrm{tr}\,[\varrho F_j]$, then

$$(\mathcal{E} \otimes \mathcal{I})(\omega) = \sum_j (\xi_j \otimes I)\mathrm{tr}_A[(F_j \otimes I)\Omega] = \sum_j q_j \xi_j \otimes Q_j\,,$$

where $q_j = \mathrm{tr}\,[(F_j \otimes I)\Omega]$ and $Q_k = q_j^{-1}\mathrm{tr}_A[\sqrt{F_j \otimes I}\,\Omega\sqrt{F_j \otimes I}]$ are positive operators with $\mathrm{tr}\,[Q_j] = q_j^{-1}\mathrm{tr}\,[(F_j \otimes I)\Omega] = 1$, i.e. $Q_j$ are density operators. $\qquad\square$

As a result we get that entanglement-breaking channels can be interpreted as *measure and prepare* procedures. In particular, they can be implemented as a particular instrument in which the measured system is replaced by a new system in the state $\xi_j$ conditioned on the observed outcome $F_j$.

**Example 83.** (*Hilbert space basis of maximally entangled states.*) Let us stress once more that maximally entangled states on $\mathcal{H}_d \otimes \mathcal{H}_d$ composite system are related by local unitary channels meaning that if $\Psi_1 = |\psi_1\rangle\langle\psi_1|$ and $\Psi_2 = |\psi_2\rangle\langle\psi_2|$ are maximally entangled states, then there exist a unitary operator $U : \mathcal{H}_d \to \mathcal{H}_d$ such that $\Psi_1 = (U \otimes I)\Psi_2(U^* \otimes I)$. Let us define unitary operators $U_1, U_2$ such that $\psi_j = (U_j \otimes I)\psi_+$, where $\psi_+$ is the vector corresponding to the canonical maximally entangled state $\Psi_+$. Due to Choi-Jamiolkowski isomorphism

$$\langle\,\psi_1\,|\,\psi_2\,\rangle = \frac{1}{d}\mathrm{tr}\,[U_1^* U_2]\,. \tag{7.28}$$

This formula implies that maximally entangled states are orthogonal if and only if the corresponding unitary operators are orthogonal in the Hilbert-Schmidt sense. As we have seen explicitly in Example 76 there do exists an orthogonal operator basis composed of unitary operators only, i.e. $\mathrm{tr}\,[U_j^* U_k] = \delta_{jk}$ for all $j, k = 1, \dots, d^2$. Consequently, the vectors $\psi_j = (U_j \otimes I)\psi_+$ are mutually orthogonal and form a complete orthonormal basis of $\mathcal{H}_d \otimes \mathcal{H}_d$ also known as *Bell basis*. An observable associated with this basis, $j \mapsto |\psi_j\rangle\langle\psi_j|$, is called *Bell observable*, or *Bell measurement*.

**Exercise 50.** Consider a composite system of a pair of two-dimensional systems, i.e. $\mathcal{H} = \mathbb{C}^2 \otimes \mathbb{C}^2$. Verify that the vectors $(\sigma_j \otimes I)\psi_+ \in \mathcal{H}$ with $\psi_+ = \frac{1}{\sqrt{2}}(\varphi \otimes \varphi + \varphi_\perp \otimes \varphi_\perp)$ are mutually orthogonal. The operators $\sigma_0 = I, \sigma_1 = \sigma_x, \sigma_2 = \sigma_y, \sigma_3 = \sigma_z$ are the Pauli operators, i.e. $\sigma_x = |\varphi\rangle\langle\varphi_\perp| + |\varphi_\perp\rangle\langle\varphi|, \sigma_y = -i(|\varphi\rangle\langle\varphi_\perp| - |\varphi_\perp\rangle\langle\varphi|), \sigma_z = |\varphi\rangle\langle\varphi| - |\varphi_\perp\rangle\langle\varphi_\perp|$.

**Example 84.** (*Superdense coding*). The existence of Bell basis is interesting *per se*, because the maximally entangled states are locally perfectly equivalent on one side (via local unitary channels), but still they are perfectly distinguishable in a single experiment only if the parties are allowed to communicate quantum systems. The superdense coding is a communication protocol in which two classical bits can be transmitted in a secure way by using a single maximally entangled state of a pair of two-dimensional systems. Suppose two partners, Alice and Bob, are separated by a large distance, but sharing a maximally entangled state $\Psi_+$ and an ideal quantum channel for transmission of a two-dimensional system. They are allowed to use the channel only once. How much information can be transferred from Alice to Bob?

The elements of the Bell basis are related by local unitary channels, i.e. $\Psi_j = (\sigma_j \otimes I)\Psi_+(\sigma_j \otimes I)$, where $\sigma_0 = I, \sigma_1 = \sigma_x, \sigma_2 = \sigma_y, \sigma_3 = \sigma_z$ are four Pauli operators forming the operator basis and satisfying $\text{tr}\,[\sigma_j \sigma_k] = 2\delta_{jk}$. Therefore, $\text{tr}\,[\Psi_j \Psi_k] = \delta_{jk}$ and Alice can choose one the four unitary channels to generate one of the basis state $\Psi_j$. If Bob receives her system, he can perfectly identify which channel has been chosen by Alice by performing the Bell measurement. Since Alice has four possibilities she can communicate two bits of information to Bob. This communication procedure is known as the superdense coding and was originally discussed in [9].

Let us note that locally the systems of Alice and Bob are described by the total mixture, i.e. $\varrho_A = \varrho_B = \frac{1}{2}I$. The total mixture is not affected by the unitary channel applied by Alice, hence the system sent through the channel is always described by the same state - the total mixture. Consequently, the transmitted system does not contain any relevant information. Therefore, the communication is perfectly secure and noone intercepting the transferred system can learn anything.

### 7.3 Entanglement detection

Introducing the concept of LOCC operations results in a division of the quantum state space of a composite system into two subsets: separable states $\mathcal{S}_{\text{sep}}$ and entangled states $\mathcal{S}_{\text{ent}}$. The main goal of *entanglement theory* is to identify whether a state is entangled or not. In particular, there are two mutually related versions of this problem:

- **experimental** - A system is given in an unknown state. Find (experimentally) whether it is entangled or not.

- **mathematical** - A complete (mathematical) description of a state is given. Determine whether it is entangled or not.

In our discussion we address both these problems, but by no means this text can be considered as an exhaustive overview of the problem. We shall describe only the basic results and for more details we refer to [63] and [65].

#### 7.3.1 Entanglement detection via linear operators

In any quantum measurement we are measuring probability distributions potentially containing a solution to our entanglement detection problem. For example, performing an informationally complete measurement each probability distribution uniquely specifies some quantum state $\varrho$. Thus, we have a complete description of quantum state and the experimental decision problem reduces to mathematical decision problem. However, the complete tomography is a difficult experimental task. Therefore, we are interested in the existence of some simpler tests.

If we assume that the number of copies of the system is unlimited, then our most general knowledge is represented by probability distributions $p_\varrho(F_k) = \mathrm{tr}\,[F_k(\varrho \otimes \cdots \otimes \varrho)]$, where $F_k$ are joint quantum effects forming a POVM on $n$ copies of the composite systems. Associating real values $x_k$ with particular outcomes $F_k$ for any observable we can introduce the mean value $f(\varrho) := \sum_k x_k p_\varrho(F_k)$ being the mean value of a selfadjoint operator $F = \sum_k x_k F_k$, i.e. $f = \mathrm{tr} F(\varrho \otimes \cdots \otimes \varrho)$. The value of $f$ separates the state space into two subsets:

(i)  $\mathcal{S}_+ = \{\varrho \in \mathcal{S}(\mathcal{H}) \mid f(\varrho) \geq 0\}$

(ii) $\mathcal{S}_- = \{\varrho \in \mathcal{S}(\mathcal{H}) \mid f(\varrho) < 0\}$

It turns our that for specific operators $F$ all the separable states are included only in one of the two subsets.

**Definition 64.** An operator $W$ is *entanglement witness* if $\mathrm{tr}\,[\varrho \otimes \cdots \otimes \varrho\, W] \geq 0$ for all $\varrho \in \mathcal{S}_{\mathrm{sep}}$ and $W$ is not positive.

The requirement that $W$ is not positive guarantees that $\mathcal{S}_-$ is not empty.

**Lemma 5.** If $\langle \Phi | A | \Phi \rangle \geq 0$ for all factorized vector states $\Phi$, then $A$ is selfadjoint.

As a consequence of the above lemma we can formulate the alternative definition of entanglement witnesses.

**Lemma 6.** An operator $W$ is an entanglement witness if and only if $\langle\, \Phi \,|\, W \Phi \,\rangle \geq 0$ for all factorized vector states $\Phi \in \mathcal{H}^{\otimes 2n}$.

**Definition 65.** An operator $W \in \mathcal{L}_s(\mathcal{H} \otimes \mathcal{H})$ is a *linear entanglement witness* if $\langle\, \psi \otimes \varphi \,|\, W \psi \otimes \varphi \,\rangle \geq 0$ for all factorized vectors $\psi \otimes \varphi \in \mathcal{H} \otimes \mathcal{H}$ and $W$ is not positive.

The definition of entanglement witness implies that if $\mathrm{tr}\,[\varrho^{\otimes n} W] < 0$, then the state $\varrho$ is entangled. As we have discussed in Chapter 3, a general state $\varrho \in \mathcal{S}(\mathcal{H} \otimes \mathcal{H})$ can be expressed in Bloch vector form as $\varrho = \frac{1}{d}(I + \vec{r} \cdot \vec{E})$, where $E_1, \ldots, E_{d^2-1}$ are mutually orthogonal traceless operators, i.e.

$$\varrho^{\otimes n} = \frac{1}{d^n}(I^{\otimes n} + \vec{r} \cdot \vec{E} \otimes I \otimes \cdots \otimes I + \cdots + \vec{r} \cdot \vec{E} \otimes \cdots \otimes \vec{r} \cdot \vec{E})\,.$$

Similarly, the entanglement witnesses $W \in \mathcal{L}_s(\mathcal{H}^{\otimes 2n})$ can be expressed as

$$\begin{aligned} W &= w_0 I^{\otimes n} + \vec{w}_1 \cdot \vec{E} \otimes I \otimes \cdots \otimes I + \cdots + I \otimes \cdots \otimes I \otimes \vec{w}_n \cdot \vec{E} \\ &\quad + \vec{w}_{12} \vec{E} \otimes \vec{E} \otimes I \otimes \cdots \otimes I + \cdots + \vec{w}_{1\ldots n} \vec{E} \otimes \cdots \otimes \vec{E}\,, \end{aligned} \tag{7.29}$$

and hence

$$\mathrm{tr}\,[\varrho^{\otimes n} W] = \frac{1}{d^2}\left[w_0 + \vec{r} \cdot (\vec{w}_1 + \cdots + \vec{w}_n) + \cdots + \sum_{j_1, \ldots, j_n} r_{j_1} \cdots r_{j_n} w^{1,\ldots,n}_{j_1,\ldots,j_n}\right]\,. \tag{7.30}$$

It follows that arbitrary polynom of $n$th degree in $\varrho$ expressed via Bloch vector $\vec{r}$ is associated with some entanglement witness and vice versa. For linear (or single copy) entanglement witnesses the expression for mean value reduces to

$$\text{tr}\left[\varrho W\right] = \frac{1}{d^2}(w_0 + \vec{w} \cdot \vec{r}) \tag{7.31}$$

and for a given witness $W$ the condition $\text{tr}\left[\varrho W\right] = 0$ defines a hyperplane in the Bloch representation of the state space separating the subsets $\mathcal{S}_+$ and $\mathcal{S}_-$.

**Definition 66.** Let $W$ and $W'$ be two entanglement witnesses. We say that $W$ is *better witness* than $W'$ ($W \succ W'$) if all the states detected by $W'$ are detected also by $W$, i.e. if for all $\varrho$ the following implication holds:

$$\text{tr}\left[W'(\varrho \otimes \cdots \otimes \varrho)\right] < 0 \Rightarrow \text{tr}\left[W(\varrho \otimes \cdots \otimes \varrho)\right] < 0\,.$$

This relation defines a partial ordering on the set of entanglement witnesses. An entanglement witness $W$ is called *optimal* if it is the maximal element with respect to partial order among witnesses, i.e. either $W \succ W'$, or $W$ and $W'$ are incomparable.

**Proposition 87.** There is no linear entanglement witness detecting all entangled states.

*Proof.* Suppose that $W$ is a linear entanglement witness detecting all entangled states. Let $\psi, \varphi \in \mathcal{H}$ be two orthogonal unit vectors, and denote $\Phi_\pm = \frac{1}{\sqrt{2}}(\psi \otimes \psi \pm \varphi \otimes \varphi)$. Both $\Phi_\pm$ are entangled vectors as shown in Example 78, and therefore, we have $\langle\,\Phi_+\,|\,W\Phi_+\,\rangle < 0$ and $\langle\,\Phi_-\,|\,W\Phi_-\,\rangle < 0$. This implies that

$$0 > \langle\,\Phi_+\,|\,W\Phi_+\,\rangle + \langle\,\Phi_-\,|\,W\Phi_-\,\rangle = \langle\,\psi \otimes \psi\,|\,W\psi \otimes \psi\,\rangle + \langle\,\varphi \otimes \varphi\,|\,W\varphi \otimes \varphi\,\rangle\,,$$

which contradicts the fact that $W$ is positive on factorized states.                                    $\square$

**Theorem 16.** (Horodecki [47]) A quantum state $\varrho$ is separable if and only if $\text{tr}\left[\varrho W\right] \geq 0$ for all linear entanglement witnesses $W$.

*Proof.* If $\varrho$ is separable, then, by the definition, $\text{tr}\left[\varrho W\right] \geq 0$ for all linear entanglement witnesses $W$.

Assume then that $\varrho$ is entangled. We need to find a linear entanglement witness $W$ such that $\text{tr}\left[\varrho W\right] < 0$. The set $\mathcal{S}^{\text{sep}}$ of separable states is a closed and convex subset of real Banach space $\mathcal{T}_s(\mathcal{H} \otimes \mathcal{H})$. It follows from the Hahn-Banach theorem (see e.g. [74, Theorem 3.4]) that there exists a linear functional $f: \mathcal{T}_s(\mathcal{H} \otimes \mathcal{H}) \rightarrow \mathbb{C}$ and $r \in \mathbb{R}$ such that

$$f(\varrho) < r < f(\eta) \tag{7.32}$$

for all $\eta \in \mathcal{S}^{\text{sep}}$. However, the dual space of $\mathcal{T}_s(\mathcal{H} \otimes \mathcal{H})$ is $\mathcal{L}_s(\mathcal{H} \otimes \mathcal{H})$, and hence there is an operator $F \in \mathcal{L}_s(\mathcal{H} \otimes \mathcal{H})$ such that $f(\varrho) = \text{tr}\left[\varrho F\right]$ for every $\varrho \in \mathcal{T}_s(\mathcal{H} \otimes \mathcal{H})$. Defining $W = F - rI$ we get the required entanglement witness, for which $\text{tr}\left[\varrho W\right] = f(\varrho) - r < 0$.    $\square$

**Corollary 4.** A quantum state is entangled if and only if there exists a linear entanglement witness $W$ such that $\text{tr}\left[\varrho W\right] < 0$.

**Exercise 51.** For a given arbitrary bipartite entangled state $\varrho_0 \in \mathcal{S}(\mathcal{H}_A \otimes \mathcal{H}_B)$, find a linear entanglement witness $W_0$ detecting this state. (Hint: see the proof of Theorem 16)

**Example 85.** (*Clauser-Horne-Shimony-Holt entanglement witness.*) Historically, an inequality based on this entanglement witness was one of the first tests of quantum entanglement [26]. Consider the following operator defined on $\mathcal{H}_2 \otimes \mathcal{H}_2$

$$W_{\text{CHSH}} = 2I \otimes I - |\vec{a} \cdot \vec{\sigma} \otimes (\vec{b} \cdot \vec{\sigma} + \vec{b}' \cdot \vec{\sigma}) + \vec{a}' \cdot \vec{\sigma} \otimes (\vec{b} \cdot \vec{\sigma} - \vec{b}' \cdot \vec{\sigma})|, \tag{7.33}$$

where $\vec{a}, \vec{a}', \vec{b}, \vec{b}'$ are unit real three-dimensional vectors. Let us use the notation $A = \vec{a} \cdot \vec{\sigma}, A' = \vec{a}' \cdot \vec{\sigma}, B = \vec{b} \cdot \vec{\sigma}, B' = \vec{b}' \cdot \vec{\sigma}$. For product states $\varrho_A \otimes \varrho_B \, \text{tr}\,[(A \otimes B)(\varrho_A \otimes \varrho_B)] = \text{tr}\,[A\varrho_A] \, \text{tr}\,[B\varrho_B]$ and $|\text{tr}\,[\varrho \vec{x} \cdot \vec{\sigma}]| \leq 1$ if $||\vec{x}|| = 1$. It follows that

$$\begin{aligned} \text{tr}\,[W_{\text{CHSH}}\varrho_A \otimes \varrho_B] &= 2 - |\text{tr}\,[A\varrho_A] \, \text{tr}\,[(B + B')\varrho_B] + \text{tr}\,[A'\varrho_A] \, \text{tr}\,[(B - B')\varrho_B]| \\ &\geq 0 \,. \end{aligned} \tag{7.34}$$

Due to convexity $W_{\text{CHSH}}$ is positive on all separable states, hence it is a linear entanglement witness. In order to see it is not trivial, let us consider the singlet state $\Psi_- = |\psi_-\rangle\langle\psi_-|$ with $\psi_- = \frac{1}{\sqrt{2}}(\varphi \otimes \varphi_\perp - \varphi_\perp \otimes \varphi)$. Let us note that the definition of $\psi_-$ is independent of the vector $\varphi \in \mathcal{H}_2$. Since $\langle\psi_+|\vec{x} \cdot \vec{\sigma} \otimes \vec{y} \cdot \vec{\sigma}|\psi_+\rangle = \frac{1}{2}\text{tr}\,[(\vec{x} \cdot \vec{\sigma})^*(\vec{y} \cdot \vec{\sigma})] = \vec{x} \cdot \vec{y}$ we obtain

$$\text{tr}\,[W_{\text{CHSH}}\Psi_+] = 2 - |\vec{a} \cdot (\vec{b} + \vec{b}') + \vec{a}' \cdot (\vec{b} - \vec{b}')| \,. \tag{7.35}$$

Setting

$$\vec{a} = (1, 0, 0)\,; \quad \vec{a}' = (0, 1, 0)\,; \quad \vec{b} = \frac{1}{\sqrt{2}}(1, 1, 0)\,; \quad \vec{b}' = \frac{1}{\sqrt{2}}(1, -1, 0)\,, \tag{7.36}$$

we find that $\text{tr}\,[W_{\text{SWAP}}\Psi_+] = 2 - 2\sqrt{2} < 0$. The inequality

$$\text{tr}\,[W_{\text{SWAP}}] \geq 0\,, \tag{7.37}$$

is known as *CHSH inequality* that was originally proposed as a test of the paradigm of local realism [33]. As we have seen due to entanglement the quantum systems violate this inequality.

### 7.3.2  Entanglement via not completely positive linear maps

In this section we shall see that the complete positivity constraints on physical transformations are intimately related with the concept of entanglement. Consider a positive linear transformation $\mathcal{F} : \mathcal{L}(\mathcal{H}) \to \mathcal{L}(\mathcal{H})$ defined on a system $A$, i.e., $\mathcal{F}[T] \geq O$ for all positive operators $T \in \mathcal{L}(\mathcal{H})$. If $\mathcal{F}$ is applied to a separable state $\varrho_{AB} = \sum_j p_j \varrho_A^{(j)} \otimes \varrho_B^{(j)}$ of a composite system $A + B$, we obtain

$$\varrho'_{AB} \equiv (\mathcal{F} \otimes \mathcal{I})[\varrho_{AB}] = \sum_j p_j \mathcal{F}[\varrho_A^{(j)}] \otimes \varrho_B^{(j)}\,, \tag{7.38}$$

which is positive. However, if $\varrho_{AB}$ is entangled, then the operator $(\mathcal{F} \otimes \mathcal{I})[\varrho_{AB}]$ need not be positive. Certainly, if $\mathcal{F}$ is completely positive then the positivity of $(\mathcal{F} \otimes \mathcal{I})[\varrho_{AB}]$ is guaranteed. As one recalls from Chapter 5, this was actually the reason why we introduced completely positive mappings. See for instance Example 50.

This observation leads us to an important application of positive, but not completely positive mappings in the detection of entanglement present in a composite state $\varrho_{AB}$. Namely, if the operator $\varrho'_{AB} = (\mathcal{F} \otimes \mathcal{I})[\varrho_{AB}]$ is negative, then the state $\varrho_{AB}$ is entangled. Moreover, the following proposition holds.

**Proposition 88.** A state $\varrho$ is separable if and only if for all positive (not necessarily completely positive) mappings $\mathcal{F}$ the operator $(\mathcal{F} \otimes \mathcal{I})(\varrho)$ is positive.

*Proof.* According to Choi-Jamiolkowski isomorphism each linear entanglement witness $W$ on a composite system $A$ and $B$ determines some linear mapping $\mathcal{F}_W$ via relation

$$\mathcal{F}_W(\xi) = \mathrm{tr}_B[(\xi^T \otimes I)W]. \tag{7.39}$$

We shall prove that linear witnesses and positive maps are in one-to-one correspondence. Consequently, the results achieved for linear entanglement witnesses can be translated into the language of positive maps. In particular, such relation makes this proposition completely equivalent to statement in Theorem 16. Let us remind that if $W$ is positive on all vectors, then the map $\mathcal{F}_W$ is completely positive.

Suppose $\mathcal{F}$ is a positive linear map, vectors $\varphi_1, \ldots, \varphi_d \in \mathcal{H}$ is an orthonormal basis in $\mathcal{H}_d$ and operators $e_{jk} = |\varphi_j\rangle\langle\varphi_k|$ form an orthonormal operator basis. Then

$$
\begin{aligned}
\langle\psi \otimes \varphi|W_{\mathcal{F}}|\psi \otimes \varphi\rangle &= \langle\psi \otimes \varphi|(\mathcal{F} \otimes \mathcal{I})[\Psi_+]|\psi \otimes \varphi\rangle \\
&= \langle\psi|\mathcal{F}[e_{jk}]|\psi\rangle\langle\varphi|e_{jk}|\varphi\rangle = \langle\psi|\mathcal{F}[\mathrm{tr}\,[e_{jk}P_\varphi]\,e_{jk}]|\psi\rangle \\
&= \langle\psi|\mathcal{F}[P_\varphi^T]|\psi\rangle \geq 0\,,
\end{aligned}
$$

where we used the notation $P_\varphi = |\varphi\rangle\langle\varphi|$ and the last inequality holds due to positivity of the transposition map $P \mapsto P^T$ and $\mathcal{F}$. That is, the operator $W_{\mathcal{F}}$ is positive on product vectors $\psi \otimes \varphi$.

Suppose $W$ is an entanglement witness. Then for the induced linear map $\mathcal{F}_W$ the following identity holds $\langle\psi|\mathcal{F}_W[X]|\psi\rangle = \langle\psi|\mathrm{tr}_A[W(X^T \otimes I)]|\psi\rangle$. Setting $X = P_\varphi^T$ we get the inequality

$$\langle\psi|\mathrm{tr}_A[W(P_\varphi \otimes I)]|\psi\rangle = \langle\varphi \otimes \psi|W(P_\varphi \otimes I)|\varphi \otimes \psi\rangle = \langle\varphi \otimes \psi|W|\varphi \otimes \psi\rangle \geq 0$$

proves that $\langle\psi|\mathcal{F}_W[P_\varphi^T]|\psi\rangle$ is positive. Since a general positive operator is a positive sum of one-dimensional projectors it follows that $\mathcal{F}[X] \geq O$ for all $X \geq O$.                    □

Although Proposition 88 is closely related to Theorem 16, there is one important difference between linear entanglement witnesses and positive, but not completely positive maps. The last ones determine *nonlinear* separability tests via the formula $(\mathcal{F} \otimes \mathcal{I})(\varrho) \geq O$.

Similarly like the entanglement witnesses, also the positive maps $\mathcal{F}$ divides the states space into two subsets: i) $\mathcal{S}_+^{\mathcal{F}} = \{\varrho : \mathcal{F} \otimes \mathcal{I}(\varrho) \geq O\}$, and ii) $\mathcal{S}_-^{\mathcal{F}} = \{\varrho : \mathcal{F} \otimes \mathcal{I}(\varrho) \ngeq O\}$. Let us note that the subset $\mathcal{S}_+^{\mathcal{F}}$ is convex and contains all separable states.

A typical example of positive, but not completely positive, mapping is the transposition map $\tau : \varrho \to \varrho^T$, where transposition is performed with respect to the basis $\varphi_1, \ldots, \varphi_d$ in which the maximally entangled state takes the form $\Psi_+ = \frac{1}{d}\sum_{jk}|\varphi_j\rangle\langle\varphi_k| \otimes |\varphi_j\rangle\langle\varphi_k|$. We use a shorthand notations $\tau_A[\varrho] = (\tau \otimes \mathcal{I})[\varrho]$ and $\tau_B[\varrho] = (\mathcal{I} \otimes \tau)[\varrho]$ for partially transposed operators. The corresponding entanglement witness is $W_\tau = \frac{1}{d}\sum_{jk}|\varphi_j\rangle\langle\varphi_k| \otimes |\varphi_k\rangle\langle\varphi_j| = \frac{1}{d}W_{\mathrm{SWAP}}$, where $W_{\mathrm{SWAP}}$ is the SWAP operator. The transformations $\tau_A \otimes \mathcal{I}$ and $\mathcal{I} \otimes \tau_B$ are called *partial transpositions* of subsystems $A$ and $B$, respectively.

**Definition 67.** A state $\varrho \in \mathcal{S}(\mathcal{H}_A \otimes \mathcal{H}_B)$ is called a *PPT state* (PPT is standing for positive partial transpose) if it remains positive under partial transposition. We denote the convex set of PPT states by $\mathcal{S}_{\mathrm{ppt}} = \mathcal{S}_+^\top$. If $\varrho \notin \mathcal{S}_{\mathrm{ppt}}$, then we say it is an *NPPT state*.

We can formulate the following separability test.

**Proposition 89.** (*PPT separability criteria.*) If a state $\varrho_{AB}$ is NPPT, then it is entangled.

The question is how strong is the entanglement criterion provided by the partial transposition. It turns out that except the simplest case of $2 \times 2$ and $2 \times 3$ dimensional composite systems the set of PPT entangled states is not empty, i.e., $\mathcal{S}_{\mathrm{ppt}} \cap \mathcal{S}_{\mathrm{ent}} \neq \emptyset$.

**Theorem 17.** (*Woronowicz [84]*) All positive linear maps $\mathcal{F} : \mathcal{L}(\mathbb{C}^2) \to \mathcal{L}(\mathbb{C}^2)$ and $\mathcal{F} : \mathcal{L}(\mathbb{C}^2) \to \mathcal{L}(\mathbb{C}^3)$ are of the form

$$\mathcal{F} = \mathcal{F}_{cp}^1 + \mathcal{F}_{cp}^2 \circ \tau \tag{7.40}$$

where $\mathcal{F}_{cp}^1, \mathcal{F}_{cp}^2$ are completely positive mappings.

If a positive mapping can be written in the form 7.40, it is called *decomposable*. As a direct consequence of Theorem 17 we obtain the following characterization of quantum entanglement for simplest composite systems.

**Theorem 18.** (*Perez-Horodecki criterion.*) For $\mathbb{C}^2 \otimes \mathbb{C}^2$ and $\mathbb{C}^2 \otimes \mathbb{C}^3$ systems a state is separable if and only if it is PPT.

*Proof.* According to Theorem 17 all the positive, but not completely positive, linear mappings on the considered types of systems are related to the transposition map via completely positive maps. Therefore, the operator $\varrho' = (\mathcal{F} \otimes \mathcal{I})(\varrho)$ is not positive if and only if $\varrho \notin \mathcal{S}_{\mathrm{ppt}}$. Hence, the positivity of $\varrho'$ is guaranteed by the positivity of partially transposed state $(\tau \otimes \mathcal{I})(\varrho)$. $\qquad\square$

In order to show that for larger dimensional systems the set of entangled PPT states is not empty we introduce the so-called *range criterion* for separability [49]. It is based on the following simple observation. By definition, if a state is separable then there exists a family of product states $\{\psi_j \otimes \phi_j\}$ spanning the whole range of $\varrho$. If we find that there is no family of product vectors $\{\psi_j \otimes \phi_j\}$ spanning the whole range of $\varrho$ we can conclude it is not separable, hence it is entangled.

**Proposition 90.** (*Range criterion.*) If the range of $\varrho \in \mathcal{S}(\mathcal{H}_A \otimes \mathcal{H}_B)$ is not spanned by a family of product vectors $\{\psi_j \otimes \phi_j\}$, then $\varrho$ is entangled. In particular, if the range of $\varrho$ does not contain any product vector, then the state $\varrho$ is entangled.

**Example 86.** (*Symmetric and antisymmetric states.*) Consider a composite Hilbert space $\mathcal{H}_d \otimes \mathcal{H}_d$ of two systems of the same type. A vector $\psi \in \mathcal{H}_d \otimes \mathcal{H}_d$ is called *symmetric* if it is preserved under the SWAP transformation, i.e., $V_{\mathrm{SWAP}} \psi = \psi$. If $V_{\mathrm{SWAP}} \psi = -\psi$, then the vector is called *antisymmetric*. Symmetric and antisymmetric vectors are mutually orthogonal and define mutually orthogonal subspaces of $\mathcal{H}_d \otimes \mathcal{H}_d$. In a similar way, we may define symmetric and antisymmetric states. In particular, if the range of $\varrho$ is contained in the symmetric (antisymmetric) subspace we say the state is symmetric (antisymmetric). Important property of antisymmetric vectors is that they cannot be of product form. In particular, applying the SWAP operator onto a product vector $\psi \otimes \phi$ we obtain $\phi \otimes \psi \neq -\psi \otimes \phi$. Thus, the antisymmetric subspace contains only entangled vectors. Consequently, the antisymmetric states have no product vector in their range and based on our previous observation they must be entangled.

In what follows we shall show that the range criterion and Peres-Horodecki criterion are different in a sense that there exist entangled states passing the PPT test, but still are identifiable by the range criterion. Recall from Section 2.2.3 the complex conjugate operator $J$ related to an orthonormal basis $\{\varphi_j\}$. We assume that the orthonormal basis is fixed and denote $\phi^* = J\phi$. Then $\tau[|\phi\rangle\langle\phi|] = |\phi^*\rangle\langle\phi^*|$. Therefore, if a product vector $\psi \otimes \phi$ belongs to the range of $\varrho$, then the product vector $\psi \otimes \phi^*$ belongs to the range of $\tau_B[\varrho]$.

**Definition 68.** A set of orthogonal product vectors $\{\psi_1 \otimes \varphi_1, \ldots, \psi_m \otimes \varphi_m\}$ with $m < d_A d_B$ is an *unextendible product basis* if there is no factorized vector $\phi \in \mathcal{H}_A \otimes \mathcal{H}_B$ such that $\phi \perp \psi_j \otimes \varphi_j$ for all $j$.

**Example 87.** (*Construction of PPT entangled state.*) Consider an unextendible product basis consisting of $m < d_A d_B$ vectors $\omega_j = \psi_j \otimes \varphi_j$. Define a projector $\Pi_{\text{upb}} = \sum_j |\omega_j\rangle\langle\omega_j|$ and a state $\varrho_{\text{upb}} = \frac{1}{d_1 d_2 - m}(I - \Pi_{\text{upb}})$. By definition the range of $\varrho$ does not contain any product state. The range criterion implies that this state cannot be separable. The vectors $\psi_j \otimes \varphi_j^*$ are also mutually orthogonal. Therefore $\tau_B[\Pi_{\text{upb}}] = \sum_j |\psi_j \otimes \phi_j^*\rangle\langle\psi_j \otimes \phi_j^*|$ is a projector, and consequently $I - \tau_B[\Pi_{\text{upb}}]$ is a projector, too. Therefore, $\tau_B[\varrho_{\text{upb}}] = \frac{1}{d_1 d_2 - m}(I - \tau_B[\Pi_{\text{upb}}])$ is a positive operator, hence, the state $\varrho_{\text{upb}}$ is a PPT entangled state. We can thus conclude that $\mathcal{S}_{\text{ent}} \cap \mathcal{S}_{\text{ppt}} \neq \emptyset$.

**Example 88.** (*Entanglement witness for detection of $\varrho_{\text{upb}}$.*) Unextendible product basis [8] can be used to design nondecomposable positive maps that can serve as novel tools for the entanglement detection. In particular, the projector $\Pi_{\text{upb}}$ onto elements of UPB is strictly positive on all factorized vectors, i.e. $\min_{\psi \otimes \varphi}\langle\psi \otimes \varphi|\Pi_{\text{upb}}|\psi \otimes \varphi\rangle = \epsilon > 0$, because of the unextendability of the product basis. Let us note that $\epsilon = 0$ implies that there exists a factorized vector orthogonal to all vectors in UPB, which is a contradiction. Let $P_\phi^\perp = |\phi\rangle\langle\phi|$ is a projector and $\phi$ is orthogonal to vectors in UPB, i.e. $\Pi_{\text{upb}}\phi = 0$. Since for the operator $W_{\text{upb}}^\phi = \Pi_{\text{upb}} - d\epsilon P_\phi^\perp$ we have

$$\langle\psi \otimes \varphi|W_{\text{upb}}^\phi|\psi \otimes \varphi\rangle \geq \epsilon(1 - |\langle\phi|\psi \otimes \varphi\rangle|^2) \geq 0\,, \qquad (7.41)$$

for all product vectors $\psi \otimes \varphi$, it follows that $W_{\text{upb}}^\phi$ is positive on all separable states. Moreover, the inequality

$$\text{tr}\left[\varrho_{\text{upb}}W_{\text{upb}}^\phi\right] = -\frac{\epsilon}{d_1 d_2 - m} < 0 \qquad (7.42)$$

implies it is an entanglement witness detecting the PPT entangled state $\varrho_{\text{upb}}$. Using the Choi-Jamiolkowski isomorphism we can define an indecomposable positive map providing an entanglement criteria qualitatively different from the PPT criteria.

**Exercise 52.** Show that vectors $\varphi_0 \otimes \varphi_{0-1}, \varphi_2 \otimes \varphi_{1-2}, \varphi_{0-1} \otimes \varphi_2, \varphi_{1-2} \otimes \varphi_0, \varphi_{0+1+2} \otimes \varphi_{0+1+2}$ form an unextendible product basis and verify that the state defined in the above example is indeed a PPT entangled state. Find the entanglement witness and associated positive, but not completely positive map. In the definition we used the shorthand notation to denote a superposition of orthonormal states $\varphi_0, \varphi_1, \varphi_2$ with equal absolute values of their amplitudes, but potentially different signs. For example, $\varphi_{0-1}$ stands for $\frac{1}{\sqrt{2}}(\varphi_0 - \varphi_1)$.

A linear mapping

$$\mathcal{R}(T) = \operatorname{tr}[T]\,I - T \tag{7.43}$$

is called *reduction*. Since for a positive operator $T$ we have

$$0 \leq \langle\,\psi\,|\,T\psi\,\rangle \leq \operatorname{tr}[T]\,,$$

this mapping is positive. However, it is not completely positive. Consider a bipartite state $\varrho_{AB} = \varrho_A \otimes \varrho_B + \Gamma_{AB}$. Let us note that $\varrho_A = \operatorname{tr}_B \varrho_{AB}$, $\varrho_B = \operatorname{tr}_A \varrho_{AB}$ and $\Gamma_{AB}$ is a traceless operator. Then

$$\begin{aligned}
(\mathcal{R} \otimes \mathcal{I})[\varrho_{AB}] &= (I - \varrho_A) \otimes \varrho_B + \operatorname{tr}_A \operatorname{tr}_B[\Gamma_{AB}]I \otimes \operatorname{tr}_A[\Gamma_{AB}] - \Gamma_{AB} \\
&= I \otimes \varrho_B - \varrho_{AB}\,.
\end{aligned} \tag{7.44}$$

Similarly, $(\mathcal{I} \otimes \mathcal{R})[\varrho_{AB}] = \varrho_A \otimes I - \varrho_{AB}$.

**Lemma 7.** A separable state $\varrho$ satisfies the following operator inequalities:

$$I \otimes \varrho_B - \varrho \geq O\,, \qquad I \otimes \varrho_A - \varrho \geq O\,. \tag{7.45}$$

*Proof.* We prove only that for separable states $\varrho \leq I \otimes \varrho_B$. It is straightforward that for each state $\varrho_A \in \mathcal{S}(\mathcal{H}_A)$ the inequality $\varrho_A \leq I$ holds. That is, factorized mixed states $\varrho_A \otimes \varrho_B$ are smaller than $I \otimes \varrho_B$. Consider a mixture $\varrho = \sum_j p_j \varrho_A^{(j)} \otimes \varrho_B^{(j)}$. Since $\sum_j p_j \varrho_B^{(j)} = \varrho_B$ we have $p_j \varrho_B^{(j)} \leq \varrho_B$. Consequently,

$$\varrho = \sum_j \varrho_A^{(j)} \otimes p_j \varrho_B^{(j)} \leq I \otimes \sum_j p_j \varrho_B^{(j)} \leq I \otimes \varrho_B\,. \tag{7.46}$$

<div align="right">□</div>

Based on this lemma we can formulate the following useful separability criterion:

**Proposition 91.** (*Reduction criterion*) Let $\varrho$ be a state of a composite system $A + B$. If either $I \otimes \varrho_B - \varrho \not\geq O$ or $\varrho_A \otimes I - \varrho \not\geq O$, then $\varrho$ is entangled.

*Proof.* It remains to show that this criterion is not trivial in a sense that there are entangled states for which $I \otimes \varrho_B - \varrho < O$. Consider, for instance, the maximally entangled state $\Psi_+$, for which by definition $\varrho_A = \varrho_B = \frac{1}{d}I$. In such case $(\mathcal{R} \otimes \mathcal{I})[\Psi_+] = \frac{1}{d}I \otimes I - \Psi_+$ and $\langle \psi_+|\frac{1}{d}I \otimes I - \Psi_+|\psi_+\rangle = (1 - d)/d < 0$, i.e. the operator $(\mathcal{R} \otimes \mathcal{I})[\Psi_+]$ is negative.     □

**Proposition 92.** If

$$\max_U \langle \Psi_+|(U^* \otimes I)\varrho(U \otimes I)|\Psi_+\rangle > \frac{1}{d}\,, \tag{7.47}$$

then the state $\varrho$ is entangled.

*Proof.* The entanglement witness associated with the reduction map reads

$$W_{\mathcal{R}} = (\mathcal{R} \otimes \mathcal{I})[\Psi_+] = \frac{1}{d}I - \Psi_+ \,. \tag{7.48}$$

It follows that if $\mathrm{tr}[W_{\mathcal{R}}\varrho] = \frac{1}{d} - \langle \Psi_+ | \varrho | \Psi_+ \rangle < 0$, then the state $\varrho$ is entangled. Maximizing these witnesses over all maximally entangled states we come to above inequality. □

The quantity defined on left side of the inequality in Proposition 92 is known as a *maximally entangled fraction* of the state $\varrho$ and can be used to quantify the degree of entanglement. In fact, it says how close in fidelity is the given state to some maximally entangled state.

### 7.3.3    Entanglement distillation

Most of the papers on quantum entanglement glorifies its importance in quantum information theory and highlights its conceptual position within the quantum theory itself. Indeed, particular entangled states exhibit very interesting features and they deserve a special attention of mathematician, physicists and computer scientists. However, it is fair to say that we are lacking an operationally clear understanding of the difference between separable and entangled states.

One class of extremely interesting states are the so-called maximally entangled states, i.e. states locally unitary equivalent to $\Psi_+$. Not only they relate the structures of linear maps and linear operators (via Choi-Jamiolkowski isomorphism), but because of their unique properties, they seem to provide us with a very rich information resource suitable for information processing. Probably the most prominent example is a procedure for *quantum state teleportation*. The goal of this communication protocol is to transfer a quantum state from side $A$ to side $B$ just by classical communication. A classical solution would be based on estimation and transmission of parameters of the unknown state, but the precision of such state transfer is limited by the accuracy of our estimation. The simplest quantum solution would be to send the system in an unknown state directly through an ideal quantum channel between the sides $A$ and $B$. However, in this case the transmission is not based on classical information exchange as it is in the definition of the problem. Moreover, the accuracy depends on our abilities to reduce the communication noise and preserve the ideality of the channel. Surprisingly, quantum theory offers one more solution based on peculiar properties of maximally entangled states.

**Example 89.** (*Quantum state teleportation* [6]) Consider a pure state of three systems of the same dimension. Two systems are in possession of Alice and one is in possession of Bob. We assume that Alice and Bob share a maximally entangled state $\Psi_+$. The total state of all three systems of the same dimension is $\Omega_{A'AB} = \varrho_{A'} \otimes \Psi_+^{AB}$, where $\varrho_A$ is a state that is going to be transmitted from Alice to Bob. The success of the state teleportation is based on the following identity. For all unitary transformations $U : \mathcal{H} \to \mathcal{H}$ and states $\varrho \in \mathcal{S}(\mathcal{H})$

$$\mathrm{tr}_{A'A}\left[(\Psi_U^{A'A} \otimes I)(\varrho_{A'} \otimes \Psi_+^{AB})(\Psi_U^{A'A} \otimes I)\right] = \frac{1}{d^2}U^*\varrho_B U \,, \tag{7.49}$$

where $\Psi_U = (U \otimes I)\Psi_+(U^* \otimes I)$, $\Psi_+ = \frac{1}{d}\sum_{jk}|\varphi_j \otimes \varphi_j\rangle\langle\varphi_k \otimes \varphi_k|$ is projection onto the maximally entangled state, and $\varrho_A = \varrho_B$. This identity is interpreted as a postselective state transformation of the system $B$ in the measurement process described by Lüder's state transformer associated with an effect $\Psi_U = \Psi_U^* = \Psi_U^2$. This effect can be observed with the

probability $p = \langle \Omega_{A'AB} | \Psi_U^{A'A} \otimes I | \Omega_{A'AB} \rangle = \frac{1}{d^2}$. Consequently, if Alice performs a local measurement observing the effect $\Psi_U$, then Bob's system is described by a unitarily transformed state $\varrho$. This fact is hidden to Bob unless this information is communicated from Alice. After receiving the information that Alice observed the wanted outcome Bob can undo the unwanted unitary transformation and retrieve the original state $\varrho$.

Luckily enough there do exists an orthonormal basis of $\mathcal{H} \otimes \mathcal{H}$ composed of maximally entangled states, hence the Alice's measurement can be composed of projections $\Psi_{U_1}, \ldots, \Psi_{U_{d^2}}$ such that $\sum_j \Psi_{U_j} = I \otimes I$. The identity

$$\langle \Psi_+ | I \otimes X | \Psi_+ \rangle = \frac{1}{d} \mathrm{tr} X \tag{7.50}$$

implies that orthogonality of $\langle \Psi_U | \Psi_V \rangle = 0$ is equivalent to orthogonality of operators $U, V$ in the Hilbert-Schmidt sense, i.e. $\mathrm{tr} U^* V = 0$. That is, the projections $\Psi_{U_j}$ forming Alice's projective measurement are associated with a orthogonal operator basis consisting of unitary transformations $U_1, \ldots, U_{d^2}$. It follows that for each outcome corresponding to some projection $\Psi_{U_j}$ Bob can perform a unitary transformation $U_j$ to retrieve the original state $\varrho$ on his side. This procedure for deterministic and perfect transmission of quantum states from Alice to Bob (just by using classical communication) is called quantum teleportation. In each run Alice is sending to Bob only $2d$ bits of classical information that themselves contain no information about the state, i.e. the sequence of outcomes is completely random. The teleportation procedure is an example of LOCC channel implementing the transformation $\varrho_{A'} \otimes \Psi_+^{AB} \mapsto \Psi_+^{A'A} \otimes \varrho_B$ providing that the local correction channel $U$ is applied also on Alice's system $A'$.

**Exercise 53.** Prove the identity in Eq. (7.49).

For perfect quantum teleportation it is crucial that the state $\Psi_+$ shared by Alice and Bob is maximally entangled. It follows from the definition of maximally entangled states that there is no LOCC channel transforming in a deterministic manner a state $\omega$ into $\Psi_+$. However, there can be an LOCC channel acting on $n$ copies of $\omega$ that asymptotically maps the state $\omega^{\otimes n}$ into $m < n$ (approximate) copies of $\Psi_+$ in a deterministic way.

**Definition 69.** Consider a sequence of LOCC channels $\mathcal{F}_1, \mathcal{F}_2, \ldots, \mathcal{F}_n$ such that each $\mathcal{F}_j$ acts on $j$ copies of bipartite states $\omega \in \mathcal{S}(\mathcal{H} \otimes \mathcal{H})$, respectively. We say that $\{\mathcal{F}_n\}_n$ is an *entanglement distillation protocol* for a state $\omega_0$ if and only if

$$\lim_{n \to \infty} ||\mathrm{tr}_{n \setminus m}[\mathcal{F}_n[\omega_0^{\otimes n}]] - \Psi_+^{\otimes m}||_{\mathrm{tr}} = 0, \tag{7.51}$$

where $\mathrm{tr}_{n \setminus m}$ stands for partial trace over $n - m$ pairs. The limiting fraction $m/n$ is called an *entanglement distillation rate*.

The entanglement distillation is a very complex task. For a general overview we refer to already mentioned review papers on entanglement theory [50] and [71]. Via the following proposition the entanglement distillation gives an interesting operational meaning to the concept of PPT entanglement criterion.

**Proposition 93.** If $\varrho \in \mathcal{S}_{\mathrm{ppt}}$, i.e. $\tau_B[\varrho] \geq O$, then it cannot be distilled, i.e. the entanglement distillation rate is zero.

*Proof.* Any LOCC channel can be written as $\varrho \mapsto \varrho' = \mathcal{F}_{\mathrm{LOCC}}[\varrho] = \sum_j A_j \otimes B_j \varrho A_j^* \otimes B_j^*$. We will show that *PPT property* is LOCC invariant, i.e. $\tau_B[\varrho] \geq O$ implies that $\tau_B[(\mathcal{F}_{\mathrm{LOCC}}[\varrho])] \geq O$. For the partial transposition of the subsystem $B$ the following identity holds $\tau_B[(A \otimes B)\varrho(C \otimes D)] = (A \otimes D^T)\tau_B[\varrho](C \otimes B^T)$. It follows that $\tau_B[\varrho'] = \sum_j A_j \otimes (B_j^*)^T \tau_B[\varrho]A_j^* \otimes B_j^T$, and it is straightforward to see that the composition of an LOCC channel with the partial transposition map results in a completely positive map. Moreover, since $\tau_B[\varrho] \geq O$, then necessarily also $\tau_B[\varrho'] \geq O$. It follows that the maximally entangled state cannot be distilled out of any PPT entangled state, because $\tau_B[\varrho^{\otimes n}] = (\tau_B[\varrho])^{\otimes n}$. □

**Definition 70.** A quantum state $\varrho \in \mathcal{S}(\mathcal{H} \otimes \mathcal{H})$ is called *bound entangled* if and only if it cannot be distilled. Let us denote by $\mathcal{S}_{\mathrm{bound}}$ the subset of bounded entangled states.

The existence of bound entangled states is indeed a strange feature of quantum entanglement that attracted lot of attention. One of the main unsolved problems of the entanglement theory is the existence, or nonexistence of so-called *NPPT bound entangled states*, i.e. the existence of states that cannot be distilled, but can be identified by means of Peres-Horodecki PPT criterion.

The following proposition provides a partial answer to another interesting problem related to the question how much entangled the bound entangled states are.

**Proposition 94.** For PPT states

$$\max_U \langle \Psi_+ | U^* \otimes I \varrho U \otimes I | \Psi_+ \rangle \leq 1/d \,. \tag{7.52}$$

*Proof.* The adjoint map $\tau^*$ to the transposition map $\tau$ is again the transposition map, i.e. $\mathrm{tr}\,[\tau[A^*]B] = \mathrm{tr}\,[A^*\tau[B]]$ and $\tau^* = \tau$. This follows from the fact that $\mathrm{tr}\,[X] = \mathrm{tr}\,[X^T]$. Setting $X = A^*B$ we get the identity $\mathrm{tr}\,[\tau[A^*]B] = \mathrm{tr}\,[\tau[\tau[A^*]B]] = \mathrm{tr}\,[\tau[B](\tau \circ \tau)[A^*]] = \mathrm{tr}\,[A^*\tau[B]]$. Using the identity $\tau \circ \tau = \mathcal{I}$ we obtain $\mathrm{tr}\,[\varrho\Psi_U] = \mathrm{tr}\,[(\tau \otimes \mathcal{I})[\varrho](\tau^* \otimes \mathcal{I})[\Psi_U]]$, where $\Psi_U = (I \otimes U)\Psi_+(I \otimes U^*)$. Let us remind that $(\tau \otimes \mathcal{I})[\Psi_+] = \frac{1}{d}V_{\mathrm{SWAP}}$, hence

$$\mathrm{tr}\,[\varrho\Psi_U] = \mathrm{tr}\,[\tau_A[\varrho](I \otimes U)\tau_A[\Psi_+](I \otimes U^*)] = \frac{1}{d}\mathrm{tr}\,[\tau_A[\varrho](I \otimes U)V_{\mathrm{SWAP}}(I \otimes U^*)] \,. \tag{7.53}$$

The operator $(I \otimes U)V_{\mathrm{SWAP}}(I \otimes U^*)$ is selfadjoint with eigenvalues $\pm 1$. The mean value of a selfadjoint operator in arbitrary state is bounded by its largest and smallest eigenvalue. If the state $\varrho$ is a PPT state, then the operator $\tau_A[\varrho]$ is also a valid quantum state. Therefore, for PPT states we find that $\mathrm{tr}\,[\varrho\Psi_U] \leq 1/d$ for all unitary operators $U$, hence also the maximum over $U$ is bounded by $1/d$. □

It follows that for PPT entangled states their maximally entangled fraction is always smaller than $1/d$. Moreover, comparing this proposition with the Proposition 92 we can conclude that the reduction criterion is weaker than PPT criterion.

## 7.4   Example: Werner states

In this section we shall investigate the properties of a one-parametric set of *Werner states*, introduced by R. Werner in [83]. In some sense their properties are generic, which makes this class of states important from the point of view of entanglement theory.

**Definition 71.** A state $\varrho \in \mathcal{S}(\mathcal{H}_d \otimes \mathcal{H}_d)$ is called *Werner state* if $[\varrho, U \otimes U] = 0$ for all unitary operators $U : \mathcal{H}_d \to \mathcal{H}_d$.

It follows that

$$\varrho_\mu = \mu \frac{1}{d_+} P_+ + (1 - \mu) \frac{1}{d_-} P_- \,, \tag{7.54}$$

where $P_\pm$ are projectors onto symmetric and antisymmetric subspaces of $\mathcal{H}_d \otimes \mathcal{H}_d$ (see Example 81) and $d_\pm = \dim P_\pm = d(d \pm 1)/2$. That is, the Werner states $\varrho_\mu$ are convex combinations of mixed states $d_\pm^{-1} P_\pm$.

We shall use the same notation as in Example 81. As usually, let $\varphi_1, \ldots, \varphi_d$ be an orthonormal basis of $\mathcal{H}_d$. The vectors $\varphi_{j \pm k}$ form an orthonormal basis of $\mathcal{H} \otimes \mathcal{H}$, too. Let us note that by definition $\varphi_{j-j}$ is the zero vector for all $j$. That is, we have $d(d+1)/2$ symmetric vectors $\varphi_{j+k}$ and $d(d-1)/2$ nonzero antisymmetric vectors $\varphi_{j-k}$ forming bases of symmetric and antisymmetric subspaces, respectively. Therefore,

$$P_\pm = \sum_{j,k=1}^{d} |\varphi_{j\pm k}\rangle\langle\varphi_{j\pm k}| = \frac{1}{2}(I \pm V_{\text{SWAP}}) \,. \tag{7.55}$$

Using the operators $I$ and $V_{\text{SWAP}}$ the Werner states can be expressed as

$$\begin{aligned}
\varrho_\mu &= \frac{\mu}{d(d+1)}(I + V_{\text{SWAP}}) + \frac{1 - \mu}{d(d-1)}(I - V_{\text{SWAP}}) \\
&= \frac{1}{d(d^2 - 1)}\left[(d + 1 - 2\mu)I + (2d\mu - d - 1)V_{\text{SWAP}}\right] \,.
\end{aligned}$$

**Proposition 95.** A Werner state $\varrho_\mu$ is entangled if and only if $\mu < 1/2$.

*Proof.* Let us start with the observation that the functional $\text{tr}\,[V_{\text{SWAP}}\omega]$ is invariant under the action of the twirling channel $\mathcal{T}(\omega) = \int_{U(d)} dU\, U \otimes U \omega U^* \otimes U^*$ described in Example 81, i.e. $\text{tr}\,[V_{\text{SWAP}}\mathcal{T}(\omega)] = \text{tr}\,[V_{\text{SWAP}}\omega]$ for all states $\omega$. In fact, as a result of the twirling channel arbitrary state $\omega$ is mapped into some Werner state $\varrho_\mu$ with some specific value of $\mu$. Therefore, as the twirling is an LOCC channel, if $\omega$ is separable, then also the corresponding Werner state $\varrho_\mu = \mathcal{T}(\omega)$ must be separable. Consequently, since for pure product states $\text{tr}\,[V_{\text{SWAP}}(|\varphi \otimes \phi\rangle\langle\varphi \otimes \phi|)] = |\langle\varphi\,|\,\phi\rangle|^2 \in [0, 1]$ it follows that for each Werner state $\varrho_\mu$ with $\text{tr}\,[V_{\text{SWAP}}\varrho_\mu] \geq 0$ there exist a separable state $\omega_0 = |\varphi \otimes \phi\rangle\langle\varphi \otimes \phi|$ such that $\varrho_\mu = \mathcal{T}[|\varphi \otimes \phi\rangle\langle\varphi \otimes \phi|]$. Therefore, if $\text{tr}\,[V_{\text{SWAP}}\varrho_\mu] \geq 0$, then the Werner state $\varrho_\mu$ must be separable.

Let us note that the selfadjoint operator $V_{\text{SWAP}}$ is proportional to the entanglement witness associated with the partial transposition criterion. Therefore, if $\text{tr}\,[V_{\text{SWAP}}\omega] < 0$, then the state $\omega$ is entangled. By definition $V_{\text{SWAP}}\psi = \pm\psi$ if $\psi$ is symmetric, or antisymmetric, respectively. Therefore, $\text{tr}\,[V_{\text{SWAP}}P_\pm] = \pm d_\pm$ and for Werner states

$$\text{tr}\,[V_{\text{SWAP}}\varrho_\mu] = \mu \frac{\text{tr}\,[V_{\text{SWAP}}P_+]}{d_+} + (1 - \mu) \frac{\text{tr}\,[V_{\text{SWAP}}P_-]}{d_-} = 2\mu - 1 \,. \tag{7.56}$$

That is, for values $\mu \geq 1/2$ the Werner states are separable and for $\mu < 1/2$ the Werner states are entangled.                                                                                      $\square$

**Proposition 96.** A Werner state is PPT if and only if $\mu \geq 1/2$.

*Proof.* Applying a partial transposition on $V_{\text{SWAP}}$ we obtain $\tau_A[V_{\text{SWAP}}] = \tau_A[\sum |\varphi_j \otimes \varphi_k\rangle \langle \varphi_k \otimes \varphi_j|] = \sum |\varphi_k \otimes \varphi_k\rangle \langle \varphi_j \otimes \varphi_j| = d\Psi_+$. For a general vector $\psi$ the mean value is between largest and smallest eigenvalue of the $\tau_A[\varrho_\mu]$. For arbitrary $\psi \perp \psi_+$ the mean value $\langle \psi | \tau_A[\varrho_\mu] | \psi \rangle = \frac{(d+1-2\mu)}{d(d^2-1)} \geq 0$, because $d > 1$. The vector $\psi_+$ is an eigenvector of $\tau_A[\varrho_\mu]$ associated with a potentially negative eigenvalue. A direct calculation gives

$$
\begin{aligned}
\langle \psi_+ | \tau_A[\varrho_\mu] | \psi_+ \rangle &= \frac{1}{d(d^2-1)} \langle \psi_+ | \left[ (d+1-2\mu)I + (2d\mu - d - 1)d\Psi_+ \right] | \psi_+ \rangle \\
&= \frac{1}{d(d^2-1)} [(2\mu-1)(d^2-1)] = \frac{2\mu-1}{d}\,.
\end{aligned}
$$

It follows that the partial transposition of a Werner state results in a positive operator only if $\mu \geq 1/2$. $\qquad\square$

**Proposition 97.** The reduction criterion does not detect the entanglement of Werner states for $d \geq 3$. For $d = 2$ the reduction criterion identifies that Werner states are entangled for $\mu < 1/2$.

*Proof.* Since $\text{tr}_A[I \otimes I] = \text{tr}_B[I \otimes I] = dI$ and $\text{tr}_A[V_{\text{SWAP}}] = \text{tr}_B[V_{\text{SWAP}}] = I$ we obtain for Werner states

$$
\text{tr}_A \varrho_\mu = \text{tr}_B \varrho_\mu = \frac{d^2 + d - 2\mu d + 2\mu d - d - 1}{d(d^2-1)}\, I = \frac{1}{d}\, I\,. \tag{7.57}
$$

The reduction criterion requires

$$
O \leq \frac{1}{d} I \otimes I - \varrho_\mu = \frac{1}{d(d^2-1)} \left[ ((d+1)(d-2) + 2\mu)I - (2d\mu - d - 1)V_{\text{SWAP}} \right] \tag{7.58}
$$

Since the eigenvalues of $V_{\text{SWAP}}$ are $\pm 1$ it follows that eigenvalues of the operator $\frac{1}{d} I \otimes I - \varrho_\mu$ are

$$
\begin{aligned}
x_\pm &= (d+1)(d-2) + 2\mu \mp (2d\mu - d - 1) = (d+1)(d-2) + 2\mu \mp 2d\mu \pm (d+1) \\
&= (d+1)(d - 2 \pm 1) + 2\mu(1 \mp d)\,.
\end{aligned}
$$

They are negative, hence the reduction criterion is violated, only if $d = 2$ and $\mu < 1/2$ as it is stated in the Proposition. $\qquad\square$

The importance of Werner states is based on the fact that an application of the twirling channel $\mathcal{T}$ to arbitrary state $\varrho$ results in a particular Werner state with some value of $\mu$. Since this channel is LOCC the state $\mathcal{T}[\varrho] \leq_{\text{LOCC}} \varrho$. Consequently, the distillability of states can be studied via distilability of Werner states. In particular, the question on the existence of bound NPPT states can be reduced the existence of bound NPPT Werner states in the following sense. If all entangled Werner states are distillable, then arbitrary NPPT entangled state is.

## Acknowledgment

Authors would like to express their thanks to Vladimír Bužek for his encouragement and also to students attending the lectures for their stimulating questions and comments.

## Appendix A: Mathematical preliminaries

This appendix is a very short refresher on some mathematical topics that are of use in quantum theory. By no means it is a complete list of all necessary mathematical tools. In what follows we shall review the definitions, basic results and provide some illustrative examples on mathematical concepts of relations and convexity. In the era of internet a common reference to some topics discussed here is the free encyclopedia - wikipedia.

### Relations

### Equivalence relation

We recall that a *relation* on a set $\Omega$ is a subset $R$ of $\Omega \times \Omega$. If $R$ is a relation on $\Omega$, we use $xRy$ to denote that $(x, y) \in R$.

An *equivalence relation* on $\Omega$ is a relation $R$ on $\Omega$ such that

- $xRx$ for every $x \in \Omega$     (reflexive);

- if $xRy$ and $yRz$, then $xRz$     (transitive);

- if $xRy$, then $yRx$     (symmetric).

We usually denote an equivalence relation by $\sim$, or some variant of this symbol containing subscripts. The *equivalence class* of an element $x$ is the set $[x] \equiv \{y \in \Omega \mid x \sim y\}$.

A collection $\mathcal{P}$ of disjoint, nonempty sets $X \subseteq \Omega$ is a *partition* of $\Omega$ if $\cup_{X \in \mathcal{P}} X = \Omega$. The equivalence classes of an equivalence relation $\sim$ form a partition, which we denote $\mathcal{P}^{\sim}$. Also, for each partition $\mathcal{P}$ of $\Omega$, we get an equivalence relation $\sim^{\mathcal{P}}$ by defining

$$x \sim^{\mathcal{P}} y \quad \text{iff} \quad x \text{ and } y \text{ belong to a same member of } \mathcal{P}. \tag{7.1}$$

If $\sim$ is an equivalence relation and we form the corresponding partition $\mathcal{P}^{\sim}$, then $\mathcal{P}^{\sim}$ determines $\sim$ through (7.1). Similarly, starting with a partition $\mathcal{P}$, the equivalence classes related to $\sim^{\mathcal{P}}$ are just the sets belonging to $\mathcal{P}$. In conclusion, equivalence relations and partitions are two different descriptions of the same structure.

**Example 90.** A *transformation group* on a set $\Omega$ is a collection $\mathcal{G}$ of bijective mappings $f : \Omega \to \Omega$ which is a group under compositions. This means that

(a) the identity mapping $1_\Omega$ belongs to $\mathcal{G}$;

(b) if $f_1, f_2 \in \mathcal{G}$, then the composition mapping $f_1 \circ f_2$ belongs $\mathcal{G}$;

(c) for each $f \in \mathcal{G}$, the inverse mapping $f^{-1}$ belongs to $\mathcal{G}$.

If $\mathcal{G}$ is a transformation group on $\Omega$, we define a relation $\sim_{\mathcal{G}}$ on $\Omega$ in the following way:

$$x \sim_{\mathcal{G}} y \text{ iff there exists } f \in \mathcal{G} \text{ such that } f(x) = y. \tag{7.2}$$

Using the properties (a)-(c), it is straightforward to verify that $\sim_{\mathcal{G}}$ is an equivalence relation. Thus, each transformation group defines an equivalence relation.

Every equivalence relation is of the form $\sim_{\mathcal{G}}$ for some transformation group $\mathcal{G}$. Indeed, let $\sim$ be an equivalence relation on $\Omega$. We define $\mathcal{G}_{\sim}$ to be the following collection of mappings:

$$\mathcal{G}_{\sim} := \{f : \Omega \to \Omega | f \text{ is a bijective and } f(x) \sim x \ \forall x \in \Omega\} \,.$$

Since $\sim$ is reflexive, transitive and symmetric, it follows that $\mathcal{G}_{\sim}$ is a group under compositions. The group $\mathcal{G}_{\sim}$ determines the original equivalence relation $\sim$ through (7.2).

### Partial order

A *partial order* on $\Omega$ is a relation $R$ on $\Omega$ which is reflexive, transitive and also has the following property:

- if $xRy$ and $yRx$, then $x = y$    (antisymmetric).

We usually denote a partial order by $\preccurlyeq$ or $\leq$, or some variants. A set with a partial order on it is a *partially ordered set*, or *poset* for short.

An element $x \in \Omega$ is

- *maximal* if $x \preccurlyeq y$ implies that $x = y$;

- *minimal* if $y \preccurlyeq x$ implies that $x = y$.

### Preordering

A relation $R$ on $\Omega$ is a *preorder* if it is reflexive and transitive. Hence, both equivalence relation and partial order are special cases of preorders.

Let $R$ be a preordering and denote

$$x \sim y \quad \text{iff} \quad xRy \text{ and } yRx \,.$$

Then $\sim$ is an equivalence relation. We denote by $[x]$ the equivalence class of an element $x$, and define

$$[x] \preccurlyeq [y] \quad \text{iff} \quad xRy \,.$$

In this way, we get a partial order $\preccurlyeq$ on the set $\mathcal{P}^{\sim}$ of all equivalence classes.

### Convex sets

A mahematical feature reflecting the statistical description of physical world is called convexity. In this section we shall introduce the basic concepts and properties of convex sets.

**Definition 72.** (*Convex set.*) A subset $S \subset V$ ($V$ is real linear space) is *convex* if for any $x, y \in S$ also the elements $\lambda x + (1 - \lambda)y \in S$ for all $\lambda \in [0, 1]$.

**Example 91.** (*Convex sets.*)

1. *Segment of a line* = $\{\vec{r} \in \mathbb{R}^n : \vec{r} = t\vec{r}_1 + (1-t)\vec{r}_2 \text{ where } \vec{r}_1, \vec{r}_2 \in \mathbb{R}^n \text{ and } t \in [0,1]\}$

2. *Square* = $\{\vec{r} \in \mathbb{R}^2 : |x| \leq 1, |y| \leq 1\}$

3. *Sphere* = $\{\vec{r} \in \mathbb{R}^n : |\vec{r}|^2 \leq 1\}$

4. *Set of probability distributions* = $\Sigma_n = \{\vec{p} \in \mathbb{R}_+^n : \sum_{j=1}^n p_j = 1\}$ is a so-called *probability simplex*. In particular, $\Sigma_2$ is a segmet of line, $\Sigma_3$ is a triangle, $\Sigma_4$ is tetrahedron, etc.

**Definition 73.** A set $S \subset V$ is called a *convex cone* if it is closed under multiplication by positive scalars ($ta \in S$ iff $a \in S$ for all $t \geq 0$) and under linear combinations, i.e. $a + b \in S$ iff $a, b \in S$.

**Example 92.** (*Convex cones.*)

1. *Positive semi-line*=$\{\vec{r} \in \mathbb{R}^n : \vec{r} = t\vec{r}_0 \text{ where } \vec{r}_0 \in \mathbb{R}^n \text{ and } t \in \mathbb{R}_+\}$

2. *Positive vectors*=$\{\vec{r} \in \mathbb{R}^n : r_j \geq 0 \quad \forall j\}$

3. *Convex angle*=$\{\vec{r} \in \mathbb{R}^n : t\vec{r}_1 + s\vec{r}_2, s, t \geq 0\}$

4. *Set of positive operators*=$\mathcal{L}_+(\mathcal{H}) = \{A : \mathcal{H} \to \mathcal{H}, \quad A \geq 0\}$

**Definition 74.** (*Convex hull.*) Let $A$ be a set of points in real vector space $V$. *Convex hull* of $A$ (denoted as $\mathrm{co}(A)$) is a minimal convex set containing $A$. Equivalently, convex hull is an intersection of all convex sets containing $A$, i.e. $\mathrm{co}(A) = \bigcap_{X : A \subset X} X$.

**Definition 75.** (*Affine (linear) combination.*) A linear combination $\sum_j a_j x_j$ is called *affine* if $\sum_j a_j = 1$. Note that $a_j$ are arbitrary real numbers.

**Definition 76.** (*Convex (linear) combination.*) A linear combination $\sum_j a_j x_j$ is called *convex* if $\sum_j a_j = 1$ and $a_j \geq 0$ for all $j$.

**Definition 77.** (*Affine linear independence.*) Set of vectors $A = \{x_1, \ldots, x_m\}$ is affinely dependent if $\sum_j a_j x_j = 0$ for some affine combination $\{a_j\}$. We say the vectors in $A$ are affinely inpenent if none of them can be expressed as affine combination of others.

Similarly, like the linear transformations of vector spaces are preserving the linear structure of vector spaces, the so-called *affine mappings* are all the transformations preserving the convex structure. That is, under affine transformations the convex sets are trasnformed into convex sets.

**Definition 78.** (*Affine map.*) We say that a mapping $F : S \to V$ ($S \subset V$ is a convex subset of some real vector space $V$) is affine if and only if for all affine linear combinations $x = \sum_j p_j x_j \in S$ the following identity holds $F(x) = \sum_j p_j F(x_j)$. It follows that linear maps are affine and the set of affine maps form a group.

A typical example of a transformation that is not linear, but still affine, is the so-called *vector translation* $T_{\vec{t}} : \mathbb{R}^n \to \mathbb{R}^n$, i.e. $\vec{r} \mapsto \vec{r}' = \vec{t} + \vec{r}$. A linear combination $\vec{r}_1 + \vec{r}_2$ is mapped into $T_{\vec{t}}\vec{r}_1 + T_{\vec{t}}\vec{r}_2 = 2\vec{t} + \vec{r}_1 + \vec{r}_2 \neq \vec{t} + \vec{r}_1 + \vec{r}_2 = T_{\vec{t}}(\vec{r}_1 + \vec{r}_2)$, hence $T_{\vec{t}}$ is not a linear transformation. However, applying the translation to an affine linear combination $\vec{r} = \sum_j a_j \vec{r}_j$ (with $\sum_j a_j = 1$) we obtain $\sum_j a_j(\vec{t} + \vec{r}_j) = \vec{t} \sum_j a_j + \sum_j a_j \vec{r}_j = \vec{t} + \vec{r}$. That is, $T_{\vec{t}}$ is the affine transformation.

The following theorem is of use in various applications. Since it is not used in the main text we shall omit the proof.

**Theorem 19.** (*Carathéodory's theorem.*) If a point $x \in \mathbb{R}^n$ lies in a convex hull of set $A$, then there is a subset $A' \subset A$ consisting of at most $n + 1$ points such that $x$ lies in the convex hull of $A'$.

**Definition 79.** (*Extremal point*) An element $x \in S$ of a convex set $S$ is called *extremal* if and only if $x = \frac{1}{2}(x_1 + x_2)$ implies $x_1 = x_2 = x$. The set of extreme points of a set $S$ will be denoted by $S_{\text{ext}}$.

**Example 93.** (*Convex sets without extremal points.*)

1. Any linear space (for example line, plane) is convex, but without extremal points.

2. Set $X$ of strictly positive functions $f : [0, 1] \to \mathbb{R}$ is obviously convex. Define a function $f_\mu \equiv \mu f$ for positive $\mu$. It is easy to check that $f(x) = \frac{1}{2}[f_\mu(x) + f_{(2-\mu)}(x)]$, i.e. each element of $X$ can be written as convex combination of other two elements.

Previous examples show that not all convex sets have some extremal points, i.e. it is possible that $S_{\text{ext}} = \emptyset$. However, if the set of extremal points is nont empty, it is of interest whether the convex hull of extremal points, or its closure, coincides with the whole convex set, or not. Since $S_{\text{ext}} \subset S \subset \mathbb{R}^n$, then according to Carathéodory's theorem each element of $\mathrm{co}(S_{\text{ext}})$ can be written as a convex combination of at most $n + 1$ extremal points. Consequently, if $\mathrm{co}(S_{\text{ext}}) = S$, then each point in $S$ can be expressed as a convex combination of at most $n + 1$ extremal points of $S$. The following theorem is roughly saying that this is the case providing that the convex sets are compact.

**Theorem 20.** (*Krein-Milman theorem.*) Every compact convex set $S$ in a locally convex space is the closed convex hull of its extreme points, i.e. $S = \overline{\mathrm{co}(S_{\text{ext}})}$

**Example 94.** (*Simplex.*) A *simplex* is a convex hull of $n + 1$ affinely independent points $e_1, \ldots, e_{n+1}$ in $n$-dimensional real vector space. It is called *regular* if $\|e_j - e_k\| = const$ for all $j \neq k$. The defining points $e_1, \ldots, e_{n+1}$ are the only extremal points of the simplex. By definition each element of the simplex can be expressed in a unique way via its extremal points, i.e. $x = \sum_j x_j e_j$ ($x_j \geq 0, \sum_j x_j = 1$). If $\sum_j x_j e_j = \sum_j a'_j e_j$ for $a_j \neq a'_j$, then it would mean that the points $e_1, \ldots, e_{n+1}$ are not affinely independent.

**Example 95.** (*Extreme points of a sphere $S^{n-1}$.*) Our task is the following: Show that an element $x \in S^{n-1}$ is an extreme point of the sphere $S^{n-1} = \{x \in \mathbb{R}^n : \|x\| \leq 1\}$ if and only if $\|x\| = 1$ and that any element $x \in S^{n-1}$ can be written as a convex combination of at most two extreme points irrespective on the dimension $n$.

We will prove something more general. Consider a unit sphere ($\|x\| \leq 1$) in a normed space. First we will show that normalized elements ($\|x\| = 1$) are extremal points of this such sphere. Assume $x = \frac{1}{2}(x_1 + x_2)$. The property of norm ($\|a + b\| \leq \|a\| + \|b\|$) implies

$$1 = \|x\| = \frac{1}{2}\|x_1 + x_2\| \leq \frac{1}{2}(\|x_1\| + \|x_2\|) \leq 1, \tag{7.3}$$

because $\|x_j\| \leq 1$ (points in the sphere). The last inequality is saturated if and only if $\|x_1\| = \|x_2\| = 1$. In order to prove that $x_1 = x_2 = x$ we will use the rectangular identity $\|a - b\|^2 +$ $\|a + b\|^2 = 2(\|a\|^2 + \|b\|^2)$ (hence we assume that the norm is induced by a scalar product!) by replacing $a \to x_1/2$ and $b \to x_2/2$. We obtain

$$\frac{1}{4}\|x_1 - x_2\|^2 + \left\|\frac{1}{2}(x_1 + x_2)\right\|^2 \quad = \quad \frac{1}{2}(\|x_1\|^2 + \|x_2\|^2) \tag{7.4}$$

$$\frac{1}{4}\|x_1 - x_2\|^2 + 1 \quad = \quad 1\,, \tag{7.5}$$

hence necessarily $\|x_1 - x_2\| = 0$ which implies that $x_1 = x_2$ and it is straightforward to see that also $x_1 = x_2 = x$. Extremality of boundary points of sphere is proved providing that the norm is generated by a scalar product, which is the case for the sphere $S^{n-1}$.

In order to prove that these are the only extremal points we shall show that arbitrary point can be expressed as a convex combination of two points from the boundary. Consider $x$ and define $x_1 = x/\|x\|$ and $x_2 = -x/\|x\|$. These points belong to boundary, hence, they are the extremal points. It is easy to see that for arbitrary $x$ the following convex decomposition holds $x = \frac{1+\|x\|}{2}x_1 + \frac{1-\|x\|}{2}x_2$. Therefore, only the elements from boundary are extremal and each nonextremal point can be written as a convex combination of two extremal points.

In conclusion, let us note that for arbitrary normed space (also not induced by some scalar product) the extremal points of a unit ball must belong to a topological boundary (in norm topology), i.e. if $x \in S_{\text{ext}}$, then $\|x\| = 1$. However, the inverse is not necessarily true. Consider for example a norm given by the maximal component, i.e. $|x| = \max_j |x_j|$. Such norm is not generated by a scalar product, i.e. the rectangular identity does not hold. The unit sphere form a cube in $\mathbb{R}^n$ and extremal points of the cube are only its corners, i.e. not arbitrary vector $|x| = 1$ is extremal.

## Convex functions

**Definition 80.** (*Convex funtion.*) A function $f : \mathbb{R}^n \to \mathbb{R} \cup \{\infty\}$ is said to be convex if

$$\lambda f(x) + (1 - \lambda)f(y) \geq f(\lambda x + (1 - \lambda)y) \tag{7.6}$$

for all $0 \leq \lambda \leq 1$.

**Definition 81.** (*Concave function.*) Function $f$ is concave if and only if $-f$ is convex, i.e.

$$\lambda f(x) + (1 - \lambda)f(y) \leq f(\lambda x + (1 - \lambda)y) \tag{7.7}$$

**Proposition 98.** (*Concave function on a convex set.*) Let $S$ be convex and closed set with at least one extreme point. A concave function $f$ attains its minimum at some extreme point of $S$.

## Appendix B: List of Symbols

### List of Hilbert space operators

- $\mathcal{L}(\mathcal{H})$ ... set of bounded linear operators

- $\mathcal{L}_s(\mathcal{H})$ ... set of bounded selfadjoint operators ($T \in \mathcal{L}(\mathcal{H})$ and $T = T^*$)

- $\mathcal{E}(\mathcal{H})$ ... set of effects ($T \in \mathcal{L}_s(\mathcal{H})$ and $O \leq T \leq I$)

- $\mathcal{P}(\mathcal{H})$ ... set of projections ($T = T^* = T^2$)

- $\mathcal{T}(\mathcal{H})$ ... set of trace class operators

- $\mathcal{T}_s(\mathcal{H})$ ... set of selfadjoint trace class operators ($T \in \mathcal{L}_s(\mathcal{H}) \cap \mathcal{T}(\mathcal{H})$)

- $\mathcal{S}(\mathcal{H})$ ... set of states ($T \in \mathcal{T}(\mathcal{H})$, $T \geq O$ and $\mathrm{tr}\,[T] = 1$)

- $\mathcal{S}^{ext}(\mathcal{H})$ ... set of pure states ($T \in \mathcal{S}(\mathcal{H})$, $\mathrm{tr}\,[T^2] = 1$)

- $\mathcal{S}^{us}(\mathcal{H})$ ... set of unnormalized states ($T \in \mathcal{T}(\mathcal{H})$, $T \geq O$ and $0 \leq \mathrm{tr}\,[T] \leq 1$)

- $\mathcal{S}^{fac}(\mathcal{H})$ ... set of factorized states

- $\mathcal{S}^{sep}(\mathcal{H})$ ... set of separable states

### List of channels and operations

- $\mathcal{M}_{cp}$ ... set of completely positive linear mappings

- $\mathcal{O}$ ... set of quantum operations ($\mathcal{E} \in \mathcal{M}_{cp}$ and $\mathcal{E}$ is trace-decreasing)

- $\mathcal{O}_c$ ... set of quantum channels ($\mathcal{E} \in \mathcal{M}_{cp}$ and $\mathcal{E}$ is trace-preserving)

- $\mathcal{O}_c^{fac}$ ... set of all local channels

- $\mathcal{O}_c^{sep}$ ... set of separable channels

- $\mathcal{O}_c^{LOCC}$ ... set of all LOCC channels